\documentclass[10pt]{article}

\usepackage{graphicx}

\usepackage{tikz}

\usepackage[all,cmtip]{xy}

\usepackage{setspace}
\usepackage{bbm,youngtab}
\usepackage{amscd,mathrsfs}
\usepackage{amssymb,amsmath,dsfont,amsfonts,amsthm}
\usepackage{euscript,enumerate}
\usepackage{stmaryrd}
\usepackage[cbgreek]{textgreek} 
\usepackage{booktabs}
\usepackage{multirow}
\usepackage{hhline} 
\usepackage{color}

\DeclareMathAlphabet{\mathpzc}{OT1}{pzc}{m}{it}

\allowdisplaybreaks 

\usepackage{authblk}
\usepackage{natbib}
\usepackage{color, soul}
\usepackage{enumerate}
\usepackage{empheq}
\usepackage{rotating}
\usepackage{ftnxtra}
\usepackage{fnpos}
\usepackage[titletoc,toc,title]{appendix}
\usepackage{euscript}
\usepackage{graphicx}
\usepackage{epsfig}
\usepackage{epstopdf}
\DeclareGraphicsExtensions{.pdf,.png,.jpg,.eps}
\usepackage{pstool}
\usepackage{upgreek}
\usepackage{mathrsfs}

\usepackage{psfrag}

\numberwithin{equation}{section}

\theoremstyle{plain}	
\newtheorem{thm}{Theorem}[section]

\newtheorem{lem}[thm]{Lemma}
\newtheorem{prop}[thm]{Proposition}
\newtheorem*{prop*}{Proposition}
\theoremstyle{definition}	
\newtheorem{defi}[thm]{Definition}
\newtheorem{remark}[thm]{Remark}
\newtheorem{example}[thm]{Example}

\usepackage{caption}
\usepackage{subcaption}

\usepackage{hyperref}
\hypersetup{colorlinks=true, linkcolor=blue}
\hypersetup{colorlinks=true,citecolor=blue}

\DeclareMathAlphabet{\mathpzc}{OT1}{pzc}{m}{it}

\usepackage{amsmath, amsthm, amssymb}

\usepackage{cleveref}

\DeclarePairedDelimiter\abs{\lvert}{\rvert}

\makeatletter
\newsavebox{\@brx}
\newcommand{\llangle}[1][]{\savebox{\@brx}{\(\m@th{#1\langle}\)}%
  \mathopen{\copy\@brx\mkern2mu\kern-0.9\wd\@brx\usebox{\@brx}}}
\newcommand{\rrangle}[1][]{\savebox{\@brx}{\(\m@th{#1\rangle}\)}%
  \mathclose{\copy\@brx\mkern2mu\kern-0.9\wd\@brx\usebox{\@brx}}}%
\let\oldabs\abs
\def\abs{\@ifstar{\oldabs}{\oldabs*}}
\makeatother

\usepackage{accents}
\newcommand{\Fe}{\accentset{e}{\mathbf{F}}}
\newcommand{\Ce}{\accentset{e}{\mathbf{C}}}

\newcommand{\Fa}{\accentset{a}{\mathbf{F}}}

\newcommand{\FeS}{\accentset{e}{\bar{\mathbf{F}}}}
\newcommand{\FaS}{\accentset{a}{\bar{\mathbf{F}}}}

\newcommand{\Sur}{\mathsf{S}}
\newcommand{\sur}{\mathsf{s}}
\newcommand{\SurE}{\mathtt{S}}
\newcommand{\surE}{\mathtt{s}}

\newcommand{\Ws}{\accentset{\mathsf{s}}{W}}

\newcommand{\Wi}{\accentset{i}{W}}
\newcommand{\Wf}{\accentset{\mathsf{f}}{W}}
\newcommand{\Wbars}{\accentset{\mathsf{s}}{\overline{W}}}

\newcommand{\Ts}{\mathcal{T}_s}
\newcommand{\Tb}{\mathcal{T}_b}
\newcommand{\Ls}{\mathcal{L}_s}
\newcommand{\Lb}{\mathcal{L}_b}
\newcommand{\Bs}{\accentset{\mathsf{s}}{\mathbf{B}}}
\newcommand{\cBs}{\accentset{\mathsf{s}}{B}}
\newcommand{\Tracs}{\accentset{\mathsf{s}}{\mathbf{T}}}

\newcommand{\FS}{\accentset{\Sur}{\boldsymbol{\mathsf{F}}}}
\newcommand{\cFS}{\accentset{\Sur}{\mathsf{F}}}
\newcommand{\Fs}{\accentset{\sur}{\boldsymbol{\mathsf{F}}}}
\newcommand{\cFs}{\accentset{\sur}{\mathsf{F}}}

\DeclareRobustCommand{\FS}{\accentset{\Sur}{\boldsymbol{\mathsf{F}}}}

\newcommand{\iSpf}{\iota_{\Sur*}}
\newcommand{\iSpb}{\iota_{\Sur}^*}
\newcommand{\ispf}{\iota_{\sur*}}

\usepackage{mathtools}

    {\end{bmatrix}}%

\usepackage[all,cmtip]{xy}

\newcommand{\df}{{\boldsymbol\omega}}                                                       
\newcommand{\mvf}{{\boldsymbol{\mu}}}                                                         

\newcommand{\N}{\accentset{1}{\boldsymbol{\mathsf{N}}}}
\newcommand{\NN}{\accentset{2}{\boldsymbol{\mathsf{N}}}}
\newcommand{\NNN}{\accentset{3}{\boldsymbol{\mathsf{N}}}}
\newcommand{\Nj}{\accentset{j}{\boldsymbol{\mathsf{N}}}}
\newcommand{\n}{\accentset{1}{\boldsymbol{\mathsf{n}}}}
\newcommand{\nn}{\accentset{2}{\boldsymbol{\mathsf{n}}}}
\newcommand{\nnn}{\accentset{3}{\boldsymbol{\mathsf{n}}}}
\newcommand{\nj}{\accentset{j}{\boldsymbol{\mathsf{n}}}}

\makeatletter
\DeclareFontFamily{OMX}{MnSymbolE}{}
\DeclareSymbolFont{MnLargeSymbols}{OMX}{MnSymbolE}{m}{n}
\SetSymbolFont{MnLargeSymbols}{bold}{OMX}{MnSymbolE}{b}{n}
\DeclareFontShape{OMX}{MnSymbolE}{m}{n}{
    <-6>  MnSymbolE5
   <6-7>  MnSymbolE6
   <7-8>  MnSymbolE7
   <8-9>  MnSymbolE8
   <9-10> MnSymbolE9
  <10-12> MnSymbolE10
  <12->   MnSymbolE12
}{}
\DeclareFontShape{OMX}{MnSymbolE}{b}{n}{
    <-6>  MnSymbolE-Bold5
   <6-7>  MnSymbolE-Bold6
   <7-8>  MnSymbolE-Bold7
   <8-9>  MnSymbolE-Bold8
   <9-10> MnSymbolE-Bold9
  <10-12> MnSymbolE-Bold10
  <12->   MnSymbolE-Bold12
}{}

\let\llangle\@undefined
\let\rrangle\@undefined
\DeclareMathDelimiter{\llangle}{\mathopen}%
                     {MnLargeSymbols}{'164}{MnLargeSymbols}{'164}
\DeclareMathDelimiter{\rrangle}{\mathclose}%
                     {MnLargeSymbols}{'171}{MnLargeSymbols}{'171}
\makeatother

\usepackage{enumitem}

\begin{document}

\title{\textbf{\vspace{-.75in}\\
A Geometric Theory of Surface Elasticity \\and Anelasticity}}

\author[1,2]{Arash Yavari\thanks{Corresponding author, e-mail: arash.yavari@ce.gatech.edu}}
\affil[1]{\small \textit{School of Civil and Environmental Engineering, Georgia Institute of Technology, Atlanta, GA 30332, USA}}
\affil[2]{\small \textit{The George W. Woodruff School of Mechanical Engineering, Georgia Institute of Technology, Atlanta, GA 30332, USA}}

\maketitle

\vskip -0.in
{\centering Dedicated to Professor David J. Steigmann on the occasion of his $70$th birthday. \par}
\vskip 0.4in

\thispagestyle{empty}

\begin{abstract}
\noindent
In this paper we formulate a geometric theory of elasticity and anelasticity for bodies containing material surfaces with their own elastic energies and distributed surface eigenstrains. Bulk elasticity is written in the language of Riemannian geometry, and the framework is extended to material surfaces by using the differential geometry of hypersurfaces in Riemannian manifolds. Within this setting, surface kinematics, surface strain measures, surface material metric, and the induced second fundamental form follow naturally from the embedding of the material surface in the material manifold. The classical theory of surface elasticity of \citet{Gurtin1975} is revisited and reformulated in this geometric framework, and then extended to anelastic bodies with anelastic material surfaces. Constitutive equations for isotropic and anisotropic material surfaces are formulated systematically, and bulk and surface anelasticity are introduced by replacing the elastic metrics with their anelastic counterparts. The balance laws are derived variationally using the Lagrange-d’Alembert principle. These include the bulk balance of linear momentum together with the surface balance of linear momentum, whose normal component gives a generalized Laplace's law. As an application, we obtain the complete solution for a spherical incompressible isotropic solid ball containing a cavity filled with a compressible hyperelastic fluid, where the cavity boundary is an anelastic material surface with distributed surface eigenstrains. The analytical and numerical results quantify the effects of surface and fluid eigenstrains on the pressure-stretch response, and residual stress.
\end{abstract}

\begin{description}
\item[Keywords:] Surface elasticity, material surfaces, eigenstrain, anelasticity, material metric, surface stress, residual stress.
\end{description}

\tableofcontents

\section{Introduction} \label{Intro}

It has been known for a long time that surface tension dominates the mechanical response of liquids at small scales. In recent years it has been observed that surface stresses have significant effects on the response of soft solids, e.g., gels \citep{Style2017,Bico2018}.
A related problem is the calculation of the effective properties of solids with small liquid/gaseous inclusions. For small enough inclusions surface stress has a significant effect on the effective properties of a composite. For example, an unexpected effect of liquid inclusions is that they can make the solid stiffer than its bulk configuration. Recall that \citet{Eshelby1957}'s classical theory predicts the opposite effect. There are some recent linear studies of this problem \citep{Style2015a,Style2015b}.  However, there is still relatively little work on the large-deformation behavior of such composites, aside from a few recent studies \citep{Liu2012,Ghosh2022}.

\citet{Gurtin1975} formulated a continuum theory of surface elasticity in the setting of finite strains by postulating a set of bulk and surface balance laws. In their formulation, material surfaces were assumed to undergo only membrane deformations. This theory was later extended to include bending deformation of material surfaces by \citet{SteigmannOgden1999}. In recent years there has been a lot of interest in the continuum theory of bodies with elastic surfaces \citep{Steigmann1999FluidFilms,Spaepen2000,Duan2005,Huang2006,Liu2012,JaviliSteinmann2010,Javili2013,Javili2018,Xu2018,Krichen2019,Heyden2022,Tamim2025}.

The classical Gurtin-Murdoch surface elasticity model is formulated in a coordinate-free manner, but it is developed within a framework that does not fully exploit the existing differential geometry of material surfaces. In the present work the theory is written directly in the language of modern differential geometry, which provides a clear and invariant description of all kinematic and kinetic quantities and avoids the need for additional assumptions or analogies.\footnote{There is an extensive literature on geometric formulations of three-dimensional elasticity and anelasticity. Representative contributions include \citet{DoyleEricksen1956,Noll1967,Wang1968,Wang1974,MaHu1983,Epstein2011GeometricalLanguage,Romano2014,Clayton2014,Sozio2019,SozioYavari2020,Segev2023}, among others.}
This geometric viewpoint also places surface elasticity in its natural setting: a surface with its own material metric. In this sense, surface elasticity is a special case of surface anelasticity, corresponding to a particular choice of surface material metric. Material surfaces in nature are almost always residually stressed, and thus a unified geometric formulation of surface anelasticity is essential. Once the surface material metric is identified as the fundamental descriptor of anelastic distortions, the overall structure of the theory becomes identical to that of classical three-dimensional elasticity, and no special constitutive hypotheses are required.\footnote{It should be emphasized that the present theory is formulated for material surfaces with purely in-plane elastic and anelastic response. No bending effects are included, and no constitutive dependence on curvature or on the second fundamental form is introduced. The surface is therefore treated as a membrane-type continuum, and all kinematic and constitutive relations are intrinsic to the surface.
}

In this paper, we first revisit the classical theory of surface elasticity formulated in \citep{Gurtin1975,Gurtin1978,Gurtin1998}. We reformulate the theory of surface elasticity in the setting of Riemannian geometry and differential geometry of embedded hypersurfaces in Riemannian manifolds.\footnote{Formulating surface elasticity in this geometric framework makes its extension to anelasticity essentially immediate. This is one of the motivations for adopting the geometric setting.} We next extend the theory to anelastic bodies with anelastic material surfaces. In surface anelasticity both the bulk and material surfaces have distributed eigenstrains. This implies that the bulk and the material surfaces have their own material metrics. The presence of bulk and surface eigenstrains induces bulk and surface residual stresses and in turn alters the mechanical properties of the body. It should be mentioned that there are previous works in the literature that have used modern differential geometry of submanifolds to formulate the mechanics of material surfaces. However, all these works are restricted to kinematics alone \citep{Betounes1986,Kadianakis2010,Kadianakis2018}. In this paper, we formulate a comprehensive geometric theory for the mechanics of material surfaces. \citet{Gurtin1975} postulated the balance laws. Here, we formulate the mechanics of surface elasticity variationally.

Recently, \citet{Eremeyev2024} formulated a finite-deformation theory of surface viscoelasticity by extending the Gurtin--Murdoch surface elasticity framework to include history-dependent surface stresses. The bulk material is assumed to be elastic, while dissipation is confined to the surface through viscoelastic constitutive relations defined on the elastic surface. As an application, the anti-plane surface waves in an elastic half-space endowed with viscoelastic surface stresses were studied. It was shown that surface viscosity modifies both the dispersion and attenuation of surface waves relative to the purely elastic surface case.
It should be emphasized that there are subtle geometric differences between anelasticity and viscoelasticity, as discussed in detail in \citep{SadikYavari2024}. In the present paper, we restrict attention to elasticity. An extension of the theory to viscoelastic material surfaces will be the subject of future work.

Elastic surfaces are inherently residually stressed, and the same is true for liquid inclusions. In this sense, within the linear or nonlinear continuum theory of soft solids, surface effects are intrinsically anelastic. The geometric framework developed in this paper therefore provides a natural setting for modeling the mechanics of such materials. By formulating surface elasticity as a continuum theory in a Riemannian geometric framework, eigenstrains are seamlessly embedded in the material metrics, and residual stresses emerge naturally from the resulting geometric frustration. This renders the formulation both conceptually transparent and mathematically consistent. As will be shown, the proposed geometric framework enables a unified treatment of coupled bulk-surface elasticity in systems where both the bulk and the surface exhibit eigenstrains and residual stresses.

Existing studies of surface elasticity with liquid inclusions have mainly focused on determining the effective elastic properties of composites containing many fluid inclusions of prescribed initial radius and capillarity number, and thus aim to compute effective moduli of a composite rather than the global mechanical response of a single finite body. In this paper we analyze the fully nonlinear spherical elasticity problem for a single solid ball containing a cavity endowed with surface elasticity and possibly filled with a compressible fluid possessing a natural volumetric state. The cavity is allowed to undergo large volumetric changes, leading to a nontrivial residual-stress field throughout the solid. As a result, the pressure-deformation response of the ball is significantly modified by either surface elasticity or the natural volume of the fluid inclusion, producing stiffening or softening effects.

This paper is organized as follows. The differential geometry of three-dimensional manifolds and their hypersurfaces is discussed in \S\ref{Sec:Cartan-Movng-Frames}. Nonlinear elasticity of bodies with material surfaces is presented in \S\ref{Sec:MaterialSurfaces}, including the bulk and surface kinematics, the bulk and surface constitutive equations in the presence of anisotropy, and the corresponding balance laws derived using the Lagrange–d'Alembert principle. The theory is then extended to surface anelasticity in \S\ref{Sec:SurfaceAnelasticity}, where a material surface is endowed with its own material metric that is not necessarily equal to its first fundamental form. As an application, in \S\ref{Sec:Example} we study a spherical ball with a concentric cavity filled with a homogeneous isotropic hyperelastic liquid, surrounded by an incompressible isotropic solid and endowed with an isotropic material surface carrying its own surface eigenstrains. The response of this system to an applied outer pressure is determined. The conclusions are summarized in \S\ref{Sec:Conclusions}.

\section{Differential Geometry via Cartan's Moving Frames} \label{Sec:Cartan-Movng-Frames}

Suppose $\mathcal{B}$ is an $n$-manifold that is equipped with a metric $\mathbf{G}$ and an affine connection $\nabla$. The triplet $(\mathcal{B},\nabla,\mathbf{G})$ is called a metric-affine manifold \citep{Gordeeva2010}. 
In our review and summary of metric-affine manifolds and hyper-surfaces of Riemannian manifolds we mainly follow \citep{Spivak1970II,Hehl2003,Sternberg2013,Nicolaescu2020}.
At $X\in\mathcal{B}$, the tangent space is an $n$-dimensional vector space that is denoted by $T_X\mathcal{B}$. Suppose $T_X\mathcal{B}$ has an orthonormal basis $\{\mathbf{e}_1(X),\hdots,\mathbf{e}_n(X)\}$. This is called a moving frame.\footnote{The method of moving frames was systematically and successfully used in formulating the mechanics of distributed point and line defects in solids in a series of papers \citep{YavariGoriely2012a,YavariGoriely2012b,YavariGoriely2013a,YavariGoriely2014}. In our formation of surface elasticity and anelasticity we use both the standard and moving frames formats of the equations of hyper-surfaces in Riemannian manifolds.}
It should be noted that a moving frame is not necessarily induced from a coordinate chart; a moving frame is a non-coordinate basis for the tangent bundle $T\mathcal{B}$ (the disjoint union of all the tangent spaces).
Every moving frame $\{\mathbf{e}_{\alpha}\}=\{\mathbf{e}_1,\hdots,\mathbf{e}_n\}$ has a corresponding moving co-frame field $\{\vartheta^{\alpha}\}=\{\vartheta^1,\hdots,\vartheta^n\}$ such that $\vartheta^{\alpha}(\mathbf{e}_{\beta})=\delta^{\alpha}_{\beta}$, $\alpha, \beta=1,\hdots,n$, where $\delta^{\alpha}_{\beta}$ is the Kronecker delta. 
With respect to an orthonormal moving frame, i.e., when $\llangle \mathbf{e}_{\alpha},\mathbf{e}_{\beta}\rrangle_{\mathbf{G}}=\delta_{\alpha\beta}$ ($\llangle .,. \rrangle_{\mathbf{G}}$ is the inner product induced by the metric $\mathbf{G}$), the metric has the following simple representation (summation over repeated indices is assumed everywhere in this paper)
\begin{equation}\label{Cartanmetric}
	\mathbf{G}=\delta_{\alpha\beta}\,\vartheta^{\alpha}\otimes\vartheta^{\beta}\,.
\end{equation}
A vector field $\mathbf{Y}$ on $\mathcal{B}$ ($\mathbf{Y}\in\mathcal{X}(\mathcal{B})$---the set of vector fields on $\mathcal{B}$) assigns a vector $\mathbf{Y}_{\!X}\in T_X\mathcal{B}$ to every $X\in\mathcal{B}$. 
At $X\in\mathcal{B}$ let us denote the space of anti-symmetric $k$-linear maps by $\Lambda^k(T_X\mathcal{B})$. The bundle of exterior $k$-forms is defined as $\Lambda^k(\mathcal{B})=\bigsqcup_{X\in\mathcal{B}}\,\Lambda^k(T_X\mathcal{B})$, where $\bigsqcup$ denotes disjoint union of sets. Smooth sections of this bundle are called differential $k$-forms. The space of differential $k$-forms is denoted as $\Omega^k(\mathcal{B})$.

The interior product between a differential $k$-form $\boldsymbol{\omega}$ and a vector field $\mathbf{W}$ is denoted as $i_{\mathbf{W}} \boldsymbol{\omega}$, and is defined as 
\begin{equation}
	i_{\mathbf{W}} \boldsymbol{\omega}(\mathbf{U}_1,\hdots,\mathbf{U}_{k-1})
	=\boldsymbol{\omega}(\mathbf{W},\mathbf{U}_1,\hdots,\mathbf{U}_{k-1})\,,\qquad \forall 
	\mathbf{U}_1,\hdots,\mathbf{U}_{k-1}\in T\mathcal{B}\,.
\end{equation}
The exterior derivative of a differential $k$-form $\boldsymbol{\omega}$ is a $(k+1)$-form that is denoted as $\mathrm d\boldsymbol\omega$.
For a $1$-form $\boldsymbol{\alpha}$, one has the following identities for arbitrary vector fields $\mathbf{U}$ and $\mathbf{W}$:
\begin{equation}\label{der-1-form}
	\iota_{\mathbf{U}} \boldsymbol{\alpha} =\langle \boldsymbol{\alpha},  \mathbf{U} \rangle
	\,,\qquad
	\mathrm d \boldsymbol{\omega} (\mathbf{U}, \mathbf{W}) =
	\langle \mathrm d \langle \boldsymbol{\omega} , \mathbf{W} \rangle   , \mathbf{U} \rangle
	-\langle \mathrm d \langle \boldsymbol{\omega}, \mathbf{U} \rangle   , \mathbf{W} \rangle
	-\langle \boldsymbol{\omega} , [\mathbf{U}, \mathbf{W}]\rangle \,,
\end{equation}
where the Lie bracket (commutator) is defined as $[\mathbf{U}, \mathbf{W}]=\mathbf{U} \mathbf{W}-\mathbf{W} \mathbf{U}$ and $\langle.,.\rangle$ is the natural pairing of one-forms and vectors.
This means that for an arbitrary scalar field $f$, $[\mathbf{U},\mathbf{W}][f]=\mathbf{U}[\mathbf{W}[f]]-\mathbf{W}[\mathbf{U}[f]]$. 
In components, $[\mathbf{U},\mathbf{W}]^A=U^B W^A{}_{,B}-W^B U^A{}_{,B}$.
Clearly, for a coordinate frame $[\partial_A,\partial_B]=0$.
Cartan's magic formula relates the Lie derivative $\mathfrak L_{\mathbf{U}}$, the interior product $i_{\mathbf{U}}$, and the exterior derivative $\mathrm d$ of a differential form $\boldsymbol{\omega}$ as
\begin{equation}\label{Cartan}
	\mathfrak L_{\mathbf{U}} \boldsymbol{\omega} =
	\mathrm d \,i_{\mathbf{U}} \boldsymbol{\omega} 
	+i_{\mathbf{U}} \mathrm d \boldsymbol{\omega} \,.
\end{equation}

Given a metric tensor $\mathbf{G}$ on an $n$-manifold $\mathcal B$, the Hodge star operator assigns to a $k$-form $\boldsymbol\omega$ the $(n-k)$-form $\star\boldsymbol\omega$ (to be more precise one should write $\star_{\mathbf{G}}$ instead of $\star$) such that
\begin{equation} \label{hodge-star}
	(\star \boldsymbol \omega)( \mathbf{U}_1, \hdots,  \mathbf{U}_k) =
	\boldsymbol \omega ( \mathbf{U}_{k+1}, \hdots,  \mathbf{U}_n) \,,
\end{equation}
for any $\mathbf{G}$-orthonormal frame $\{\mathbf{U}_1,\hdots,\mathbf{U}_n\}$.
Note that $\star\star\df=(-1)^{k(n-k)}\df$.
For the Riemannian volume $n$-form $\boldsymbol{\mu}_{\mathbf{G}}$,\footnote{For a positively oriented set of vectors $\{\mathbf{W}_1,\hdots,\mathbf{W}_n\}$, $\boldsymbol{\mu}_{\mathbf{G}}(\mathbf{W}_1,\hdots,\mathbf{W}_n)=\sqrt{\det(\mathbf{G}(\mathbf{W}_i,\mathbf{W}_j))}$. For a $\mathbf{G}$-orthonormal frame $\{\mathbf{U}_1,\hdots,\mathbf{U}_n\}$, $\boldsymbol{\mu}_{\mathbf{G}}(\mathbf{U}_1,\hdots,\mathbf{U}_n)=1$.} $\star \boldsymbol{\mu}_{\mathbf{G}}=1$. With respect to an orthonormal moving coframe $\{\vartheta^1,\hdots,\vartheta^n\}$ Hodge star has a simple representation. It acts on a basis element of $k$-forms as follows 
\begin{equation} 
	\star\left(\vartheta^{\gamma_1}\wedge\hdots\wedge\vartheta^{\gamma_k}\right)
	=\epsilon^{\gamma_1\hdots \gamma_k}{}_{\eta_{k+1}\hdots\eta_n}\,
	\vartheta^{\eta_{k+1}}\wedge\hdots\wedge\vartheta^{\eta_n}\,,
\end{equation}
where summation is over $\eta_{k+1}<\hdots < \eta_{n}$, and $\boldsymbol{\epsilon}$ is the totally anti-symmetric tensor defined as
\begin{equation} 
	\epsilon_{\eta_1\hdots\eta_n}=
	\begin{dcases}
	+1 & (\eta_1\hdots\eta_n)~\text{is~an~even~permutation~of~}(1\hdots n)\,, \\
	-1 & (\eta_1\hdots\eta_n)~\text{is~an~odd~permutation~of~}(1\hdots n)\,, \\
	0 & \text{otherwise} \,,
	\end{dcases}
\end{equation}
and indices are raised by $\delta^{\alpha\beta}$.
Let $\boldsymbol{\alpha}$ and $\boldsymbol{\beta}$ be $k$-forms. The pointwise inner product induced by the metric $\mathbf{G}$ is defined by
\begin{equation}
	\langle \boldsymbol{\alpha}, \boldsymbol{\beta} \rangle_{\mathbf{G}} = \sum_{I} \alpha_I\,\beta_I,
\end{equation}
where the sum is over all increasing multi-indices $I = (i_1 < \cdots < i_k)$ and $\boldsymbol{\alpha} = \sum_I \alpha_I\, dx^{i_1} \wedge \cdots \wedge dx^{i_k}$, $\boldsymbol{\beta} = \sum_I \beta_I\, dx^{i_1} \wedge \cdots \wedge dx^{i_k}$, with the coefficients $\alpha_I$ and $\beta_I$ taken in an orthonormal coframe with respect to $\mathbf{G}$.

The raised Hodge operator is defined as $\star^{\sharp} \boldsymbol\omega= (\star \boldsymbol\omega)^{\sharp}$, i.e., raising all the indices of the Hodge star operator. This gives an alternating tensor of type $(k,0)$.
For a $k$-form $\boldsymbol{\alpha}$ and an $(n-k)$-form $\boldsymbol{\gamma}$ one has
\begin{equation} \label{wedge-hodge}
	\boldsymbol{\alpha} \wedge \boldsymbol{\gamma}=
	\langle \boldsymbol{\alpha}  , \star^{\sharp}  \boldsymbol{\gamma} \rangle \,\mvf=
	\langle \boldsymbol{\gamma} , \star^{\sharp}  \boldsymbol{\alpha} \rangle \,\mvf \,,
\end{equation}
where $\langle,\rangle$ is the natural pairing of forms with alternating multi-vectors.
The natural pairing $\langle\, , \, \rangle$ is the canonical contraction between a $k$-form and a $k$-vector (alternating contravariant tensor of rank $k$). Given a $k$-form $\boldsymbol{\alpha} = \sum_I \alpha_I\, dx^{i_1} \wedge \cdots \wedge dx^{i_k}$ and a $k$-vector $\mathbf{V} = \sum_I V^I\, \partial_{i_1} \wedge \cdots \wedge \partial_{i_k}$, their pairing is defined pointwise by
\begin{equation}
	\langle \boldsymbol{\alpha}, \mathbf{V} \rangle = \sum_I \alpha_I\, V^I\,,
\end{equation}
where the sum is over all increasing multi-indices $I = (i_1 < \cdots < i_k)$. In particular, the pairing $\langle \boldsymbol{\alpha}, \star^{\sharp} \boldsymbol{\gamma} \rangle$ in \eqref{wedge-hodge} contracts a $k$-form with an $(n-k)$-form that has been transformed into a $k$-vector using the raised Hodge star.

Let $S(k,n)$ be the set of permutations $\tau$ of the set $\{1,\hdots,n\}$ such that $\tau(1)<\hdots<\tau(k)$ and $\tau(k+1)<\hdots<\tau(n)$. With respect to the orthonormal basis $\{\mathbf{e}_1,\hdots,\mathbf{e}_n\}$, 
\begin{equation} 
	(\star\boldsymbol{\omega})(\mathbf{e}_{\tau(k+1)},\hdots,\mathbf{e}_{\tau(n)})
	=(\mathsf{sgn}\,\tau)\,\boldsymbol{\omega}(\mathbf{e}_{\tau(1)},\hdots,\mathbf{e}_{\tau(k)})\,,
\end{equation}
where $\mathsf{sgn}\,\tau$ is the sign of the permutation $\tau$ \citep{Schwarz2006}.
For any vector $\mathbf{V}$, with respect to the orthonormal frame one has
\begin{equation} 
	(\star\mathbf{V}^{\flat})(\mathbf{e}_{\tau(2)},\hdots,\mathbf{e}_{\tau(n)})
	=(\mathsf{sgn}\,\tau)\,\mathbf{V}^{\flat}(\mathbf{e}_{\tau(1)})\,.
\end{equation}
On the other hand
\begin{equation} 
\begin{aligned}
	(i_{\mathbf{V}}\mvf_{\mathbf{G}})(\mathbf{e}_{\tau(2)},\hdots,\mathbf{e}_{\tau(n)})
	& =\mvf_{\mathbf{G}}(\mathbf{V},\mathbf{e}_{\tau(2)},\hdots,\mathbf{e}_{\tau(n)}) \\
	& =\mvf_{\mathbf{G}}(V^{\tau(1)}\mathbf{e}_{\tau(1)},\mathbf{e}_{\tau(2)},\hdots,\mathbf{e}_{\tau(n)})
	\\
	&=(\mathsf{sgn}\,\tau)\,V^{\tau(1)}\\
	&=(\mathsf{sgn}\,\tau)\,\mathbf{V}^{\flat}(\mathbf{e}_{\tau(1)})
	\,.
\end{aligned}
\end{equation}
Thus, we have proved the following classical result 
\begin{equation} \label{vector-volumeform}
	i_{\mathbf{V}}\mvf_{\mathbf{G}}=\star\mathbf{V}^{\flat} \,.
\end{equation}

Given a volume form $\mvf$, the divergence of a vector field $\mathbf{V}$ is defined as
$(\operatorname{div} \mathbf{V})\, \mvf=\mathfrak L_{\mathbf{V}}\mvf$. Using Cartan's magic formula and the fact that  $\mathrm d \mvf=0$ one writes $(\operatorname{div} \mathbf{V})\, \mvf= \mathrm d (i_{ \mathbf{V}} \mvf)$.
For the special volume form $\boldsymbol{\mu}_{\mathbf{G}}$---the Riemannian volume form, $\operatorname{div}_{\mathbf{G}} \mathbf{V} = \operatorname{tr}_{\mathbf{G}} (\nabla^{\mathbf{G}} \mathbf{V})$, where $\nabla^{\mathbf{G}}$ is the Levi-Civita connection associated with $\mathbf{G}$. Using $\left(\operatorname{div}_{\mathbf{G}} \!\mathbf{V}\right) \mvf_{\mathbf{G}}= \mathrm d( i_{ \mathbf{V}} \mvf_{\mathbf{G}})$ and \eqref{vector-volumeform} one finds the following classic result
\begin{equation} \label{d-div}
	\left(\operatorname{div}_{\mathbf{G}} \!\mathbf{V}\right) \mvf_{\mathbf{G}}
	=\mathrm d\star\!\mathbf{V}^{\flat} \,.
\end{equation}

In a manifold it does not make sense to differentiate vector fields unless an extra structure is provided---an affine connection. An affine (or linear) connection is an operation $\nabla:\mathcal{X}(\mathcal{B})\times\mathcal{X}(\mathcal{B})\to\mathcal{X}(\mathcal{B})$ that has the following three properties: 
\begin{equation} 
\begin{aligned}
	& (i) && \nabla_{f_1\mathbf{X}_1+f_2\mathbf{X}_2}\mathbf{Y}
	=f_1\nabla_{\mathbf{X}_1}\mathbf{Y}+f_2\nabla_{\mathbf{X}_2}\mathbf{Y}\,, \\
	& (ii) &&  \nabla_{\mathbf{X}}(a_1\mathbf{Y}_1+a_2\mathbf{Y}_2)
	=a_1\nabla_{\mathbf{X}}\mathbf{Y}_1+a_2\nabla_{\mathbf{X}}\mathbf{Y}_2\,,\\
	& (iii) &&  \nabla_{\mathbf{X}}(f\mathbf{Y})=f\nabla_{\mathbf{X}}\mathbf{Y}+(\mathbf{X}[f])\mathbf{Y} 
	\,,
\end{aligned}
\end{equation}
where $\mathbf{X}$, $\mathbf{Y}$, $\mathbf{X}_1$, $\mathbf{X}_2$, $\mathbf{Y}_1$, and $\mathbf{Y}_2$ are arbitrary vector fields, $f,f_1,f_2$ are arbitrary functions, and $a_1,a_2$ are arbitrary scalars. 
The vector field $\nabla_{\mathbf{X}}\mathbf{Y}$ is called the covariant derivative of $\mathbf{Y}$ along $\mathbf{X}$. 
For a given connection $\nabla$, the covariant derivative of the moving frame defines the connection $1$-forms:
\begin{equation}
    \nabla\mathbf{e}_{\alpha}=\mathbf{e}_{\gamma}\otimes\omega^{\gamma}{}_{\alpha}\,.
\end{equation}
One defines the connection coefficients as $\nabla_{\mathbf{e}_{\beta}}\mathbf{e}_{\alpha}=\left\langle \omega^{\gamma}{}_{\alpha},\mathbf{e}_{\beta} \right\rangle \mathbf{e}_{\gamma}=\omega^{\gamma}{}_{\beta\alpha}\,\mathbf{e}_{\gamma}$.
The connection $1$-forms are represented with respect to the moving co-frame as $\omega^{\gamma}{}_{\alpha}=\omega^{\gamma}{}_{\beta\alpha}\,\vartheta^{\beta}$. 
For a vector $\mathbf{Y}=Y^{\alpha}\,\mathbf{e}_{\alpha}$, one has
\begin{equation}
\begin{aligned}
    \nabla\mathbf{Y} &=\nabla(Y^{\alpha}\mathbf{e}_{\alpha})\\
    &=\mathbf{e}_{\alpha}\otimes dY^{\alpha}+Y^{\alpha}\,\nabla\mathbf{e}_{\alpha}\\
    &=\mathbf{e}_{\alpha}\otimes dY^{\alpha}
    +Y^{\alpha}\,\mathbf{e}_{\gamma}\otimes\omega^{\gamma}{}_{\alpha}\\
     &=\mathbf{e}_{\alpha}\otimes dY^{\alpha}
     +\mathbf{e}_{\alpha}\otimes \omega^{\alpha}{}_{\gamma}Y^{\gamma}\\
     &=\mathbf{e}_{\alpha}\otimes \left(dY^{\alpha}+\omega^{\alpha}{}_{\gamma}Y^{\gamma}\right)
    \,.
\end{aligned}
\end{equation}
It is observed that $\nabla\mathbf{Y}$ is a vector-valued $1$-fom. 
The covariant derivative of $\mathbf{Y}$ along another vector $\mathbf{X}$ is defined as
\begin{equation}
    \nabla_{\mathbf{X}}\mathbf{Y} = \big\langle dY^{\alpha}+\omega^{\alpha}{}_{\gamma}Y^{\gamma} , \mathbf{X}
    \big \rangle \,\mathbf{e}_{\alpha}
    \,.
\end{equation}
The covariant derivative of the moving co-frame field can be calculated by differentiating the identity $\vartheta^{\alpha}(\mathbf{e}_{\beta})=\delta^{\alpha}_{\beta}$. It gives us $\nabla\vartheta^{\alpha}=-\omega^{\alpha}{}_{\gamma}\,\vartheta^{\gamma}$, and $\nabla_{\mathbf{e}_{\beta}}\vartheta^{\alpha}=-\omega^{\alpha}{}_{\beta\gamma}\,\vartheta^{\gamma}$.

Any coordinate chart $\{X^A\}$ for $\mathcal{B}$ defines a coordinate basis $\left\{\partial_A=\frac{\partial}{\partial X^A}\right\}$ for $T_X\mathcal{B}$.
The coordinate chart and the moving frame field $\{\mathbf{e}_{\alpha}\}$ are related as $\mathbf{e}_{\alpha}=\mathsf{F}_{\alpha}{}^A\,\partial_A$, where $\boldsymbol{\mathsf{F}}\in GL(n,\mathbb{R})$ is an invertible linear transformation (it is assumed that $\det[\mathsf{F}_{\alpha}{}^A]>0$ in order to preserve orientation).
The analogous relation between the moving and coordinate co-frames is $\vartheta^{\alpha}=\mathsf{F}^{\alpha}{}_A\,dX^A$, where $[\mathsf{F}^{\alpha}{}_A]$ is the inverse of $[\mathsf{F}_{\alpha}{}^A]$.\footnote{Metric has the coordinate components $G_{A B}=\mathsf{F}_{A}{}^{\alpha}\, \delta_{\alpha\beta}\, \mathsf{F}_{B}{}^{\beta}$.}
For the moving frame field the \emph{object of anhonolomy} is a $2$-form defined as $c^{\gamma}=d\vartheta^{\gamma}$, which has the following representation  
\begin{equation}
    c^{\gamma}=d\left(\mathsf{F}^{\gamma}{}_B\,dX^B\right)
    =\sum_{\alpha<\beta}c^{\gamma}{}_{\alpha\beta} \,\vartheta^{\alpha}\wedge\vartheta^{\beta}\,,
    \qquad c^{\gamma}{}_{\alpha\beta}=\mathsf{F}_{\alpha}{}^A\,\mathsf{F}_{\beta}{}^B
	\left(\partial_A\mathsf{F}^{\gamma}{}_B-\partial_B\mathsf{F}^{\gamma}{}_A\right)\,.
\end{equation}
For the moving frame one has $[\mathbf{e}_{\alpha},\mathbf{e}_{\beta}]=-c^{\gamma}{}_{\alpha\beta}\,\mathbf{e}_{\gamma}$.
In a coordinate chart $\{X^A\}$, $\nabla_{\partial_A}\partial_B=\Gamma^C{}_{AB}\,\partial_C$, where $\Gamma^C{}_{AB}$ are the Christoffel symbols of the connection. 

In a metric-affine manifold there are three important tensors, namely, non-metricity, torsion, and curvature, which are discussed next.

\subsection{Non-metricity}

Non-metricity in a metric-affine manifold quantifies how far $\nabla\mathbf{G}$ is from zero. More specifically, non-metricity $\boldsymbol{\mathcal{Q}}:\mathcal{X}(\mathcal{B})\times\mathcal{X}(\mathcal{B})\times\mathcal{X}(\mathcal{B})\to \mathcal{X}(\mathcal{B})$ is defined as
\begin{equation}
     \boldsymbol{\mathcal{Q}}(\mathbf{X},\mathbf{Y},\mathbf{Z})
     =\llangle \nabla_{\mathbf{X}}\mathbf{Y},\mathbf{Z} \rrangle_{\mathbf{G}}
     +\llangle \mathbf{Y},\nabla_{\mathbf{X}}\mathbf{Z} \rrangle_{\mathbf{G}}
     -\mathbf{X}\big[\llangle\mathbf{Y},\mathbf{Z}\rrangle_{\mathbf{G}}\big]\,,
     \qquad \forall \mathbf{X},\mathbf{Y},\mathbf{Z}\in T\mathcal{B}  \,,
\end{equation}
where $\llangle .,. \rrangle_{\mathbf{G}}$ is the inner product of vectors induced from $\mathbf{G}$.
With respect to the moving frame $\{\mathbf{e}_{\alpha}\}$ non-metricity coefficients are defined as $\mathcal{Q}_{\gamma\alpha\beta}=\boldsymbol{\mathcal{Q}}(\mathbf{e}_{\gamma},\mathbf{e}_{\alpha},\mathbf{e}_{\beta})$.
The non-metricity $1$-forms are defined as $\mathcal{Q}_{\alpha\beta}=\mathcal{Q}_{\gamma\alpha\beta}\,\vartheta^{\gamma}$. 
Following the definition of non-metricity $\mathcal{Q}_{\gamma\alpha\beta}
     =\omega^{\xi}{}_{\gamma\alpha}\,G_{\xi\beta}+\omega^{\xi}{}_{\gamma\beta}\,G_{\xi\alpha}
     -\langle dG_{\alpha\beta},\mathbf{e}_{\gamma} \rangle
     =\omega_{\beta\gamma\alpha}+\omega_{\alpha\gamma\beta}
     -\langle dG_{\alpha\beta},\mathbf{e}_{\gamma} \rangle$, where $d$ is the exterior derivative. Thus, $\mathcal{Q}_{\alpha\beta}=\omega_{\alpha\beta}+\omega_{\beta\alpha}-dG_{\alpha\beta}$. Note that $dG_{\alpha\beta}=d\delta_{\alpha\beta}=0$, and hence we have obtained \emph{Cartan's zeroth structural equations}:
\begin{equation}
     \mathcal{Q}_{\alpha\beta}=\omega_{\alpha\beta}+\omega_{\beta\alpha}.
\end{equation}
In solids, non-metricity describes the geometry of the reference configuration of a body with distributed point defects \citep{Falk1981,DeWit1981,Grachev1989,Kroner1990,MiriRivier2002,YavariGoriely2012b,YavariGoriely2014,Golgoon2018b}.

The connection $\nabla$ is $\mathbf{G}$-compatible if non-metricity vanishes, i.e., 
\begin{equation}
    \nabla_{\mathbf{X}}\llangle \mathbf{Y},\mathbf{Z}\rrangle_{\mathbf{G}}
    =\llangle \nabla_{\mathbf{X}}\mathbf{Y},\mathbf{Z} \rrangle_{\mathbf{G}}
    +\llangle \mathbf{Y},\nabla_{\mathbf{X}}\mathbf{Z}\rrangle_{\mathbf{G}}\,.
\end{equation}
This is equivalent to $\nabla\mathbf{G}=\mathbf{0}$, and in coordinates reads $G_{AB|C}=G_{AB,C}-\Gamma^D{}_{CA}G_{DB}-\Gamma^D{}_{CB}G_{AD}=0$. 
For a metric-compatible connection with respect to the moving frame one has $\omega_{\alpha\beta}+\omega_{\beta\alpha}=0$, i.e., the matrix of connection $1$-forms of a metric-compatible connection is anti-symmetric.

\subsection{Torsion}

Torsion $\mathbf{T}:\mathcal{X}(\mathcal{B})\times\mathcal{X}(\mathcal{B})\to\mathcal{X}(\mathcal{B})$ of the connection $\nabla$ is defined as
\begin{equation}
    \mathbf{T}(\mathbf{X},\mathbf{Y})=\nabla_{\mathbf{X}}\mathbf{Y}-\nabla_{\mathbf{Y}}\mathbf{X}
    -[\mathbf{X},\mathbf{Y}]\,.
\end{equation}
Note that $\mathbf{T}(\mathbf{Y},\mathbf{X})=-\mathbf{T}(\mathbf{X},\mathbf{Y})$.
With respect to a coordinate chart $\{X^A\}$ and the moving frame $\{\mathbf{e}_{\alpha}\}$, torsion has components $T^A{}_{BC}=\Gamma^A{}_{BC}-\Gamma^A{}_{CB}$, and $T^{\alpha}{}_{\beta\gamma}=\omega^{\alpha}{}_{\beta\gamma}-\omega^{\alpha}{}_{\gamma\beta}+c^{\alpha}{}_{\beta\gamma}$, respectively. 
The torsion $2$-forms $\mathcal{T}^{\alpha}$ are defined as 
\begin{equation} 
    \mathcal{T}^{\alpha}(\mathbf{X},\mathbf{Y})\,\mathbf{e}_{\alpha}
    =\mathbf{T}(\mathbf{X},\mathbf{Y})\,,
    \qquad \forall~\mathbf{X}, \mathbf{Y}\in\mathcal{X}(\mathcal{B})
    \,.
\end{equation}
Thus, $\mathcal{T}^{\alpha}=\frac{1}{2}\,T^{\alpha}{}_{\beta\gamma}\,\vartheta^{\beta}\wedge\vartheta^{\gamma}$.
\emph{Cartan's first structural equations} relate the torsion $2$-forms and the connection $1$-forms:
\begin{equation} \label{First-Structural-Equations}
    \mathcal{T}^{\alpha} = d\vartheta^{\alpha}+\omega^{\alpha}{}_{\beta}\wedge\vartheta^{\beta}\,.
\end{equation}
In solids, torsion describes the geometry of the material manifold of a body with distributed dislocations \citep{Bilby1955,Bilby1956,Bilby1968,Kondo1955,Kroner1959,KronerSeeger1959,YavariGoriely2012a,OzakinYavari2014,YavariGoriely2014,Golgoon2018b}.
A  torsion-free connection is called symmetric, for which $\nabla_{\mathbf{X}}\mathbf{Y}-\nabla_{\mathbf{Y}}\mathbf{X}=[\mathbf{X},\mathbf{Y}]$, and with respect to the moving frame $d\vartheta^{\alpha}+\omega^{\alpha}{}_{\beta}\wedge\vartheta^{\beta}=0$.

\subsection{Curvature}

The curvature tensor $\mathbf{R}:\mathcal{X}(\mathcal{B})\times\mathcal{X}(\mathcal{B})\times\mathcal{X}(\mathcal{B})\to\mathcal{X}(\mathcal{B})$ of the affine connection $\nabla$ is defined as
\begin{equation}
    \mathbf{R}(\mathbf{X},\mathbf{Y})\mathbf{Z}
    =[\nabla_{\mathbf{X}},\nabla_{\mathbf{Y}}]\mathbf{Z}-\nabla_{[\mathbf{X},\mathbf{Y}]}\mathbf{Z}
    =\nabla_{\mathbf{X}}\nabla_{\mathbf{Y}}\mathbf{Z}
    -\nabla_{\mathbf{Y}}\nabla_{\mathbf{X}}\mathbf{Z}-\nabla_{[\mathbf{X},\mathbf{Y}]}\mathbf{Z}\,.
\end{equation}
Notice that $\mathbf{R}(\mathbf{Y},\mathbf{X})\mathbf{Z}=-\mathbf{R}(\mathbf{X},\mathbf{Y})\mathbf{Z}$.
In a coordinate chart $\{X^A\}$, $R^A{}_{BCD}=\Gamma^A{}_{CD,B}-\Gamma^A{}_{BD,C}+\Gamma^A{}_{BM}\,\Gamma^M{}_{CD}-\Gamma^A{}_{CM}\,\Gamma^M{}_{BD}$.
With respect to a moving frame, the curvature tensor has the components $R^{\alpha}{}_{\beta\lambda\mu}
    =\partial_{\beta}\omega^{\alpha}{}_{\lambda\mu}-\partial_{\lambda}\omega^{\alpha}{}_{\beta\mu}
    +\omega^{\alpha}{}_{\beta\xi}\,\omega^{\xi}{}_{\lambda\mu}
    -\omega^{\alpha}{}_{\lambda\xi}\,\omega^{\xi}{}_{\beta\mu}
    +\omega^{\alpha}{}_{\xi\mu}\,c^{\xi}{}_{\beta\lambda}$.
The curvature $2$-forms $\mathcal{R}^{\alpha}{}_{\beta}$ are defined as 
\begin{equation} 
    \mathcal{R}^{\alpha}{}_{\beta}(\mathbf{X},\mathbf{Y})\,\mathbf{e}_{\alpha}
    =\boldsymbol{R}(\mathbf{X},\mathbf{Y})\, \mathbf{e}_{\beta}\,,
    \qquad \forall~\mathbf{X}, \mathbf{Y}\in\mathcal{X}(\mathcal{B})
    \,.
\end{equation}
Thus, $R^{\alpha}{}_{\beta\xi\eta}=\mathcal{R}^{\alpha}{}_{\beta}(\mathbf{e}_{\xi},\mathbf{e}_{\eta})$.
It is straightforward to show that $\mathcal{R}^{\alpha}{}_{\beta}=\frac{1}{2}R^{\alpha}{}_{\beta\xi\eta}\,     \vartheta^{\xi}\wedge\vartheta^{\eta}$.
Note that
\begin{equation} \label{Curvature-Forrmula}
	 \llangle \mathbf{R}(\mathbf{X},\mathbf{Y})\, \mathbf{e}_{\beta},
	 \mathbf{e}_{\alpha}\rrangle_{\mathbf{G}}
	 = \mathcal{R}_{\alpha\beta}(\mathbf{X},\mathbf{Y}) \,,
\end{equation}
where $ \mathcal{R}_{\alpha\beta}= \mathcal{R}^{\gamma}{}_{\beta}\,\delta_{\alpha\gamma}$.
\emph{Cartan's second structural equations} relate the curvature $2$-forms and the connection $1$-forms:
\begin{equation} \label{Second-Structural-Equations}
    \mathcal{R}^{\alpha}{}_{\beta}=d\omega^{\alpha}{}_{\beta}
    +\omega^{\alpha}{}_{\gamma}\wedge\omega^{\gamma}{}_{\beta}\,.
\end{equation}

There is a unique connection that is both metric compatible and torsion free---the Levi-Civita connection.
With respect to a coordinate chart $\{X^A\}$ it has the connection coefficients (Christoffel symbols) $\Gamma^C{}_{AB}=\frac{1}{2}G^{CD}(G_{BD,A}+G_{AD,B}-G_{AB,D})$.

\begin{lem}
For a $1$-form $\boldsymbol{\alpha}$ on a Riemannian manifold $(\mathcal{M},\mathbf{G})$ we have the following identity
\begin{equation} \label{d-nabla-identity}
	d \boldsymbol{\alpha} (\mathbf{v}, \mathbf{w}) 
	= \llangle \nabla_{\mathbf{v}} \boldsymbol{\alpha}, \mathbf{w} \rrangle_{\mathbf{G}} 
	- \llangle \nabla_{\mathbf{w}} \boldsymbol{\alpha}, \mathbf{v} \rrangle_{\mathbf{G}} \,,
\end{equation}
where $\nabla$ is the Levi-Civita connection and $\mathbf{v}, \mathbf{w}$ are arbitrary vector fields.
\end{lem}
\begin{proof}
The exterior derivative of a $1$-form $\boldsymbol{\alpha}$ is defined by
\begin{equation}
	d\boldsymbol{\alpha}(\mathbf{v}, \mathbf{w}) 
	= \mathbf{v}[\boldsymbol{\alpha}(\mathbf{w})] 
	- \mathbf{w}[\boldsymbol{\alpha}(\mathbf{v})] 
	- \boldsymbol{\alpha}([\mathbf{v}, \mathbf{w}])\,,
\end{equation}
where $\mathbf{v}, \mathbf{w}$ are arbitrary vector fields. This is a purely algebraic definition and does not depend on any connection.
We can express the directional derivatives of the scalar functions $\boldsymbol{\alpha}(\mathbf{w})$ and $\boldsymbol{\alpha}(\mathbf{v})$ in terms of covariant derivatives:
\begin{equation}
	\mathbf{v}[\boldsymbol{\alpha}(\mathbf{w})] 
	= \llangle \nabla_{\mathbf{v}}\boldsymbol{\alpha}, \mathbf{w} \rrangle_{\mathbf{G}} 
	+ \llangle \boldsymbol{\alpha}, \nabla_{\mathbf{v}}\mathbf{w} \rrangle_{\mathbf{G}}\,,
\end{equation}
\begin{equation}
	\mathbf{w}[\boldsymbol{\alpha}(\mathbf{v})] 
	= \llangle \nabla_{\mathbf{w}}\boldsymbol{\alpha}, \mathbf{v} \rrangle_{\mathbf{G}} 
	+ \llangle \boldsymbol{\alpha}, \nabla_{\mathbf{w}}\mathbf{v} \rrangle_{\mathbf{G}}\,.
\end{equation}
Also, the Lie bracket $[\mathbf{v},\mathbf{w}]$ can be expressed in terms of the Levi-Civita connection as
\begin{equation}
	[\mathbf{v}, \mathbf{w}] = \nabla_{\mathbf{v}}\mathbf{w} - \nabla_{\mathbf{w}}\mathbf{v}\,,
\end{equation}
because the torsion of $\nabla$ vanishes: $\mathcal{T}(\mathbf{v}, \mathbf{w}) = \nabla_{\mathbf{v}}\mathbf{w}  - \nabla_{\mathbf{w}}\mathbf{v} - [\mathbf{v},\mathbf{w}] = 0$.
Substituting these expressions into the definition of $d\boldsymbol{\alpha}$, we obtain:
\begin{equation}
\begin{aligned}
	d\boldsymbol{\alpha}(\mathbf{v}, \mathbf{w}) 
	&= \llangle \nabla_{\mathbf{v}}\boldsymbol{\alpha}, \mathbf{w} \rrangle_{\mathbf{G}} 
	+ \llangle \boldsymbol{\alpha}, \nabla_{\mathbf{v}}\mathbf{w} \rrangle_{\mathbf{G}} 
	- \llangle \nabla_{\mathbf{w}}\boldsymbol{\alpha}, \mathbf{v} \rrangle_{\mathbf{G}} 
	- \llangle \boldsymbol{\alpha}, \nabla_{\mathbf{w}}\mathbf{v} \rrangle_{\mathbf{G}} 
	- \boldsymbol{\alpha}([\mathbf{v}, \mathbf{w}]) \\
	&= \llangle \nabla_{\mathbf{v}}\boldsymbol{\alpha}, \mathbf{w} \rrangle_{\mathbf{G}} 
	- \llangle \nabla_{\mathbf{w}}\boldsymbol{\alpha}, \mathbf{v} \rrangle_{\mathbf{G}} 
	+ \llangle \boldsymbol{\alpha}, \nabla_{\mathbf{v}}\mathbf{w} - \nabla_{\mathbf{w}}\mathbf{v} - [\mathbf{v}, \mathbf{w}] \rrangle_{\mathbf{G}}\,.
\end{aligned}
\end{equation}
The last term vanishes due to the torsion-free condition, proving \eqref{d-nabla-identity}.
\end{proof}

\subsection{Differential Geometry of Embedded Submanifolds} \label{Sec:GeometrySurfaces}

In this section we review the geometry of embedded submanifolds. More specifically, we consider hyper-surfaces in a Riemannian manifold, and mainly follow \citet{do1992riemannian, CapovillaGuven1995, Spivak1970III} and \citet{Kuchar1976}.
The body is denoted by $\mathcal{B}$. The natural configuration of the body is a Riemannian manifold with a metric $\mathbf{G}$.\footnote{In elasticity $\mathbf{G}$ is a flat metric. However, in anelastic bodies $\mathbf{G}$ is non-flat and quantifies the natural distances in the presence of eigenstrains. In the material manifold the bulk and material surfaces have their own Riemannian metrics. In general, the material metric need not be continuous, see \citep{YavariGoriely2013Inclusions,Golgoon2018a} for examples of discontinuous material metrics.} The ambient space is another Riemannian manifold $(\mathcal{S},\mathbf{g})$. We denote the inner products induced by the metrics $\mathbf{G}$ and $\mathbf{g}$ by $\llangle \cdot,\cdot\rrangle_{\mathbf{G}}$ and $\llangle \cdot,\cdot\rrangle_{\mathbf{g}}$, respectively.
Consider a Riemannian manifold $\Sur$ embedded in the material manifold $(\mathcal{B},\mathbf{G})$ such that $\dim \Sur=\dim \mathcal{B}-1$.
$\Sur$ can be viewed in two ways: (i) as a subset of $\mathcal{B}$ or (ii) as an abstract $(n-1)$-dimensional manifold. These two perspectives are related via the inclusion map  $\iota_{\Sur}:\Sur\hookrightarrow\mathcal{B}$, $\iota_{\Sur}(X)=X$. 
We distinguish between $\Sur$ and $\SurE=\iota_{\Sur}(\Sur)\subset \mathcal{B}$.\footnote{It is assumed that either $\partial\SurE=\emptyset$ or $\partial\SurE\subset\partial \mathcal{B}$.}

The metric $\mathbf{G}$ on $\mathcal{B}$ induces a metric $\bar{\mathbf{G}}=\iota_{\Sur}^*\mathbf{G}=\FS^\star\mathbf{G}\FS=\mathbf{G}\big|_{\mathsf{S}}$ on $\mathsf{S}$,\footnote{In surface anelasticity this is not the material metric of the surface, in general.} where $\FS=T\iota_{\Sur}$. This is the \emph{first fundamental form} of the hyper-surface.
We denote the local coordinates on $\mathcal{B}$, $\Sur$, and $\mathcal{S}$ by $\{X^A\}$, $\{\bar{X}^{\bar{A}}\}$, and $\{x^a\}$, respectively, where $A,a=1,\hdots, n$ and $\bar{A}=1,\hdots, n-1$.
In particular, this implies that $\Sur$ is locally represented as $X^A=X^A(\bar{X}^{\bar{A}})$. Therefore \citep{Yano1970}\footnote{The first fundamental form has components $\bar{G}_{\bar{A}\bar{B}}=\cFS^A{}_{\bar{A}}\,G_{AB}\,\cFS^B{}_{\bar{B}}$.}
\begin{equation} 
	\cFS^A{}_{\bar{A}}=\frac{\partial X^A}{\partial \bar{X}^{\bar{A}}} \,,\qquad
	\FS= \frac{\partial X^A}{\partial \bar{X}^{\bar{A}}}\, \frac{\partial}{\partial X^A}
	\otimes d\bar{X}^{\bar{A}}
	\,.
\end{equation}
Let $\dim\mathcal{S}=\dim\mathcal{B}= n$.\footnote{In surface elasticity applications $n=3$.}

Let us define the induced bundle $\iSpb T\mathcal{B}$ over $\Sur$ whose fiber at $X\in\Sur$ is $T_X\mathcal{B}$. One has the following orthogonal decomposition \citep{Dajczer2019}
\begin{equation}
   \iSpb T\mathcal{B} = \iSpf T\Sur \oplus \left(\iSpf T\Sur\right)^{\perp} = T\SurE \oplus \left(T\SurE\right)^{\perp}\,,
\end{equation}
where $\left(T\SurE\right)^{\perp}$ is the normal bundle of the embedding $\iota_{\Sur}$ whose fiber at $X\in\Sur$ is the orthogonal complement of $T_{X}\SurE$ in $T_X\mathcal{B}$.\footnote{For a vector $\bar{\mathbf{W}}\in T_X\Sur$, $\iSpf\bar{\mathbf{W}}=\FS\bar{\mathbf{W}}$, and in coordinates $(\iSpf \bar{W})^A=\cFS^A{}_{\bar{A}}\bar{W}^{\bar{A}}$.}
Therefore, at $X\in\Sur$, the tangent space $T_{X}\SurE$ has an orthogonal complement $\left(T_{X}\SurE\right)^{\perp}\subset T_{X}\mathcal{B}$ such that
\begin{equation}\label{T_decomp}
	T_X\mathcal{B} = T_{X}\SurE \oplus \left(T_{X}\SurE\right)^{\perp}\,.
\end{equation}
In other words, any vector field $\mathbf{W}\in T_X\mathcal{B}$ is uniquely written as the sum of a vector $\mathbf{W}_{\parallel}\in T_X\SurE$ (that is tangent to $\SurE$) and a vector $\mathbf{W}_\perp:=\mathbf{W}-\mathbf{W}_\parallel$ (that is normal to $\SurE$, i.e., $\mathbf{W}_\perp\in (T_X\SurE)^\perp$).
For $\mathbf{W}\in T_X\mathcal{B}$, one writes $\mathbf{W} = \mathbf{W}_\parallel + \mathbf{W}_\perp$.\footnote{We say that $\mathbf{W}$ is a vector field \emph{along} $\SurE$. If $\mathbf{W} = \mathbf{W}_\parallel$, we say that $\mathbf{W}$ is a vector field \emph{on} $\SurE$. Obviously, a vector field on $\SurE$ is a vector field along $\SurE$ \citep{Tu2017}. \label{footnote-surface}} 
Note that $\left(T_{X}\SurE\right)^{\perp}$ is one dimensional. We assume it is spanned by the unit vector $\mathbf{N}(X)$.
This means that every vector field $\mathbf{W}$ on $\mathcal{B}$ along $\SurE$ can be written as $\mathbf{W} = \mathbf{W}_\parallel + W_n\, \mathbf{N}$.
Note that $\llangle \mathbf{N},\mathbf{W}_\parallel \rrangle_{\mathbf{G}} = 0$.

Let $\{X^A\}_{A=1,...,n}$ be a local coordinate chart for $\mathcal{B}$ such that at any point of $\SurE$, $\{X^1,...,X^{n-1}\}$ is a local coordinate chart for $\SurE$, and such that the unit normal vector field $\mathbf{N}$ is tangent to the $X^n$-coordinate curve. We call such coordinates \emph{foliation coordinates} \citep{Sozio2019}.
With respect to a foliation coordinate chart the inclusion map $\iota_{\Sur}:\Sur\hookrightarrow\mathcal{B}$ has the following representation
\begin{equation}
	\iota_{\Sur}:
	\begin{Bmatrix} X^1 \\ \vdots \\ X^{n-1} \end{Bmatrix} \mapsto 
	\begin{Bmatrix} X^1 \\ \vdots \\ X^{n-1} \\ 0 \end{Bmatrix}\,.
\end{equation}
Therefore, the tangent map of $\iota_{\Sur}$, $\FS=T\iota_{\Sur}:T\Sur\to T\mathcal{B}$ has the following representation\footnote{$\FS$ is what \citet{Gurtin1975} call ``inclusion map" and denote it by $\boldsymbol{\mathsf{I}}$.}
\begin{equation} \label{FS-Matrix}
	\big[\FS\big]=\big[\cFS^A{}_{\bar{A}}\big]
	= \begin{bmatrix} \mathbf{I}_{n-1} \\ \mathbf{0}_{1\times (n-1)} \end{bmatrix}\,,
\end{equation}
where $\mathbf{I}_{n-1} $ is the $n-1$ by $n-1$ identity matrix and $\mathbf{0}_{1\times (n-1)}$ is the $(n-1)$-dimensional zero row vector. 
In components, $\cFS^A{}_{\bar{A}}=\delta^A_{\bar{A}}$.
The dual of $\FS$, $\FS^\star:T^*\mathcal{B}\to T^*\Sur$ is defined such that
\begin{equation}
	\langle\boldsymbol\alpha,\FS \bar{\mathbf{U}} \rangle
	=\langle\FS^\star \boldsymbol\alpha,\bar{\mathbf{U}} \rangle\,, \qquad\forall\, 
	\bar{\mathbf{U}} \in T_X\Sur\,,~ \boldsymbol\alpha \in T^*_X\mathcal B\,.
\end{equation}
$\FS^\star$ has the following matrix representation
\begin{equation}
	\big[\FS^\star\big]= \begin{bmatrix} \mathbf{I}_{n-1} & \mathbf{0}_{(n-1)\times 1} \end{bmatrix}\,.
\end{equation}
In components, $(\cFS^\star)_{\bar{A}}{}^A=\delta^A_{\bar{A}}$.
Obviously, $[\FS^\star]=[\FS]^{\mathsf{T}}$. Also note that $[\FS^\star][\FS]=\mathbf{I}_{n-1}$.
With respect to arbitrary coordinate charts
\begin{equation} 
	\FS^\star= \frac{\partial X^A}{\partial \bar{X}^{\bar{A}}}\, d\bar{X}^{\bar{A}}\otimes
	\frac{\partial}{\partial X^A}
	\,.
\end{equation}

\begin{defi}[Projection Map]
For $X\in\SurE$, the projection map $\pi_{\SurE}: T_X\mathcal{B}\big|_{\SurE} \to T_X\SurE$ is defined such that $\pi_{\SurE}(\mathbf{W}) = \mathbf{W}_\parallel$. The explicit form of the projection map is
\begin{equation} \label{Referential-Projection}
	\pi_{\SurE}=\operatorname{id}_{T\mathcal{B}} - \mathbf{N}\otimes \mathbf{N}^{\flat}\,.
\end{equation}
\end{defi}

One can easily check that $\pi_{\SurE}(\mathbf{W})=\mathbf{W}- \mathbf{N}\langle  \mathbf{N}^{\flat} ,\mathbf{W} \rangle=\mathbf{W}- W_n\,\mathbf{N} =\mathbf{W}-\mathbf{W}^{\perp}=\mathbf{W}_\parallel$. The matrix representation of the projection map with respect to foliation coordinates is
\begin{equation}
	\big[\pi_{\SurE}\big]= \begin{bmatrix} \mathbf{I}_{n-1} & \mathbf{0}_{(n-1)\times 1} \\
	\mathbf{0}_{1\times (n-1)}  & 0
	\end{bmatrix}\,.
\end{equation}

\begin{figure}[h] \centerline{%
\xymatrix@C=+1.50cm{
    T_X\mathcal{B}\big|_{\SurE}  \ar[d]^{\bar{\pi}_{\Sur}} \ar[dr]^{\pi_{\SurE}} \\
    T_X\Sur  \ar[r]^{\FS}  
    &  T_X\SurE
    }}
\caption{The commutative diagram of projection and surface projection maps.}\label{Fig:ProjectionMaps}
\end{figure}
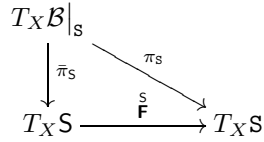

\begin{defi}[Surface Projection Map]
Given $\mathbf{W}_\parallel\in T\SurE$, there is a unique $\bar{\mathbf{W}}\in T\Sur$ such that
$\mathbf{W}_\parallel=\iSpf \bar{\mathbf{W}}=\FS\bar{\mathbf{W}}$.
The surface projection $\bar{\pi}_{\Sur}:T\mathcal{B}\big|_{\SurE}\to T\Sur$ is defined such that $\bar{\pi}_{\Sur}(\mathbf{W})=\bar{\mathbf{W}}$. This implies that (see Fig.~\ref{Fig:ProjectionMaps})
\begin{equation} \label{Pi-bar-Pi}
	\pi_{\SurE}=\iSpf \bar{\pi}_{\Sur}=\FS\circ\bar{\pi}_{\Sur}
	\,.
\end{equation}
\end{defi}

Note that $\bar{\pi}_{\Sur}(\mathbf{W})=\bar{\pi}_{\Sur}(\mathbf{W}_\parallel + \mathbf{W}_\perp)=\bar{\pi}_{\Sur}(\mathbf{W}_\parallel)=\bar{\mathbf{W}}$. Therefore, when restricted to $T\SurE$, from the commutative diagram of Fig.~\ref{Fig:ProjectionMaps} we see that $\FS\circ\bar{\pi}_{\Sur}\big|_{T\SurE}=\operatorname{id}_{T\SurE}$.
From $\bar{\pi}_{\Sur}(\mathbf{W}_\parallel)=\bar{\mathbf{W}}$, one has $\bar{\pi}_{\Sur}(\mathbf{W}_\parallel)=\bar{\pi}_{\Sur}(\FS \bar{\mathbf{W}})=(\bar{\pi}_{\Sur}\circ\FS)\bar{\mathbf{W}}=\bar{\mathbf{W}}$, and hence, $\bar{\pi}_{\Sur}\circ\FS=\operatorname{id}_{T\Sur}$.
Therefore, in summary 
\begin{equation} \label{FS-Identities}
	\bar{\pi}_{\Sur}\circ\FS=\operatorname{id}_{T\Sur}\,,\qquad
	\FS\circ\bar{\pi}_{\Sur}\big|_{T\SurE}=\operatorname{id}_{T\SurE}
	\,.
\end{equation}
For an arbitrary $\bar{\mathbf{U}}\in T\Sur$ and $\mathbf{W}\in T\mathcal{B}\big|_{\SurE}$,
\begin{equation}
\begin{aligned}
	\left\llangle \bar{\pi}_{\Sur}(\mathbf{W}) , \bar{\mathbf{U}} \right\rrangle_{\bar{\mathbf{G}}}
	& =\left\llangle \FS \,\bar{\pi}_{\Sur}(\mathbf{W}),\FS\bar{\mathbf{U}} \right\rrangle_{\mathbf{G}} \\
	&=\left\llangle \pi_{\Sur}(\mathbf{W}) , \FS\bar{\mathbf{U}} \right\rrangle_{\mathbf{G}} \\
	&=\left\llangle \mathbf{W}- W_n\,\mathbf{N} , \FS\bar{\mathbf{U}} \right\rrangle_{\mathbf{G}} \\
	&=\left\llangle \mathbf{W} , \FS\bar{\mathbf{U}} \right\rrangle_{\mathbf{G}} \\
	& =\left\llangle \FS^{\mathsf{T}}\mathbf{W} , \bar{\mathbf{U}} \right\rrangle_{\mathbf{G}}
	\,,
\end{aligned}
\end{equation}
where in the second equality \eqref{Pi-bar-Pi} was used.
This implies that\footnote{$\FS^{\mathsf{T}}$ is what \citet{Gurtin1975} call ``perpendicular projection" and denote by $\mathbf{P}$.}
\begin{equation} \label{Projection-Formula}
	\bar{\pi}_{\Sur}=\FS^{\mathsf{T}}	\,.
\end{equation}
From \eqref{Pi-bar-Pi} and \eqref{Projection-Formula}, it follows that it follows that $\pi_{\Sur} = \FS\,\FS^{\mathsf{T}}$.  
Also, from \eqref{FS-Identities} and \eqref{Projection-Formula}, one obtains the following identities
\begin{equation}
	\FS^{\mathsf{T}}\,\FS = \operatorname{id}_{T\Sur}\,,\qquad
	\FS\,\FS^{\mathsf{T}}\big|_{T\SurE} = \operatorname{id}_{T\SurE}
	\,.
\end{equation}

With respect to foliation coordinates $\{X^1,...,X^n\}$, we have
\begin{equation} 
	\pi_{\SurE}:
	\begin{Bmatrix} W^1 \\ \vdots \\ W^{n-1} \\ W^n \end{Bmatrix} \mapsto 
	\begin{Bmatrix} W^1 \\ \vdots \\ W^{n-1} \\ 0 \end{Bmatrix}\,,\qquad
	\bar{\pi}_{\Sur}:
	\begin{Bmatrix} W^1 \\ \vdots \\ W^{n-1} \\ W^n \end{Bmatrix} \mapsto 
	\begin{Bmatrix} W^1 \\ \vdots \\ W^{n-1}  \end{Bmatrix}
	\,.
\end{equation}
This implies that 
\begin{equation}
	\big[\bar{\pi}_{\Sur}\big] = \begin{bmatrix} \mathbf{I}_{n-1} & \mathbf{0}_{(n-1)\times 1} \end{bmatrix}\,.
\end{equation}

\paragraph[The metric-compatible left inverse of FS]{The metric-compatible left inverse of \texorpdfstring{$\FS$}{FS}.}
The induced (first fundamental) form on $\Sur$ is the pullback of $\mathring{\mathbf{G}}$ by the inclusion $\iota_{\Sur}$, i.e., $\mathring{\bar{\mathbf{G}}} \;=\; \FS^{\star}\,\mathring{\mathbf{G}}\,\FS$.
Its inverse $\mathring{\bar{\mathbf{G}}}^{\sharp}:T_X^{\!*}\Sur\to T_X\Sur$ is characterized by
\begin{equation}
	\mathring{\bar{\mathbf{G}}}^{\sharp}\,\mathring{\bar{\mathbf{G}}}
	= \operatorname{id}_{T\Sur}\,,
	\qquad
	\mathring{\bar{\mathbf{G}}}\,\mathring{\bar{\mathbf{G}}}^{\sharp}
	= \operatorname{id}_{T^{\!*}\Sur}\,.
\end{equation}
One can show that
\begin{equation} \label{Surface-G-Inverse}
	\FS\,\mathring{\bar{\mathbf{G}}}^{\sharp}\,\FS^{\star}\,\mathring{\mathbf{G}}=
	\operatorname{id}_{T\mathcal{B}} - \mathbf{N}\otimes \mathbf{N}^{\flat}\,.
\end{equation}
Let $\bar{\mathbf{V}}\in T_X\Sur$ and set $\mathbf{V}_{\parallel}=\FS\,\bar{\mathbf{V}}\in T_X\mathcal{B}$. Then
\begin{equation}
\begin{aligned}
	\FS\,\mathring{\bar{\mathbf{G}}}^{\sharp}\,\FS^{\star}\,\mathring{\mathbf{G}}\,\mathbf{V}_{\parallel}
	&= \FS\,\mathring{\bar{\mathbf{G}}}^{\sharp}\,\FS^{\star}\,\mathring{\mathbf{G}}\,\FS\,\bar{\mathbf{V}} \\
	&= \FS\,\mathring{\bar{\mathbf{G}}}^{\sharp}\,\big(\FS^{\star}\,\mathring{\mathbf{G}}\,\FS\big)\,\bar{\mathbf{V}} \\
	&= \FS\,\mathring{\bar{\mathbf{G}}}^{\sharp}\,\mathring{\bar{\mathbf{G}}}\,\bar{\mathbf{V}} \\
	&= \FS\,\bar{\mathbf{V}} \\
	&= \mathbf{V}_{\parallel}\,.
\end{aligned}
\end{equation}
Hence, the left-hand side acts as the identity on tangential vectors. 
Note that for all $\bar{\mathbf{W}}\in T_X\Sur$,
\begin{equation}
	\llangle \FS\,\bar{\mathbf{W}},\mathbf{N}\rrangle_{\mathring{\mathbf{G}}}
	=\langle \FS^{\star}\mathbf{N}^\flat,\bar{\mathbf{W}}\rangle
	= 0\,,
\end{equation}
and hence $\FS^{\star}\mathbf{N}^\flat=\mathbf{0}$.
Thus
\begin{equation}
	\FS\,\mathring{\bar{\mathbf{G}}}^{\sharp}\,\FS^{\star}\,\mathbf{N}^\flat
	= \FS\,\mathring{\bar{\mathbf{G}}}^{\sharp}\,\mathbf{0}
	= \mathbf{0}\,.
\end{equation}
Therefore $\FS\,\mathring{\bar{\mathbf{G}}}^{\sharp}\,\FS^{\star}\,\mathring{\mathbf{G}}$ is the $\mathring{\mathbf{G}}$-orthogonal tangential projector, i.e., \eqref{Surface-G-Inverse} holds.
The metric-compatible left inverse of $\FS$ is the map\footnote{Any linear map $\boldsymbol{\mathsf{L}}:T_X\mathcal{B}\to T_X\Sur$ satisfying $\boldsymbol{\mathsf{L}}\,\FS=\operatorname{id}_{T\Sur}$ is a left inverse of $\FS$. However, such a left inverse is not unique. The map $\FS^{\dagger}$ is distinguished because it is defined using the Riemannian metrics on $T_X\Sur$ and $T_X\mathcal{B}$, i.e., it is the adjoint of $\FS$ with respect to these metrics. This is why it is called the \emph{metric-compatible} left inverse.}
\begin{equation}
	\FS^{\dagger} 
	= \mathring{\bar{\mathbf{G}}}^{\sharp}\,\FS^{\star}\,\mathring{\mathbf{G}}:T_X\mathcal{B}\to T_X\Sur
	\,.
\end{equation}
Notice that
\begin{equation}
	\FS^{\dagger}\,\FS
	= \mathring{\bar{\mathbf{G}}}^{\sharp}\,\FS^{\star}\,\mathring{\mathbf{G}}\,\FS
	= \mathring{\bar{\mathbf{G}}}^{\sharp}\,\big(\FS^{\star}\,\mathring{\mathbf{G}}\,\FS\big)
	= \mathring{\bar{\mathbf{G}}}^{\sharp}\,\mathring{\bar{\mathbf{G}}}
	= \operatorname{id}_{T\Sur}\,,
\end{equation}
i.e., $\FS^{\dagger}$ is a left inverse of $\FS$ compatible with the metrics.
The metric-compatible left inverse can also be explicitly expressed using the normal vector. Define the $\mathring{\mathbf{G}}$-orthogonal tangential projector
\begin{equation}
	\pi_{\SurE} = \operatorname{id}_{T\mathcal{B}} - \mathbf{N}\otimes \mathbf{N}^\flat \,,
\end{equation}
so that $\pi_{\SurE}:T_X\mathcal{B}\to T_X\SurE$ and $\pi_{\Sur}\!\circ\FS=\FS$. Let $(\FS)^{-1}:T_X\SurE \to T_X\Sur$ denote the inverse of the restriction of $\FS$ to $T_X\Sur$. Then the left inverse can be written as
\begin{equation}
	\FS^{\dagger}
	=\big(\FS\big)^{-1}\circ\pi_{\Sur}
	=\big(\FS\big)^{-1}\circ\big(\operatorname{id}_{T\mathcal{B}}-\mathbf{N}\otimes \mathbf{N}^\flat\big)
	\,.
\end{equation}
For any $\mathbf{V}\in T_X\mathcal{B}$, this gives the explicit formula
\begin{equation}
	\FS^{\dagger}(\mathbf{V})
	=\big(\FS\big)^{-1}\!\left[\mathbf{V}-\langle \mathbf{N}^\flat,\mathbf{V}\rangle\,\mathbf{N}\right]
	\,.
\end{equation}

Let us denote the connection coefficients for the Levi-Civita connections $\nabla^{\mathbf{G}}$, $\nabla^{\bar{\mathbf{G}}}$ and $\nabla^{\mathbf{g}}$ corresponding to the metrics $\mathbf{G}$, $\bar{\mathbf{G}}$, and $\mathbf{g}$ by $\Gamma^{A}{}_{BC}\,$, $\bar{\Gamma}^{\bar{A}}{}_{\bar{B}\bar{C}}\,$, and $\gamma^a{}_{bc}\,$, respectively. Note that with respect to foliation coordinates $\Gamma^{\bar{A}}{}_{\bar{B}\bar{C}}=\bar{\Gamma}^{\bar{A}}{}_{\bar{B}\bar{C}}$.

\begin{prop}\label{LieDerivative-Extension}
Let $\bar{\mathbf{X}}$ be a vector field on $\Sur$. $\mathbf{X}$ is an extension of $\iota_{\Sur*}\bar{\mathbf{X}}$ if $\mathbf{X}\big|_{T\SurE}=\iota_{\Sur*}\bar{\mathbf{X}}$.
Let $\bar{\mathbf{X}}$ and $\bar{\mathbf{Y}}$ be vector fields on $\Sur$, and let $\mathbf{X}$ and $\mathbf{Y}$ be arbitrary extensions of $\bar{\mathbf{X}}$ and $\bar{\mathbf{Y}}$, respectively. One can show that 
\begin{equation} 
	[\mathbf{X},\mathbf{Y}]\big|_{\SurE} = [\bar{\mathbf{X}}, \bar{\mathbf{Y}}]
	\,,
\end{equation}
i.e., $[\bar{\mathbf{X}}, \bar{\mathbf{Y}}]$ is independent of the extensions.
\end{prop}
\noindent For a proof, see \citep{Chen2019}.

\subsubsection{Geometry of hyper-surfaces via Cartan's moving frames}

Next we derive the fundamental equations of surfaces in Riemannian manifolds using the machinery of Cartan's moving frames.\footnote{A similar treatment for space-like hyper-surfaces in Lorentzian manifolds can be found in \citep{Straumann2012}. See also \citep{Nicolaescu2020}.}
For any $X\in\Sur$, in the orthonormal frame field $\{\mathbf{e}_1(X),\hdots,\mathbf{e}_n(X)\}$ for $T_X\mathcal{B}$, suppose that the first $n-1$ vectors, i.e., $\{\mathbf{e}_1(X),\hdots,\mathbf{e}_{n-1}(X)\}$ is a basis for $T_X\SurE$. 
We call $\{\mathbf{e}_1(X),\hdots,\mathbf{e}_n(X)\}$ a foliation moving frame.
The corresponding moving coframe field is denoted as $\{\vartheta^1(X),\hdots,\vartheta^n(X)\}$. In particular, metric has the simple representation $\mathbf{G}=\delta_{\alpha\beta}\,\vartheta^{\alpha}\otimes\vartheta^{\beta}$.
Thus, $\{\bar{\mathbf{e}}_{\bar{\alpha}}\}$, $\bar{\alpha}=1,\hdots n-1$, where $\bar{\mathbf{e}}_{\bar{\alpha}}=\mathbf{e}_{\bar{\alpha}}\big|_{\SurE}$, is a moving frame for $\SurE$.
This moving frame is dual to the restrictions $\bar{\vartheta}^{\bar{\alpha}}=\vartheta^{\bar{\alpha}}|_{\SurE}$, $\bar{\alpha}=1,\hdots n-1$.
The induced metric has the simple representation $\bar{\mathbf{G}}=\delta_{\bar{\alpha}\bar{\beta}}\,\vartheta^{\bar{\alpha}}\otimes\vartheta^{\bar{\beta}}$.
Note that $\langle \vartheta^n,\bar{\mathbf{e}}_{\bar{\alpha}} \rangle=0$, $\bar{\alpha}=1,\hdots n-1$, and hence $\vartheta^n\big|_{T\SurE}=0$. In the absence of torsion, the first structural equations \eqref{First-Structural-Equations} restricted to $T\Sur$ (and $T\SurE$) read 
\begin{equation} \label{First-Structural-Equations-S}
\begin{dcases}
    0= d\bar{\vartheta}^{\bar{\alpha}}
    +\omega^{\bar{\alpha}}{}_{\bar{\beta}}\wedge\bar{\vartheta}^{\bar{\beta}}\,,
    \qquad \bar{\alpha}=1,\hdots n-1\,,\\
    0 = \omega^{n}{}_{\bar{\beta}}\wedge\bar{\vartheta}^{\bar{\beta}} \,.
\end{dcases}
\end{equation}
The first structural equations of $(\Sur,\bar{\mathbf{G}},\bar{\nabla})$ with connection $1$-forms $\bar{\omega}^{\bar{\alpha}}{}_{\bar{\beta}}$ compatible with the induced metric $\bar{\mathbf{G}}$ are: $0 = d\bar{\vartheta}^{\bar{\alpha}}+\bar{\omega}^{\bar{\alpha}}{}_{\bar{\beta}}\wedge\bar{\vartheta}^{\bar{\beta}}$. Knowing that the solution for the connection $1$-forms is unique (even in the presence of torsion \citep{YavariGoriely2012a}), one concludes that on $T\SurE=\iota_{\Sur*}T\Sur$, $\bar{\omega}^{\bar{\alpha}}{}_{\bar{\beta}}=\omega^{\bar{\alpha}}{}_{\bar{\beta}}$.

The Levi-Civita connection $\nabla^{\mathbf{G}}$ of $(\mathcal{B},\mathbf{G})$ induces a unique connection $\hat{\nabla}^{\mathbf{G}}$ on $\iSpb T\mathcal{B}$ defined as 
\begin{equation} \label{Induced-Connection-Definition}
	\hat{\nabla}^{\mathbf{G}}_{\bar{\mathbf{X}}} (\mathbf{Y}\circ\iota_{\Sur})
	= \nabla^{\mathbf{G}}_{\iSpf\bar{\mathbf{X}}} \mathbf{Y}\,,\qquad \forall X\in\Sur\,,
	~\bar{\mathbf{X}}\in T_X\Sur\,,~\mathbf{Y}\in T_X\mathcal{B}
	\,.
\end{equation}
We follow \citep{Dajczer2019} and identify $\hat{\nabla}^{\mathbf{G}}$ and $\nabla^{\mathbf{G}}$ and simply write $\nabla^{\mathbf{G}}$.
For $\bar{\mathbf{X}},\bar{\mathbf{Y}}\in T_X\Sur$, one writes
\begin{equation}
	 \nabla^{\mathbf{G}}_{\bar{\mathbf{X}}} (\iSpf\bar{\mathbf{Y}})
	 =\left( \nabla^{\mathbf{G}}_{\bar{\mathbf{X}}} (\iSpf\bar{\mathbf{Y}})\right)_{\parallel}
	 +\left( \nabla^{\mathbf{G}}_{\bar{\mathbf{X}}} (\iSpf\bar{\mathbf{Y}})\right)_{\perp}
	\,.
\end{equation}
Consider vector fields $\mathbf{Y}$ and $\bar{\mathbf{X}}$ defined on $\mathcal{B}$ and $\Sur$, respectively, such that $\mathbf{Y}$ is everywhere tangent to $\SurE$ (in this case we say that ``$\mathbf{Y}$ is a vector field on $\SurE$", see Footnote \ref{footnote-surface}). 
The vector field $\mathbf{Y}$ can be written as $\mathbf{Y}=\iSpf\bar{\mathbf{Y}}$ such that $\bar{\mathbf{Y}}\in T_X\Sur$.
Thus, $\bar{\mathbf{X}}=\bar{X}^{\bar{\alpha}}\,\bar{\mathbf{e}}_{\bar{\alpha}}$ and $\bar{\mathbf{Y}}=\bar{Y}^{\bar{\alpha}}\,\bar{\mathbf{e}}_{\bar{\alpha}}$.
Note that $\mathbf{Y}=\iSpf\bar{\mathbf{Y}}=Y^{\bar{\alpha}}\,\mathbf{e}_{\bar{\alpha}}$ but also $\mathbf{Y}=Y^{\alpha}\,\mathbf{e}_{\alpha}=\cFS^{\alpha}{}_{\bar{\alpha}}\,\bar{Y}^{\bar{\alpha}}\,\mathbf{e}_{\alpha}=\delta^{\alpha}_{\bar{\alpha}}\,\bar{Y}^{\bar{\alpha}}\,\mathbf{e}_{\alpha}=\bar{Y}^{\bar{\alpha}}\,(\delta^{\alpha}_{\bar{\alpha}}\,\mathbf{e}_{\alpha})=\bar{Y}^{\bar{\alpha}}\,\mathbf{e}_{\bar{\alpha}}$. This implies that $Y^{\bar{\alpha}}=\bar{Y}^{\bar{\alpha}}$. Thus, one writes (note that $Y^n=0$ as $\mathbf{Y}$ is a tangent vector)
\begin{equation} \label{Covariant-Derivative-Surface}
\begin{aligned}
    \nabla^{\mathbf{G}}_{\bar{\mathbf{X}}}\mathbf{Y} 
    &=\langle dY^{\alpha}+\omega^{\alpha}{}_{\gamma}Y^{\gamma}, 
    \bar{\mathbf{X}}\rangle \,\mathbf{e}_{\alpha}\\
    & = \langle dY^{\bar{\alpha}}+\omega^{\bar{\alpha}}{}_{\bar{\gamma}}Y^{\bar{\gamma}}, 
    \bar{\mathbf{X}}\rangle \,\mathbf{e}_{\bar{\alpha}}
    +\langle \omega^{n}{}_{\bar{\gamma}}Y^{\bar{\gamma}}, 
    \bar{\mathbf{X}}\rangle \,\mathbf{e}_{n}
     \\
    & = \langle d\bar{Y}^{\bar{\alpha}}+\bar{\omega}^{\bar{\alpha}}{}_{\bar{\gamma}}
    \bar{Y}^{\bar{\gamma}}, \bar{\mathbf{X}}\rangle \,\bar{\mathbf{e}}_{\bar{\alpha}}
    +\langle \omega^{n}{}_{\bar{\gamma}}Y^{\bar{\gamma}}, 
    \bar{\mathbf{X}}\rangle \,\mathbf{e}_{n}\\
    & =\iSpf \nabla^{\bar{\mathbf{G}}}_{\bar{\mathbf{X}}}\bar{\mathbf{Y}} 
    +\langle \omega^{n}{}_{\bar{\gamma}}Y^{\bar{\gamma}}, 
    \bar{\mathbf{X}}\rangle \,\mathbf{e}_{n}
    \,.
\end{aligned}
\end{equation}
Thus, $(\nabla^{\mathbf{G}}_{\bar{\mathbf{X}}}\mathbf{Y})_{\parallel}=\iSpf \nabla^{\bar{\mathbf{G}}}_{\bar{\mathbf{X}}}\bar{\mathbf{Y}}$.

The second structural equation \eqref{First-Structural-Equations-S}$_2$ and Cartan's lemma\footnote{Let $\vartheta^1,...,\vartheta^p$ be $p$ linearly independent $1$-forms in $\mathcal{B}$ ($p\leq n$). Now suppose that the $1$-forms $\xi_1,...,\xi_p$ satisfy the relation $\xi_{\alpha}\wedge\vartheta^{\alpha}=0$. Then, \emph{Cartan's Lemma} tells us that $\xi_{\alpha}=\xi_{\alpha\beta}\,\vartheta^{\beta}$, such that $\xi_{\alpha\beta}=\xi_{\beta\alpha}$ \citep{Sternberg1999}.} imply that on $\Sur$
\begin{equation}
    \omega^{n}{}_{\bar{\beta}}=-K_{\bar{\beta}\bar{\alpha}}\,\bar{\vartheta}^{\bar{\alpha}}\,,\qquad 
    K_{\bar{\beta}\bar{\alpha}}=K_{\bar{\alpha}\bar{\beta}}
    \,.
\end{equation}
The tensor $\mathbf{K}=K_{\bar{\alpha}\bar{\beta}}\,\bar{\vartheta}^{\bar{\alpha}}\otimes\bar{\vartheta}^{\bar{\beta}}=K_{\bar{\alpha}\bar{\beta}}\,\vartheta^{\bar{\alpha}}\otimes\vartheta^{\bar{\beta}}$ is called the \emph{second fundamental form} of $\SurE$. 
From \eqref{Covariant-Derivative-Surface} the normal component of the covariant derivative is written as $(\nabla^{\mathbf{G}}_{\bar{\mathbf{X}}}\mathbf{Y})^{\perp}= \langle \omega^{n}{}_{\bar{\gamma}}Y^{\bar{\gamma}}, \bar{\mathbf{X}}\rangle \,\mathbf{e}_n$. Thus, choosing $\bar{\mathbf{X}}=\bar{\mathbf{e}}_{\bar{\alpha}}$ and $\mathbf{Y}=\mathbf{e}_{\bar{\beta}}=\bar{\mathbf{e}}_{\bar{\beta}}$ we have
\begin{equation}
    (\nabla^{\mathbf{G}}_{\bar{\mathbf{e}}_{\bar{\alpha}}}\bar{\mathbf{e}}_{\bar{\beta}})^{\perp}
    = \big\langle \omega^{n}{}_{\bar{\gamma}}\,\delta^{\bar{\gamma}}_{\bar{\beta}}, 
    \bar{\mathbf{e}}_{\bar{\alpha}} \big\rangle \,\mathbf{e}_n
    = \big\langle \omega^{n}{}_{\bar{\beta}}, \bar{\mathbf{e}}_{\bar{\alpha}} \big\rangle \,\mathbf{e}_n 
    = \big\langle -K_{\bar{\beta}\bar{\gamma}}\,\bar{\vartheta}^{\bar{\gamma}}, 
    \bar{\mathbf{e}}_{\bar{\alpha}} \big\rangle \,\mathbf{e}_n 
    = -K_{\bar{\beta}\bar{\alpha}}\,\mathbf{e}_n \,.
\end{equation}
Therefore, \eqref{Covariant-Derivative-Surface} can be simplified as
\begin{equation} \label{Gauss-formula}
\begin{aligned}
    \nabla^{\mathbf{G}}_{\bar{\mathbf{X}}}\mathbf{Y} 
    & =\iSpf \nabla^{\bar{\mathbf{G}}}_{\bar{\mathbf{X}}}\bar{\mathbf{Y}} 
    + \big\langle \omega^{n}{}_{\bar{\gamma}}Y^{\bar{\gamma}}, \bar{\mathbf{X}} \big\rangle 
    \,\mathbf{e}_{n} \\
    & = \iSpf \nabla^{\bar{\mathbf{G}}}_{\bar{\mathbf{X}}}\bar{\mathbf{Y}} 
    + \big\langle -K_{\bar{\gamma}\bar{\beta}}\,\bar{\vartheta}^{\bar{\beta}} Y^{\bar{\gamma}}, 
    \bar{\mathbf{X}} \big\rangle \,\mathbf{e}_{n} \\
    & = \iSpf \nabla^{\bar{\mathbf{G}}}_{\bar{\mathbf{X}}}\bar{\mathbf{Y}} 
    -K_{\bar{\gamma}\bar{\beta}}\,Y^{\bar{\gamma}}\, \big\langle \bar{\vartheta}^{\bar{\beta}} , 
    \bar{\mathbf{X}} \big\rangle \,\mathbf{e}_{n} \\
    & = \iSpf \nabla^{\bar{\mathbf{G}}}_{\bar{\mathbf{X}}}\bar{\mathbf{Y}} 
    -K_{\bar{\gamma}\bar{\beta}}\,Y^{\bar{\gamma}}\, \bar{X}^{\bar{\beta}} \,\mathbf{e}_{n} \\
    & = \iSpf \nabla^{\bar{\mathbf{G}}}_{\bar{\mathbf{X}}}\bar{\mathbf{Y}} 
    -K_{\bar{\gamma}\bar{\beta}}\,Y^{\bar{\gamma}}\, X^{\bar{\beta}} \,\mathbf{e}_{n} \\
    & = \iSpf \nabla^{\bar{\mathbf{G}}}_{\bar{\mathbf{X}}}\bar{\mathbf{Y}} 
    -\mathbf{K}(\mathbf{X}, \mathbf{Y}) \,\mathbf{e}_{n} 
    \,,
\end{aligned}
\end{equation}
where $\mathbf{X}=\iSpf\bar{\mathbf{X}}$. 
Using the definition and identification given in \eqref{Induced-Connection-Definition}, the above identity can be rewritten as
\begin{equation} \label{Gauss-Formula}
	\nabla^{\mathbf{G}}_{\iSpf\bar{\mathbf{X}}} (\iSpf\bar{\mathbf{Y}})
	= \iSpf \nabla^{\bar{\mathbf{G}}}_{\bar{\mathbf{X}}}\bar{\mathbf{Y}} 
	-\mathbf{K}(\iSpf\bar{\mathbf{X}}, \iSpf\bar{\mathbf{Y}}) \,\mathbf{e}_{n}     \,.
\end{equation}
This is the \emph{Gauss formula}, which implies that $\llangle  \nabla^{\mathbf{G}}_{\bar{\mathbf{X}}}\mathbf{Y} , \mathbf{e}_n \rrangle_{\mathbf{G}}=-\mathbf{K}(\iSpf\bar{\mathbf{X}}, \iSpf\bar{\mathbf{Y}})$. Knowing that  $\llangle  \mathbf{Y} , \mathbf{e}_n \rrangle_{\mathbf{G}}=0$, one concludes that $\llangle  \nabla^{\mathbf{G}}_{\bar{\mathbf{X}}}\mathbf{Y} , \mathbf{e}_n \rrangle_{\mathbf{G}}=-\llangle  \nabla^{\mathbf{G}}_{\bar{\mathbf{X}}}\mathbf{e}_n, \mathbf{Y}  \rrangle_{\mathbf{G}}$. Thus, 
\begin{equation} \label{Weingarten-equation}
	\mathbf{K}(\mathbf{X}, \mathbf{Y})
	=\llangle \mathbf{Y} , \nabla^{\mathbf{G}}_{\mathbf{X}}\mathbf{e}_n 
	\rrangle_{\mathbf{G}} \,, \quad \text{or} \qquad 
	\mathbf{K}= \left(\nabla^{\mathbf{G}}\mathbf{e}_n \right)^\flat \,,
\end{equation}
which is the \emph{Weingarten equation}.

\begin{remark}
The expression $\mathbf{K} = \nabla^{\mathbf{G}} \mathbf{e}_n$ requires careful interpretation, as $\mathbf{e}_n$ is defined only along the surface $\SurE \subset \mathcal{B}$ and not in a neighborhood of it. Consequently, the full covariant derivative $\nabla^{\mathbf{G}} \mathbf{e}_n$ is not defined as a tensor field on $\mathcal{B}$. 
What is well-defined is the map
\begin{equation}
	\mathbf{X} \mapsto \nabla^{\mathbf{G}}_{\mathbf{X}} \mathbf{e}_n \in T_X\SurE \,,
	\qquad \mathbf{X} \in T_X\SurE \,,
\end{equation}
which defines the \textit{shape operator} $\mathbf{S} = \mathbf{G}^\sharp \mathbf{K}$ (not to be confused with the second Piola--Kirchhoff stress). The second fundamental form is then given by
\begin{equation}
	\mathbf{K}(\mathbf{X}, \mathbf{Y}) 
	= \llangle \nabla^{\mathbf{G}}_{\mathbf{X}} \mathbf{Y}, \mathbf{e}_n \rrangle_{\mathbf{G}} 
	= -\llangle \mathbf{Y}, \nabla^{\mathbf{G}}_{\mathbf{X}} \mathbf{e}_n \rrangle_{\mathbf{G}} \,, 
	\qquad \forall\, \mathbf{X}, \mathbf{Y} \in T_X\SurE \,.
\end{equation}
In this light, the shorthand notation $\mathbf{K} = \nabla^{\mathbf{G}} \mathbf{e}_n$, which is sometimes used in the literature, must be understood as purely symbolic: it refers only to the \textit{restriction of the covariant derivative to tangential directions} and to the identification
\begin{equation}
	\mathbf{K}(\mathbf{X}, \cdot) = \llangle \nabla^{\mathbf{G}}_{\mathbf{X}} \mathbf{e}_n, 
	\cdot \rrangle_{\mathbf{G}} \,, 
	\qquad \mathbf{X} \in T_X\SurE \,,
\end{equation}
where the right-hand side is interpreted as a one-form on $T_X\SurE$ and the pairing is with tangent vectors only. Without this restriction, the notation $\nabla^{\mathbf{G}} \mathbf{e}_n$ has no intrinsic meaning.
\end{remark}

Using \eqref{Gauss-formula}, one obtains
\begin{equation} 
\begin{aligned}
    \nabla^{\mathbf{G}}_{\mathbf{X}}\mathbf{Y}-\nabla^{\mathbf{G}}_{\mathbf{Y}}\mathbf{X} 
    & = \iSpf \big(\nabla^{\bar{\mathbf{G}}}_{\bar{\mathbf{X}}}\bar{\mathbf{Y}}
    -\nabla^{\bar{\mathbf{G}}}_{\bar{\mathbf{X}}}\bar{\mathbf{Y}} \big) 
    +\big(\mathbf{K}(\mathbf{Y}, \mathbf{X}) -\mathbf{K}(\mathbf{X}, \mathbf{Y})\big) \,\mathbf{e}_{n} 
    \,.
\end{aligned}
\end{equation}
Because $\nabla^{\mathbf{G}}$ and $\nabla^{\bar{\mathbf{G}}}$ are torsion-free one has $\nabla^{\mathbf{G}}_{\mathbf{X}}\mathbf{Y}-\nabla^{\mathbf{G}}_{\mathbf{Y}}\mathbf{X}=[\mathbf{X},\mathbf{Y}]=\iSpf [\bar{\mathbf{X}},\bar{\mathbf{Y}}]$ and $\nabla^{\bar{\mathbf{G}}}_{\bar{\mathbf{X}}}\bar{\mathbf{Y}}-\nabla^{\bar{\mathbf{G}}}_{\bar{\mathbf{X}}}\bar{\mathbf{Y}}=[\bar{\mathbf{X}},\bar{\mathbf{Y}}]$.
Therefore, $\mathbf{K}(\mathbf{Y}, \mathbf{X}) = \mathbf{K}(\mathbf{X}, \mathbf{Y})$, i.e., the second fundamental form $\mathbf{K}:T\SurE \times T\SurE \to\mathbb{R}$ is a symmetric bilinear form.

The second structural equations \eqref{Second-Structural-Equations} restricted to $T\SurE$ read 
\begin{equation} \label{Second-Structural-Equations-S}
\begin{dcases}
    \mathcal{R}^{\bar{\alpha}}{}_{\bar{\beta}} =d\omega^{\bar{\alpha}}{}_{\bar{\beta}}
    +\omega^{\bar{\alpha}}{}_{\gamma}\wedge\omega^{\gamma}{}_{\bar{\beta}}
    =d\omega^{\bar{\alpha}}{}_{\bar{\beta}}
    +\omega^{\bar{\alpha}}{}_{\bar{\gamma}}\wedge\omega^{\bar{\gamma}}{}_{\bar{\beta}}
    +\omega^{\bar{\alpha}}{}_{n}\wedge\omega^{n}{}_{\bar{\beta}}\,,
    \qquad \bar{\alpha}=1,\hdots n-1\,,\\
    \mathcal{R}^{n}{}_{\bar{\beta}} =d\omega^{n}{}_{\bar{\beta}}
    +\omega^{n}{}_{\gamma}\wedge\omega^{\gamma}{}_{\bar{\beta}}
    =d\omega^{n}{}_{\bar{\beta}}
    +\omega^{n}{}_{\bar{\gamma}}\wedge\omega^{\bar{\gamma}}{}_{\bar{\beta}} \,.
\end{dcases}
\end{equation}
Eq.~\eqref{Second-Structural-Equations-S}$_1$ is simplified as follows
\begin{equation} \label{Second-Structural-Equations-S1}
\begin{aligned}
    \mathcal{R}^{\bar{\alpha}}{}_{\bar{\beta}} 
    &=d\omega^{\bar{\alpha}}{}_{\bar{\beta}}
    +\omega^{\bar{\alpha}}{}_{\bar{\gamma}}\wedge\omega^{\bar{\gamma}}{}_{\bar{\beta}}
    +\omega^{\bar{\alpha}}{}_{n}\wedge\omega^{n}{}_{\bar{\beta}} \\
    &=d\omega^{\bar{\alpha}}{}_{\bar{\beta}}
    +\omega^{\bar{\alpha}}{}_{\bar{\gamma}}\wedge\omega^{\bar{\gamma}}{}_{\bar{\beta}}
    +\omega_{\bar{\gamma}}{}^n\,G^{\bar{\gamma}\bar{\alpha}}\wedge\omega^{n}{}_{\bar{\beta}} \\
    &= d\bar{\omega}^{\bar{\alpha}}{}_{\bar{\beta}}
    +\bar{\omega}^{\bar{\alpha}}{}_{\bar{\gamma}}\wedge\bar{\omega}^{\bar{\gamma}}{}_{\bar{\beta}}
    -\omega^n{}_{\bar{\gamma}}\,G^{\bar{\gamma}\bar{\alpha}}\wedge\omega^{n}{}_{\bar{\beta}} \\
    &= \bar{\mathcal{R}}^{\bar{\alpha}}{}_{\bar{\beta}}
    -G^{\bar{\gamma}\bar{\alpha}}\,K_{\bar{\gamma}\bar{\xi}} \,K_{\bar{\beta}\bar{\eta}}
    \, \vartheta^{\bar{\xi}} \wedge \vartheta^{\bar{\eta}}    \\
    &= \bar{\mathcal{R}}^{\bar{\alpha}}{}_{\bar{\beta}}
    -K^{\bar{\alpha}}{}_{\bar{\xi}} \,K_{\bar{\beta}\bar{\eta}}
    \, \vartheta^{\bar{\xi}} \wedge \vartheta^{\bar{\eta}} \,.
\end{aligned}
\end{equation}
From \eqref{Second-Structural-Equations-S1} and \eqref{Curvature-Forrmula} for $\mathbf{X},\mathbf{Y}\in T\SurE$ we have
\begin{equation} 
\begin{aligned}
	 \llangle \mathbf{R}(\mathbf{X},\mathbf{Y})\,\mathbf{e}_{\bar{\beta}},\mathbf{e}_{\bar{\alpha}}
	 \rrangle_{\mathbf{G}}
	 &=\mathcal{R}_{\bar{\alpha}\bar{\beta}}(\mathbf{X},\mathbf{Y}) \\
	 &=\bar{\mathcal{R}}_{\bar{\alpha}\bar{\beta}}(\mathbf{X},\mathbf{Y})
	 -K_{\bar{\alpha}\bar{\xi}} \,K_{\bar{\beta}\bar{\eta}}\, (\vartheta^{\bar{\xi}} 
	 \wedge \vartheta^{\bar{\eta}})
	 (\mathbf{X},\mathbf{Y}) \\
	 &=\bar{\mathcal{R}}_{\bar{\alpha}\bar{\beta}}(\mathbf{X},\mathbf{Y})
	 -K_{\bar{\alpha}\bar{\xi}}\,\vartheta^{\bar{\xi}}(\mathbf{X})
	 \,K_{\bar{\beta}\bar{\eta}}\,\vartheta^{\bar{\eta}}(\mathbf{Y}) 
	 +K_{\bar{\alpha}\bar{\xi}}\,\vartheta^{\bar{\xi}}(\mathbf{Y})
	 \,K_{\bar{\beta}\bar{\eta}} \,\vartheta^{\bar{\eta}}(\mathbf{X}) \\
	 &=\bar{\mathcal{R}}_{\bar{\alpha}\bar{\beta}}(\mathbf{X},\mathbf{Y})
	 -\mathbf{K}(\bar{\mathbf{e}}_{\bar{\alpha}},\mathbf{X})
	 \,\mathbf{K}(\bar{\mathbf{e}}_{\bar{\beta}},\mathbf{Y}) 
	 +\mathbf{K}(\bar{\mathbf{e}}_{\bar{\alpha}},\mathbf{Y})
	 \,\mathbf{K}(\bar{\mathbf{e}}_{\bar{\beta}},\mathbf{X})  \\
	 &= \llangle \bar{\mathbf{R}}(\mathbf{X},\mathbf{Y})\mathbf{e}_{\bar{\beta}},
	 \mathbf{e}_{\bar{\alpha}}\rrangle_{\mathbf{G}}
	 -\mathbf{K}(\bar{\mathbf{e}}_{\bar{\alpha}},\mathbf{X})
	 \,\mathbf{K}(\bar{\mathbf{e}}_{\bar{\beta}},\mathbf{Y}) 
	 +\mathbf{K}(\bar{\mathbf{e}}_{\bar{\alpha}},\mathbf{Y})
	 \,\mathbf{K}(\bar{\mathbf{e}}_{\bar{\beta}},\mathbf{X})  	    \,.
\end{aligned}
\end{equation}
Therefore
\begin{equation} \label{Gauss-Theorema-Egregium}
\begin{aligned}
	 \llangle \mathbf{R}(\mathbf{X},\mathbf{Y})\mathbf{Z},\mathbf{W}\rrangle_{\mathbf{G}}
	 &=\llangle \bar{\mathbf{R}}(\mathbf{X},\mathbf{Y})\mathbf{Z},\mathbf{W}\rrangle_{\mathbf{G}}
	 -\mathbf{K}(\mathbf{W},\mathbf{X}) \,\mathbf{K}(\mathbf{Z},\mathbf{Y}) 
	 +\mathbf{K}(\mathbf{W},\mathbf{Y}) \,\mathbf{K}(\mathbf{Z},\mathbf{X})  \\
	 &=\llangle \bar{\mathbf{R}}(\mathbf{X},\mathbf{Y})\mathbf{Z},\mathbf{W}\rrangle_{\mathbf{G}}
	 -\left[\mathbf{K}(\mathbf{X},\mathbf{W}) \,\mathbf{K}(\mathbf{Y},\mathbf{Z}) 
	 -\mathbf{K}(\mathbf{Y},\mathbf{W}) \,\mathbf{K}(\mathbf{X},\mathbf{Z})\right] \\
	 &=\llangle \bar{\mathbf{R}}(\mathbf{X},\mathbf{Y})\mathbf{Z},\mathbf{W}\rrangle_{\mathbf{G}}
	 -\begin{vmatrix}
	 \mathbf{K}(\mathbf{X},\mathbf{Z}) & \mathbf{K}(\mathbf{X},\mathbf{W}) \\
	 \mathbf{K}(\mathbf{Y},\mathbf{Z}) & \mathbf{K}(\mathbf{Y},\mathbf{W}) 
	 \end{vmatrix}
	 \,,\qquad \forall \mathbf{X},\mathbf{Y},\mathbf{Z},\mathbf{W} \in T\SurE	    \,.
\end{aligned}
\end{equation}
This is \emph{Gauss's Theorema Egregium}  (the Remarkable Theorem).

Eq.~\eqref{Second-Structural-Equations-S}$_2$ is simplified as follows
\begin{equation} \label{Second-Structural-Equations-S2}
\begin{aligned}
    \mathcal{R}^{n}{}_{\bar{\beta}} &=d\big(-K_{\bar{\beta}\bar{\xi}}\,\bar{\vartheta}^{\bar{\xi}}\big)
    -K_{\bar{\gamma}\bar{\eta}}\,\bar{\vartheta}^{\bar{\eta}} \wedge\bar{\omega}^{\bar{\gamma}}{}_{\bar{\beta}} \\
    &=-d K_{\bar{\beta}\bar{\xi}}\wedge \bar{\vartheta}^{\bar{\xi}}-K_{\bar{\beta}\bar{\xi}}\,d\bar{\vartheta}^{\bar{\xi}}
    -K_{\bar{\gamma}\bar{\eta}}\,\bar{\vartheta}^{\bar{\eta}} \wedge\bar{\omega}^{\bar{\gamma}}{}_{\bar{\beta}} \\
    &=-d K_{\bar{\beta}\bar{\xi}}\wedge \bar{\vartheta}^{\bar{\xi}}-K_{\bar{\beta}\bar{\xi}}
    \,\big(-\omega^{\bar{\xi}}{}_{\bar{\eta}}\wedge\bar{\vartheta}^{\bar{\eta}}\big)
    -K_{\bar{\gamma}\bar{\xi}}\,\bar{\vartheta}^{\bar{\xi}} \wedge\bar{\omega}^{\bar{\gamma}}{}_{\bar{\beta}} \\
    &=-d K_{\bar{\beta}\bar{\xi}}\wedge \bar{\vartheta}^{\bar{\xi}}
    +K_{\bar{\beta}\bar{\xi}}\,\omega^{\bar{\xi}}{}_{\bar{\eta}}\wedge\bar{\vartheta}^{\bar{\eta}}
    +K_{\bar{\gamma}\bar{\xi}}\,\bar{\omega}^{\bar{\gamma}}{}_{\bar{\beta}} \wedge \bar{\vartheta}^{\bar{\xi}}\\
    &=-d K_{\bar{\beta}\bar{\xi}}\wedge \bar{\vartheta}^{\bar{\xi}}
    +K_{\bar{\beta}\bar{\eta}}\,\omega^{\bar{\eta}}{}_{\bar{\xi}}\wedge\bar{\vartheta}^{\bar{\xi}}
    +K_{\bar{\gamma}\bar{\xi}}\,\bar{\omega}^{\bar{\gamma}}{}_{\bar{\beta}} \wedge \bar{\vartheta}^{\bar{\xi}}\\
    & =-\left(d K_{\bar{\beta}\bar{\xi}}
    -K_{\bar{\beta}\bar{\eta}}\,\omega^{\bar{\eta}}{}_{\bar{\xi}}
    -K_{\bar{\gamma}\bar{\xi}}\,\bar{\omega}^{\bar{\gamma}}{}_{\bar{\beta}}\right) \wedge \bar{\vartheta}^{\bar{\xi}} 
    \,.
\end{aligned}
\end{equation}
Using \eqref{Curvature-Forrmula} and \eqref{Second-Structural-Equations-S2} one can write (note that $\mathcal{R}_{n\beta}=\mathcal{R}^n{}_{\beta}$)
\begin{equation} 
\begin{aligned}
	 \llangle \mathbf{R}(\mathbf{X},\mathbf{Y})\mathbf{e}_{\bar{\beta}},\mathbf{e}_{n}\rrangle_{\mathbf{G}}
	 &= \mathcal{R}_{n\bar{\beta}} (\mathbf{X},\mathbf{Y})  \\
	 &= -\left[\left(d K_{\bar{\beta}\bar{\xi}}
	 -K_{\bar{\beta}\bar{\eta}}\,\omega^{\bar{\eta}}{}_{\bar{\xi}}
	 -K_{\bar{\gamma}\bar{\xi}}\,\bar{\omega}^{\bar{\gamma}}{}_{\bar{\beta}}\right) 
	 \wedge \bar{\vartheta}^{\bar{\xi}}\right] (\mathbf{X},\mathbf{Y}) \\
	 &= -\langle d K_{\bar{\beta}\bar{\xi}}
	 -K_{\bar{\beta}\bar{\eta}}\,\omega^{\bar{\eta}}{}_{\bar{\xi}}
	 -K_{\bar{\gamma}\bar{\xi}}\,\bar{\omega}^{\bar{\gamma}}{}_{\bar{\beta}},\mathbf{X} \rangle
	 \,\bar{\vartheta}^{\bar{\xi}}(\mathbf{Y})
	 +\langle d K_{\bar{\beta}\bar{\xi}}
	 -K_{\bar{\beta}\bar{\eta}}\,\omega^{\bar{\eta}}{}_{\bar{\xi}}
	 -K_{\bar{\gamma}\bar{\xi}}\,\bar{\omega}^{\bar{\gamma}}{}_{\bar{\beta}},\mathbf{Y} \rangle
	 \,\bar{\vartheta}^{\bar{\xi}}(\mathbf{X})\\
	 & = \langle d K_{\bar{\beta}\bar{\xi}}
	 -K_{\bar{\beta}\bar{\eta}}\,\omega^{\bar{\eta}}{}_{\bar{\xi}}
	 -K_{\bar{\gamma}\bar{\xi}}\,\bar{\omega}^{\bar{\gamma}}{}_{\bar{\beta}},\mathbf{Y} \rangle 
	 \,X^{\bar{\xi}}
	 -\langle d K_{\bar{\beta}\bar{\xi}}
	 -K_{\bar{\beta}\bar{\eta}}\,\omega^{\bar{\eta}}{}_{\bar{\xi}}
	 -K_{\bar{\gamma}\bar{\xi}}\,\bar{\omega}^{\bar{\gamma}}{}_{\bar{\beta}},\mathbf{X} \rangle 
	 \,Y^{\bar{\xi}}\\
	 & = \left(\nabla^{\bar{\mathbf{G}}}_{\mathbf{Y}}\mathbf{K}\right)(\mathbf{X},\bar{\mathbf{e}}
	 _{\bar{\beta}})
	 - \left(\nabla^{\bar{\mathbf{G}}}_{\mathbf{X}}\mathbf{K}\right)(\mathbf{Y},\bar{\mathbf{e}}
	 _{\bar{\beta}})
	  \,.
\end{aligned}
\end{equation}
Therefore
\begin{equation}
	 \llangle \mathbf{R}(\mathbf{X},\mathbf{Y})\mathbf{Z},\mathbf{e}_{n}\rrangle_{\mathbf{G}}
	 = (\nabla^{\bar{\mathbf{G}}}_{\mathbf{Y}}\mathbf{K})(\mathbf{X},\mathbf{Z})
	 - (\nabla^{\bar{\mathbf{G}}}_{\mathbf{X}}\mathbf{K})(\mathbf{Y},\mathbf{Z})
	  \,.
\end{equation}
This is the \emph{Codazzi-Mainardi equation}.

The main theorems of surface theory are summarized in Table \ref{Table-Summary}.
\begin{table}[h]
\centering
\vskip 0.2in
\renewcommand{\arraystretch}{1.6} 
\setlength{\tabcolsep}{10pt} 
\begin{tabular}{@{} l l @{}}
    \toprule
    \textbf{Gauss Formula} &
    $\nabla^{\mathbf{G}}_{\iSpf\bar{\mathbf{X}}} (\iSpf\bar{\mathbf{Y}})
    = \iSpf \nabla^{\bar{\mathbf{G}}}_{\bar{\mathbf{X}}}\bar{\mathbf{Y}} 
    - \mathbf{K}(\iSpf\bar{\mathbf{X}}, \iSpf\bar{\mathbf{Y}}) \,\mathbf{e}_{n}$ \\
    
    \textbf{Weingarten Equation} & 
    $\mathbf{K}(\mathbf{X}, \mathbf{Y})
    =\llangle \mathbf{Y} , \nabla^{\mathbf{G}}_{\mathbf{X}}\mathbf{e}_n 
    \rrangle_{\mathbf{G}}, \quad \text{or} \quad 
    \mathbf{K}=\nabla^{\mathbf{G}}\mathbf{e}_n$ \\
    
    \textbf{Gauss's Theorema Egregium} & 
    $\llangle \mathbf{R}(\mathbf{X},\mathbf{Y})\mathbf{Z},\mathbf{W}\rrangle_{\mathbf{G}}
    =\llangle \bar{\mathbf{R}}(\mathbf{X},\mathbf{Y})\mathbf{Z},\mathbf{W}\rrangle_{\mathbf{G}}
    - \begin{vmatrix}
    \mathbf{K}(\mathbf{X},\mathbf{Z}) & \mathbf{K}(\mathbf{X},\mathbf{W}) \\
    \mathbf{K}(\mathbf{Y},\mathbf{Z}) & \mathbf{K}(\mathbf{Y},\mathbf{W}) 
    \end{vmatrix}$ \\

    \textbf{Codazzi-Mainardi Equation} & 
    $\llangle \mathbf{R}(\mathbf{X},\mathbf{Y})\mathbf{Z},\mathbf{e}_{n}\rrangle_{\mathbf{G}}
    = (\nabla^{\bar{\mathbf{G}}}_{\mathbf{Y}}\mathbf{K})(\mathbf{X},\mathbf{Z})
    - (\nabla^{\bar{\mathbf{G}}}_{\mathbf{X}}\mathbf{K})(\mathbf{Y},\mathbf{Z})$ \\
    \bottomrule
\end{tabular}
\caption{Summary of the main equations in surface theory. Here, $\bar{\mathbf{X}},\bar{\mathbf{Y}}\in T\Sur$ and $\mathbf{X},\mathbf{Y},\mathbf{Z},\mathbf{W}\in T\SurE$.}
\label{Table-Summary}
\end{table}

\subsubsection{Coordinate representation of surface equations}

Consider a local coordinate chart $\{X^1,X^2,X^3\}$ for $\mathcal{B}$ such that $\{X^1,X^2\}$ is a local chart for $\SurE$ and at any point of the hypersurface $\mathbf{N}$ is in the direction $\partial/\partial X^3$. In this foliation coordinate chart, the metric of $\mathcal{B}$ has the representation
\begin{equation}
	\mathbf{G}(X)
	=\begin{bmatrix}
	G_{11}(X) & G_{12}(X)  & 0 \\
	G_{12}(X) & G_{22}(X)  & 0  \\
	0 & 0  & 1  
\end{bmatrix}\,,\qquad \forall ~ X\in\Sur\,.
\end{equation}
The induced metric on $\Sur$ (the first fundamental form) is given by
\begin{equation}
	\bar{\mathbf{G}}(X)=\left(\iota_{\Sur}^*\mathbf{G}\right)(X)
	=\begin{bmatrix}
	G_{11}(X) & G_{12}(X) \\
	G_{12}(X) & G_{22}(X)  
\end{bmatrix}\,, \qquad \forall ~ X\in\Sur\,.
\end{equation}
The Christoffel symbols of the induced connection $\bar{\nabla}$ read
\begin{equation}\label{key-me1}
	\bar{\Gamma}^A{}_{BC}=\frac{1}{2}\bar{G}^{AK}\left(\bar{G}_{KB,C}
	+\bar{G}_{KC,B}-\bar{G}_{BC,K}\right)\,,\qquad A,B,C,K=1,2.
\end{equation}
The coordinate components of the second fundamental form are calculated as folloows.
Recall that $\mathbf{K}=(\nabla^{\mathbf{G}}\mathbf{e}_n)^\flat$. Thus
\begin{equation}
	K_{AB} = G_{AC} \,(\mathbf{e}_n)^C{}_{|B} 
	\,.
\end{equation}
Note that $(\mathbf{e}_n)^C{}_{|B}  = (\mathbf{e}_n)^C{}_{,B} +\Gamma^C{}_{DB} \,(\mathbf{e}_n)^D$. Knowing that $(\mathbf{e}_n)^D=\delta_3^D$, we obtain $(\mathbf{e}_n)^C{}_{|B}  =\Gamma^C{}_{3B}$.
From $\Gamma^{C}{}_{3B}= \frac{1}{2}\,G^{CD}\left(G_{BD,3} + G_{3D,B} - G_{3B,D}\right)$, and knowing that $G_{3D}$ is constant (either $0$ or $1$), $(\mathbf{e}_n)^C{}_{|B} =\frac{1}{2}\,G^{CD}\,G_{BD,3}$. 
Thus, $G_{AC} \,(\mathbf{e}_n)^C{}_{|B}=\frac{1}{2}\,G_{AC} \,G^{CD}\,G_{BD,3} =\frac{1}{2}\,G_{BA,3} =\frac{1}{2}\,G_{AB,3}$.
Therefore
\begin{equation} \label{Second-Fundamental-Form-Coordinates}
	K_{AB}(X)= \frac{1}{2}\frac{\partial G_{AB}}{\partial X^3}\bigg|_{\SurE}(X), \qquad A,B=1,2\,,
	\quad \forall X\in \SurE\,. 
\end{equation}

The fundamental theorem of surface theory tells us that the geometry of the hypersurface $\Sur$ is fully described by its first and second fundamental forms $\bar{\mathbf{G}}$ and $\bar{\mathbf{K}}$. 
The Gauss equation in the local coordinate chart $\{X^1,X^2,X^3\}$ reads
\begin{equation}
	R_{1212}=\bar{R}_{1212}+K^2_{12}-K_{11}K_{22}\,.
\end{equation}
The Codazzi-Mainardi equations given in the local coordinate chart $\{X^1,X^2,X^3\}$ are written as
\begin{equation}
	R_{1213}=K_{11|2}-K_{12|1}\,,\qquad
	R_{2123}=K_{22|1}-K_{12|2}\,,
\end{equation}
where the covariant derivatives correspond to the Levi-Civita connection $\bar{\nabla}$ of $(\Sur,\bar{\mathbf{G}})$ with Christoffel symbols $\bar{\Gamma}^C{}_{AB}$ ($A,B,C=1,2$). Note that in components: $K_{AB|C}=K_{AB,C}-\bar{\Gamma}^K{}_{AC}K_{KB}-\bar{\Gamma}^K{}_{BC}K_{AK}$.

\subsubsection{Stokes' theorem, divergence theorem, and Nanson's formula}

Stokes' theorem is metric independent and states that for an $(n-1)$-form $\boldsymbol\omega$ on an $n$-manifold $\mathcal{B}$ one has
\begin{equation} \label{Stokes-Theorem}
	\int_{\mathcal B} \mathrm d \boldsymbol\omega 
	= \int_{\partial\mathcal B} \iota^* \boldsymbol\omega
	\,,
\end{equation}
where $\iota:\partial\mathcal{B} \hookrightarrow \mathcal{B}$ is the inclusion map.
Divergence theorem is the metric-dependent variant of Stokes' theorem on a Riemannian manifold.
Let $\mathbf{N}$ be the $\mathbf{G}$-unit normal of the boundary $\partial\mathcal{B}$.
Then, $\star\mathbf{N}^{\flat}$ is an $(n-1)$-form, and the induced Riemannian volume form of $\partial\mathcal{B}$ is given as
\begin{equation} 
	\boldsymbol{\mu}_{\bar{\mathbf{G}}}=
	\iota^* (\star\,\mathbf{N}^{\flat})=
	\iota^*\!\left(i_{\mathbf{N}} \boldsymbol{\mu}_{\mathbf{G}}\right) \,.
\end{equation}

The inclusion map satisfies the following identity
\begin{equation} \label{N-area-element}
	\iota^*\!\left( i_{\mathbf{W}} \boldsymbol{\mu}_{\mathbf{G}} \right)=
	\langle \mathbf{N}^{\flat} , \mathbf{W}\rangle \, \boldsymbol{\mu}_{\bar{\mathbf{G}}}
	\,,\qquad \forall \mathbf{W}\in T\mathcal{B}\,,
\end{equation}
where $i_{\mathbf{W}} \boldsymbol{\mu}_{\mathbf{G}}$ is the interior product of the $n$-form $\boldsymbol{\mu}_{\mathbf{G}}$ and the vector field $\mathbf{W}$, and is an $(n-1)$-form. A coordinate proof of this result is given in \citep{MaHu1983}. Here we give a coordinate-free proof.
Recall that $\mathbf{W}=\mathbf{W}_{\parallel}+\langle \mathbf{N}^{\flat} , \mathbf{W}\rangle\,\mathbf{N}$. Thus
\begin{equation} \label{N-area-element-bulk}
	i_{\mathbf{W}} \boldsymbol{\mu}_{\mathbf{G}} = 
	i_{\mathbf{W}_{\parallel}} \boldsymbol{\mu}_{\mathbf{G}}
	+ \langle \mathbf{N}^{\flat} , \mathbf{W}\rangle\,i_{\mathbf{N}} \boldsymbol{\mu}_{\mathbf{G}} \,.
\end{equation}
By definition of interior product
\begin{equation} 
	\iota^*(i_{\mathbf{W}_{\parallel}} \boldsymbol{\mu}_{\mathbf{G}})
	(\mathbf{V}_1,\hdots,\mathbf{V}_{n-1})
	=\boldsymbol{\mu}_{\mathbf{G}}(\mathbf{W}_{\parallel},\iota_*\mathbf{V}_1,\hdots,
	\iota_*\mathbf{V}_{n-1})\,,
	\qquad \forall~ \mathbf{V}_1,\hdots,\mathbf{V}_{n-1} \in T\SurE\,.
\end{equation}
However, notice that the set $\big\{ \mathbf{W}_{\parallel},\iota_*\mathbf{V}_1,\hdots,\iota_*\mathbf{V}_{n-1} \big\}$ is linearly dependent, and hence, the right-hand side is identically zero. Pulling back \eqref{N-area-element-bulk} to $\partial\mathcal{B}$ and using $\iota^*(i_{\mathbf{W}_{\parallel}} \boldsymbol{\mu}_{\mathbf{G}})=\mathbf{0}$, proves the identity \eqref{N-area-element}.

Let $\boldsymbol\omega=\star\mathbf{V}^{\sharp}$. From \eqref{d-div}, we have $d \boldsymbol\omega=\left(\operatorname{div}_{\mathbf{G}} \!\mathbf{V}\right) \mvf_{\mathbf{G}}$.
Also, using \eqref{vector-volumeform} and \eqref{N-area-element} one writes $\iota^*\boldsymbol\omega=\iota^*(\star\mathbf{V}^{\sharp})=\iota^*\left(i_{\mathbf{V}}\mvf_{\mathbf{G}}\right)=\langle \mathbf{N}^{\flat} , \mathbf{V}\rangle \, \boldsymbol{\mu}_{\bar{\mathbf{G}}}$. Therefore, the metric-dependent variant of \eqref{Stokes-Theorem}--- the Divergence Theorem---reads
\begin{equation} \label{Stokes-Theorem-Metric}
	\int_{\mathcal B} \left(\operatorname{div}_{\mathbf{G}} \!\mathbf{V}\right) \mvf_{\mathbf{G}}
	= \int_{\partial\mathcal B} 
	\langle \mathbf{N}^{\flat} , \mathbf{V}\rangle \, \boldsymbol{\mu}_{\bar{\mathbf{G}}}
	\,.
\end{equation}

\begin{remark}
If $\SurE\subset\mathcal{B}$ is a hypersurface, applying Stokes' theorem to an $(n-2)$-form $\boldsymbol\alpha$ on $\SurE$, yields
\begin{equation} 
	\int_{\SurE} \mathrm d \boldsymbol\alpha
	= \int_{\partial\SurE} \iota^* \boldsymbol\alpha
	\,,
\end{equation}
where $\iota:\partial\SurE \hookrightarrow \SurE$ is the inclusion map. 
Repeating the same reasoning as in the proof of the bulk divergence theorem, but now with respect to the induced metric on $\SurE$, one immediately obtains the metric-dependent variant
\begin{equation} \label{Surface-Stokes-Theorem-Metric}
	\int_{\SurE} \left(\operatorname{div}_{\parallel} \!\mathbf{W}\right) \mvf_{\parallel}
	= \int_{\partial\SurE} 
	\langle \boldsymbol{\nu}^{\flat} , \mathbf{W}\rangle \,  \bar{\mvf}_{\parallel}	\,,
\end{equation}
where $\operatorname{div}_{\parallel}$ is the surface divergence with respect to the first fundamental form of $\SurE$, $\mvf_{\parallel}$ is the induced surface volume form, $\boldsymbol{\nu}$ is the unit normal to $\partial\SurE$ within $\SurE$, and $\bar{\mvf}_{\parallel}$ is the induced volume form on $\partial\SurE$. 
Thus, \eqref{Surface-Stokes-Theorem-Metric} is a direct corollary of the bulk divergence theorem and will be referred to as the \textit{Surface Divergence Theorem}.
\end{remark}

Let us denote the area elements on $\partial\mathcal{B}$ and $\partial\mathcal{C}$ by $dA$ and $da$, respectively. They are related as $da=\bar{J}\,dA$, where $\bar{J}$ is the surface Jacobian. 
Given $\mathbf{w}\in\mathcal{X}(\mathcal{S})$, its Piola transform is defined as $\mathbf{W}=J\varphi^*\mathbf{w}$ \citep{MaHu1983}. It can be shown that
\begin{equation} 
	\varphi^*\left(i_{\mathbf{w}}\boldsymbol{\mu}_{\mathbf{g}} \right)
	= i_{\mathbf{W}}\boldsymbol{\mu}_{\mathbf{G}}
	\,.
\end{equation}
From \eqref{N-area-element}, we have 
\begin{equation} 
\begin{aligned}
	i_{\mathbf{W}} \boldsymbol{\mu}_{\mathbf{G}}  &=
	\langle \mathbf{N}^{\flat} , \mathbf{W}\rangle \, \iota_* \boldsymbol{\mu}_{\bar{\mathbf{G}}}\,, 
	&&& \forall \mathbf{W}\in T\mathcal{B}\,, \\
	i_{\mathbf{w}} \boldsymbol{\mu}_{\mathbf{g}}  &=
	\langle \mathbf{n}^{\flat} , \mathbf{w}\rangle \, \iota_* \boldsymbol{\mu}_{\bar{\mathbf{g}}}\,, 
	&&& \forall \mathbf{w}\in T\mathcal{C}\,.
\end{aligned}	
\end{equation}
Let us substitute $\mathbf{W}=J\varphi^*\mathbf{w}$ in the first identity and note that $\varphi^*\!\left( \iota_* \boldsymbol{\mu}_{\bar{\mathbf{g}}} \right)=\bar{J}\,\iota_* \boldsymbol{\mu}_{\bar{\mathbf{G}}}$. Thus
\begin{equation} 
	\big \langle \mathbf{N}^{\flat} , J\mathbf{F}^{-1}\mathbf{w} \big\rangle =
	\bar{J} \, \langle \mathbf{n}^{\flat} , \mathbf{w}\rangle\,,\qquad \forall \mathbf{w}\in T\mathcal{C}
	\,.
\end{equation}
Thus, $\bar{J}\,\mathbf{n}^{\flat}=J\,\mathbf{F}^{-\star}\,\mathbf{N}^{\flat}$. 
Therefore, noticing that $da=\bar{J}\,dA$, one obtains
\begin{equation} \label{Nanson}
	\mathbf{n}^{\flat}\,da=J\,\mathbf{F}^{-\star}\,\mathbf{N}^{\flat}\,dA\,,\quad
	\text{or~in~components}\qquad n_ada=J\,F^{-A}{}_a\,N_AdA
	\,.
\end{equation}
This is called Nanson's formula. It can be rewritten as $n^a da=J g^{ab} F^{-B}{}_b \,G_{AB} \,N^B dA$, or 
\begin{equation} \label{Nanson-Vector}
	\mathbf{n} \,da=J\mathbf{F}^{-\mathsf{T}} \mathbf{N} \,dA
	\,.
\end{equation}
Thus
\begin{equation} \label{Nanson-J}
	\bar{J}\,\mathbf{n}= J\mathbf{F}^{-\mathsf{T}} \mathbf{N} 
	\,.
\end{equation}
This implies that 
\begin{equation} \label{Surface-Jacobian}
	\bar{J}= J \,
	\sqrt{\llangle \mathbf{F}^{-\mathsf{T}} \mathbf{N},\mathbf{F}^{-\mathsf{T}} \mathbf{N}\rrangle_{\mathbf{g}}}		\,.
\end{equation}
From \eqref{Nanson-J} we have $\bar{J}\,\mathbf{n}= J\mathbf{F}^{-\mathsf{T}} \mathbf{N}$. 
Taking $\mathbf{g}$-norms and using $\llangle \mathbf{n},\mathbf{n}\rrangle_{\mathbf{g}}=1$ gives
\begin{equation}
	\bar{J} = J \,\big\| \mathbf{F}^{-\mathsf{T}} \mathbf{N} \big\|_{\mathbf{g}}
	= J  \sqrt{	\llangle \mathbf{F}^{-\mathsf{T}} \mathbf{N},\,\mathbf{F}^{-\mathsf{T}} \mathbf{N}\rrangle_{\mathbf{g}} }
	\,.
\end{equation}
Equivalently (when convenient), since $\mathbf{C}=\mathbf{F}^{\mathsf{T}}\mathbf{g}\,\mathbf{F}$ relative to $\mathbf{G}$, one may write
\begin{equation}
	\bar{J} = J \sqrt{\llangle \mathbf{N},\,\mathbf{C}^{-1}\mathbf{N}\rrangle_{\mathbf{G}} }	\,.
\end{equation}

\section{Nonlinear Elasticity with Material Surfaces} \label{Sec:MaterialSurfaces}

In this section we formulate a continuum theory of elastic bodies with material surfaces using the modern differential geometry of surfaces that was reviewed in the previous section. This will be a prelude to our theory of surface anelasticity that will be formulated in \S\ref{Sec:SurfaceAnelasticity}.

\begin{figure}[t!]
\centering
\includegraphics[width=0.65\textwidth]{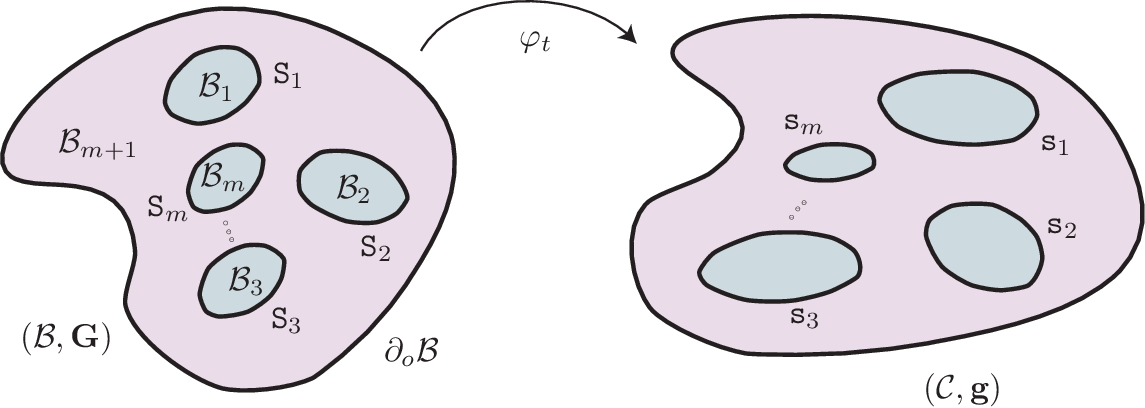}
\vspace*{0.10in}
\caption{\footnotesize  
Schematic representation of a body $\mathcal{B}$ containing $m$ material surfaces.  
The material surfaces form an abstract $2$-manifold $\mathsf{S}=\bigsqcup_{i=1}^{m}\mathsf{S}_i$, and their inclusion in the body manifold is denoted by $\SurE=\iota_{\Sur}(\mathsf{S})\subset \mathcal{B}$, where $\iota_{\Sur}:\mathsf{S}\hookrightarrow\mathcal{B}$ is the inclusion map.  
The bulk region is $\mathring{\mathcal{B}}=\mathcal{B}\setminus\SurE$, with connected components $\mathring{\mathcal{B}}=\bigsqcup_{i=1}^{m+1}\mathcal{B}_i$.  
For $i=1,\cdots,m$, the boundary of the $i$th inclusion is $\partial\mathcal{B}_i=\SurE_i$, whereas  
$\partial\mathcal{B}_{m+1}=\partial_o\mathcal{B}\,\sqcup\,\bigsqcup_{i=1}^{m}(-\SurE_i)$, where $-\SurE_i$ denotes the orientation-reversed surface.  
A material surface $\SurE_i$ moves under the deformation $\varphi_t$ to the deformed surface $\surE_i=\varphi_t(\SurE_i)$ in the ambient space.  }
\label{Reference-Current-Configurations}
\end{figure}

A body is denoted by $\mathcal{B}$. The set of material surfaces in the body is denoted by $\Sur$.
As submanifolds of the body manifold, the material surfaces are given by $\SurE=\iota_{\Sur}(\Sur) \subset \mathcal{B}$, where $\iota_{\Sur}: \Sur \hookrightarrow \mathcal{B}$ denotes the inclusion map.
We assume that $\mathsf{S}$ has $m$ connected components, i.e., $\mathsf{S}=\bigsqcup_{i=1}^{m}\mathsf{S}_i$, where $\bigsqcup$ denotes disjoint union of sets, and $\mathsf{S}_i$ is the $i$th connected component of $\mathsf{S}$. 
The body is partitioned into the bulk and material surfaces: 
$\mathcal{B}=\mathring{\mathcal{B}}\sqcup \SurE$, where $\mathring{\mathcal{B}}=\mathcal{B}\setminus \SurE$.
The bulk body $\mathring{\mathcal{B}}$ has $m+1$ connected components
\begin{equation} 
	\mathring{\mathcal{B}}= \bigsqcup_{i=1}^{m+1}  \mathcal{B}_i	\,,
\end{equation}
where, for $i=1,\cdots,m$, $\partial \mathcal{B}_i=\SurE_i$. Notice that
\begin{equation} 
	\partial \mathcal{B}_{m+1}=
	\partial_o\mathcal{B} \sqcup \bigsqcup_{i=1}^{m} (-\SurE_i)\,,
\end{equation}
where $-\SurE_i$ denotes the surface $\SurE_i$ with its orientation reversed.
We call $\partial_o\mathcal{B}$ the outer boundary of the body, see Fig.~\ref{Reference-Current-Configurations}.
We assume that for $i=1,\cdots,m$, either $\partial \SurE_i=\emptyset$ or $\partial \SurE_i \subset \partial_o\mathcal{B}$.

Deformation is a time-dependent map $\varphi_t:\mathcal{B}\to \mathcal{S}$, where $\mathcal{S}$ is the Euclidean ambient space, see Fig.~\ref{Reference-Current-Configurations}.
We denote the time-dependent current configuration by $\mathcal{C}_t=\varphi_t(\mathcal{B})\subset\mathcal{S}$. For a fixed value of $t$  we denote $\varphi=\varphi_t$, and $\mathcal{C}=\varphi(\mathcal{B})$. 
Deformation is sometimes written as $\varphi_t:(\mathcal{B},\mathring{\mathbf{G}})\to (\mathcal{C}_t,\mathbf{g})$, where $\mathbf{g}$ is a background (fixed) metric in the ambient space.
This notation explicitly highlights the metrics associated with the reference and current configurations.
However, note that $\mathbf{g}\neq \varphi_* \mathring{\mathbf{G}}$, in general.
We assume that $\mathcal{B}$ is an embedded submanifold of $\mathcal{S}$, and hence $\mathring{\mathbf{G}}=\mathbf{g}\big|_{\mathcal{B}}$.\footnote{We reserve $\mathbf{G}$ for the non-flat material metric in the presence of eigenstrains, which will be discussed in \S\ref{Sec:SurfaceAnelasticity}.} 
The inner products induced by the metrics $\mathring{\mathbf{G}}$ and $\mathbf{g}$ are denoted by
$\llangle , \rrangle_{\mathring{\mathbf{G}}}$ and $\llangle , \rrangle_{\mathbf{g}}$, respectively.
Under the deformation map, the surface $\SurE$ is mapped to $\surE=\sqcup_{i=1}^{m} \surE_i \subset\mathcal{C}$, where $\surE_i=\varphi (\SurE_i)$. 
The deformed surface $\surE$ is a submanifold of $\mathcal{C}$, which can be viewed as the inclusion of an abstract surface $\sur$ into the ambient space. The corresponding inclusion map is denoted as $\iota_{\sur}:\sur \hookrightarrow\mathcal{S}$. 

The first and second fundamental forms of $\SurE$ are denoted by $\mathring{\mathbf{G}}$ and $\mathring{\mathbf{K}}$, respectively. Those of $\Sur$ are denoted by $\mathring{\bar{\mathbf{G}}}$ and $\mathring{\bar{\mathbf{K}}}$, respectively. Similarly, the first and second fundamental forms of $\sur$ are denoted by $\mathbf{g}\big|_{\surE}$ and $\mathbf{k}$, respectively. We first review the standard bulk kinematics. This is followed by a formulation of the kinematics of material surfaces in terms of their initial geometries and the deformation mapping.

\paragraph{Volume forms.}
A volume form on the $3$-manifold $\mathcal{B}$ is any non-vanishing $3$-form \citep{Nakahara2003}. 
In an orthonormal coframe field $\{\vartheta^{\alpha}\}$ a volume form is written as $\boldsymbol{\mu}=h\,\vartheta^1\wedge\vartheta^2\wedge\vartheta^3$, for some function $h>0$. 
The Riemannian volume form corresponding to the metric $\mathring{\mathbf{G}}=\vartheta^1\otimes\vartheta^1+\vartheta^2\otimes\vartheta^2+\vartheta^3\otimes\vartheta^3$ is $\boldsymbol{\mu}_{\mathring{\mathbf{G}}}=\vartheta^1\wedge\vartheta^2\wedge\vartheta^3$.
If this is a foliation moving frame for $\SurE$, then $\mathring{\bar{\mathbf{G}}}=\vartheta^1\otimes\vartheta^1+\vartheta^2\otimes\vartheta^2$ and $\boldsymbol{\mu}_{\mathring{\bar{\mathbf{G}}}}=\vartheta^1\wedge\vartheta^2$.
In a foliation coordinate chart $\{X^A\}$ the Riemannian volume elements are written as
\begin{equation}
	\boldsymbol{\mu}_{\mathring{\mathbf{G}}}=\sqrt{\det\mathring{\mathbf{G}}}\,
	dX^1\wedge dX^2 \wedge dX^3\,,\qquad
	\boldsymbol{\mu}_{\mathring{\bar{\mathbf{G}}}}=\sqrt{\det\mathring{\bar{\mathbf{G}}}}\,
	dX^1\wedge dX^2
	\,.
\end{equation}
Similarly, the Riemannian volume elements in the deformed configuration have the following coordinate representations with respect to a foliation coordinate chart $\{x^a\}$:
\begin{equation}
	\boldsymbol{\mu}_{\mathbf{g}}=\sqrt{\det\mathbf{g}}~
	dx^1\wedge dx^2\wedge dx^3\,,\qquad
	\boldsymbol{\mu}_{\bar{\mathbf{g}}}=\sqrt{\det\bar{\mathbf{g}}}~
	dx^1\wedge dx^2
	\,.
\end{equation}

\subsection{Bulk kinematics}

For $\mathcal{B}$ and $\mathcal{C}$, we consider coordinate charts $\{X^A\}:\mathcal{B}\to\mathbb{R}^n$ and $\{x^a\}:\mathcal{C}\to\mathbb{R}^n$, respectively ($n=2$ or $3$). 
The material velocity $\mathbf{V}_t:\mathcal{B}\to T\mathcal{S}$ is defined as 
\begin{equation}
	\mathbf{V}_t(X)=\frac{\partial}{\partial t}\varphi_t(X)\,, 
	\qquad
	V^a(X,t)=\frac{\partial \varphi^a}{\partial t}(X,t) 
	\,.
\end{equation}
The spatial velocity is defined as $\mathbf{v}=\mathbf{V}_t\circ\varphi_t^{-1}$. 
The material acceleration is defined as 
\begin{equation} \label{Material-Acceleration}
	\mathbf{A}(X,t)=D^{\mathbf{g}}_{t}\mathbf{V}(X,t)
	\in T_{\varphi_t(X)}\mathcal{S}
	\,,
\end{equation}
where $D^{\mathbf{g}}_{t}$ denotes the covariant derivative along the curve $\varphi_t(X)$ in $\mathcal{S}$.\footnote{Let $\mathbf{U}(X,t)$ and $\mathbf{W}(X,t)$ be vector fields along the motion $\varphi_t(X)$, i.e., $\mathbf{U}(X,t),\mathbf{W}(X,t)\in T_{\varphi_t(X)}\mathcal{S}$ for each $t$.  
The covariant time derivative $D_t^{\mathbf g}$ is defined as the unique operator satisfying the Leibniz rule
\begin{equation}
	\frac{d}{dt}\llangle \mathbf{U},\mathbf{W}\rrangle_{\mathbf g}
	=\llangle D_t^{\mathbf g}\mathbf{U},\mathbf{W}\rrangle_{\mathbf g}
	+\llangle \mathbf{U},D_t^{\mathbf g}\mathbf{W}\rrangle_{\mathbf g}\,,
\end{equation}
together with compatibility with the Levi-Civita connection of $\mathbf g$.  
In coordinates, for $\mathbf{U}=U^a\,\partial/\partial x^a$,
\begin{equation}
	(D_t^{\mathbf g}\mathbf{U})^a
	=\frac{\partial U^a}{\partial t}
	+\gamma^a{}_{bc}\,U^b\,V^c\,,
\end{equation}
where $\mathbf{V}=\partial\varphi_t/\partial t$ and $\gamma^a{}_{bc}$ are the Christoffel symbols of $\mathbf g$.  
The covariant time derivative of a vector field can equivalently be obtained by using the local coordinate representation $\mathbf{U}(X,t)=U^a(X,t)\,\mathbf{e}_a(\varphi_t(X))$. Thus,
\begin{equation}
	D_t^{\mathbf g}\mathbf{U}
	=\frac{\partial U^a}{\partial t}\,\mathbf{e}_a +U^a\, \nabla_{\mathbf{e}_b} \mathbf{e}_a\,V^b
	=\frac{\partial U^a}{\partial t}\,\mathbf{e}_a +U^a\, \gamma^c{}_{ba} \,\mathbf{e}_c\,V^b
	=\left(\frac{\partial U^a}{\partial t} + \gamma^a{}_{bc} \,V^b\,U^c\right)\mathbf{e}_a
	\,,
\end{equation}
and recall that $\gamma^a{}_{bc} =\gamma^a{}_{cb} $.
} 
In components, 
\begin{equation}
	A^a=\frac{\partial V^a}{\partial t}+\gamma^a{}_{bc}\,V^b\,V^c
	\,,
\end{equation}
where $\gamma^a{}_{bc}$ are the Levi-Civita connection coefficients of $\mathbf{g}$.\footnote{In \citep{MaHu1983}, the material acceleration is written formally as $\mathbf{A}=\frac{\partial}{\partial t}\mathbf{V}$. 
This notation should be interpreted with care: although the material point $X$ is fixed, the vector $\mathbf{V}(X,t)$ takes values in different tangent spaces $T_{\varphi_t(X)}\mathcal{S}$ as $t$ varies. A precise definition therefore requires a covariant time derivative along the curve $\varphi_t(X)$.} 
The spatial acceleration is defined as $\mathbf{a}_t=\mathbf{A}_t\circ\varphi_t^{-1}=\dfrac{\partial \mathbf{v}}{\partial t}+\nabla_{\mathbf{v}}^{\mathbf{g}}\mathbf{v}$, and has components $a^a=\dfrac{\partial v^a}{\partial t}+\dfrac{\partial v^a}{\partial x^b}v^b+\gamma^a{}_{bc}v^bv^c$.

The derivative map of $\varphi$---the deformation gradient---at $X\in\mathcal{B}$, is denoted by $\mathbf{F}(X)$ and is a linear mapping that maps a vector $\mathbf{U}\in T_X\mathcal{B}$ to $\mathbf{F}(X)\mathbf{U}\in T_{\varphi(X)}\mathcal{C}$. The deformation gradient is a two-point tensor and has the following coordinate representation 
\begin{equation}
	\mathbf{F}(X)=\frac{\partial \varphi^a(X)}{\partial X^A}\, \frac{\partial}{\partial x^a}\otimes dX^A
	=F^a{}_A(X)\, \frac{\partial}{\partial x^a}\otimes dX^A\,.
\end{equation}
Its dual $\mathbf{F}^\star:T_{\varphi(X)}^*\mathcal{C}\to T_X^*\mathcal{B}$, where $T_{\varphi(X)}^*\mathcal{C}$ and $T_X^*\mathcal{B}$ denote the cotangent spaces of $T_{\varphi(X)}\mathcal{C}$ and $T_X\mathcal{B}$, respectively, is defined such that
\begin{equation}
	\langle\boldsymbol\alpha,\mathbf F \mathbf U\rangle
	=\langle\mathbf F^\star \boldsymbol\alpha,\mathbf U\rangle\,, \qquad\forall\, 
	\mathbf U \in T_X\mathcal B\,,~ \boldsymbol\alpha \in T^*_{\varphi(X)}\mathcal S\,,
\end{equation}
where $\langle , \rangle$ is the natural pairing of $1$-forms and vectors, e.g., $\langle\boldsymbol{\alpha},\mathbf{u}\rangle=\alpha_a\,u^a$.
$\mathbf{F}^\star$ has the following coordinate representation
\begin{equation}
	\mathbf{F}^{\star}\circ\varphi(X)=F^a{}_A(X) \,dX^A \otimes \frac{\partial}{\partial x^a}\,.
\end{equation}
The transpose of the deformation gradient, which is metric dependent, is defined as
\begin{equation}
	\mathbf{F}^{\textsf{T}}:T_{x}\mathcal{C}  
	\to T_{X}\mathcal{B},\qquad \llangle \mathbf{FU}\,,	\mathbf{u} \rrangle_{\mathbf{g}}    
	=\llangle\mathbf{U},\mathbf{F}^{\textsf{T}}\mathbf{u}\rrangle_{\mathring{\mathbf{G}}}\,, 	
	\qquad
	\forall \mathbf{U} \in T_{X}\mathcal{B}\,,~\mathbf{u} \in T_{x} \mathcal{C}\,,
\end{equation}
and has the following coordinate representation 
\begin{equation}
	\mathbf{F}^{\textsf{T}}\circ\varphi(X)=(F^{\textsf{T}}(X))^A{}_{a}\, \frac{\partial}{\partial X^A}
	\otimes dx^a=g_{ab}\circ\varphi(X)\,F^b{}_{B}(X)\,\mathring{G}^{AB}(X) \, 
	\frac{\partial}{\partial X^A}\otimes dx^a	\,.
\end{equation}

In addition to the deformation gradient, there are several other measures of strain that have direct applications in nonlinear elasticity and anelasticity \citep{MaHu1983,Ogden1984,Goriely2017,YavariSozio2023}. 
The dot product of two vectors $\mathbf{u}, \mathbf{w}\in T_x\mathcal{C}$ is calculated via the ambient space metric $\mathbf{g}$: 
\begin{equation} 
	\llangle \mathbf{u},\mathbf{w} \rrangle_{\mathbf{g}}
	=\llangle \mathbf{F}\mathbf{U},\mathbf{F}\mathbf{W} \rrangle_{\mathbf{g}}
	=\llangle \mathbf{U},\mathbf{W} \rrangle_{\mathbf{F}^*\mathbf{g}}
	\,,
\end{equation}
where $\mathbf{F}^*\mathbf{g}=\mathbf{F}^{\star}\mathbf{g}\mathbf{F}=\varphi^*\mathbf{g}=\mathbf{C}^{\flat}$ is the pulled-back metric or the right Cauchy-Green strain. 
In components, this is written as $\llangle \mathbf{u},\mathbf{w} \rrangle_{\mathbf{g}}=u^aw^bg_{ab}=(F^a{}_AF^b{}_B\,g_{ab})\,U^AW^B=C_{AB}\,U^AW^B$, and hence $C_{AB}=F^a{}_A\,g_{ab}\,F^b{}_B=(\mathbf{F}^*\mathbf{g})_{AB}$.
Note that $\mathbf{C}=\mathbf{F}^{\mathsf{T}}\mathbf{F}$ because $C^A{}_B=G^{AM}C_{MB}=(G^{AM}F^a{}_M\,g_{ab})F^b{}_B=\big(F^{\mathsf{T}}\big)^A{}_b\,F^b{}_B$.
The dot product of two vectors $\mathbf{U}, \mathbf{W}\in T_X\mathcal{B}$ is calculated via the material metric $\mathbf{G}$ as
\begin{equation} 
	\llangle \mathbf{U},\mathbf{W} \rrangle_{\mathbf{G}}
	=\llangle \mathbf{F}^{-1}\mathbf{u},\mathbf{F}^{-1}\mathbf{w} \rrangle_{\mathbf{G}}
	=\llangle \mathbf{u},\mathbf{w} \rrangle_{\mathbf{F}_*\mathbf{G}}
	\,,
\end{equation}
where $\mathbf{c}^{\flat}=\mathbf{F}_*\mathbf{G}=\mathbf{F}^{-\star}\mathbf{G}\mathbf{F}^{-1}$ is the push-forward of the material metric, and is the spatial analogue of the right Cauchy-Green strain. It has components $c_{ab}=F^{-A}{}_a\,G_{AB}\,F^{-B}{}_b$, where $F^{-A}{}_a$ are the components of $\mathbf{F}^{-1}$.

Considering $1$-forms (covectors) instead of vectors two other measures of strain are defined: the left Cauchy-Green strain $\mathbf{B}^{\sharp}=\varphi^*\mathbf{g}^{\sharp}$, and its spatial analogue $\mathbf{b}^{\sharp}=\varphi_*\mathbf{G}^{\sharp}=\mathbf{F}\mathbf{G}^{\sharp}\mathbf{F}^{\star}$. They have components $B^{AB}=F^{-A}{}_a\,F^{-B}{}_b\,g^{ab}$, and $b^{ab}=F^a{}_AF^b{}_B\,G^{AB}$, respectively.
The tensors $\mathbf{b}$ and $\mathbf{c}$ are defined as $\mathbf{b}=\mathbf{b}^{\sharp}\mathbf{g}$, and $\mathbf{c}=\mathbf{g}^{\sharp}\mathbf{c}^{\flat}$. 
Notice that $\mathbf{c}\mathbf{b}=\mathbf{g}^{\sharp}\mathbf{c}^{\flat}\mathbf{b}^{\sharp}\mathbf{g}=\mathbf{g}^{\sharp}\mathbf{F}^{-\star}\mathbf{G}\mathbf{F}^{-1}  
\mathbf{F}\mathbf{G}^{\sharp}\mathbf{F}^{\star}\mathbf{g}=\mathbf{g}^{\sharp}\mathbf{F}^{-\star}\mathbf{G}\mathbf{G}^{\sharp}\mathbf{F}^{\star}\mathbf{g}=\mathbf{g}^{\sharp}\mathbf{F}^{-\star}\mathbf{F}^{\star}\mathbf{g}=\mathbf{g}^{\sharp}\mathbf{g}=\operatorname{id}_{\mathcal{S}}$, i.e., $\mathbf{b} = \mathbf{c}^{-1}$. Similarly, $\mathbf{B} = \mathbf{C}^{-1}$.

The \textit{polar decomposition} of the deformation gradient can be expressed as
\begin{equation}\label{Polar-Decomposition}
    \mathbf{F}=\mathbf{R}\mathbf{U}=\mathbf{V}\mathbf{R} \,,
\end{equation}
where $\mathbf{U}$ and $\mathbf{V}$ (which should not be confused with the material velocity) represent the material and spatial stretch tensors, respectively, and $\mathbf{R}:T\mathcal{B}\to T\mathcal{C}$ is a $(\mathbf{G},\mathbf{g})$-orthogonal tensor field \citep{SimoMarsden1984}, i.e.,\footnote{This can be written as $\mathbf{G}^\sharp \mathbf{R}^\star (\mathbf{g}\circ\varphi)\, \mathbf{R}=\mathbf{R}^{\mathsf{T}}\mathbf{R}=\operatorname{id}_{T\mathcal{B}}$.}
\begin{equation} \label{G-g-Orthogonal}
	\mathbf{R}^\star (\mathbf{g}\circ\varphi)\, \mathbf{R}=\mathbf{G}    \,.
\end{equation}
In component form, this is written as $R^a{}_A\,(g_{ab}\circ\varphi)\,R^b{}_B=G_{AB}$.
The component-wise expression of the polar decomposition is given by
\begin{equation}
	F^a{}_A= R^a{}_B\, U^B{}_A =V^a{}_b\,R^b{}_A    \,.
\end{equation}
Eq.~\eqref{G-g-Orthogonal} implies that $(\det\mathbf{R})^2\det\mathbf{g}=\det\mathbf{G}$, and from \eqref{Polar-Decomposition}, it follows that $\det\mathbf{U}=\det\mathbf{V}$. Recall that the Jacobian of deformation is defined by the relation $dv=J\,dV$ and is given by
\begin{equation} \label{Jacobian-Anelasticity}
	J=\sqrt{\frac{\det\mathbf{g}}{\det\mathbf{G}}}\,\det\mathbf{F}=\det\mathbf{U}=\det\mathbf{V}  \,.
\end{equation}
The material stretch tensor $\mathbf{U}:T_X\mathcal{B}\to T\mathcal{B}$ and the spatial stretch tensor $\mathbf{V}:T_x\mathcal{C}\to T\mathcal{C}$ are related to the right and left Cauchy-Green deformation tensors through the relations
\begin{equation} \label{C-U-b_v}
\begin{aligned}
	\mathbf{C} &=\mathbf{F}^{\textsf{T}} \mathbf{F}
	=(\mathbf{R}\mathbf{U})^{\textsf{T}}\mathbf{R}\mathbf{U}
	=\mathbf{G}^\sharp(\mathbf{R}\mathbf{U})^\star \mathbf{g}\mathbf{R}\mathbf{U}
	=\mathbf{G}^\sharp \mathbf{U}^\star \mathbf{R}^\star \mathbf{g}\mathbf{R}\mathbf{U}
	=\mathbf{G}^\sharp \mathbf{U}^\star \mathbf{G} \mathbf{U}=\mathbf{U}^2 \,,\\
	\mathbf{b} &=\mathbf{F}\mathbf{F}^{\textsf{T}}
	=\mathbf{V}\mathbf{R} (\mathbf{V}\mathbf{R})^{\textsf{T}}
	=\mathbf{V}\mathbf{R}\mathbf{G}^\sharp(\mathbf{V}\mathbf{R})^\star\mathbf{g}
	=\mathbf{V}\mathbf{R}\mathbf{G}^\sharp \mathbf{R}^\star\mathbf{V}^\star \mathbf{g}
	=\mathbf{V} \mathbf{g}^\sharp \mathbf{V}^\star \mathbf{g}
	=\mathbf{V}^2	    \,.
\end{aligned}
\end{equation}
Alternatively, this can be expressed as
\begin{equation}
	\mathbf{C}^\flat=\mathbf{U}^\star \mathbf{G} \mathbf{U}\,,\qquad 
	\mathbf{b}^\sharp = \mathbf{V} \mathbf{g}^\sharp \mathbf{V}^\star
	\,,
\end{equation}
which, in component form, are given by $C_{AB} = U^M{}_A\,G_{MN}\,U^N{}_B$ and $b^{ab}=V^a{}_m\,g^{mn}\,V^b{}_n$.
The relations \eqref{C-U-b_v} are typically expressed as $\mathbf{U}=\sqrt{\mathbf{C}}$ and $\mathbf{V}=\sqrt{\mathbf{b}}$.

The right Cauchy-Green deformation tensor and the material stretch tensor admit the following spectral representations \citep{Ogden1984}:
\begin{equation} \label{C-Spectral}
	\mathbf{C}^\sharp= \lambda_1^2 \,\N \otimes\N	+\lambda_2^2 \,\NN \otimes\NN
	+\lambda_3^2 \,\NNN \otimes\NNN
	\,,\qquad
	\mathbf{U}^\sharp= \lambda_1 \,\N \otimes\N+\lambda_2 \,\NN \otimes\NN
	+\lambda_3 \,\NNN \otimes\NNN\,,
\end{equation}
where $\lambda_1, \lambda_2, \lambda_3$ and $\N, \NN, \NNN$ represent the principal stretches and their associated principal directions.
It is important to note that $\N\otimes\N+ \NN\otimes\NN+ \NNN\otimes\NNN=\mathbf{G}^\sharp$.
The representation \eqref{C-Spectral}$_2$ is equivalent to $\mathbf{U}= \lambda_1 \,\N \otimes\N^\flat+\lambda_2 \,\NN \otimes\NN^\flat+\lambda_3 \,\NNN \otimes\NNN^\flat$, which leads to $\mathbf{U}\Nj=\lambda_j\Nj$ (no summation).
Additionally, $\mathbf{F}\Nj=\mathbf{R}\mathbf{U}\Nj=\lambda_j \mathbf{R}\Nj=\mathbf{V}\mathbf{R}\Nj$.
This implies that if we denote the eigenbasis of $\mathbf{V}$ by $\nj$, for $j=1,2,3$, then we have $\nj=\mathbf{R}\Nj$, for $j=1,2,3$ \citep{Ogden1984}.
As a result, the Finger and spatial stretch tensors take the following spectral representations:
\begin{equation}
	\mathbf{b}^\sharp= 
	\lambda_1^2 \,\n \otimes\n+\lambda_2^2 \,\nn \otimes\nn+\lambda_3^2 \,\nnn \otimes\nnn\,,\qquad
	\mathbf{V}^\sharp= 
	\lambda_1 \,\n \otimes\n+\lambda_2 \,\nn \otimes\nn+\lambda_3 \,\nnn \otimes\nnn
	\,.
\end{equation}
Observe that
\begin{equation} 
	\sum_{j=1}^3 \nj\otimes \nj = \sum_{j=1}^3 \mathbf{R}\Nj\otimes \mathbf{R}\Nj
	= \mathbf{R} \Bigg(\sum_{j=1}^3 \Nj\otimes \Nj \Bigg) \mathbf{R}^\star
	= \mathbf{R} \mathbf{G}^\sharp \mathbf{R}^\star =\mathbf{g}^\sharp \,.
\end{equation}

\subsection{Surface kinematics}

Let us consider a motion $\varphi_t:\mathcal{B}\to\mathcal{S}$ (see Fig.~\ref{Reference-Current-Configurations}).

\begin{figure}[t!]
\centering
\includegraphics[width=0.6\textwidth]{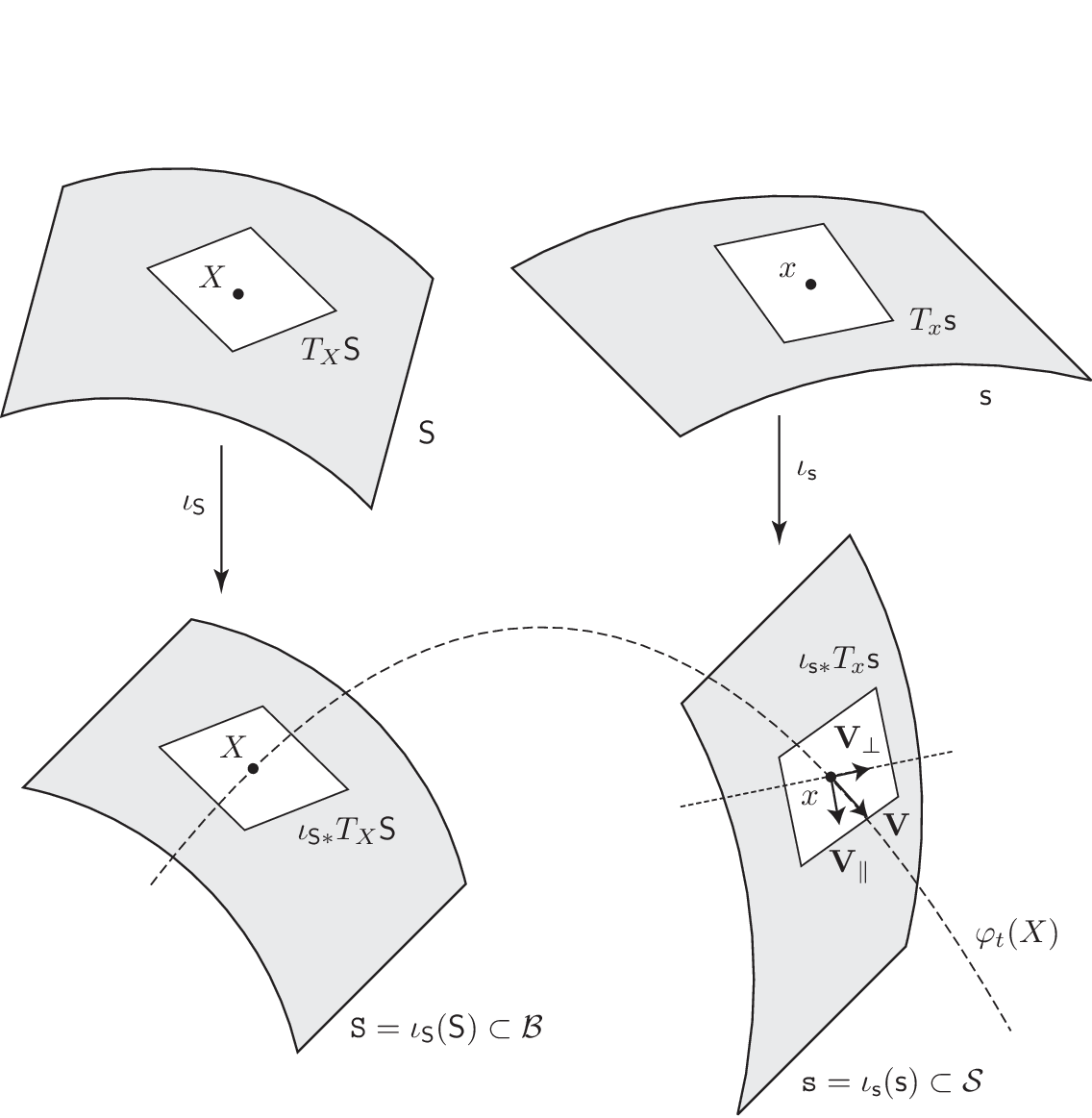}
\vspace*{0.10in}
\caption{Motion of a material surface $\Sur$, which is an abstract $2$-manifold. Its inclusion in the material manifold $\SurE=\iota_{\Sur}\Sur\subset\mathcal{B}$ is the undeformed material surface. For a fixed $X\in \SurE$, $\varphi(X,t)$ is a curve in the ambient space. Tangent to this curve at $x=\varphi(X,t)$ is the material velocity $\mathbf{V}(X,t)$, which has parallel $\mathbf{V}_\parallel(X,t)$ and normal $\mathbf{V}_\perp(X,t)$ components with respect to the deformed material surface $\surE$.} 
\label{Surface-Motion}
\end{figure}

\subsubsection{Surface velocity and acceleration}

For $X\in\Sur$, material velocity has the following decomposition (see Fig.~\ref{Surface-Motion})
\begin{equation} 
   \mathbf{V}(X)=\mathbf{V}_\parallel(X)+V_n(X)\,\mathbf{n}\circ\varphi(X) \,.
\end{equation}
Material acceleration is given in \eqref{Material-Acceleration}. 
Velocity and acceleration fields are continuous everywhere in the body. One simply has $V_n=\llangle \mathbf{V},\mathbf{n}\rrangle_{\mathbf{g}}$, and $\mathbf{V}_\parallel=\mathbf{V}-V_n\,\mathbf{n}$. Similarly, for acceleration: $A_n=\llangle \mathbf{A},\mathbf{n}\rrangle_{\mathbf{g}}$, and $\mathbf{A}_\parallel=\mathbf{A}-A_n\,\mathbf{n}$.

\subsubsection{Surface-restricted, tangent, and surface deformation gradients}

Let us choose a foliation coordinate chart $\{X^A\}=\{X^1,X^2,X^3\}$ for $\mathcal{B}$ such that at any point of $\SurE$, $\{X^1,X^2\}$ is a local coordinate chart for $\SurE$ and the unit normal vector field $\mathbf{N}$ is tangent to the coordinate curve $X^3$. 
Let us also choose a foliation coordinate chart $\{x^a\}=\{x^1,x^2,x^3\}$ for $\mathcal{C}$ such that at any point of $\surE=\iota_{\sur}(\sur)\subset \mathcal{C}$, $\{x^1,x^2\}$ is a local coordinate chart for $\surE$ and the unit normal vector field $\mathbf{n}$ is tangent to the $x^3$-coordinate curve. 

\begin{defi}
The \textit{surface restricted deformation gradient} is defined as $\mathbf{F}_\SurE(X):=\mathbf{F}\big |_{T_X\SurE}(X)=\mathbf{F}\big |_{\SurE}(X)\circ \pi_{\SurE}:T_X\SurE \to T_x\mathcal{C}$, where the material projection $\pi_{\SurE}:T\mathcal{B}\to T_X\SurE$ is defined as\footnote{Recall that $T_X\SurE=\iSpf T_X\Sur$.}
\begin{equation}
	\pi_{\SurE}=\operatorname{id}_{T\mathcal{B}} - \mathbf{N} \otimes \mathbf{N}^{\flat}\,.
\end{equation}
\end{defi}

Given a vector $\bar{\mathbf{W}}(X)\in T_X\Sur$, define $\mathbf{W}_\parallel(X)=\iota_{\Sur*}\bar{\mathbf{W}}(X)=\FS(X) \bar{\mathbf{W}}(X) \in T_X\SurE \subset T_X\mathcal{B}$. The surface restricted deformation gradient maps it to $\mathbf{w}(x)=\mathbf{F}_{\SurE}(X) \mathbf{W}_\parallel(X)\in T_x\mathcal{C}$. 
Note that $\mathbf{w}(x)$ does not necessarily lie in $T_x\surE \subset T_x\mathcal{C}$;\footnote{Recall that $T_x\surE=\ispf T_x\mathsf{s}$.} it has a normal component, in general. Let us denote the unit normal vector  to $\surE$ at $x=\varphi(X)$ by $\mathbf{n}(x)$. Thus
\begin{equation} 
   \mathbf{w}(x)=\mathbf{F}_{\SurE}(X) \,\mathbf{W}_\parallel(X)
   =\mathbf{w}_\parallel(x)+\mathbf{w}_\perp(x)
   =\mathbf{w}_\parallel(x)+w_n(x)\,\mathbf{n}(x)
   \,,
\end{equation}
where $w_n(x)=\llangle\mathbf{w}(x),\mathbf{n}(x)\rrangle_{\mathbf{g}}$.

\begin{defi}
The \textit{tangential deformation gradient} $\mathbf{F}_{\parallel}(X): T_X\SurE \to T_x\surE$ is defined such that $\mathbf{w}_\parallel\circ\varphi=\mathbf{F}_{\parallel} \mathbf{W}_\parallel$. Note that
\begin{equation} \label{F-Parallel}
\begin{aligned}
  \mathbf{F}_{\parallel}(X)\mathbf{W}_\parallel(X)
  &=\mathbf{F}_{\SurE}(X)\, \mathbf{W}_\parallel(X) -w_n(x)\,\mathbf{n}(x) \\
  & =\mathbf{F}_{\SurE}(X)\, \mathbf{W}_\parallel(X) 
   -\llangle \mathbf{F}_{\SurE}(X)\,\mathbf{W}_\parallel(X),
   \mathbf{n}(x)\rrangle_{\mathbf{g}}\,\mathbf{n}(x)
   \,,
\end{aligned}
\end{equation}
where $x=\varphi(X)$.
In coordinates, 
\begin{equation} 
\begin{aligned}
   (F_{\parallel})^a{}_{A}\,W_{\parallel}^{A}
   &=(F_\SurE)^a{}_{A}\, W_{\parallel}^{A}-(F_\SurE)^b{}_{A}\,W_{\parallel}^{A} \,n^c\,g_{cb}  \, n^a \\
   &=\left[(F_\SurE)^a{}_{A}-n^a\,n_b\,(F_\SurE)^b{}_{A}\right] W_{\parallel}^{A} \\
   &=\left(\delta^a_b-n^a\,n_b\right)\,(F_\SurE)^b{}_{A}\,W_{\parallel}^{A} \\
   &=\left(\delta^a_b-n^a\,n_b\right) F^b{}_{B} \left(\delta^B_A-N^B N_A\right) W_{\parallel}^{A} 
   \,.
\end{aligned}
\end{equation}
Hence, $(F_{\parallel})^a{}_{A}=\left(\delta^a_b-n^a\,n_b\right) (F_\SurE)^b{}_{A}=\left(\delta^a_b-n^a\,n_b\right) F^b{}_{B} \left(\delta^B_A-N^B N_A\right)$, or in a coordinate-independent form
\begin{equation} 
 	\mathbf{F}_{\parallel}(X)= \pi_{\surE} \circ \mathbf{F}_{\SurE}(X)
	= \pi_{\surE} \circ \mathbf{F}(X) \circ \pi_{\SurE}   \,,
\end{equation}
where $\pi_{\surE}:T\mathcal{C}\to T\surE$ is defined as\footnote{With respect to the material and spatial foliation coordinates the projection maps have the following representations
\begin{equation}
	[\pi_{\Sur}]=[\pi_{\sur}]=
	\begin{bmatrix}
		1 & 0 & 0  \\
		0 & 1 & 0 \\
		0 & 0 & 0 
	\end{bmatrix}
	\,.
\end{equation}
}
\begin{equation}
	\pi_{\surE}=\operatorname{id}_{T\mathcal{C}} - \mathbf{n}\circ\varphi \otimes \mathbf{n}^{\flat}\circ\varphi\,.
\end{equation}
Note that 
\begin{equation} \label{Tangent-Parallel-Velocities}
	\mathbf{W}_\parallel(X)=\iota_{\Sur*}\bar{\mathbf{W}}(X)=\FS \bar{\mathbf{W}}(X)\,,\qquad
	\mathbf{w}_\parallel(x)=\iota_{\sur*}\bar{\mathbf{w}}(x)=\Fs \bar{\mathbf{w}}(x)
	\,.
\end{equation}
Thus, $\iota_{\sur*}\bar{\mathbf{w}}(x)=\mathbf{F}_{\parallel}(X)\,\iota_{\Sur*}\bar{\mathbf{W}}(X)$. 
In components, $W_\parallel^A=\cFS^A{}_{\bar{A}} \,\bar{W}^{\bar{A}}$ and $w_\parallel^a=\cFs^a{}_{\bar{a}} \,\bar{w}^{\bar{a}}$.
\end{defi}

\begin{remark}
It is important to note that the tangential deformation gradient $\mathbf{F}_{\parallel}$ is not the restriction of the deformation gradient $\mathbf{F}$ to the tangent bundle of the material surface $T\SurE$. Instead, it is defined by projecting the image of $\mathbf{F}$ onto the tangent bundle of the spatial surface $T\surE$. This distinction is crucial since, in general, $\mathbf{F}$ maps tangent vectors of the material surface to vectors that do not necessarily lie in $T\surE$.
\end{remark}

\begin{figure}[t!]
\centering
\includegraphics[width=0.35\textwidth]{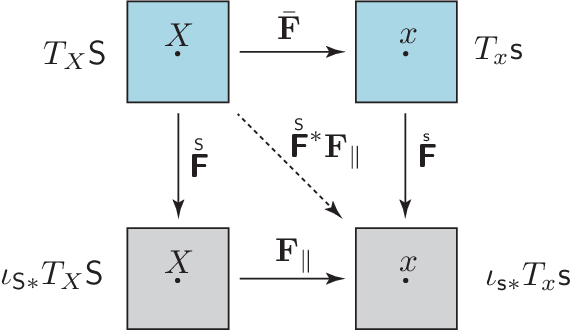}
\vspace*{0.10in}
\caption{The schematic relationship between tangential and surface deformation gradients and the tangents of inclusion maps.} 
\label{Deformation-Gradients}
\end{figure}

\begin{defi}
The \emph{surface deformation gradient} $\bar{\mathbf{F}}(X):T_X\Sur\to T_x\sur$ is defined such that $\bar{\mathbf{w}}(x)=\bar{\mathbf{F}}(X)\bar{\mathbf{W}}(X)$.
From $\mathbf{w}_\parallel=\mathbf{F}_{\parallel}\mathbf{W}_\parallel$, $\mathbf{W}_\parallel=\FS\bar{\mathbf{W}}$, $\mathbf{w}_\parallel=\Fs\bar{\mathbf{w}}$, and $\bar{\mathbf{w}}=\bar{\mathbf{F}}\bar{\mathbf{W}}$, one concludes that $\Fs\bar{\mathbf{F}}\bar{\mathbf{W}}=\mathbf{F}_{\parallel}\FS\bar{\mathbf{W}}$. Therefore (see Fig.~\ref{Deformation-Gradients})\footnote{Recall that $\Fs=T\iota_{\sur}:T\sur\to T\mathcal{C}$ and has the following representation with respect to foliation coordinates
\begin{equation} \label{Fs-Matrix}
	\big[\Fs\big]=\big[\cFs^a{}_{\bar{a}}\big]
	= \begin{bmatrix} \mathbf{I}_{n-1} \\ \mathbf{0}_{1\times (n-1)} \end{bmatrix}\,.
\end{equation}
}
\begin{equation} \label{Surface-Deformation-Gradient}
	\Fs\bar{\mathbf{F}} = \mathbf{F}_{\parallel} \FS\,, \quad \text{or} \qquad
	\Fs_*\bar{\mathbf{F}} = \FS^*\mathbf{F}_{\parallel} \,.
\end{equation}
\end{defi}

This defining relation can be rewritten as
\begin{equation} 
	\Fs\,\bar{\mathbf{F}} =  \pi_{\surE} \circ \mathbf{F}\big |_{\SurE} \circ \FS
	\,.
\end{equation}
Note that $\pi_{\surE}:T\mathcal{C}\big|_{\surE} \to T\surE$ is the spatial projection, and $\bar{\pi}_{\sur}:T\mathcal{C}\big|_{\surE} \to T\sur$ is the intrinsic projection onto the tangent bundle of the surface. Similar to \eqref{Pi-bar-Pi}, one has
\begin{equation} 
	\pi_{\sur} = \Fs \circ \bar{\pi}_{\sur}
	\,,
\end{equation}
where $\Fs = T\iota_{\sur}$ is the inclusion map. Thus, $\bar{\mathbf{F}}$ maps $T\Sur$ to $T\sur$, and the composition with $\Fs$ gives the tangential projection of the deformation gradient. This identity emphasizes that $\bar{\mathbf{F}}$ is not the restriction of $\mathbf{F}$ to $T\Sur$, but a projection of its action onto $T\sur$.

With respect to foliation coordinates, the surface restricted, the parallel, and surface deformation gradients have the following representations
\begin{equation}
	\left[ (F_\SurE)^a{}_A \right] =
	\begin{bmatrix}
		F^1{}_1 & F^1{}_2 & 0  \\
		F^2{}_1 & F^2{}_2 &0 \\
		F^3{}_1 & F^3{}_2 & 0
	\end{bmatrix}\,,\qquad
	\left[ (F_\parallel)^a{}_A \right] =
	\begin{bmatrix}
		F^1{}_1 & F^1{}_2 & 0  \\
		F^2{}_1 & F^2{}_2 &0 \\
		0 & 0 & 0
	\end{bmatrix}\,,\qquad
	\left[ \bar{F}^{\,\bar{a}}{}_{\bar{A}} \right] =
	\begin{bmatrix}
		F^1{}_1 & F^1{}_2  \\
		F^2{}_1 & F^2{}_2 
	\end{bmatrix}	
	 \,.
\end{equation}

\paragraph{The first fundamental forms.}
The induced metric $\bar{\mathbf{g}}$ on $\sur$---the first fundamental form of $\sur$---is defined as $\bar{\mathbf{g}}=\mathbf{g}\big|_{\mathsf{s}}=\iota_{\sur}^*\mathbf{g}=\Fs^\star \mathbf{g} \,\Fs$.
Similarly, the induced metric $\mathring{\bar{\mathbf{G}}}$ on $\Sur$---the first fundamental form of $\Sur$---is defined as $\mathring{\bar{\mathbf{G}}}=\mathring{\mathbf{G}}\big|_{\Sur}=\iota_{\Sur}^*\mathring{\mathbf{G}}=\FS^\star \mathring{\mathbf{G}} \,\FS$.
Recall that with respect to foliation coordinates $\{X^1,X^2,X^3\}$ and $\{x^1,x^2,x^3\}$ 
for $\SurE$ and $\surE$, respectively, the metrics $\mathring{\mathbf{G}}$ and $\mathbf{g}$ when restricted to $\SurE$ and $\surE$, respectively, have the following representations 
\begin{equation}
	\mathring{\mathbf{G}}\big|_{\SurE}(X)=\begin{bmatrix}
	\mathring{G}_{11}(X) & \mathring{G}_{12}(X)  & 0 \\
	\mathring{G}_{12}(X) & \mathring{G}_{22}(X)  & 0  \\
	0 & 0  & 1  
	\end{bmatrix}\,,\qquad
	\mathbf{g}\big|_{\surE}(x)=\begin{bmatrix}
	g_{11}(x) & g_{12}(x)  & 0 \\
	g_{12}(x) & g_{22}(x)  & 0  \\
	0 & 0  & 1  
	\end{bmatrix}\,,\quad \forall ~ X\in\SurE\,,~ x\in\surE\,.
\end{equation}
Thus, the first fundamental forms have the following foliation coordinate representations:
\begin{equation}
	\mathring{\bar{\mathbf{G}}}(X) = \left[\mathring{\bar{G}}_{\bar{A}\bar{B}}(X) \right]=
	\begin{bmatrix}
		\mathring{G}_{11}(X) & \mathring{G}_{12}(X)  \\
		\mathring{G}_{12}(X) & \mathring{G}_{22}(X) 
	\end{bmatrix}\,,\qquad
	\bar{\mathbf{g}}(x) = \left[\bar{g}_{\bar{a}\bar{b}}(x) \right]=
	\begin{bmatrix}
		g_{11}(x) & g_{12}(x)  \\
		g_{12}(x) & g_{22}(x) 
	\end{bmatrix}
	 \,.
\end{equation}

The dual of the surface deformation gradient is denoted by 
 $\bar{\mathbf{F}}^{\star}:T_x^*\mathsf{s}\to T_X^*\Sur$, where $T_x^*\mathsf{s}$ and $T_X^*\Sur$ denote the cotangent spaces of $\mathsf{s}$ at $x$ and $\Sur$ at $X$, respectively.
 $\bar{\mathbf{F}}^{\star}$ has the following coordinate representation
\begin{equation}
	\bar{\mathbf{F}}^{\star}(X)= \bar{F}^{\,\bar{a}}{}_{\bar{A}}(X) \,
	dX^{\bar{A}} \otimes \frac{\partial}{\partial x^{\bar{a}}}\,.
\end{equation}
The transpose of the surface deformation gradient is defined as
\begin{equation}
	\bar{\mathbf{F}}^{\textsf{T}}: T_x\mathsf{s} \to T_X\Sur\,,
	\qquad \llangle \bar{\mathbf{F}}\bar{\mathbf{V}}\,, \bar{\mathbf{v}} \rrangle_{\bar{\mathbf{g}}}    
	=\llangle \bar{\mathbf{V}}, \bar{\mathbf{F}}^{\textsf{T}} \bar{\mathbf{v}}
	\rrangle_{\,\mathring{\bar{\mathbf{G}}}}\,, \qquad
	\forall \,\bar{\mathbf{V}} \in T_{X}\Sur,~\bar{\mathbf{v}} \in T_{x} \mathsf{s}\,.
\end{equation}
Thus, $\bar{\mathbf{F}}^{\textsf{T}}=\mathring{\bar{\mathbf{G}}}^\sharp \,\bar{\mathbf{F}}^\star \,\bar{\mathbf{g}}$. In coordinates, $(\bar{F}^{\mathsf{T}})^{\bar{A}}{}_{\bar{a}}=\mathring{\bar{G}}^{\bar{A}\bar{B}} \,\bar{F}^{\bar{b}}{}_{\bar{B}}\,\bar{g}_{\bar{a}\bar{b}}$.
The (intrinsic) surface right Cauchy-Green strain is defined as $\bar{\mathbf{C}}^{\flat}=\bar{\mathbf{F}}^{\,*}\bar{\mathbf{g}}=\bar{\mathbf{F}}^{\,\star}\bar{\mathbf{g}}\bar{\mathbf{F}}$, which has components $\bar{C}_{\bar{A}\bar{B}}=\bar{F}^{\,\bar{a}}{}_{\bar{A}}\,\bar{F}^{\,\bar{b}}{}_{\bar{B}}\,\bar{g}_{\bar{a}\bar{b}}$. 

The left surface Cauchy-Green deformation tensor is defined as $\bar{\mathbf{B}}^{\,\sharp}=\bar{\mathbf{F}}^*\,\bar{\mathbf{g}}^{\,\sharp}$, and has components $\bar{B}^{\bar{A}\bar{B}}=\bar{F}^{\,-\bar{A}}{}_{\bar{a}}\,\bar{F}^{\,-\bar{B}}{}_{\bar{b}}\,\,\bar{g}^{\,\bar{a}\bar{b}}$, where $\bar{\mathbf{g}}^{\sharp}$ is the inverse of $\bar{\mathbf{g}}$.
The spatial analogues of $\bar{\mathbf{C}}^{\,\flat}$ and $\bar{\mathbf{B}}^{\,\sharp}$ are defined as $\bar{\mathbf{c}}^{\,\flat}=\bar{\mathbf{F}}_*\mathring{\bar{\mathbf{G}}}$ and $\bar{\mathbf{b}}^{\,\sharp}=\bar{\mathbf{F}}_*\mathring{\bar{\mathbf{G}}}^{\sharp}$, respectively.
They have components $\bar{c}_{\bar{a}\bar{b}}=\bar{F}^{\,-\bar{A}}{}_{\bar{a}}\,\bar{F}^{\,-\bar{B}}{}_{\bar{b}}\,\mathring{\bar{G}}_{\bar{A}\bar{B}}$, and $\bar{b}^{\,\bar{a}\bar{b}}=\bar{F}^{\bar{a}}{}_{\bar{A}}\,\bar{F}^{\bar{b}}{}_{\bar{B}}\,\mathring{\bar{G}}^{\bar{A}\bar{B}}$.

\begin{remark}
These measures of strain are all related to the in-plane (membrane) deformations of the material surfaces. The out-of-plane (bending) deformations are described by the pull-back of the second fundamental form $\bar{\mathbf{k}}$. 
The extrinsic defamation tensor is defined as $\bar{\mathbf{H}}=\bar{\mathbf{F}}^{*}\bar{\mathbf{k}}$, which has components $\bar{H}_{\bar{A}\bar{B}}=\bar{F}^{\,\bar{a}}{}_{\bar{A}}\,\bar{F}^{\,\bar{b}}{}_{\bar{B}}\,\bar{k}_{\bar{a}\bar{b}}$ \citep{AngoshtariYavari2015,Sadik2016}. In this paper, we only consider membrane deformations for the elastic surfaces.
\end{remark}

\paragraph{Polar decomposition of surface deformation gradient.}
The \textit{surface polar decomposition} of the surface deformation gradient is expressed as
\begin{equation}\label{Surface-Polar-Decomposition}
    \bar{\mathbf{F}}=\bar{\mathbf{R}} \bar{\mathbf{U}}= \bar{\mathbf{V}} \bar{\mathbf{R}} \,,
\end{equation}
where $\bar{\mathbf{U}}:T\Sur\to T\Sur$ and $\bar{\mathbf{V}}:T\sur\to T\sur$ are the material and spatial surface stretch tensors, respectively, and $\bar{\mathbf{R}}:T\Sur\to T\sur$ is a $(\bar{\mathbf{G}},\bar{\mathbf{g}})$-orthogonal tensor field, defined in the sense of \citep{SimoMarsden1984}:
\begin{equation} \label{Surface-G-g-Orthogonal}
	\bar{\mathbf{R}}^\star (\bar{\mathbf{g}}\circ\varphi)\, \bar{\mathbf{R}}=\bar{\mathbf{G}}    \,.
\end{equation}
In component form, this is written as $\bar{R}^{\bar{a}}{}_{\bar{A}}\,(\bar{g}_{\bar{a}\bar{b}}\circ\varphi)\,\bar{R}^{\bar{b}}{}_{\bar{B}}=\bar{G}_{\bar{A}\bar{B}}$.

\begin{remark}
The condition \eqref{Surface-G-g-Orthogonal} implies that $\bar{\mathbf{R}}$ preserves the inner product structure between $\Sur$ and $\sur$, i.e., $\bar{\mathbf{R}}$ maps orthonormal bases in $(T\Sur,\bar{\mathbf{G}})$ to orthonormal bases in $(T\sur,\bar{\mathbf{g}})$.
\end{remark}

The component-wise expression of the polar decomposition is given by
\begin{equation}
	\bar{F}^{\bar{a}}{}_{\bar{A}} 
	= \bar{R}^{\bar{a}}{}_{\bar{B}}\, \bar{U}^{\bar{B}}{}_{\bar{A}} 
	= \bar{V}^{\bar{a}}{}_{\bar{b}}\, \bar{R}^{\bar{b}}{}_{\bar{A}}
   \,.
\end{equation}
Eq.~\eqref{Surface-G-g-Orthogonal} implies that $(\det\bar{\mathbf{R}})^2 \det\bar{\mathbf{g}} = \det\bar{\mathbf{G}}$, and from \eqref{Surface-Polar-Decomposition}, it follows that $\det\bar{\mathbf{U}} = \det\bar{\mathbf{V}}$. The surface Jacobian of deformation is defined by the relation $da_s=\bar{J}\,dA_s$ and is given as
\begin{equation}
	\bar{J} = \sqrt{\frac{\det\bar{\mathbf{g}}}{\det\bar{\mathbf{G}}}}\,\det\bar{\mathbf{F}} = \det\bar{\mathbf{U}} = \det\bar{\mathbf{V}}  \,.
\end{equation}

The material stretch tensor $\bar{\mathbf{U}}:T\Sur \to T\Sur$ and the spatial stretch tensor $\bar{\mathbf{V}}:T\sur \to T\sur$ are related to the right and left Cauchy-Green surface deformation tensors through the following relations
\begin{equation} \label{Surface-C-U-b_v}
\begin{aligned}
	\bar{\mathbf{C}} &= \bar{\mathbf{F}}^{\textsf{T}} \bar{\mathbf{F}}
	= (\bar{\mathbf{R}} \bar{\mathbf{U}})^{\textsf{T}} \bar{\mathbf{R}} \bar{\mathbf{U}}
	= \bar{\mathbf{G}}^\sharp (\bar{\mathbf{R}} \bar{\mathbf{U}})^\star \bar{\mathbf{g}} 
	\bar{\mathbf{R}} \bar{\mathbf{U}}
	= \bar{\mathbf{G}}^\sharp \bar{\mathbf{U}}^\star \bar{\mathbf{R}}^\star \bar{\mathbf{g}} 
	\bar{\mathbf{R}} \bar{\mathbf{U}}
	= \bar{\mathbf{G}}^\sharp \bar{\mathbf{U}}^\star \bar{\mathbf{G}} \bar{\mathbf{U}} 
	= \bar{\mathbf{U}}^2 \,,\\
	\bar{\mathbf{b}} &= \bar{\mathbf{F}} \bar{\mathbf{F}}^{\textsf{T}}
	= \bar{\mathbf{V}} \bar{\mathbf{R}} (\bar{\mathbf{V}} \bar{\mathbf{R}})^{\textsf{T}}
	= \bar{\mathbf{V}} \bar{\mathbf{R}} \bar{\mathbf{G}}^\sharp (\bar{\mathbf{V}} 
	\bar{\mathbf{R}})^\star \bar{\mathbf{g}}
	= \bar{\mathbf{V}} \bar{\mathbf{R}} \bar{\mathbf{G}}^\sharp \bar{\mathbf{R}}^\star 
	\bar{\mathbf{V}}^\star \bar{\mathbf{g}}
	= \bar{\mathbf{V}} \bar{\mathbf{g}}^\sharp \bar{\mathbf{V}}^\star \bar{\mathbf{g}} 
	= \bar{\mathbf{V}}^2 \,.
\end{aligned}
\end{equation}
Alternatively, this can be expressed as
\begin{equation}
	\bar{\mathbf{C}}^\flat = \bar{\mathbf{U}}^\star \bar{\mathbf{G}} \bar{\mathbf{U}} \,,\qquad 
	\bar{\mathbf{b}}^\sharp = \bar{\mathbf{V}} \bar{\mathbf{g}}^\sharp \bar{\mathbf{V}}^\star
	\,,
\end{equation}
which, in component form, are given by $\,\bar{C}_{\bar{A}\bar{B}} = \bar{U}^{\bar{M}}{}_{\bar{A}}\, \bar{G}_{\bar{M}\bar{N}}\, \bar{U}^{\bar{N}}{}_{\bar{B}}\,$ and $\,\bar{b}^{\bar{a}\bar{b}} = \bar{V}^{\bar{a}}{}_{\bar{m}}\, \bar{g}^{\bar{m}\bar{n}}\, \bar{V}^{\bar{b}}{}_{\bar{n}}$. The relations \eqref{Surface-C-U-b_v} are equivalently expressed as $\bar{\mathbf{U}} = \sqrt{\bar{\mathbf{C}}}$ and $\bar{\mathbf{V}} = \sqrt{\bar{\mathbf{b}}}$.
The square roots of tensors are interpreted in the spectral sense (i.e., defined via diagonalization for symmetric positive-definite tensors).

Let us define the \textit{normal stretch} as
\begin{equation}
	\lambda_n = \llangle \mathbf{F}\mathbf{N}, \mathbf{n} \rrangle_{\mathbf{g}} = F^a{}_A\,N^A\,n_a\,.
\end{equation}
One can show that $\lambda_n = J/\bar{J}$. This follows from the dual form of Nanson’s formula:
\begin{equation}
	\mathbf{F}^{\mathsf{T}}\mathbf{n}^\flat = \dfrac{J}{\bar{J}}\,\mathbf{N}^\flat\,,
\end{equation}
which implies that
\begin{equation}
	\llangle \mathbf{F}\mathbf{N}, \mathbf{n} \rrangle_{\mathbf{g}} 
	= \llangle \mathbf{N}, \mathbf{F}^{\mathsf{T}}\mathbf{n} \rrangle_{\mathbf{G}}
	= \Big \llangle \mathbf{N}, \frac{J}{\bar{J}}\,\mathbf{N} \Big\rrangle_{\mathbf{G}}
	= \frac{J}{\bar{J}}\,.
\end{equation}

\paragraph{Polar decomposition of the surface-restricted deformation gradient \citep{ManCohen1986}.}
The surface-restricted deformation gradient $\mathbf{F}_\SurE(X)=\mathbf{F}\big|_{\SurE}(X)\circ \pi_{\SurE}:T_X\SurE \to T_x\mathcal{C}$ admits the following polar decomposition. This is the membrane polar decomposition introduced by \citet{ManCohen1986}, rewritten here in our notation:
\begin{equation} \label{Polar-Decomposition-ManCohen}
	\mathbf{F}_\SurE(X)=\boldsymbol{\mathsf{R}}\circ\varphi(X)\,\FS(X)\,\bar{\mathbf{U}}(X)\,,
\end{equation}
where $\bar{\mathbf{U}}$ is the surface stretch tensor and $\boldsymbol{\mathsf{R}}:T\mathcal{B}\big|_{\SurE}\to T\mathcal{C}\big|_{\surE}$ is the unique $(\mathbf{G},\mathbf{g})$-orthogonal tensor field satisfying
\begin{equation}
	\mathbf{n}\circ\varphi(X)=\boldsymbol{\mathsf{R}}\circ\varphi(X)\cdot\mathbf{N}(X)\,.
\end{equation}
Note that 
\begin{equation}
	\mathbf{F}_\SurE^{\mathsf{T}} \mathbf{F}_\SurE
	= \bar{\mathbf{U}}^{\mathsf{T}} \FS^{\mathsf{T}} \boldsymbol{\mathsf{R}}^{\mathsf{T}} 
	\boldsymbol{\mathsf{R}}\,\FS\,\bar{\mathbf{U}}
	= \bar{\mathbf{U}}^{\mathsf{T}} \FS^{\mathsf{T}} \operatorname{id}_{T\mathcal{B}}\,\FS\,\bar{\mathbf{U}}
	= \bar{\mathbf{U}}^{\mathsf{T}} \FS^{\mathsf{T}} \,\FS\,\bar{\mathbf{U}}
	= \bar{\mathbf{U}}^{\mathsf{T}} \operatorname{id}_{T\SurE}\,\bar{\mathbf{U}}
	= \bar{\mathbf{U}}^{\mathsf{T}} \,\bar{\mathbf{U}}
	=\bar{\mathbf{C}} 
	\,.
\end{equation}

\subsection{Constitutive equations}

In surface elasticity, the bulk and material surfaces have different constitutive equations, in general. 
We assume that both the bulk and material surfaces are made of hyper-elastic materials, i.e., each has an elastic energy function.\footnote{One can extend the results of this paper to Cauchy elasticity, i.e., elasticity without a stored-energy function \citep{Cauchy1828,Truesdell1952,YavariGoriely2025Cauchy}. In this paper we restrict attention to hyperelasticity.}
Let us denote the energy function of the bulk and the material surfaces by $W$ and $\Ws$, respectively.

\subsubsection{Bulk constitutive equations}

\paragraph{Measures of stress.}
Here we briefly review three commonly used measures of stress in nonlinear elasticity. It should be emphasized that there are infinitely many measures of stress in continuum mechanics, and the three discussed here are simply more convenient in many applications.
Consider an element of area $da$ in the deformed configuration $\mathcal{C}$ with $\mathbf{g}$-unit normal $\mathbf{n}$. The traction on this element is $\mathbf{t}=\boldsymbol{\sigma}\mathbf{n}^\flat$, where $\boldsymbol{\sigma}$ is the \textit{Cauchy stress}, and $\mathbf{n}^\flat=\mathbf{g}\mathbf{n}$. The force on this element is $\mathbf{f}=\mathbf{t}\,da$. In components, $t^a=\sigma^{ab}n_b$, where $n_b=g_{bc}n^c$.
Let the corresponding area element in the reference configuration $\mathcal{B}$ be $dA$ with $\mathbf{G}$-unit normal $\mathbf{N}$. The \textit{first Piola--Kirchhoff stress tensor} $\mathbf{P}$ is defined such that
\begin{equation}
	\mathbf{t}\,da = \mathbf{P}\,\mathbf{N}^\flat\,dA\,.
\end{equation}
Using Nanson's formula \eqref{Nanson}, one finds
\begin{equation}
	\mathbf{P} = J\,\boldsymbol{\sigma}\,\mathbf{F}^{-\star}\,,
\end{equation}
where
\begin{equation} \label{Jacobian-Elasticity}
	J=\sqrt{\frac{\det\mathbf{g}}{\det\mathring{\mathbf{G}}}}\,\det\mathbf{F}  \,.
\end{equation}
In components, $P^{aA}=J\,\sigma^{ab}\,F^{-A}{}_b$. Now if one pulls back the force $\mathbf{f}$, the \textit{second Piola-Kirchhoff stress tensor} $\mathbf{S}$ is defined such that $\mathbf{F}^{-1}\mathbf{t}\,da=\mathbf{S}\,\mathbf{N}^\flat\,dA$. Therefore,
\begin{equation}
	\mathbf{S} = \mathbf{F}^{-1}\mathbf{P} = J\,\mathbf{F}^{-1}\,\boldsymbol{\sigma}\,\mathbf{F}^{-\star}\,.
\end{equation}
In components, $S^{AB}=F^{-A}{}_a\,P^{aB}=J\,F^{-A}{}_a\,\sigma^{ab}\,F^{-B}{}_b$.

Assuming an inhomogeneous bulk material, for $X\in\mathring{\mathcal{B}}$, $W=W(X,\mathbf{F},\mathring{\mathbf{G}},\mathring{\boldsymbol{\Lambda}},\mathbf{g})$, where $\mathring{\boldsymbol{\Lambda}}$ is a set of structural tensors that characterize the bulk anisotropy of the material \citep{liu1982,boehler1987,zheng1993,zheng1994theory,lu2000,MazzucatoRachele2006}. 
Objectivity implies that $W=\hat{W}(X,\mathbf{C}^{\flat},\mathring{\mathbf{G}},\mathring{\boldsymbol{\Lambda}})$. When structural tensors are included as arguments of the energy function, the energy function becomes an isotropic (or materially covariant) function of its arguments \citep{Boehler1979}.
Thus, the energy function can be written as a function of its isotropic invariants (or integrity basis) \citep{Spencer1971}.
Denoting the integrity basis by $I_j, ~j=1,\dots,m$, we have $W=\overline{W}(X,I_1,...,I_m)$. 
The second Piola-Kirchhoff stress tensor has the following representation \citep{DoyleEricksen1956, MaHu1983, Yavari2006}
\begin{equation} \label{Second-PK-Stress-Representation}
	\mathbf{S}=2\frac{\partial \hat{W}}{\partial\mathbf{C}^\flat}
	=\sum_{j=1}^{m}2W_j\frac{\partial I_j}{\partial\mathbf{C}^\flat}\,, \qquad 
	W_j=W_j(X,I_1,...,I_m):=\frac{\partial \overline{W}}{\partial I_j}\,, \qquad j=1,\dots,m\,.
\end{equation}
Recall that $\mathbf{P}=\mathbf{F}\mathbf{S}$ and $\boldsymbol{\sigma}=J^{-1} \mathbf{F}\mathbf{S}\mathbf{F}^\star$.

\paragraph{Isotropic solids.}
For an isotropic solid, the energy function depends only on the principal invariants of $\mathbf{C}^{\flat}$, i.e., $W=\overline{W}(X,I_1,I_2,I_3)$, where
\begin{equation} 
\begin{aligned}
	I_1 &=\operatorname{tr}_{\mathring{\mathbf{G}}}\mathbf{C}^{\flat}
	=\mathbf{C}^{\flat}\!:\!\mathring{\mathbf{G}}^{\sharp}
	=C_{AB}\,\mathring{G}^{AB}\,, \\
	I_2 &=\frac{1}{2}\left[I_1^2-\operatorname{tr}_{\mathring{\mathbf{G}}}\mathbf{C}^2\right]
	=\frac{1}{2}\left(I_1^2-C_{MB}\,C_{NA}\,\mathring{G}^{AM}\mathring{G}^{BN}\right)\,, \\
	I_3 &=\frac{\det\mathbf{C}^{\flat}}{\det\mathring{\mathbf{G}}}
	\,.
\end{aligned}
\end{equation}
For an isotropic solid the Cauchy stress has the following representation \citep{DoyleEricksen1956}
\begin{equation} \label{Cauchy-Representation-Isotropic-Comp}
	\boldsymbol{\sigma} = \frac{2}{\sqrt{I_3}} \left[ \left(I_2\,\overline{W}_2+I_3\,\overline{W}_3\right)
	\mathbf{g}^{\sharp} 
	+\overline{W}_1\,\mathbf{b}^{\sharp}-I_3\,\overline{W}_2\,\mathbf{c}^{\sharp} \right]\,.
\end{equation}
For an incompressible isotropic solid $I_3=1$, and $W=\overline{W}(X,I_1,I_2)$. Thus
\begin{equation}
	\boldsymbol{\sigma} = \left(-p+2I_2\,\overline{W}_2\right)
	\mathbf{g}^{\sharp} 
	+2\overline{W}_1\,\mathbf{b}^{\sharp}-2\,\overline{W}_2\,\mathbf{c}^{\sharp} \,,
\end{equation}
where $p$ is the Lagrange multiplier associated with the incompressibility constraint $J=\sqrt{I_3}=1$. As $p$ is an unknown a priori, $-p+2I_2\,\overline{W}_2$ can be replaced by $-p$, and hence
\begin{equation}  \label{Cauchy-Representation-Isotropic-Incomp}
	\boldsymbol{\sigma} = -p\,\mathbf{g}^{\sharp} 
	+2\overline{W}_1\,\mathbf{b}^{\sharp}-2\,\overline{W}_2\,\mathbf{c}^{\sharp}\,.
\end{equation}

\paragraph{Transversely isotropic solids.} In transversely isotropic materials, each point has a single material preferred direction, orthogonal to the isotropy plane at that point. A unit vector $\mathbf{N}(X)$ denotes the material preferred direction at $X \in \mathcal{B}$. For inhomogeneous transversely isotropic materials, the energy function is expressed as $W=W(X,\mathring{\mathbf{G}},\mathbf{C}^\flat, \mathbf{A})$, where $\mathbf{A}=\mathbf{N}\otimes\mathbf{N}$ is a structural tensor \citep{DoyleEricksen1956,spencer1982formulation,lu2000covariant}.
The energy function $W$ depends on the following five independent invariants:
\begin{equation} \label{invartr}
\begin{aligned}
	& I_1=\mathrm{tr}\,\mathbf{C}=C^A{}_A\,,\quad 
	I_2=\mathrm{det}\,\mathbf{C}~\mathrm{tr}~\mathbf{C}^{-1}
	=\mathrm{det}(C^A{}_B)(C^{-1})^D{}_D\,,\quad 
	I_3=\mathrm{det}\mathbf{C}=\mathrm{det}(C^A{}_B)\,, \\
	& I_4=\mathbf{N}\cdot\mathbf{C}\cdot\mathbf{N}=N^AN^B\,C_{AB}\,,\quad 
	I_5=\mathbf{N}\cdot\mathbf{C}^2\cdot\mathbf{N}=N^AN^B\,C_{BM}\,C^M{}_A\,.
\end{aligned}
\end{equation}
The second Piola-Kirchhoff stress tensor is given by
\begin{equation}
	\mathbf{S}=\sum_{j=1}^{5}2W_j\frac{\partial I_j}{\partial \mathbf{C}^\flat}\,, \qquad
	W_j=W_j(X,I_1,...,I_5):=\frac{\partial W}{\partial I_j}\,,\, \qquad j=1,\dots,5\,.
\end{equation}
This can be simplified to read
\begin{equation} \label{piolas}
\begin{aligned}
	\mathbf{S} &=2W_1\,\mathring{\mathbf{G}}^\sharp+2W_2\left(I_2\,\mathbf{C}^{-1}
	-I_3\,\mathbf{C}^{-2}\right)+2W_3\,I_3\,\mathbf{C}^{-1} \\
	&\quad +2W_4\left(\mathbf{N}\otimes\mathbf{N}\right)
	+2W_5\left[\mathbf{N}\otimes(\mathbf{C}\cdot\mathbf{N})
	+(\mathbf{C}\cdot\mathbf{N})\otimes\mathbf{N} \right] \,.
\end{aligned}
\end{equation} 
The Cauchy stress tensor is given by \citep{Ericksen1954Anisotropic,Golgoon2018a,Golgoon2018b}:
\begin{equation}\label{Cauchy-Compressible-Transversely-Isotropic}
	\sigma^{ab}=\frac{2}{\sqrt{I_3}}\left[W_1b^{ab}+\left(I_2W_2+I_3W_3 \right)g^{ab}-I_3W_2\,c^{ab}
	+W_4\, n^an^b+W_5\,\ell^{ab}
	\right]\,,
\end{equation}  
where $n^a=F^a{}_A N^A$, and $\ell^{ab}=n^a\,b^{bc}\,n_c+n^b\,b^{ac}\,n_c$.
For incompressible transversely isotropic materials ($I_3=1$), $W=W(\mathbf{X},I_1,I_2,I_4,I_5)$, and hence \citep{Ericksen1954Anisotropic,Spencer1986,Golgoon2018a,Golgoon2018b}
\begin{equation}\label{Cauchy-Incompressible-Transversely-Isotropic}
	\sigma^{ab}=-p\,g^{ab}+2W_1\,b^{ab}-2W_2\,c^{ab}
	+2W_4\, n^a\,n^b+2W_5 \left(n^a\,b^{bc}\,n^d\,g_{cd}+n^b\,b^{ac}\,n^d\,g_{cd}\right) \,.
\end{equation}  

\paragraph{Hyperelastic fluids.}
The free energy function for hyperelastic fluids takes the form 
\begin{equation}\label{Fluid-Energy}
	\Wf=\Wf(X,T,J) \,,
\end{equation}  
where $\Wf$ is a smooth and strictly convex function of $J$, diverging to infinity as $J \to 0$ \citep{Podio1985}. Consequently, the Cauchy, the first Piola-Kirchhoff, and the second Piola-Kirchhoff stress tensors are expressed as:
\begin{equation} \label{Fluid-Stress}
	\boldsymbol{\sigma}= \frac{\partial \Wf}{\partial J} \,\mathbf{g}^\sharp  
	\,,\qquad 
	\mathbf{P}=J \frac{\partial \Wf}{\partial J} \,\mathbf{g}^\sharp \mathbf{F}^{-\star}
	\,,\qquad 
	\mathbf{S}=J \frac{\partial \Wf}{\partial J} \,\mathbf{F}^{-1}\mathbf{g}^\sharp \mathbf{F}^{-\star}\,.
\end{equation}
It is important to note that $\frac{\partial \Wf}{\partial J} < 0$, as hydrostatic stresses in fluids are inherently compressive.

\subsubsection{Surface constitutive equations}

\paragraph{Measures of surface stress.}
One can use different measures of surface stress. Here we describe three such measures that are the surface analogues of their corresponding bulk stresses. Let us consider a line element $d\ell_0$ in the reference configuration with $\bar{\mathbf{G}}$-unit normal vector $\bar{\mathbf{N}}$. In the deformed configuration, the corresponding line element $d\ell$ has the $\bar{\mathbf{g}}$-unit normal vector $\bar{\mathbf{n}}$. The surface traction on this line element is $\bar{\mathbf{t}}=\bar{\boldsymbol{\sigma}}\bar{\mathbf{n}}^\flat$, where $\bar{\boldsymbol{\sigma}}$ is the \textit{surface Cauchy stress}, $\bar{\mathbf{n}}^\flat=\bar{\mathbf{g}}\bar{\mathbf{n}}$, and the force on this element is $\bar{\mathbf{f}}=\bar{\mathbf{t}}\,d\ell$. 

Similar to the bulk, the \textit{first Piola-Kirchhoff surface stress} $\bar{\mathbf{P}}$ and the \textit{second Piola-Kirchhoff surface stress} $\bar{\mathbf{S}}$ are defined as
\begin{equation} \label{Surface-Stresses}
	\bar{\mathbf{P}} = \bar{J}\,\bar{\boldsymbol{\sigma}}\,\bar{\mathbf{F}}^{-\star}\,, \qquad
	\bar{\mathbf{S}} = \bar{\mathbf{F}}^{-1}\bar{\mathbf{P}} 
	= \bar{J}\,\bar{\mathbf{F}}^{-1}\,\bar{\boldsymbol{\sigma}}\,\bar{\mathbf{F}}^{-\star}\,.
\end{equation}

A material surface has mechanical properties different from those of the bulk. 
For a general hyperelastic surface, elastic energy (per unit undeformed surface area) depends on the restricted surface deformation gradient $\mathbf{F}_\SurE$, i.e., 
\begin{equation} 
	\Ws=W_{\!s}(X,\mathbf{F}_\SurE,\mathring{\mathbf{G}},\mathbf{g})\,.
\end{equation}
This energy includes both membrane and bending deformations.
Assuming that only membrane deformations induce stress in the elastic surface, the energy depends on the deformation only through in-surface stretches. Consequently, any dependence on the restricted surface deformation gradient $\mathbf{F}_\SurE$ can be reduced to a dependence on the tangential deformation gradient $\mathbf{F}_{\parallel}$, and equivalently on the surface deformation gradient $\bar{\mathbf{F}}$, using the identity $\Fs\,\bar{\mathbf{F}}=\mathbf{F}_{\parallel}\,\FS$.

The (membrane) constitutive equations of a material surface are given in terms of its intrinsic (surface) coordinates.
The surface right Cauchy-Green strain is defined as $\bar{\mathbf{C}}^{\flat}=\bar{\mathbf{F}}^*\bar{\mathbf{g}}:T_X\Sur\to T_X\Sur$ and quantifies the membrane (in-plane) deformations of the material surface. Using \eqref{Surface-Deformation-Gradient}, it is straightforward to see that
\begin{equation} \label{Surface-Strain}
	\bar{\mathbf{C}}^{\flat}=\bar{\mathbf{F}}^*\bar{\mathbf{g}}
	=\bar{\mathbf{F}}^*\Fs^*\mathbf{g}
	= (\Fs \bar{\mathbf{F}})^*\mathbf{g}
	= ( \mathbf{F}_{\parallel} \FS)^*\mathbf{g}
	=\FS^* \mathbf{F}_{\parallel}^{*}\,\mathbf{g}
	=\FS^* \mathbf{C}_{\parallel}^{\flat} \,.
\end{equation}
Assuming only membrane deformations and isotropic response, one has\footnote{For an anisotropic material surface $\Ws = W_{\!s}(X,\mathbf{F}_\SurE,\mathring{\mathbf{G}},\mathring{\bar{\boldsymbol{\Lambda}}},\mathbf{g})$ for $X\in\SurE$, where $\mathring{\bar{\boldsymbol{\Lambda}}}$ is a set of surface structural tensors that characterize the surface anisotropy of the material.} 
\begin{equation}  \label{Surface-Energy-Isotropic-Reduction}
	\Ws
	= W_{\!s}(X,\mathbf{F}_\SurE,\mathring{\mathbf{G}},\mathbf{g})
	= W_{\!s}(X,\mathbf{F}_{\parallel},\mathring{\mathbf{G}},\mathbf{g})
	= \hat{\Ws}(X,\mathbf{C}^\flat_{\parallel},\mathring{\mathbf{G}})
	= \hat{\Ws}(X,\FS^*\mathbf{C}^\flat_{\parallel},\FS^*\mathring{\mathbf{G}})
	= \hat{\Ws}(X,\bar{\mathbf{C}}^\flat,\mathring{\bar{\mathbf{G}}})
	\,,
\end{equation}
where $\mathbf{C}^\flat_{\parallel}=\mathbf{F}_{\parallel}^{*}\,\mathbf{g}=\mathbf{F}_{\parallel}^{\star}\,\mathbf{g}\mathbf{F}_{\parallel}$.
In \eqref{Surface-Energy-Isotropic-Reduction}, the second equality follows from the assumption of membrane deformations, the third from objectivity, the fourth from isotropy (material covariance), and the last from the definitions of the surface strain and the first fundamental form.

\begin{defi}[Projected Metric]
Let us consider arbitrary $\mathbf{U}, \mathbf{W}\in T\mathcal{B}\big|_{\mathsf{S}}$. The \textit{projected metric} $\mathring{\mathbf{G}}_\parallel$ is defined such that
\begin{equation}
\begin{aligned}
	\left\llangle \mathbf{U} , \mathbf{W} \right\rrangle_{\mathring{\mathbf{G}}_\parallel}
	&=\left\llangle \bar{\pi}_{\Sur}(\mathbf{U}) , 
	\bar{\pi}_{\Sur}(\mathbf{W}) \right\rrangle_{\mathring{\bar{\mathbf{G}}}} \\
	&=\left\llangle \bar{\pi}_{\Sur}(\mathbf{U}) , 
	\bar{\pi}_{\Sur}(\mathbf{W}) \right\rrangle_{\iota_{\Sur}^*\mathring{\mathbf{G}}}\\
	&=\left\llangle \iota_{\Sur*}\bar{\pi}_{\Sur}(\mathbf{U}) 
	, \iota_{\Sur*}\bar{\pi}_{\Sur}(\mathbf{W}) \right\rrangle_{\mathring{\mathbf{G}}} \\
	&=\left\llangle \pi_{\SurE}(\mathbf{U}) , \pi_{\SurE}(\mathbf{W}) \right\rrangle_{\mathring{\mathbf{G}}}\\
	&=\left\llangle \mathbf{U}-\langle \mathbf{N}^\flat,\mathbf{U} \rangle  \mathbf{N}
	, \mathbf{W}-\langle \mathbf{N}^\flat,\mathbf{W} \rangle  \mathbf{N} \right\rrangle_{\mathring{\mathbf{G}}} \\
	&=\left\llangle \mathbf{U} , \mathbf{W} \right\rrangle_{\mathring{\mathbf{G}}}
	-\langle \mathbf{N}^\flat,\mathbf{U} \rangle \langle \mathbf{N}^\flat,\mathbf{W} \rangle \\
	&=\left\llangle \mathbf{U} , \mathbf{W} \right\rrangle_{\mathring{\mathbf{G}}
	-\mathbf{N}^\flat\otimes \mathbf{N}^\flat}
	\,.
\end{aligned}
\end{equation}
Therefore,
\begin{equation} \label{Projected-Metric}
	\mathring{\mathbf{G}}_\parallel = \mathring{\mathbf{G}}-\mathbf{N}^\flat\otimes \mathbf{N}^\flat
	\,.
\end{equation}
\end{defi}

\begin{remark}
Since $\mathbf{C}_\parallel^\flat$ defined in \eqref{Projected-Metric} vanishes whenever one of its arguments is normal to the surface, all metric contractions in the invariants of $\mathbf{C}_\parallel^\flat$ involve only tangential components. Therefore, replacing $\mathring{\mathbf{G}}$ by its projected counterpart $\mathring{\mathbf{G}}_\parallel$ leaves these contractions unchanged. This justifies using $\mathring{\mathbf{G}}_\parallel$ in place of $\mathring{\mathbf{G}}$ in the surface energy function, i.e., $\Ws=\hat{\Ws}(X,\mathbf{C}_\parallel^\flat,\mathring{\mathbf{G}}_\parallel)$.  
\end{remark}

For an anisotropic material surface 
\begin{equation}
\begin{aligned}
	\Ws
	&= W_{\!s}(X,\mathbf{F}_\SurE,\mathring{\mathbf{G}},\mathring{\boldsymbol{\Lambda}},\mathbf{g}) \\
	&= W_{\!s}(X,\mathbf{F}_{\parallel},\mathring{\mathbf{G}},\mathring{\boldsymbol{\Lambda}},\mathbf{g}) \\
	&= \hat{\Ws}(X,\mathbf{C}^\flat_{\parallel},\mathring{\mathbf{G}},\mathring{\boldsymbol{\Lambda}}) \\
	&= \hat{\Ws}(X,\FS^*\mathbf{C}^\flat_{\parallel},\FS^*\mathring{\mathbf{G}},
	\FS^*\mathring{\boldsymbol{\Lambda}}) \\
	&= \hat{\Ws}(X,\bar{\mathbf{C}}^\flat,\mathring{\bar{\mathbf{G}}},\mathring{\bar{\boldsymbol{\Lambda}}})
	\,,
\end{aligned}
\end{equation}
where $\mathring{\bar{\boldsymbol{\Lambda}}}=\FS^*\mathring{\boldsymbol{\Lambda}}$ is a set of surface structural tensors that characterize the surface anisotropy of the material. 

Let us consider only membrane deformations and denote the surface integrity basis by $\bar{I}_j, ~j=1,\dots,m_s$. Thus, $\Ws=\Wbars(X,\bar{I}_1,...,\bar{I}_{m_s})$. The second Piola-Kirchhoff surface stress tensor is written as
\begin{equation}
	\bar{\mathbf{S}}=2\frac{\partial \Ws}{\partial \bar{\mathbf{C}}^\flat}
	=\sum_{j=1}^{m}2\Ws_j\frac{\partial \bar{I}_j}{\partial\bar{\mathbf{C}}^\flat}\,,\qquad 
	\Ws_j=\Ws_j(X,\bar{I}_1,...,\bar{I}_{m_s}):=\frac{\partial \Wbars}{\partial \bar{I}_j}\,, 
	\qquad j=1,\dots,m_s\,.
\end{equation}
Recall that $\bar{\mathbf{P}}=\bar{\mathbf{F}}\bar{\mathbf{S}}$ and $\bar{\boldsymbol{\sigma}}=\bar{J}^{-1} \bar{\mathbf{F}}\bar{\mathbf{S}}\bar{\mathbf{F}}^\star$.

\paragraph{Isotropic material surfaces.}
For an isotopic material surface, the surface energy function depends on the two principal invariants of $\bar{\mathbf{C}}^{\flat}$, i.e., $\Ws=\Wbars(X,\bar{I}_1,\bar{I}_2)$, where 
\begin{equation} 
	\bar{I}_1 =\operatorname{tr}_{\mathring{\bar{\mathbf{G}}}}\bar{\mathbf{C}}^{\flat}
	=\bar{C}_{\bar{A}\bar{B}}\,\mathring{\bar{G}}^{\bar{A}\bar{B}}\,, \qquad
	\bar{I}_2 =\frac{\det\bar{\mathbf{C}}^{\flat}}{\det\mathring{\bar{\mathbf{G}}}}
	\,.
\end{equation}

First, we show that $\bar{I}_1 = I_1(\mathbf{C}^\flat_\parallel)$.
Recall that $\mathbf{C}_{\parallel}^{\flat}=\mathbf{F}_{\parallel}^{\star}\,\mathbf{g}\,\mathbf{F}_{\parallel}$ is a symmetric $(0,2)$ tensor on $T\SurE \subset T\mathcal{B}$, while $\bar{\mathbf{C}}^{\flat}=\bar{\mathbf{F}}^{\star}\bar{\mathbf{g}}$ is a symmetric $(0,2)$ tensor on $T\Sur$. They are related as $\bar{\mathbf{C}}^{\flat} = \FS^{\star}\mathbf{C}_{\parallel}^{\flat}$.
In components, relative to material surface coordinates $\{\bar{X}^{\bar{A}}\}$ and bulk coordinates $\{X^A\}$, this reads $\bar{C}_{\bar{A}\bar{B}}= \cFS^{A}{}_{\bar{A}}\,(C_{\parallel})_{AB}\,\cFS^{B}{}_{\bar{B}}$.
The reference surface metric $\mathring{\bar{\mathbf{G}}}$ is related to the bulk reference metric $\mathring{\mathbf{G}}$ by $\mathring{\bar{\mathbf{G}}} = \FS^{\star}\mathring{\mathbf{G}}\FS$, or in components $\mathring{\bar{G}}_{\bar{A}\bar{B}}
	= \cFS^{A}{}_{\bar{A}}\,\mathring{G}_{AB}\,\cFS^{B}{}_{\bar{B}}$.
The first invariant of $\bar{\mathbf{C}}^{\flat}$ is defined as
\begin{equation}
	\bar{I}_1
	=\operatorname{tr}_{\mathring{\bar{\mathbf{G}}}}\bar{\mathbf{C}}^{\flat}
	=\bar{G}^{\bar{A}\bar{B}}\,\bar{C}_{\bar{A}\bar{B}}\,.
\end{equation}
Substituting the above component relations gives us $\bar{I}_1 = \bar{G}^{\bar{A}\bar{B}}\,	\cFS^{A}{}_{\bar{A}}\,(C_{\parallel})_{AB}\,\cFS^{B}{}_{\bar{B}}$.
Using the identity \eqref{Surface-G-Inverse}, in components we have $\cFS^{A}{}_{\bar{A}}\,\bar{G}^{\bar{A}\bar{B}}\,\cFS^{B}{}_{\bar{B}}=\mathring{G}^{AB}-N^{A}N^{B}$. Hence
\begin{equation}
	\bar{I}_1
	=(\mathring{G}^{AB}-N^{A}N^{B})\,C^{\parallel}_{AB}\,.
\end{equation}
This is precisely the definition of the first invariant of $\mathbf{C}_{\parallel}^{\flat}$ with respect to the projected metric $\mathring{\mathbf{G}}_\parallel$ defined in \eqref{Projected-Metric}, i.e.,
\begin{equation}
	I_1(\mathbf{C}^\flat_\parallel)
	=\operatorname{tr}_{\mathring{\mathbf{G}}_\parallel}\mathbf{C}_{\parallel}^{\flat}
	=(\mathring{G}^{AB}-N^{A}N^{B})\,C^{\parallel}_{AB}\,.
\end{equation}
Therefore, $\bar{I}_1 = I_1(\mathbf{C}^\flat_\parallel)$.

Second, we show that $\bar{I}_2=\bar{J}^{\,2}$, where surface Jacobian was defined in \eqref{Surface-Jacobian}.
Fix $X\in\Sur$ and let $\{\bar{X}^{\bar{A}}\}$, $\bar{A}=1,2$, be a local coordinate chart on the material surface $\Sur$.  
Let $\{\bar{\mathbf{E}}_{\bar{A}}\}$ denote the associated tangent basis of $T_X\Sur$, i.e., $\bar{\mathbf{E}}_{\bar{A}}=\partial/\partial\bar{X}^{\bar{A}}|_X$.  
Also, let $\{\bar{x}^{\bar{a}}\}$, $\bar{a}=1,2$, be a local coordinate chart on the deformed surface $\sur$, with associated tangent basis $\{\bar{\mathbf{e}}_{\bar{a}}\}$, $\bar{\mathbf{e}}_{\bar{a}}=\partial/\partial \bar{x}^{\bar{a}}|_x$.
In the chosen coordinates, the reference area element on $\Sur$ is written as
\begin{equation}
	dA_s = \sqrt{\det[\mathring{\bar{\mathbf{G}}}_{\bar{A}\bar{B}}(X)]}\;
	d\bar{X}^1 d\bar{X}^2= A_0(X)\,d\bar{X}^1 d\bar{X}^2\,,
\end{equation}
where $A_0^2(X)=\det\big[\mathring{\bar{\mathbf{G}}}(\bar{\mathbf{E}}_{\bar{\alpha}},\bar{\mathbf{E}}_{\bar{\beta}})\big]
=\det[\mathring{\bar{\mathbf{G}}}_{\bar{A}\bar{B}}(X)]$ is the squared area density in the reference configuration, expressed in the chosen local coordinates.  
Similarly, the deformed area element, expressed in the same material coordinates, is obtained by pulling back the spatial surface metric via $\bar{\mathbf{F}}$:
\begin{equation}
	da_s = \sqrt{\det[(\bar{\mathbf{F}}^{*}\bar{\mathbf{g}})_{\bar{A}\bar{B}}(X)]}\;
	d\bar{X}^1 d\bar{X}^2= A(X)\,d\bar{X}^1 d\bar{X}^2\,,
\end{equation}
where $A^2(X)=\det\big[(\bar{\mathbf{F}}^{*}\bar{\mathbf{g}})(\bar{\mathbf{E}}_{\bar{\alpha}},\bar{\mathbf{E}}_{\bar{\beta}})\big]
=\det[(\bar{\mathbf{F}}^{*}\bar{\mathbf{g}})_{\bar{A}\bar{B}}(X)]$.
For any oriented $2$-frame, the squared area equals the determinant of the metric matrix on that frame. This leads to the surface Jacobian, defined as the ratio of the deformed to undeformed area elements:
\begin{equation}
	\bar{J}(X)=\frac{da_s}{dA_s}=\frac{A(X)}{A_0(X)}
	=\left(\frac{\det[\bar{\mathbf{F}}^{*}\bar{\mathbf{g}}]}{\det[\mathring{\bar{\mathbf{G}}}]}\right)^{\tfrac12}\,.
\end{equation}
By definition of the surface right Cauchy-Green strain tensor 
$\bar{\mathbf{C}}^{\flat}=\bar{\mathbf{F}}^{*}\bar{\mathbf{g}}:T_X\Sur\to T_X\Sur$,
we can write
\begin{equation}
	\bar{J}(X)=\left(\frac{\det[\bar{\mathbf{C}}^{\flat}(X)]}{\det[\mathring{\bar{\mathbf{G}}}(X)]}\right)^{\tfrac12}\,.
\end{equation}
This expression is independent of the particular choice of basis $\{\bar{\mathbf{E}}_{\bar{A}}\}$ and hence is an intrinsic scalar associated with the surface deformation.  
Therefore
\begin{equation}
	\bar{J}^{\,2}(X) =\frac{\det[\bar{\mathbf{C}}^{\flat}(X)]}{\det[\mathring{\bar{\mathbf{G}}}(X)]}	=\bar{I}_2(X)\,.
\end{equation}

For an isotropic material the Cauchy surface stress has the following representation\footnote{This follows from
\begin{equation}
	\bar{\mathbf{S}}=2\frac{\partial \hat{\Ws}}{\partial {\bar{\mathbf{C}}^\flat}}
	= 2\Wbars_1 \frac{\partial \bar{I}_1}{\partial {\bar{\mathbf{C}}^\flat}}
	+2\Wbars_1 \frac{\partial \bar{I}_2}{\partial {\bar{\mathbf{C}}^\flat}}
	= 2\Wbars_1 \mathring{\bar{\mathbf{G}}}+2 \bar{I}_2 \bar{\mathbf{C}}^{-\sharp}\,,
\end{equation}
and $\bar{\boldsymbol{\sigma}}=\frac{2}{\sqrt{\bar{I}_2}} \,\bar{\mathbf{F}}\,\bar{\mathbf{S}}\,\bar{\mathbf{F}}^{\star}$.} 
\begin{equation} \label{Surface-Stress-Isotropic}
	\bar{\boldsymbol{\sigma}} = \frac{2}{\sqrt{\bar{I}_2}} 
	\left[ \bar{I}_2\,\Wbars_2 \,\bar{\mathbf{g}}^{\sharp} 
	+\Wbars_1\,\bar{\mathbf{b}}^{\sharp} \right]\,,
\end{equation}
where $\bar{\mathbf{b}}^{\sharp}=\bar{\mathbf{F}}\,\bar{\mathbf{G}}^\sharp \,\bar{\mathbf{F}}^\star$.
For an incompressible material surface $\bar{I}_2=1$, and $\Ws=\Wbars(X,\bar{I}_1)$. Hence
\begin{equation} \label{Surface-Cauchy-Representation-Comp}
	\bar{\boldsymbol{\sigma}} = \big(-\bar{p}+2\Wbars_2\big) \,\bar{\mathbf{g}}^{\sharp} 
	+2\Wbars_1\,\bar{\mathbf{b}}^{\sharp} \,,
\end{equation}
where $\bar{p}$ is the Lagrange multiplier associated with the incompressibility constraint $\bar{J}=\sqrt{\bar{I}_2}=1$. Because $\bar{p}$ is an unknown a priori, $-\bar{p}+2\Wbars_2$ can be replaced by $-\bar{p}$, and hence
\begin{equation} \label{Surface-Cauchy-Representation-Incomp}
	\bar{\boldsymbol{\sigma}} = -\bar{p}\,\bar{\mathbf{g}}^{\sharp} 
	+2\Wbars_1\,\bar{\mathbf{b}}^{\sharp}\,.
\end{equation}

\begin{example} A compressible neo-Hookean elastic surface has the following energy function and constitutive equations
\begin{equation}
	\Wbars(\bar{I}_1,\bar{I}_2) = \frac{\mu_s}{2}\,\big(\bar{I}_1 - 2-\ln \bar{I}_2 \big)
	+ \frac{\kappa_s}{2}\left( \bar{I}_2^{\,\frac{1}{2}} - 1\right)^2\,,\qquad
	\bar{\boldsymbol{\sigma}} 
	= \mu_s \bar{I}_2^{\,-\frac{1}{2}}\big(\bar{\mathbf{b}}^\sharp -\bar{\mathbf{g}}^\sharp\big)
	+ \kappa_s \big(\bar{I}_2^{\,\frac{1}{2}}-1 \big)\,\bar{\mathbf{g}}^\sharp \,,
\end{equation}
where $\bar{\mu}$ and $\bar{\kappa}$ are, respectively, the surface shear and bulk moduli---two material constants.
\end{example}

\begin{example} An incompressible neo-Hookean elastic surface has the following energy function and constitutive equations
\begin{equation}
	\Wbars(\bar{I}_1)=\frac{\bar{\mu}}{2}(\bar{I}_1-2)\,,\qquad
	\bar{\boldsymbol{\sigma}} = -\bar{p}\,\bar{\mathbf{g}}^{\sharp} 
	+\bar{\mu}\,\bar{\mathbf{b}}^{\sharp} \,,
\end{equation}
where $\bar{\mu}$ is the surface shear modulus---a material constant.
\end{example}

\subsection{Balance laws of surface elasticity}

In this section we derive the governing equations of surface elasticity variationally using the Lagrange-d'Alembert principle. We will need the following results in our calculations.

\begin{defi}[Induced Bundle and Connection]
Given a deformation mapping $\varphi:\mathcal{B}\rightarrow\mathcal{C}$, one defines an induced vector bundle $\varphi^{-1}T\mathcal{C}$, which is a vector bundle over $\mathcal{B}$ whose fiber over $X\in\mathcal{B}$ is $T_{\varphi(X)}\mathcal{C}$ \citep{Nishikawa2002}. 
The Levi-Civita connection $\nabla^{\mathbf{g}}$ induces a unique connection $\nabla^{\varphi}$ on $\varphi^{-1}T\mathcal{C}$ defined as
\begin{equation} \label{Induced-Connection}
	\nabla^{\varphi}_{\mathbf{W}}(\mathbf{y}\circ\varphi)
	=\nabla^{\mathbf{g}}_{\varphi_*\mathbf{W}}\mathbf{y}
	\,,\qquad \forall \mathbf{W}\in T_X\mathcal{B},~\mathbf{y}\in \Gamma(T\mathcal{C})\,. 
\end{equation}	
The induced connection has connection coefficients $F^b{}_A\,\gamma^a{}_{bc}=\dfrac{\partial\varphi^b}{\partial X^A}\,\gamma^a{}_{bc}$ with respect to the coordinate charts $\{X^A\}$ and $\{x^a\}$ for $\mathcal{B}$ and $\mathcal{C}$, respectively. 
In coordinates, $(\nabla^{\varphi}(\mathbf{y}\circ\varphi))^a{}_{|A}=(\nabla^{\mathbf{g}}\mathbf{y})^a{}_{|b}\,F^b{}_A$.
\end{defi}

\paragraph{Covariant derivative of $\mathbf{F}$.}
The covariant derivative of the deformation gradient has the following coordinate representation
\begin{equation}
\begin{aligned}
	& \nabla^{\varphi}\mathbf{F}=F^a{}_{A|B}\,dX^B\otimes dX^A\otimes \frac{\partial}{\partial x^a}\,,\\
	& F^a{}_{A|B}=\frac{\partial F^a{}_A}{\partial X^B}+(F^b{}_B\gamma^a{}_{bc})F^c{}_A
	-\Gamma^C{}_{AB}F^a{}_C=\frac{\partial F^a{}_A}{\partial X^B}+\gamma^a{}_{bc}F^b{}_BF^c{}_A
	-\Gamma^C{}_{AB}F^a{}_C\,. 
\end{aligned}
\end{equation}	
It is straightforward to show that \citep[Lemma 3.3]{Nishikawa2002}
\begin{equation} \label{F-identities}
	(\nabla^{\varphi}\mathbf{F})(\mathbf{X},\mathbf{Y})
	=\nabla^{\varphi}_{\mathbf{X}}(\varphi_*\mathbf{Y})
	-\varphi_*\nabla^{\mathbf{G}}_{\mathbf{X}}\mathbf{Y},\qquad
	\nabla^{\varphi}_{\mathbf{X}}(\varphi_*\mathbf{Y})-\nabla^{\varphi}_{\mathbf{Y}}(\varphi_*\mathbf{X})
	=\varphi_*[\mathbf{X},\mathbf{Y}]
	\,. 
\end{equation}	
One can also show that \citep{Nishikawa2002} 
\begin{equation}
    \nabla^{\varphi}_{\mathbf{V}}\mathbf{F}= \nabla^{\varphi}_{\frac{\partial}{\partial t}}{\mathbf{F}}
    =\nabla^{\varphi}\mathbf{V} \,.
\end{equation}
This is proved as follows.
Deformation gradient is given as $\mathbf{F}_{t}=T \varphi_{t}$. Let us consider the vector fields $\Big(\dfrac{\partial}{\partial X^A},0\Big)$ and $\left(0,\dfrac{\partial}{\partial t}\right)$ on $\mathcal{B}\times I$, where $I$ is some interval. 
Notice that $\Big[\Big(\dfrac{\partial}{\partial X^A},0\Big),\Big(0,\dfrac{\partial}{\partial t}\Big) \Big]=0$. Therefore, using \eqref{F-identities} one writes
\begin{equation}
	\nabla_{\left(0,\frac{\partial}{\partial t}\right)}\, \varphi_{t*}
	\left(\frac{\partial}{\partial X^A},0\right)
	=\nabla_{\left(\frac{\partial}{\partial X^A},0\right)} \,\varphi_{t*}
	\left(0,\frac{\partial}{\partial t}\right).
\end{equation}
Thus
\begin{equation}
	\nabla_{\frac{\partial}{\partial t}} \,\frac{\partial \varphi_{t}^a}{\partial X^A}
	=\nabla_{\frac{\partial}{\partial X^A}} \,\frac{\partial \varphi_{t}^a}{\partial t}.
\end{equation}
This implies that $(\nabla^{\varphi}_{\mathbf{V}}\mathbf{F})^a{}_A=V^a{}_{|A}=F^b{}_A V^a{}_{|b}$, where $V^a{}_{|b}=\dfrac{\partial V^b}{\partial x^b}+\gamma^a{}_{bc}V^c$.

\subsubsection{The Lagrange-d'Alembert principle}

We will derive the governing equations variationally using the Lagrange-d'Alembert principle. 
The bulk Lagrangian density is defined as $\Lb = \Tb -W$, where $\Tb = \frac{1}{2}\rho_0\Vert\boldsymbol V\Vert^2_{\mathbf{g}}= \frac{1}{2}\rho_0 \llangle \mathbf{V},\mathbf{V}\rrangle_{\mathbf{g}}$ is the bulk kinetic energy density.
The surface Lagrangian density is defined as $\Ls = \Ts -\Ws$, where $\Ts = \frac{1}{2}\bar{\rho}_0\Vert\boldsymbol V\Vert^2_{\mathbf{g}}= \frac{1}{2}\bar{\rho}_0 \llangle \mathbf{V},\mathbf{V}\rrangle_{\mathbf{g}}$ is the surface kinetic energy density. Notice that on $\surE$, $\llangle \mathbf{V},\mathbf{V}\rrangle_{\mathbf{g}}=\llangle \mathbf{V}_\parallel,\mathbf{V}_\parallel\rrangle_{\mathbf{g}}+V_n^2$.

Variation of the deformation map $\varphi_t:\mathcal{B}\to\mathcal{S}$ is a map $\varphi_{t,\epsilon}:\mathcal{B}\times I\to\mathcal{S}$, where $I$ is some interval, such that $\varphi_{t,0}=\varphi_t$. 
For $X$ and $t$ fixed, let us denote with $\delta\varphi_t(X)$ the vector field tangent to the curve $\epsilon\mapsto\varphi_{t,\epsilon}(X)$ in $\mathcal S$ and evaluated at $\epsilon=0$:
\begin{equation} \label{Deformation-Variation}
	\delta \varphi_t(X) = \left[ (\varphi_t(X))_*  \tfrac{\partial}{\partial \epsilon} \right] \Big\vert_{\epsilon=0}
	=\frac{d}{d\epsilon}\Bigg|_{\epsilon=0}\varphi_{t,\epsilon}(X) \,.
\end{equation}

The Lagrange-d'Alembert principle tells us that the physical configuration of the body satisfies the following identity \citep{Lanczos1962,Goldstein2002,MarsRat2013}:
\begin{equation} \label{LD-Principle}
\begin{aligned}
	& \delta \int_{t_1}^{t_2}  \int_{\mathcal{B}\setminus\SurE} \Lb \, dV \, \mathrm dt 
	+\delta \int_{t_1}^{t_2}  \int_{\SurE} \Ls \, dA_s \, \mathrm dt 
	+ \int_{t_1}^{t_2} \int_{\mathcal{B}\setminus\SurE} \rho_0 \llangle \mathbf{B}, 
	\delta\varphi \rrangle_{\mathbf{g}} \, dV\, \mathrm dt +
	\int_{t_1}^{t_2} \int_{\SurE} \bar{\rho}_0 \llangle \Bs, 
	\delta\varphi \rrangle_{\mathbf{g}} \, dA_s\, \mathrm dt \\
	& +\int_{t_1}^{t_2} \int_{\partial_o\mathcal{B} } \llangle \mathbf{T} , 
	\delta\varphi \rrangle_{\mathbf{g}} \, dA \,\mathrm dt 
	+\int_{t_1}^{t_2} \int_{\partial \SurE} \llangle \Tracs , 
	\delta\varphi \rrangle_{\mathbf{g}} \, d\ell_s \,\mathrm dt  = 0 \,,
\end{aligned}
\end{equation}
for any variation field $\delta\varphi$,\footnote{It is assumed that 
\begin{equation} \label{Variation-t1-t2}
	\delta\varphi(X,t_1)=\delta\varphi(X,t_2)=0\,.
\end{equation}
}
where $\rho_0$ and $\bar{\rho}_0$ are the bulk and surface mass densities, respectively, $\mathbf{B}$ and $\mathbf{T}$ are, respectively, the body force per unit mass and the boundary traction per unit undeformed area, and $\Bs$ and $\Tracs$ are, respectively, the surface body force per unit mass and the surface boundary traction per unit undeformed length, and $dA_s$ and $d\ell_s$ are the elements of area and length on the material surface $\SurE$.
It is assumed that $\Tracs$ is tangential and is specified on part of the surface boundary that is denoted by $\partial_N \SurE$.
Note that $\llangle \Bs,  \delta\varphi \rrangle_{\mathbf{g}} = \llangle \Bs_\parallel,  \delta\varphi_\parallel \rrangle_{\mathbf{g}}+\cBs_n\,\delta \varphi_n$, where $\Bs_\parallel$ and $\cBs_n$ are the tangential and normal parts of the surface body force.

The variation of the material velocity is given by
\begin{equation} 
	\mathbf{V}_{\!\epsilon}(X,t) 
	= \dfrac{\partial \varphi_{t,\epsilon}(X)}{\partial t} \in T{\varphi_{t,\epsilon}(X)}\mathcal{S}\,,
\end{equation}
meaning that for different values of $\epsilon$, the velocity lies in different tangent spaces. Consequently, to obtain the velocity variation field, a covariant derivative along the curve $\epsilon \mapsto \varphi_{t,\epsilon}(X)$ must be used \citep{MaHu1983}. Thus,
\begin{equation}
	\delta\mathbf{V}(X,t)
	=\nabla_{\!\frac{\partial}{\partial \epsilon}}\frac{\partial \varphi_{t,\epsilon}(X)}{\partial t}\Big|_{\epsilon=0}
	=\nabla_{\!\frac{\partial}{\partial t}}\frac{\partial \varphi_{t,\epsilon}(X)}{\partial \epsilon}\Big|_{\epsilon=0}
	=D^{\mathbf{g}}_t \delta\varphi(X,t)
	\,,
\end{equation}
where $D^{\mathbf{g}}_t$ is the covariant time derivative.
In the above calculation, the second equality follows from the symmetry lemma of Riemannian geometry \citep{Lee1997}. In components, $\left(\delta V\right)^a=\partial \delta\varphi^a/\partial t+\gamma^a{}_{bc}\,V^b\,\delta\varphi^c$. 

Note that
\begin{equation} 
\begin{aligned}
	\delta \int_{t_1}^{t_2}  \int_{\mathcal{B}\setminus\SurE} \Tb \, dV \, \mathrm dt 
	& = \int_{t_1}^{t_2}  \int_{\mathcal{B}\setminus\SurE} \rho_0 
	\llangle \mathbf{V},\delta\mathbf{V}\rrangle_{\mathbf{g}} \, dV \, \mathrm dt  \\
	& = \int_{t_1}^{t_2}  \int_{\mathcal{B}\setminus\SurE}  
	\left[ \frac{d}{dt} \left(\rho_0  \llangle \mathbf{V},\delta\varphi\rrangle_{\mathbf{g}} \right)-
	\llangle \rho_0\mathbf{A},\delta\varphi \rrangle_{\mathbf{g}} \right] dV \, \mathrm dt  \\
	& = -\int_{t_1}^{t_2}  \int_{\mathcal{B}\setminus\SurE}  
	\llangle \rho_0\mathbf{A},\delta\varphi \rrangle_{\mathbf{g}} \, dV \, \mathrm dt
	\,,
\end{aligned}
\end{equation}
where in the third equality \eqref{Variation-t1-t2} was used. Similarly,
\begin{equation} 
\begin{aligned}
	\delta \int_{t_1}^{t_2}  \int_{\SurE} \Ts \, dA_s \, \mathrm dt 
	& = \int_{t_1}^{t_2}  \int_{\SurE} \bar{\rho}_0 
	\llangle \mathbf{V},\delta\mathbf{V}\rrangle_{\mathbf{g}} \, dA_s \, \mathrm dt  \\
	& = \int_{t_1}^{t_2}  \int_{\SurE}  
	\left[ \frac{d}{dt} \left(\bar{\rho}_0  \llangle \mathbf{V},\delta\varphi\rrangle_{\mathbf{g}} \right)-
	\llangle \bar{\rho}_0 \mathbf{A},\delta\varphi \rrangle_{\mathbf{g}} \right] dA_s \, \mathrm dt  \\
	& = -\int_{t_1}^{t_2}  \int_{\SurE}  
	\llangle \bar{\rho}_0\mathbf{A},\delta\varphi \rrangle_{\mathbf{g}} \, dA_s \, \mathrm dt \\
	& = -\int_{t_1}^{t_2}  \int_{\SurE}  
	\left( \llangle \bar{\rho}_0 \mathbf{A}_\parallel,\delta\varphi_\parallel \rrangle_{\mathbf{g}} 
	+\bar{\rho}_0 A_n \delta\varphi_n \right) dA_s \, \mathrm dt
	\,,
\end{aligned}
\end{equation}
where on $\surE$, $ \delta\varphi = \delta\varphi_\parallel + \delta\varphi_\perp= \delta\varphi_\parallel + \delta\varphi_n\,\mathbf{n}$, and $\mathbf{A}_\parallel$ and $A_n$ are the tangential and normal accelerations, respectively.

The deformation gradient of the motion $\varphi_{t,\epsilon}$ is given by $\mathbf{F}_{t,\epsilon}=\dfrac{\partial \varphi_{t,\epsilon}}{\partial X}$. Consider the vector fields $\left(\partial/\partial X^A,0\right)$ and $\left(0,\partial/\partial \epsilon \right)$ on $\mathcal{B} \times I$. Since these vector fields commute, i.e., $\left[\left(\partial/\partial X^A,0\right),\left(0,\partial/\partial \epsilon \right) \right] = 0$, we can use \eqref{F-identities} to obtain the following relation: 
\begin{equation}
	\nabla_{\left(0,\frac{\partial}{\partial \epsilon}\right)} \,\varphi_{t,\epsilon*}
	\left(\frac{\partial}{\partial X^A},0\right)
	=\nabla_{\left(\frac{\partial}{\partial X^A},0\right)} \,\varphi_{t,\epsilon*}
	\left(0,\frac{\partial}{\partial \epsilon}\right)\,.
\end{equation}
Thus
\begin{equation}
	\nabla_{\frac{\partial}{\!\partial \epsilon}} \,\frac{\partial \varphi_{\epsilon}^a}{\partial X^A}
	=\nabla_{\frac{\partial}{\partial X^A}} \,\frac{\partial \varphi_{\epsilon}^a}{\partial \epsilon}\,.
\end{equation}
Therefore
\begin{equation}
	\delta\mathbf{F} =\nabla^\varphi \delta\varphi \,.
\end{equation}
Hence, in components, we have $\delta F^a{}_A = \delta\varphi^a{}_{|A} =  \delta\varphi^a{}_{|b}\,F^b{}_A$, where $\delta\varphi^a{}_{|b} = \dfrac{\partial \delta\varphi^b}{\partial x^b} + \gamma^a{}_{bc}\, \delta\varphi^c$. Thus,
\begin{equation}
	\delta\mathbf{F} =\nabla \delta\varphi \cdot\mathbf{F}\,.
\end{equation}

For the perturbed motion $\varphi_{t,\epsilon}$, $W_{\epsilon}:=W(X,\mathbf{F}_{\epsilon},\mathring{\mathbf{G}},\mathbf{g}\circ\varphi_{\epsilon})$, and hence,
\begin{equation}
	\delta W
	= \frac{\partial W_{\epsilon}}{\partial \mathbf{F}_{\epsilon}}
	\!:\! \nabla_{\!\frac{\partial}{\partial \epsilon}}\mathbf{F}_{\epsilon}\Big|_{\epsilon=0}
	+\frac{\partial W_{\epsilon}}{\partial \mathbf{g}\circ\varphi_{\epsilon}}
	\!:\! \nabla_{\frac{\partial}{\partial \epsilon}}\mathbf{g}\circ\varphi_{\epsilon}\Big|_{\epsilon=0}
	=\frac{\partial W}{\partial \mathbf{F}}\!:\! \nabla \delta\varphi \,,
\end{equation}
where, in the second equality the geometric $\omega$-lemma \citep{MaHu1983}, and in the last equality the metric compatibility of the Levi-Civita connection ($\nabla^{\mathbf{g}}\mathbf{g}=\mathbf{0}$) was used.
Thus
\begin{equation} \label{Bulk-L-Variation}
\begin{aligned}
	\delta \int_{t_1}^{t_2}  \int_{\mathcal{B}\setminus\SurE} -W \, dV \, \mathrm dt 
	& = -\int_{t_1}^{t_2}  \int_{\mathcal{B}\setminus\SurE} \delta W \, dV \, \mathrm dt  \\
	& = -\int_{t_1}^{t_2}  \int_{\mathcal{B}\setminus\SurE}  
	\frac{\partial W}{\partial \mathbf{F}}\!:\! \nabla \delta\varphi \, dV \, \mathrm dt  \\
	& = \int_{t_1}^{t_2}  \int_{\mathcal{B}\setminus\SurE}  \left[ 
	\operatorname{Div} \left( \frac{\partial W}{\partial \mathbf{F}} \right)\cdot \delta\varphi
	-\operatorname{Div} \left( \frac{\partial W}{\partial \mathbf{F}}\cdot \delta\varphi  \right)
	\right] dV \, \mathrm dt
	\,.
\end{aligned}
\end{equation}
Recall that $\SurE=\sqcup_{i=1}^{m}\SurE_i$ (see Fig.~\ref{Reference-Current-Configurations}). The bulk body $\mathring{\mathcal{B}}$ has $m+1$ connected components: 
\begin{equation} 
	\mathring{\mathcal{B}}= \bigsqcup_{i=1}^{m+1}  \mathcal{B}_i
	\,.
\end{equation}
For $i=1,\cdots,m$, $\partial \mathcal{B}_i=\SurE_i$. We assume that $\partial \SurE_i = \emptyset$, $i=1,\cdots,m$.
These $m$ connected components are called \textit{inclusions}. Inclusions are assumed to have the energy functions $\Wi$, $i=1,\cdots m$. In a particular case, the inclusions are made of the same hyperelastic fluid. The $(m+1)$-th connected component of $\mathring{\mathcal{B}}$, i.e., $\mathcal{B}_{m+1}$ has the energy function $W$. Also, note that
\begin{equation} 
	\partial \mathcal{B}_{m+1}=
	\partial_o\mathcal{B} \sqcup \bigsqcup_{i=1}^{m} (-\SurE_i)\,,
\end{equation}
where $-\SurE_i$ denotes the surface $\SurE_i$ with its orientation reversed and $\partial_o\mathcal{B}$ is the outer boundary of $\mathcal{B}$ (see Fig.~\ref{Reference-Current-Configurations}).
Using the divergence theorem we have
\begin{equation} 
\begin{aligned}
	\int_{\mathcal{B}\setminus\SurE}  
	\operatorname{Div} \left( \frac{\partial W}{\partial \mathbf{F}}\cdot \delta\varphi  \right) dV 
	& =\sum_{i=1}^{m+1} \int_{\mathcal{B}_i}  
	\operatorname{Div}\! \left( \frac{\partial W}{\partial \mathbf{F}}\cdot \delta\varphi  \right) dV \\
	 & =\sum_{i=1}^{m} \int_{\SurE_i}  
	 \left \llangle \mathbf{g}^\sharp\frac{\partial \Wi}{\partial \mathbf{F}} \mathbf{N}, 
	 \delta\varphi   \right \rrangle_{\mathbf{g}} \, dA_s \\
	 & \qquad +\int_{\partial_o\mathcal{B} }  
	 \left \llangle \mathbf{g}^\sharp\frac{\partial W}{\partial \mathbf{F}} \mathbf{N}, 
	 \delta\varphi   \right \rrangle_{\mathbf{g}} \, dA
	 -\sum_{i=1}^{m} \int_{\SurE_i}  
	 \left \llangle \mathbf{g}^\sharp\frac{\partial W}{\partial \mathbf{F}} \mathbf{N}, 
	 \delta\varphi   \right \rrangle_{\mathbf{g}} \, dA_s \\
	 & =  \int_{\partial_o\mathcal{B} }  
	 \left \llangle \mathbf{g}^\sharp\frac{\partial W}{\partial \mathbf{F}} \mathbf{N}, 
	 \delta\varphi   \right \rrangle_{\mathbf{g}} \, dA
	 +\sum_{i=1}^{m} \int_{\SurE_i}  
	 \left \llangle \mathbf{g}^\sharp\left(\frac{\partial \Wi}{\partial \mathbf{F}}\mathbf{N}
	 -\frac{\partial W}{\partial \mathbf{F}} \mathbf{N} \right) , 
	 \delta\varphi   \right \rrangle_{\mathbf{g}} \, dA_s \\
	 & =  \int_{\partial_o\mathcal{B} }  
	 \left \llangle \mathbf{g}^\sharp\frac{\partial W}{\partial \mathbf{F}} \mathbf{N}, 
	 \delta\varphi   \right \rrangle_{\mathbf{g}} \, dA
	 -\sum_{i=1}^{m} \int_{\SurE_i}  
	 \left \llangle \mathbf{g}^\sharp
	 \left\llbracket \frac{\partial W}{\partial \mathbf{F}}\mathbf{N} \right\rrbracket, 
	 \delta\varphi   \right \rrangle_{\mathbf{g}} \, dA_s
	\,,
\end{aligned}
\end{equation}
where on $\SurE_i$\footnote{$\llbracket A \rrbracket$ denotes the jump of the enclosed quantity $A$ across the interface, with $\mathbf{N}$ the unit normal pointing toward the exterior. The jump is defined as $\llbracket A \rrbracket = A^{+}-A^{-}$, where $A^{+}$ and $A^{-}$ are the exterior and interior limiting values, respectively.}
\begin{equation} 
	\left\llbracket \frac{\partial W}{\partial \mathbf{F}} \mathbf{N}\right\rrbracket
	=\frac{\partial W}{\partial \mathbf{F}}\mathbf{N}-\frac{\partial \Wi}{\partial \mathbf{F}}\mathbf{N}
	\,,
\end{equation}
where on $\SurE_i$, $\mathbf{N}$ points from the inclusion toward the bulk.
It should be noted that the material metric $\mathbf{G}$ is, in general, discontinuous across the interface between the inclusion and the surrounding bulk. Consequently, the unit normal $\mathbf{N}$, defined as a $\mathbf{G}$-unit vector, is not necessarily continuous across the interface.

Therefore, \eqref{Bulk-L-Variation} is now simplified as
\begin{equation} 
\begin{aligned}
	\delta \int_{t_1}^{t_2}  \int_{\mathcal{B}\setminus\SurE} -W \, dV \, \mathrm dt 
	& =  \int_{t_1}^{t_2}  \int_{\mathcal{B}\setminus\SurE}  \left[ 
	\operatorname{Div} \left( \frac{\partial W}{\partial \mathbf{F}} \right)\cdot \delta\varphi
	-\operatorname{Div} \left( \frac{\partial W}{\partial \mathbf{F}}\cdot \delta\varphi  \right)
	\right] dV \, \mathrm dt \\
	& = \int_{t_1}^{t_2}  \int_{\mathcal{B}\setminus\SurE}   
	\operatorname{Div} \left( \frac{\partial W}{\partial \mathbf{F}} \right)\cdot \delta\varphi
	 \,dV \, \mathrm dt
	 -\int_{t_1}^{t_2} \int_{\partial_o\mathcal{B} }  
	 \left \llangle \mathbf{g}^\sharp\frac{\partial W}{\partial \mathbf{F}} \mathbf{N}, 
	 \delta\varphi   \right \rrangle_{\mathbf{g}} \, dA \, \mathrm dt \\
	 & \qquad +  \int_{t_1}^{t_2}  \sum_{i=1}^{m} \int_{\SurE_i}  
	 \left \llangle \mathbf{g}^\sharp
	 \left\llbracket \frac{\partial W}{\partial \mathbf{F}} \mathbf{N} \right\rrbracket, 
	 \delta\varphi   \right \rrangle_{\mathbf{g}} \, dA_s\, \mathrm dt	\,.
\end{aligned}
\end{equation}
Note that on $\surE_i$, $\delta\varphi = \delta\varphi_\parallel + \delta\varphi_\perp= \delta\varphi_\parallel + \delta\varphi_n\,\mathbf{n}$, and hence
\begin{equation} 
\begin{aligned}
	\delta \int_{t_1}^{t_2}  \int_{\mathcal{B}\setminus\SurE} -W \, dV \, \mathrm dt 
	& = \int_{t_1}^{t_2}  \int_{\mathcal{B}\setminus\SurE}    
	 \left \llangle \operatorname{Div}\mathbf{P}, \delta\varphi   \right \rrangle_{\mathbf{g}} 
	 \,dV \, \mathrm dt
	 -\int_{t_1}^{t_2} \int_{\partial_o\mathcal{B} }  
	 \left \llangle \mathbf{P} \mathbf{N}, \delta\varphi   \right \rrangle_{\mathbf{g}} \, dA \, \mathrm dt \\
	 & \qquad +  \int_{t_1}^{t_2}  \sum_{i=1}^{m} \int_{\SurE_i}  
	 \left( 
	 \left \llangle \left\llbracket \mathbf{P}  \mathbf{N} \right\rrbracket,  \delta\varphi_\parallel   \right \rrangle_{\mathbf{g}}
	 +\left \llangle \left\llbracket \mathbf{P}  \mathbf{N} \right\rrbracket, \mathbf{n} \right \rrangle_{\mathbf{g}} \delta\varphi_n
	 \right)
	 \, dA_s\, \mathrm dt	\,,
\end{aligned}
\end{equation}
where $\mathbf{P}=\mathbf{g}^\sharp \dfrac{\partial W}{\partial \mathbf{F}}$ is the first Piola-Kirchhoff stress tensor.\footnote{For an incompressible bulk material ($J=1$), $\mathbf{P}= -p\,\mathbf{F}^{-1}+\mathbf{g}^\sharp \dfrac{\partial W}{\partial \mathbf{F}}$, where $p$ is the Lagrange multiplier corresponding to the internal constraint $J=1$.}

\paragraph{Flows and Lie derivatives.}  
Let $\chi : \mathcal{B} \times \mathbb{R} \to \mathcal{B}$ be a smooth one-parameter family of diffeomorphisms satisfying $\chi(X,0) = X$. The vector field $\mathbf{U}(X,t)$ generating the flow is defined as the velocity of the curve $t \mapsto \chi(X,t)$, i.e., its components are given by $U^A(X,t) = \partial \chi^A(X,t) / \partial t $. For fixed $t$, we write $\chi_t = \chi(\cdot, t)$, and define the relative flow map
\begin{equation}
	\chi^t_s = \chi_s \circ \chi_t^{-1} : \chi_t(\mathcal{B}) \to \chi_s(\mathcal{B}) \,,
\end{equation}
so that $\chi_s = \chi^t_s \circ \chi_t$ and $\chi^s_s = \operatorname{id}_{\mathcal{B}}$.  
Let $\mathbf{T}_t$ be a time-dependent tensor field defined on $\chi_t(\mathcal{B})$. The \emph{non-autonomous Lie derivative} of $\mathbf{T}_t$ along $\mathbf{U}_t$ is defined by
\begin{equation} \label{non-lie-def}
	\mathbf{L}_{\mathbf{U}_t} \mathbf{T}_t =
	\left. \frac{d}{ds} \left[ \left( \chi^t_s \right)^* \mathbf{T}_s \right] \right|_{s=t} \,,
\end{equation}
where $(\chi^t_s)^* \mathbf{T}_s$ denotes the pullback of $\mathbf{T}_s$ by $\chi^t_s$.  
The \emph{autonomous Lie derivative} of the family $\mathbf{T}_t$, viewed as a fixed tensor field along the flow, is defined by
\begin{equation} \label{lie-def}
	\mathfrak{L}_{\mathbf{U}_t} \mathbf{T}_t =
	\left. \frac{d}{ds} \left[ \left( \chi^t_s \right)^* \mathbf{T}_t \right] \right|_{s=t} \,.
\end{equation}
Using the chain rule, one obtains the following identity:
\begin{equation} \label{non-aut-lie}
	\mathbf{L}_{\mathbf{U}_t} \mathbf{T}_t =
	\partial_t \mathbf{T}_t + \mathfrak{L}_{\mathbf{U}_t} \mathbf{T}_t \,.
\end{equation}

Next we need to simplify the contribution of the surface energy to the variational principle:
\begin{equation} 
\begin{aligned}
	\delta \int_{t_1}^{t_2}  \int_{\SurE} -W_{\!s} \, dA_s \, \mathrm dt 
	& = -\int_{t_1}^{t_2}  \int_{\SurE} \delta W_{\!s} \, dA_s \, \mathrm dt 
	& = -\int_{t_1}^{t_2}  \sum_{i=1}^{m} \int_{\SurE_i} \delta W_{\!s} \, dA_s \, \mathrm dt 	
	\,.
\end{aligned}
\end{equation}
Recalling that $\Ws=W_{\!s}(X,\mathbf{F}_\SurE,\mathring{\mathbf{G}},\mathbf{g})$, the variation of the surface elastic energy (a scalar field) is calculated using either Lie derivative or covariant derivative
\begin{equation}
	\delta W_{\!s}
	= \frac{\partial W_{\!s}}{\partial {\mathbf{F}_\SurE}}\!:\!\mathbf{L}_{\delta\varphi}{\mathbf{F}_\SurE}
	+ \frac{\partial W_{\!s}}{\partial \mathbf{g}}\!:\!\mathbf{L}_{\delta\varphi}\mathbf{g}
	= \frac{\partial W_{\!s}}{\partial {\mathbf{F}_\SurE}}\!:\!\nabla_{\delta\varphi}{\mathbf{F}_\SurE}
	\,,
\end{equation}
where metric compatibility of the Levi-Civita was used.

For an arbitrary time-independent material vector field $\mathbf{U}$, one has
\begin{equation}\label{LF=0}
	\mathbf{L}_{\delta\varphi}{\mathbf{F}}\mathbf{U}
	= \left[\frac{d}{d\epsilon}\left(\varphi_{\epsilon}\circ\varphi_{s}^{-1}\right)^* 
	{\mathbf{F}}_{\epsilon} \mathbf{U}\right]_{s=\epsilon}
	=\left[\frac{d}{d\epsilon} \varphi_{s*} \varphi_{\epsilon}^* \varphi_{\epsilon*} 
	\mathbf{U}\right]_{s=\epsilon}
	=\left[\frac{d}{d\epsilon}\varphi_{s*}\mathbf{U}\right]_{s=\epsilon}=\mathbf{0}\,.
\end{equation}
This implies that $\mathbf{L}_{\delta\varphi}{\mathbf{F}} = \mathbf{0}$.  
Recall that $\mathbf{F}_\SurE(X)= \mathbf{F}\big |_{\SurE}(X)\circ \pi_{\SurE}$, where $\pi_{\SurE}:T\mathcal{B}\to T\SurE$ is defined as $\pi_{\SurE}=\operatorname{id}_{T\mathcal{B}} - \mathbf{N} \otimes \mathbf{N}^{\flat}$.
Using the product rule for the Lie derivative, and noticing that $\mathbf{L}_{\delta\varphi} \pi_{\SurE}=\mathbf{0}$, we obtain
\begin{equation}
	\mathbf{L}_{\delta\varphi} \mathbf{F}_\SurE
	= \mathbf{F}\circ \left( \mathbf{L}_{\delta\varphi} \pi_{\SurE} \right)
	+ \mathbf{L}_{\delta\varphi} \mathbf{F} \circ \pi_{\SurE}
	= \mathbf{0} \,.
\end{equation}
Therefore, 
\begin{equation} \label{Surface-Energy-Variation}
	\delta W_{\!s} =  \frac{\partial W_{\!s}}{\partial \mathbf{g}}\!:\!\mathbf{L}_{\delta\varphi}\mathbf{g}
	\,.
\end{equation}

\begin{lem}
The Lie derivative of the spatial metric along the variation field has the following form
\begin{equation}
	\mathbf{L}_{\delta\varphi}\mathbf{g} 
	= \mathbf{L}_{\delta\varphi_\parallel}\mathbf{g} + 2\delta\varphi_n\, \mathbf{k}
	= \mathfrak{L}_{\delta\varphi_\parallel}\mathbf{g} + 2\delta\varphi_n\, \mathbf{k} \,.
\end{equation}
\end{lem}
\begin{proof}
Using the bilinearity of the Lie derivative and the Leibniz rule, we write (note that the background metric $\mathbf{g}$ is time independent, and hence its autonomous and non-autonomous Lie derivatives coincide)
\begin{equation}
	\mathbf{L}_{\delta\varphi} \mathbf{g} 
	= \mathbf{L}_{\delta\varphi_\parallel} \mathbf{g}
	+ \mathbf{L}_{(\delta\varphi_n \mathbf{n})} \,\mathbf{g}
	= \mathfrak{L}_{\delta\varphi_\parallel} \mathbf{g}
	+ \mathfrak{L}_{(\delta\varphi_n \mathbf{n})} \,\mathbf{g} \,.
\end{equation}
The first term is the standard Lie derivative of $\mathbf{g}$ along a surface tangent vector field.
To compute the second term, we note that for two tangent vectors $\mathbf{X}_\parallel, \mathbf{Y}_\parallel \in T\surE$,
\begin{equation}
	 \left( \mathfrak{L}_{(\delta\varphi_n \mathbf{n})} \,\mathbf{g} \right)(\mathbf{X}_\parallel, \mathbf{Y}_\parallel)
	= \delta\varphi_n \, \left( \mathfrak{L}_{\mathbf{n}} \mathbf{g} \right)(\mathbf{X}_\parallel, \mathbf{Y}_\parallel)
	+ \mathbf{X}_\parallel [\delta\varphi_n]\, \llangle \mathbf{n}, \mathbf{Y}_\parallel \rrangle_{\mathbf{g}}
	+ \mathbf{Y}_\parallel [\delta\varphi_n]\, \llangle \mathbf{X}_\parallel, \mathbf{n} \rrangle_{\mathbf{g}} \,.
\end{equation}
But $\llangle \mathbf{n}, \mathbf{Y}_\parallel \rrangle_{\mathbf{g}} = \llangle \mathbf{X}_\parallel, \mathbf{n} \rrangle_{\mathbf{g}} = 0$ since $\mathbf{X}_\parallel, \mathbf{Y}_\parallel \in T\surE$, so we obtain
\begin{equation}
	\left( \mathfrak{L}_{\delta\varphi_n \mathbf{n}} \mathbf{g} \right)(\mathbf{X}_\parallel, \mathbf{Y}_\parallel)
	= \delta\varphi_n \, \left( \mathfrak{L}_{\mathbf{n}} \mathbf{g} \right)(\mathbf{X}_\parallel, \mathbf{Y}_\parallel) \,.
\end{equation}
Now, the Lie derivative of the metric along $\mathbf{n}$ applied to two tangent vectors gives twice the second fundamental form:
\begin{equation}
	\left( \mathfrak{L}_{\mathbf{n}} \mathbf{g} \right)(\mathbf{X}_\parallel, \mathbf{Y}_\parallel)
	= \llangle \nabla^{\mathbf{g}}_{\mathbf{X}_\parallel} \mathbf{n}, \mathbf{Y}_\parallel \rrangle_{\mathbf{g}}
	+ \llangle \mathbf{X}_\parallel, \nabla^{\mathbf{g}}_{\mathbf{Y}_\parallel} \mathbf{n} \rrangle_{\mathbf{g}}
	= 2\, \llangle \nabla^{\mathbf{g}}_{\mathbf{X}_\parallel} \mathbf{n}, \mathbf{Y}_\parallel \rrangle_{\mathbf{g}}
	= 2\, \mathbf{k}(\mathbf{X}_\parallel, \mathbf{Y}_\parallel) \,.
\end{equation}
Therefore, $\mathfrak{L}_{\delta\varphi_n \mathbf{n}} \mathbf{g} = 2 \,\delta\varphi_n\, \mathbf{k}$, and so the total Lie derivative becomes:
\begin{equation} \label{Metric-Lie-Derivative}
	\mathbf{L}_{\delta\varphi} \mathbf{g}
	= \mathfrak{L}_{\delta\varphi_\parallel} \mathbf{g} + 2\,\delta\varphi_n\, \mathbf{k} \,.
\end{equation}
\end{proof}

Now, let us assume only membrane energy, and hence $\Ws=\hat{\Ws}(X,\bar{\mathbf{C}}^{\flat},\mathring{\bar{\mathbf{G}}})$. 
We know that $\bar{\mathbf{C}}^{\flat}=\bar{\mathbf{F}}^\star \,\bar{\mathbf{g}} \,\bar{\mathbf{F}}$, $\bar{\mathbf{g}}=\Fs^\star \mathbf{g} \Fs$, and hence $\bar{\mathbf{C}}^{\flat}= (\Fs\bar{\mathbf{F}})^\star \,\mathbf{g} \, \Fs\bar{\mathbf{F}}$. Therefore,
\begin{equation} \label{Surface-Energy-Derivative}
	\frac{\partial W_{\!s}}{\partial \mathbf{g}} 
	=  \frac{\partial \hat{\Ws}}{\partial \bar{\mathbf{C}}^\flat}\!:\!\frac{\partial \bar{\mathbf{C}}^\flat}{\partial \mathbf{g}} 
	= \Fs\bar{\mathbf{F}}  \frac{\partial \hat{\Ws}}{\partial \bar{\mathbf{C}}^\flat} (\Fs\bar{\mathbf{F}})^\star
	= \Fs\bar{\mathbf{F}}  \frac{\partial \hat{\Ws}}{\partial \bar{\mathbf{C}}^\flat} \bar{\mathbf{F}}^\star \Fs^\star
	= \iota_{\sur*} \Big(\bar{\mathbf{F}} \frac{\partial \hat{\Ws}}{\partial \bar{\mathbf{C}}^\flat}
	\bar{\mathbf{F}}^\star \Big)
	= \iota_{\sur*} \Big(\frac{\partial \hat{\Ws}}{\partial \bar{\mathbf{g}}} \Big)
	\,.
\end{equation}
Substituting \eqref{Metric-Lie-Derivative} into \eqref{Surface-Energy-Variation} and using \eqref{Surface-Energy-Derivative} we obtain
\begin{equation} 
\begin{aligned}
	\delta \Ws
	&= \frac{\partial W_{\!s}}{\partial \mathbf{g}}\!:\! \mathfrak{L}_{\delta\varphi_\parallel} \mathbf{g}
	+2\,\delta\varphi_n\, \frac{\partial W_{\!s}}{\partial \mathbf{g}}\!:\! \mathbf{k} \\
	& = \iota_{\sur*}\Big( \frac{\partial \hat{\Ws}}{\partial \bar{\mathbf{g}}} \Big)\!:\! 
	\mathfrak{L}_{\delta\varphi_\parallel} \mathbf{g}
	+2\,\delta\varphi_n\,\iota_{\sur*}\Big( \frac{\partial \hat{\Ws}}{\partial \bar{\mathbf{g}}} \Big)\!:\! \mathbf{k}\\
	&= \frac{\partial \hat{\Ws}}{\partial \bar{\mathbf{g}}} \!:\! 
	\mathfrak{L}_{\iota_{\sur}^*\delta\varphi_\parallel} \bar{\mathbf{g}}
	+2\,\delta\varphi_n\, \frac{\partial \hat{\Ws}}{\partial \bar{\mathbf{g}}} \!:\! \bar{\mathbf{k}}\\
	&= 2\frac{\partial \hat{\Ws}}{\partial \bar{\mathbf{g}}} \!:\! \bar{\nabla}(\iota_{\sur}^*\delta\varphi_\parallel)
	+2\,\delta\varphi_n\, \frac{\partial \hat{\Ws}}{\partial \bar{\mathbf{g}}} \!:\! \bar{\mathbf{k}}
	\,,
\end{aligned}
\end{equation}
where $\bar{\mathbf{k}}=\iota_{\sur}^*\mathbf{k}$ and the fourth equality follows from $\mathfrak{L}_{\bar{\mathbf{u}}}\bar{\mathbf{g}}=\bar{\nabla}\bar{\mathbf{u}}+(\bar{\nabla}\bar{\mathbf{u}})^{\star}$ for the Levi-Civita connection $\bar{\nabla}$ corresponding to the metric $\bar{\mathbf{g}}$.
Note that 
\begin{equation}
	2\frac{\partial \hat{\Ws}}{\partial \bar{\mathbf{g}}} 
	= 2 \bar{\mathbf{F}} \,\frac{\partial \hat{\Ws}}{\partial \bar{\mathbf{C}}^{\flat}} \,\bar{\mathbf{F}}^\star
	= \bar{\mathbf{F}} \bar{\mathbf{S}} \,\bar{\mathbf{F}}^\star
	= \bar{\mathbf{P}} \bar{\mathbf{F}}^\star
	\,.
\end{equation}
Thus,
\begin{equation} 
\begin{aligned}
	\delta \Ws
	&= \bar{\mathbf{P}} \bar{\mathbf{F}}^\star \!:\! \bar{\nabla}(\iota_{\sur}^*\delta\varphi_\parallel)
	+ \delta\varphi_n\, \bar{\mathbf{P}} \bar{\mathbf{F}}^\star \!:\! \bar{\mathbf{k}}\\
	&= \bar{\mathbf{P}} \!:\! \bar{\nabla}_0(\iota_{\sur}^*\delta\varphi_\parallel)
	+ \delta\varphi_n\, \bar{\mathbf{P}} \bar{\mathbf{F}}^\star \!:\! \bar{\mathbf{k}}\\
	&= \overline{\operatorname{Div}}\Big( \bar{\mathbf{P}} \cdot\iota_{\sur}^*\delta\varphi_\parallel \Big)
	-\overline{\operatorname{Div}} \bar{\mathbf{P}} 	\cdot\iota_{\sur}^*\delta\varphi_\parallel
	+ \delta\varphi_n\, \bar{\mathbf{P}} \bar{\mathbf{F}}^\star \!:\! \bar{\mathbf{k}}
	\,,
\end{aligned}
\end{equation}
where $\bar{\nabla}_0$ is the Levi-Civita connection corresponding to the metric $\mathring{\bar{\mathbf{G}}}$ and $\overline{\operatorname{Div}}$ is the surface material divergence.
Therefore,
\begin{equation} 
\begin{aligned}
	& \delta \int_{t_1}^{t_2}  \int_{\SurE}  -W_{\!s} \, dA_s \, \mathrm dt 
	= \int_{t_1}^{t_2}  \sum_{i=1}^{m} \int_{\SurE_i} -\delta W_{\!s} \, dA_s \, \mathrm dt  \\
	& = \int_{t_1}^{t_2}  \sum_{i=1}^{m} \int_{\SurE_i} \left[
	-\overline{\operatorname{Div}}\Big( \bar{\mathbf{P}} \cdot\iota_{\sur}^*\delta\varphi_\parallel \Big)
	+\overline{\operatorname{Div}} \bar{\mathbf{P}} 	\cdot\iota_{\sur}^*\delta\varphi_\parallel
	- \delta\varphi_n\, \bar{\mathbf{P}} \bar{\mathbf{F}}^\star \!:\! \bar{\mathbf{k}}
	\right] dA_s \, \mathrm dt \\
	& = \int_{t_1}^{t_2}  \sum_{i=1}^{m}  \left\{
	-\int_{\partial \SurE_i} \left \llangle \iota_{\sur*}\big(\bar{\mathbf{P}}\bar{\mathbf{N}}^\flat\big), 
	\delta\varphi_\parallel
	\right\rrangle_{\mathbf{g}} \, d\ell
	+\int_{\SurE_i} \Big[ \left \llangle \iota_{\sur*}\left(\overline{\operatorname{Div}} \bar{\mathbf{P}} \right), 
	\delta\varphi_\parallel \right\rrangle_{\mathbf{g}}
	- \delta\varphi_n\, \iota_{\sur*}\!\left(\bar{\mathbf{P}} \bar{\mathbf{F}}^\star\right) \!:\! \mathbf{k}	\Big] dA_s \, 
	\right\}\mathrm dt    
	\,.
\end{aligned}
\end{equation}

The variational principle \eqref{LD-Principle} is now simplified to read
\begin{equation} \label{LD-Principle-Simplified}
\begin{aligned}
	& \int_{t_1}^{t_2}  \int_{\mathcal{B}\setminus\SurE}    
	 \left \llangle \operatorname{Div}\mathbf{P}+\rho_0(\mathbf{B}
	 -\mathbf{A}), \delta\varphi   \right \rrangle_{\mathbf{g}} 
	 \,dV \, \mathrm dt 
	 +\int_{t_1}^{t_2} \int_{\partial_o\mathcal{B} } \llangle \mathbf{T}-\mathbf{P} \mathbf{N} , 
	\delta\varphi \rrangle_{\mathbf{g}} \, dA \,\mathrm dt \\
	 & \qquad +  \int_{t_1}^{t_2}  \sum_{i=1}^{m} \int_{\SurE_i}  
	 \left \llangle \iota_{\sur*}\!\left(\overline{\operatorname{Div}} \bar{\mathbf{P}} \right)
	 +\left\llbracket \mathbf{P}  \mathbf{N} \right\rrbracket
	 +\bar{\rho}_0(\Bs_\parallel- \mathbf{A}_\parallel),  \delta\varphi_\parallel   \right \rrangle_{\mathbf{g}}
	 \, dA_s\, \mathrm dt	\\
	 & \qquad +  \int_{t_1}^{t_2}  \sum_{i=1}^{m} \int_{\SurE_i}  
	\left( \left \llangle \left\llbracket \mathbf{P}  \mathbf{N} \right\rrbracket, \mathbf{n} \right \rrangle_{\mathbf{g}} 
	- \iota_{\sur*}\!\left(\bar{\mathbf{P}} \bar{\mathbf{F}}^\star\right) \!:\! \mathbf{k}	+\bar{\rho}_0(\cBs_n-A_n)
	\right)\delta\varphi_n
	 \, dA_s\, \mathrm dt	\\
	 & \qquad+ \int_{t_1}^{t_2}  \sum_{i=1}^{m}  \left\{
	\int_{\partial \SurE_i} \left \llangle \Tracs - \iota_{\sur*}\big(\bar{\mathbf{P}}\bar{\mathbf{N}}^\flat\big) ,
	\delta\varphi_\parallel \right\rrangle_{\mathbf{g}}\, d\ell 
	\right\}\mathrm dt   	=0 \,.
\end{aligned}
\end{equation}

The outer boundary of the body is assumed to be the disjoint union of the Dirichlet and Neumann boundaries, expressed as $\partial_o\mathcal{B} = \partial_D\mathcal{B} \sqcup \partial_N\mathcal{B}$, where $\partial_D \mathcal{B}$ is the Dirichlet boundary on which motion is specified, and hence $\delta\varphi|_{\partial_D \mathcal{B}}=0$, and $\partial_N\mathcal{B}$ denotes the Neumann boundary.

\subsubsection{The bulk and surface governing equations}

The variational principle \eqref{LD-Principle-Simplified} gives us the following bulk and surface Euler-Lagrange equations and Neumann boundary conditions:
\begin{empheq}[left={\empheqlbrace }]{align} 
	\label{EL-Compressible-1}
	& \operatorname{Div} \mathbf{P}+\rho_0\mathbf{B} = \rho_0\mathbf{A}\,, && 
	\text{in~}\mathcal{B}\setminus\SurE\,,\\[2pt]
	\label{EL-Compressible-2}
	& \mathbf{P}\mathbf{N} = \mathbf{T}\,, && \text{on~}\partial_N \mathcal{B}\,, \\[2pt]
	\label{EL-Compressible-3}
	& \iota_{\sur*}\!\left(\overline{\operatorname{Div}} \bar{\mathbf{P}} \right)
	+ \left\llbracket \mathbf{P} \mathbf{N} \right\rrbracket_\parallel
	+ \bar{\rho}_0 \Bs_\parallel = \bar{\rho}_0 \mathbf{A}_\parallel\,, && \text{on~}\SurE_i\,,~ i=1,\cdots,m\,, \\[2pt]
	\label{EL-Compressible-4}
	& \left \llangle \left\llbracket \mathbf{P} \mathbf{N} \right\rrbracket, \mathbf{n} \right \rrangle_{\mathbf{g}}
	- \iota_{\sur*}\!\left(\bar{\mathbf{P}} \bar{\mathbf{F}}^\star\right)\!:\!\mathbf{k}
	+ \bar{\rho}_0 \cBs_n = \bar{\rho}_0 A_n\,, && \text{on~}\SurE_i\,,~ i=1,\cdots,m\,, \\[2pt]
	\label{EL-Compressible-5}
	& \iota_{\sur*}\big(\bar{\mathbf{P}}\bar{\mathbf{N}}^\flat\big) = \Tracs\,, && 
	\text{on~} \partial_N \SurE_i\,,~ i=1,\cdots,m\,.
\end{empheq}
where $\left\llbracket \mathbf{P} \mathbf{N} \right\rrbracket_\parallel$ is the tangential bulk traction jump across the material surface.
Eqs.~\eqref{EL-Compressible-1} and \eqref{EL-Compressible-2} are the standard bulk balance of linear momentum and the associated natural boundary condition, respectively. Eq.~\eqref{EL-Compressible-3} represents the tangential surface balance of linear momentum, Eq.~\eqref{EL-Compressible-4} is the normal surface balance of linear momentum, and Eq.~\eqref{EL-Compressible-5} specifies the surface natural boundary condition.\

Recall that $\bar{\mathbf{P}} \bar{\mathbf{F}}^\star=\bar{J} \bar{\boldsymbol{\sigma}}$, and hence Eq.~\eqref{EL-Compressible-4}  can be rewritten as\footnote{For a spherical interface with inner pressure $p_{\text{in}}$ and outer pressure $p_{\text{out}}$, one has $\left\llangle\left\llbracket\mathbf{P}\mathbf{N}\right\rrbracket,\mathbf{n}\right\rrangle_{\mathbf{g}}=p_{\text{in}}-p_{\text{out}}$. For an isotropic surface tension $\gamma_0$ ($\bar{\boldsymbol{\sigma}}=\gamma_0\,\bar{\mathbf{g}}^\sharp$) on a sphere of radius $r_i$ ($\bar{\mathbf{k}}=\bar{\mathbf{g}}/r_i$), one has $\bar{J}\,\bar{\boldsymbol{\sigma}}\!:\!\bar{\mathbf{k}}=2\gamma_0/r_i$. Substituting into \eqref{Normal-Equilibrium-Equation} with $\bar{\rho}_0\cBs_n=\bar{\rho}_0 A_n=0$ gives $p_{\text{in}}-p_{\text{out}}=2\gamma_0/r_i$, i.e., \eqref{Normal-Equilibrium-Equation} is consistent with Laplace’s law \citep{Laplace1805,DeGennes2003}. Physically, a tensile surface stress pulls inward on each surface patch, and this inward resultant together with the outer pressure balances the outward action of the inner pressure.}
\begin{equation} \label{Normal-Equilibrium-Equation}
	\left \llangle \left\llbracket \mathbf{P} \mathbf{N} \right\rrbracket, \mathbf{n} \right \rrangle_{\mathbf{g}}
	- \bar{J} \bar{\boldsymbol{\sigma}} \!:\! \bar{\mathbf{k}}
	+ \bar{\rho}_0 \cBs_n = \bar{\rho}_0 A_n\,,\qquad \text{on~}\SurE_i\,,~ i=1,\cdots,m
	\,.
\end{equation}

\section{Bulk and Surface Anelasticity} \label{Sec:SurfaceAnelasticity}

In this section, the surface elasticity theory of the previous section is extended to take into account anelastic effects, both in the bulk and on the material surfaces. When elasticity is formulated in a Riemannian geometric setting, extending it to anelasticity is straightforward, as all the anelastic effects are encoded in the geometry of the material manifold. We will see that this is the case for surface anelasticity as well.

\subsection{Bulk and surface material metrics} 

Let us consider a body $\mathcal{B}$ with elastic surfaces $\Sur$. The bulk body $\mathring{\mathcal{B}}=\mathcal{B}\setminus \SurE$ is assumed to have a distribution of finite \textit{eigenstrains}.\footnote{\emph{Eigenstrain} is a hybrid German–English term originating in the pioneering paper of Hans Reissner \citep{Reissner1931} (Eigenspannung is the German term for self stress). This notion was later adopted and widely used by Mura \citep{Kinoshita1971,Mura1982}. In the mechanics literature the same idea appears under several equivalent names, e.g., \emph{initial strain} \citep{Kondo1949}, \emph{nuclei of strain} \citep{Mindlin1950}, \emph{transformation strain} \citep{Eshelby1957}, \emph{inherent strain} \citep{Ueda1975}, and \emph{residual strain} \citep{ambrosi2019growth}. For infinite bodies in linear elasticity, the first systematic treatment of eigenstrains and their induced stresses is due to \citet{Eshelby1957}. Extensions to nonlinear elasticity are due to \citet{DianiParks2000,Yavari2013}, \citet{Golgoon2018a}, and \citet{Yavari2021}.} Eigenstrains are modeled by anelastic distortions $\Fa$ as follows. The deformation gradient at $X\in\mathcal{B}$ is assumed to be multiplicatively decomposed as $\mathbf{F}(X)=\Fe(X)\Fa(X)$,\footnote{For a historical account of this multiplicative decomposition and other possibilities see \citep{Sadik2017,YavariSozio2023}} where $\Fe(X):T_X\mathcal{B}\to T_x\mathcal{C}$ is the elastic distortion and $\Fa:T_X\mathcal{B}\to T_X\mathcal{B}$ is the anelastic distortion. The natural stress-free distances are measured by the metric $\mathbf{G}=\Fa^*\mathring{\mathbf{G}}=\Fa^\star\mathring{\mathbf{G}}\Fa$, which is called the \textit{material metric} \citep{YavariGoriely2013Inclusions,Yavari2021}.
In the absence of external forces, the natural distances within the body are determined by its material metric. In general, a global stress-free configuration cannot be realized in the Euclidean ambient space, since the Riemannian material manifold cannot be isometrically embedded in it. This geometric incompatibility arises from the non-flatness of the material metric and leads to the presence of residual stresses induced by the eigenstrain distribution.\footnote{This idea is due to \citet{Eckart1948}. He argued that classical elasticity rests on two restrictive assumptions: a fixed, load-independent relaxed configuration and the existence of a globally Euclidean stress-free state. Motivated by earlier geometric insights \citep{Eisenhart1926}, he proposed replacing global relaxability by a local notion of relaxability. In doing so he recognized the natural role of Riemannian geometry in anelasticity and modeled anelastic strains by introducing a Riemannian material metric.}

The material surfaces are endowed with their own eigenstrain distributions.
The surface deformation gradient is assumed to have the multiplicative decomposition $\bar{\mathbf{F}}=\FeS\FaS$, where $\FeS:T_X\Sur\to T_x\sur$ and $\FaS:T_X\Sur\to T_X\Sur$ are the elastic and anelastic surface distortions, respectively.
The surface material metric is defined as 
\begin{equation}
	\bar{\mathbf{G}}
	=\FaS^*\mathring{\bar{\mathbf{G}}}
	=\FaS^\star\mathring{\bar{\mathbf{G}}}\FaS
	=\FaS^\star \FS^\star \mathring{\mathbf{G}} \,\FS \FaS
	=(\FS \FaS)^\star \mathring{\mathbf{G}} \,\FS \FaS
	\,.
\end{equation}
In components, $\bar{G}_{\bar{A}\bar{B}}=\bar{F}^{\bar{M}}{}_{\bar{A}} \,\cFS^A{}_{\bar{M}} \,\mathring{G}_{AB}\,\cFS^B{}_{\bar{N}} \, \bar{F}^{\bar{N}}{}_{\bar{B}}$.
It should be emphasized that $\bar{\mathbf{G}}\neq \mathring{\bar{\mathbf{G}}}$, i.e., the surface material metric and the first fundamental form are different, in general, in the presence of surface eigenstrains.

\begin{remark}
In an anelastic body with anelastic material surfaces, the material metric may have discontinuities across such surfaces. This is not surprising, as several examples of discontinuous material metrics have already appeared in the literature, particularly in the study of inclusions \citep{YavariGoriely2013Inclusions,Golgoon2018a}.
There are also examples of discontinuous ambient space metrics in the literature \citep{YavariOzakinSadik2016}. Moreover, time-dependent ambient space metrics have been used in the modeling of lipid membrane mechanics, for example in \citep{Arroyo2009}.
\end{remark}

\subsection{Bulk and surface constitutive equations in anelasticity} 

In this section we first briefly review the formulation of constitutive equations of bulk hyper-anelasticity. This is followed by formulating the surface constitutive equations in the presence of surface eigenstrains.

\subsubsection{Bulk constitutive equations in the presence of eigenstrains} 

In hyper-anelasticity, it is assumed that in the bulk at every point there is an energy function that explicitly depends on the elastic distortion, i.e., for $X\in\mathring{\mathcal{B}}$, $W=W(X,\Fe,\mathring{\mathbf{G}},\mathring{\boldsymbol{\Lambda}},\mathbf{g})$, 
where $\mathring{\boldsymbol{\Lambda}}$ is a set of structural tensors that characterize the bulk anisotropy of the material and $\mathring{\mathbf{G}}$ is the flat metric of the reference configuration. 
Objectivity implies that $W=\hat{W}(X,\Ce^{\flat},\mathring{\mathbf{G}},\mathring{\boldsymbol{\Lambda}})$, where $\Ce^\flat=\Fe^*\mathbf{g}=\Fe^\star\mathbf{g}\Fe$.
When structural tensors are included as arguments of the energy function, the resulting energy function is isotropic, that is, materially covariant, with respect to its arguments. This, in particular, implies that
\begin{equation} 
	W=W(X,\Fe,\mathring{\mathbf{G}},\mathring{\boldsymbol{\Lambda}},\mathbf{g})
	=W(X,\Fa^*\Fe,\Fa^*\mathring{\mathbf{G}},\Fa^*\mathring{\boldsymbol{\Lambda}},\mathbf{g})
	=W(X,\mathbf{F},\mathbf{G},\boldsymbol{\Lambda},\mathbf{g})
	=\hat{W}(X,\mathbf{C}^{\flat},\mathbf{G},\boldsymbol{\Lambda})
	\,,
\end{equation}
where $\boldsymbol{\Lambda}=\Fa^*\mathring{\boldsymbol{\Lambda}}$ denotes the set of \textit{bulk anelastic structural tensors}.
In other words, the energy function depends on the total deformation gradient when the material metric and the anelastic structural tensors are used \citep{YavariSozio2023}.
The second Piola–Kirchhoff stress tensor admits the representation given in \eqref{Second-PK-Stress-Representation}, with the only difference that the isotropic invariants are computed using the material metric $\mathbf{G}$ rather than the flat metric $\mathring{\mathbf{G}}$.
Again, $\mathbf{P}=\mathbf{F}\mathbf{S}$ and $\boldsymbol{\sigma}=J^{-1} \mathbf{F}\mathbf{S}\mathbf{F}^\star$ but in this case $J$ explicitly depends on the material metric, see \eqref{Jacobian-Anelasticity}.

\paragraph{Isotropic solids.}
If in the absence of eigenstrain a solid is isotropic, its energy function depends only on the principal invariants of $\mathbf{C}^{\flat}$, i.e., $W=\overline{W}(X,I_1,I_2,I_3)$, where
\begin{equation} 
\begin{aligned}
	I_1 &=\operatorname{tr}_{\mathbf{G}}\mathbf{C}^{\flat}
	=\mathbf{C}^{\flat}\!:\!\mathbf{G}^{\sharp}
	=C_{AB}\,G^{AB}\,, \\
	I_2 &=\frac{1}{2}\left[I_1^2-\operatorname{tr}_{\mathbf{G}}\mathbf{C}^2\right]
	=\frac{1}{2}\left(I_1^2-C_{MB}\,C_{NA}\,G^{AM} G^{BN}\right)\,, \\
	I_3 &=\frac{\det\mathbf{C}^{\flat}}{\det \mathbf{G}}
	\,.
\end{aligned}
\end{equation}
It should be emphasized that the principal invariants are computed with respect to the material metric $\mathbf{G}$ rather than the flat reference configuration metric $\mathring{\mathbf{G}}$.  
The Cauchy stress tensor has the representations given in \eqref{Cauchy-Representation-Isotropic-Comp} and \eqref{Cauchy-Representation-Isotropic-Incomp} for the compressible and incompressible cases, respectively, except that the principal invariants are now evaluated using the material metric $\mathbf{G}$ instead of $\mathring{\mathbf{G}}$.  
Other symmetry classes can be treated analogously to their treatment in standard nonlinear elasticity.

\subsubsection{Surface constitutive equations in the presence of eigenstrains} 

We neglect bending deformations and consider only membrane deformations for anelastic surfaces.
For a hyper-anelastic surface, surface energy explicitly depends on the elastic surface distortion, i.e., 
\begin{equation} 
	\Ws=\Ws(X,\FeS,\mathring{\bar{\mathbf{G}}},\mathring{\bar{\boldsymbol{\Lambda}}},\bar{\mathbf{g}})\,.
\end{equation}
With structural tensors included, the surface energy function is an isotropic function of its argument (materially covariant). In particular, one can write
\begin{equation} 
	\Ws
	=\Ws(X,\FaS^*\FeS,\FaS^*\mathring{\bar{\mathbf{G}}},\FaS^*\mathring{\bar{\boldsymbol{\Lambda}}},\bar{\mathbf{g}})
	=\Ws(X,\bar{\mathbf{F}},\bar{\mathbf{G}},\bar{\boldsymbol{\Lambda}},\bar{\mathbf{g}})
	\,,
\end{equation}
where $\bar{\boldsymbol{\Lambda}}=\FaS^*\mathring{\bar{\boldsymbol{\Lambda}}}$ is the set of \textit{surface anelastic structural tensors}.
Objectivity implies that
\begin{equation} 
	\Ws=\hat{\Ws}(X,\bar{\mathbf{C}}^{\flat},\bar{\mathbf{G}},\bar{\boldsymbol{\Lambda}})
	\,.
\end{equation}

For isotropic solids, the surface Cauchy stress representations given in \eqref{Surface-Cauchy-Representation-Comp} and \eqref{Surface-Cauchy-Representation-Incomp} for compressible and incompressible cases, respectively, still hold for hyper-anelasticity as long as the two principal invariants are calculated using the surface material metric, i.e.,
\begin{equation} 
	\bar{I}_1 =\operatorname{tr}_{\bar{\mathbf{G}}}\bar{\mathbf{C}}^{\flat}
	=\bar{C}_{\bar{A}\bar{B}}\,\bar{G}^{\bar{A}\bar{B}}\,, \qquad
	\bar{I}_2 =\frac{\det\bar{\mathbf{C}}^{\flat}}{\det\bar{\mathbf{G}}}
	\,.
\end{equation}

\begin{defi}[Projected Material Metric]
Let us consider arbitrary $\mathbf{U}, \mathbf{W}\in T\mathcal{B}\big|_{\SurE}$. The \textit{projected material metric} $\mathring{\mathbf{G}}_\parallel$ is defined such that
\begin{equation}
\begin{aligned}
	\left\llangle \mathbf{U} , \mathbf{W} \right\rrangle_{\mathbf{G}_\parallel}
	&=\left\llangle \bar{\pi}_{\Sur}(\mathbf{U}) , \bar{\pi}_{\Sur}(\mathbf{W}) \right\rrangle_{\bar{\mathbf{G}}} \\
	&=\left\llangle \bar{\pi}_{\Sur}(\mathbf{U}) , \bar{\pi}_{\Sur}(\mathbf{W}) \right\rrangle_{\iota_{\Sur}^*\mathbf{G}}\\
	&=\left\llangle \iota_{\Sur*}\bar{\pi}_{\Sur}(\mathbf{U}) 
	, \iota_{\Sur*}\bar{\pi}_{\Sur}(\mathbf{W}) \right\rrangle_{\mathbf{G}} \\
	&=\left\llangle \pi_{\SurE}(\mathbf{U}) , \pi_{\SurE}(\mathbf{W}) \right\rrangle_{\mathbf{G}}\\
	&=\left\llangle \mathbf{U}-\langle \mathbf{N}^\flat,\mathbf{U} \rangle  \mathbf{N}
	, \mathbf{W}-\langle \mathbf{N}^\flat,\mathbf{W} \rangle  \mathbf{N} \right\rrangle_{\mathbf{G}} \\
	&=\left\llangle \mathbf{U} , \mathbf{W} \right\rrangle_{\mathbf{G}}
	-\langle \mathbf{N}^\flat,\mathbf{U} \rangle \langle \mathbf{N}^\flat,\mathbf{W} \rangle \\
	&=\left\llangle \mathbf{U} , \mathbf{W} \right\rrangle_{\mathbf{G}-\mathbf{N}^\flat\otimes \mathbf{N}^\flat}
	\,.
\end{aligned}
\end{equation}
Therefore
\begin{equation} \label{Projected-Material-Metric}
	\mathbf{G}_\parallel = \mathbf{G}-\mathbf{N}^\flat\otimes \mathbf{N}^\flat
	\,.
\end{equation}
Note that $\mathbf{N}$ is a $\mathbf{G}$-unit vector, i.e, $\left\llangle \mathbf{N} , \mathbf{N} \right\rrangle_{\mathbf{G}}=1$.
The projected material metric \eqref{Projected-Material-Metric} is the anelastic analogue of the projected metric defined in \eqref{Projected-Metric}.
\end{defi}

Similar to surface elasticity, it is straightforward to show that 
\begin{equation} 
	\bar{I}_1 = I_1(\mathbf{C}^\flat_\parallel)=\operatorname{tr}_{\mathbf{G}_\parallel}\mathbf{C}_{\parallel}^{\flat}\,,
	\qquad
	\bar{I}_2 = I_2(\mathbf{C}^\flat_\parallel)=\frac{\det\bar{\mathbf{C}}^{\flat}}{\det\bar{\mathbf{G}}}
	\,.
\end{equation}

\begin{remark}
It has been known that the surface stress in solids is strain dependent \citep{Shuttleworth1950,Cammarata1994,Spaepen2000,Huang2006,Xu2018,Krichen2019}.
In a deformable solid, the appropriate material input is a \emph{surface constitutive equation}, that is, a relation between the surface stress and the surface deformation (and also surface eigenstrains). 
The classical notion of a constant surface tension is only an idealization, corresponding to a very particular choice of surface energy and to a restricted class of deformations. 
In the present framework, the surface Cauchy stress is obtained by differentiating the surface energy density $\Wbars$ with respect to the surface metric, and the \emph{effective} surface tension depends on the current geometry and the surface eigenstrain. 
Thus, surface stress is an outcome of the surface constitutive law, not a prescribed scalar material parameter.
\end{remark}

\subsection{Balance laws of surface anelasticity} 

We assume that both the bulk and the material surfaces carry prescribed eigenstrains. In the geometric formulation adopted here, these eigenstrains are encoded directly in the bulk and surface material metrics. Once the material metrics are specified, the configuration of the body is kinematically incompatible in general, and the resulting residual stresses follow from the balance laws written with respect to these modified metrics.
The governing equations of anelasticity are obtained by replacing the geometric descriptors of the stress-free state---namely, the induced bulk metric and the surface first fundamental forms---with the corresponding bulk and surface material metrics in the variational formulation.
No additional assumptions are required: the structure of the theory is exactly the same as that of classical nonliear elasticity. In particular, the bulk and surface equilibrium equations, the balance of linear and angular momenta, and the natural boundary conditions follow from the variational principle, where the constitutive equations are written in terms of the bulk and surface material metrics. 
As a consequence, the balance laws of surface anelasticity retain the same form as those of surface elasticity, with the sole difference that all metric-dependent quantities are evaluated using the bulk and surface material metrics. In this sense, anelasticity simply replaces the geometric descriptors of the stress-free state. The resulting field equations are therefore identical to \eqref{EL-Compressible-1}-\eqref{EL-Compressible-5}, with the understanding that these equations are now written on a bulk–surface system endowed with its own intrinsic material metrics.

\begin{remark}
In surface anelasticity the material metric is, in general, discontinuous across a material surface. This should be kept in mind when interpreting the jump condition in \eqref{EL-Compressible-3}, where the term $\mathbf{P}\mathbf{N}$ appears. The normal vector $\mathbf{N}$ is a unit normal with respect to the corresponding material metric $\mathbf{G}$. 
Both the normal on the solid part $\mathbf{N}^{+}$ and the normal from the inclusion side $\mathbf{N}^{-}$ point away from the inclusion and into the bulk solid.
Since $\mathbf{N}^{+}$ and $\mathbf{N}^{-}$ are normalized with respect to different material metrics (those of the bulk solid and the inclusion, respectively), they are not necessarily equal, i.e., $\mathbf{N}^{+}\neq \mathbf{N}^{-}$. 
Consequently, the two sides of the jump in \eqref{EL-Compressible-3} involve normals that are both metric-dependent and, in general, not equal.
\end{remark}

\section{Radial Deformations of an Incompressible Spherical Shell Filled with a Compressible Hyperelastic Liquid} \label{Sec:Example}

Consider a spherical body that, in its undeformed and unstressed state, has an outer radius $R_o$. In the reference configuration, a concentric spherical cavity of radius $R_i$ is embedded at the center. The surrounding material is an incompressible isotropic elastic solid. The interface between the cavity and the surrounding material is modeled as a compressible isotropic elastic surface endowed with $2$D dilatational eigenstrains. In this problem $\Sur$ is a sphere of radius $R_i$ and $\partial\Sur=\emptyset$. We assume that in the absence of surface eigenstrains the material surface is a hyperelastic membrane.
We consider both a dry cavity and a cavity filled with a compressible hyperelastic fluid. There are no bulk or surface body forces. The spherical ball is under uniform pressure on its outer boundary, i.e., $\sigma^{rr}(R_o)=-p_0$.

\subsection{The spherical shell bulk body}

In the reference configuration we use spherical coordinates $(R,\Theta,\Phi)$, where $R \ge 0\,, \, 0 \le \Theta \le \pi\,, \, 0 \le \Phi < 2\pi$ (recall that $X = R \sin\Theta \cos\Phi\,, \, Y = R \sin\Theta \sin\Phi\,, \, Z = R \cos\Theta$). Here $\Theta$ is the polar angle measured from the positive $Z$-axis, and $\Phi$ is the azimuthal angle measured in the $XY$-plane from the positive $X$-axis. In the current configuration, spherical coordinates $(r,\theta,\phi)$ are used. The Euclidean spatial metric is written as
$\mathbf{g} = dr \otimes dr + r^{2}\, d\theta \otimes d\theta + r^{2}\sin^{2}\theta\, d\phi \otimes d\phi.$
Hence, $[g_{ab}] = \mathrm{diag}(1,\,r^{2},\,r^{2}\sin^{2}\theta).$
In the reference configuration, and in the absence of eigenstrains, the Euclidean reference metric is writren as
$\mathring{\mathbf{G}} = dR \otimes dR + R^{2}\, d\Theta \otimes d\Theta + R^{2}\sin^{2}\Theta\, d\Phi \otimes d\Phi.$
Hence, $[\mathring{G}_{AB}] = \mathrm{diag}(1,\,R^{2},\,R^{2}\sin^{2}\Theta)$. Thus
\begin{equation} \label{Metrics-Sphere}
	\mathbf{g} =
	\begin{bmatrix}
	1 & 0 & 0 \\
	0 & r^{2} & 0 \\
	0 & 0 & r^{2}\sin^{2}\theta
	\end{bmatrix}\,,\qquad
	\mathring{\mathbf{G}} =
	\begin{bmatrix}
	1 & 0 & 0 \\
	0 & R^2 & 0 \\
	0 & 0 & R^{2}\sin^{2}\Theta
	\end{bmatrix}
	\,.
\end{equation}

We consider radial deformations of the form\footnote{\label{Footote:Universal-Solid}
This corresponds to Family $4$ universal deformations \citep{Ericksen1954}. A universal deformation is one that can be maintained, in the absence of body forces, by boundary tractions alone for every member of a given material class. In this case, radial deformations can be maintained for any incompressible isotropic solid. \citet{Ericksen1954} established this result for homogeneous bodies, and the analysis was later extended to inhomogeneous \citep{Yavari2021Universal} and anisotropic bodies \citep{YavariGoriely2021Universal,Yavari2023universal}.
}
\begin{equation} \label{Deformation-Family4}
	r(R,\Theta,\Phi)=r(R)\,,\qquad \theta(R,\Theta,\Phi)=\Theta\,,\qquad \phi(R,\Theta,\Phi)=\Phi
	\,.
\end{equation}
The deformation gradient is written as
\begin{equation}
	\mathbf{F}=
	\begin{bmatrix}
	r'(R) & 0 & 0 \\
	0 & 1 & 0 \\
	0 & 0 & 1
	\end{bmatrix}
	\,.
\end{equation}
For $R_i\leq R\leq R_o$, the body is incompressible and hence $J=\dfrac{r^2(R)r'(R)}{R^2}=1$. Therefore, $r^3(R)=R^3+r_o^3-R_o^3$, where $r_o=r(R_o)$. Thus 
\begin{equation}
	r(R)=\left[R^3+R_o^3(\lambda_o^3-1)\right]^{\frac{1}{3}}
	\,,\qquad R_i\leq R \leq R_o\,,
\end{equation}
where $\lambda_o=\dfrac{r_o}{R_o}$ is the radial stretch. Let $r_i=r(R_i)$. Hence, $r_i=\left[R_i^3+R_o^3(\lambda_o^3-1)\right]^{\frac{1}{3}}$.\footnote{One can equivalently write
\begin{equation} \label{Radial-Kinematics}
	r(R)=\left[R^3+r_i^3-R_i^3\right]^{\frac{1}{3}}
	\,,\qquad R_i\leq R \leq R_o\,.
\end{equation}
} The principal invariants read
\begin{equation}
	I_1 = \frac{R^6 + 2 r^6(R)}{R^2 r^4(R)}\,,\qquad
	I_2 = \frac{2 R^6 + r^6(R)}{R^4 r^2(R)}\,.
\end{equation}

The non-zero Cauchy stress components are
\begin{equation}
\begin{aligned}
	\sigma^{rr}(R) &= -p + \frac{2 R^{4} W_1}{r^{4}(R)} - \frac{2 W_2\, r^{4}(R)}{R^{4}}\,,\\[4pt]
	\sigma^{\theta\theta}(R) &= -\frac{p}{r^{2}(R)} + \frac{2 W_1}{R^{2}} - \frac{2 R^{2} W_2}{r^{4}(R)}\,,\\[4pt]
	\sigma^{\phi\phi}(R) &= \left[-\frac{p}{r^{2}(R)} + \frac{2 W_1}{R^{2}} - \frac{2 R^{2} W_2}{r^{4}(R)}\right] \csc^{2}\Theta\,.
\end{aligned}
\end{equation}
The only non-trivial equilibrium equation is 
\begin{equation} \label{Radial-Equilibrium}
	\frac{\partial \sigma^{rr}}{\partial r}
	+\frac{2}{r}\sigma^{rr}-r\,\sigma^{\theta\theta}-(r\sin^2\theta)\,\sigma^{\phi\phi}=0 \,,
\end{equation}
which is simplified to read
\begin{equation}
\begin{aligned}
	\frac{\partial \sigma^{rr}(R)}{\partial R}
	&= - 2r'(R) \left[\frac{1}{r(R)}\sigma^{rr}(R)-r(R)\,\sigma^{\theta\theta}(R)\right] \\
	& = - \frac{2R^2}{r^2(R)} \left[\frac{1}{r(R)}\sigma^{rr}(R)-r(R)\,\sigma^{\theta\theta}(R)\right]\\
	&= \frac{4\left(r^6(R) - R^6\right)}{r^5(R)}\left[\frac{W_1}{r^2(R)} + \frac{W_2}{R^2}\right]
	\,.
\end{aligned}
\end{equation}
Thus,
\begin{equation} \label{Radial-Stress-1}
	\sigma^{rr}(R)= \sigma_i+\int_{R_i}^R f(\xi)\,d\xi\,,
\end{equation}
where 
\begin{equation}
	\sigma_i=\sigma^{rr}(R_i^+)\,,\qquad f(R)=\frac{4\left(r^6(R) - R^6\right)}{r^5(R)}\left[\frac{W_1}{r^2(R)} + \frac{W_2}{R^2}\right]\,.
\end{equation}
If $\sigma^{rr}(R_o)=-p_o$, we have
\begin{equation}  \label{Radial-Stress-2}
	\sigma_i = -p_o -\int_{R_i}^{R_o} f(\xi)\,d\xi
	\,.
\end{equation}
Note that $\mathbf{N}=\{ 1,0,0\}$, and hence
\begin{equation}\label{Traction-Bulk}
	\mathbf{P}\mathbf{N}=J\boldsymbol{\sigma}\mathbf{F}^{-\star}\mathbf{N}
	= \begin{bmatrix} 
	\dfrac{r_i^2}{R_i^2} \,\sigma_i  \\
	0\\
	0
	\end{bmatrix}
	\,.
\end{equation}

\subsection{The liquid inclusion}

The spherical cavity is assumed to be either dry or filled with a compressible hyperelastic fluid subject to internal pressure.
Consequently, a fluid-filled cavity can be regarded as an inclusion with a purely dilatational eigenstrain. The inclusion is assumed to be homogeneous and composed of the same liquid. As the temperature is taken to be fixed, \eqref{Fluid-Energy} reduces to $\Wf=\Wf(J)$, where
\begin{equation} \label{Jacobian-Liquid}
	J=\sqrt{\frac{\det\mathbf{g}}{\det\mathbf{G}}}\,\det\mathbf{F}\,,
\end{equation}
and $\mathbf{G}$ denotes the material metric of the liquid, expressed as
\begin{equation} \label{Metrics-liquid}
	\mathbf{G} = e^{2\Omega_l}
	\begin{bmatrix}
	1 & 0 & 0 \\
	0 & R^2 & 0 \\
	0 & 0 & R^2 \sin^2\Theta
	\end{bmatrix}
	\,,
\end{equation}
where $\Omega_l>0$ represents the isotropic eigenstrain parameter of the liquid inclusion\footnote{
The material metric \eqref{Metrics-liquid} encodes a purely dilatational eigenstrain through the factor $e^{2\Omega_l}$. For $\Omega_l>0$, the natural distances prescribed by the material metric are larger than the Euclidean distances in the reference configuration. Thus, when the inclusion is placed in the Euclidean reference geometry of radius $R_i$, it is forced to occupy a smaller volume than its natural (stress-free) volume, inducing a compressive residual stress. Because this incompatibility is purely volumetric, the resulting residual stress is hydrostatic, i.e., proportional to $\mathbf{g}^{\sharp}$.} (see \citep{YavariGoriely2013Inclusions,Yavari2021} for more details). 
It should be emphasized that the fluid inclusion is modeled as a compressible phase with a natural volumetric state $J_{\mathrm{nat}} = e^{3\Omega_l}$. Deviations of the actual volume $J$ from this natural state generate hydrostatic pressure.
For the liquid, we adopt the same kinematic ansatz \eqref{Deformation-Family4}, from which
\begin{equation} 
	J(R)=\frac{r^2(R)\,r'(R)}{R^2}\,e^{-3\Omega_l}\,.
\end{equation}
The Cauchy stress inside the inclusion is written as\footnote{The initial (residual) stress inside the inclusion is then given by
\begin{equation} 	
	\boldsymbol{\sigma}_r= \Wf'(1)\,\mathbf{g}^\sharp\,.
\end{equation}
}
\begin{equation} 	
	\boldsymbol{\sigma}(R)= \Wf'(J(R))\,\mathbf{g}^\sharp\circ\varphi(R)\,.
\end{equation}
For $R \leq R_i$ we have a compressible hyperelastic fluid whose deformation $r=r(R)$ is not known \textit{a priori}. 
In this case, the equilibrium equation \eqref{Radial-Equilibrium} simplifies to 
$\dfrac{\partial \sigma^{rr}}{\partial r} = 0$, 
indicating that the radial stress $\sigma^{rr}(R)=\Wf'(J(R))$ is uniform within the fluid inclusion. 
If $\Wf’$ is a monotone function of $J$ (as is typical for compressible hyperelastic fluids),\footnote{Strict monotonicity of $\Wf'(J)$ follows from the requirement that the bulk modulus be positive, ensuring material stability under volumetric deformations. For an isotropic compressible fluid, the radial Cauchy stress is $\sigma^{rr}=\Wf'(J)$. The bulk modulus is defined as $K = J\,\Wf''(J)$, and thermodynamic stability requires $K>0$, i.e., $\Wf''(J) > 0$. Therefore, $\Wf'(J)$ is strictly increasing, and the condition $\dfrac{\partial\sigma^{rr}}{\partial r}=0$ implies that $J$ must be constant within the fluid inclusion.} it follows that
$J(R) = J_0$, a constant.
Thus
\begin{equation}
	r^3(R)=r_i^3+J_0 \,e^{3\Omega_l} (R^3-R_i^3)	\,,\qquad 0\leq R \leq R_i\,.
\end{equation}
Knowing that $r(0)=0$,\footnote{A regular deformation of a simply-connected spherical inclusion requires $r(0)=0$. Choosing $r(0)>0$ implies that the material point at $R=0$ is mapped to a sphere of finite radius and the region $0\le r < r(0)$ remains empty. This corresponds to a cavitated configuration \citep{Ball1982, HorganAbeyaratne1986}. Since the present model assumes that the compressible fluid fills the entire cavity, only the branch with $r(0)=0$ is physically admissible.} we obtain $J_0 \,e^{3\Omega_l} =\dfrac{r_i^3}{R_i^3}$, and hence\footnote{This solution is similar to what was found in \citep{Pradhan2024} in a solidification problem.}
\begin{equation}
	r(R)= \frac{r_i}{R_i} R	\,,\qquad 0\leq R \leq R_i\,.
\end{equation}
We thus have
\begin{equation}
	\mathbf{F}=
	\begin{bmatrix}
	\dfrac{r_i}{R_i} & 0 & 0 \\
	0 & 1 & 0 \\
	0 & 0 & 1
	\end{bmatrix}
	\,.
\end{equation}
The constant stress inside the liquid inclusion is $\boldsymbol{\sigma}(R)= p_f\,\mathbf{g}^\sharp$, where 
\begin{equation} \label{Inclusion-Pressure}
	p_f=\Wf'(J_0)<0\,,\qquad J_0 =\frac{r_i^3}{R_i^3} \,e^{-3\Omega_l}
	\,.
\end{equation}
Note that $\mathbf{N}=\{ e^{-\Omega_l},0,0\}$,\footnote{Recall that $\mathbf{N}=\{ N^1,0,0\}$ is a $\mathbf{G}$-unit vector, i.e., $\llangle\mathbf{N},\mathbf{N}\rrangle_{\mathbf{G}}=(N^1)^2 \,e^{2\Omega_l}=1$, and hence $N^1=e^{-\Omega_l}$.} and hence
\begin{equation} \label{Traction-Inclusion}
	\mathbf{P}\mathbf{N}=J\boldsymbol{\sigma}\mathbf{F}^{-\star}\mathbf{N}
	= \begin{bmatrix} 
	e^{-4\Omega_l} \dfrac{r_i^2}{R_i^2}\,p_f \\
	0\\
	0
	\end{bmatrix}
	\,.
\end{equation}

Therefore, from \eqref{Traction-Inclusion} and \eqref{Traction-Bulk} we have
\begin{equation}
	\left\llbracket \mathbf{P}\mathbf{N} \right\rrbracket
	= \frac{r_i^2}{R_i^2}
	\begin{bmatrix}
		\sigma_i - e^{-4\Omega_l} p_f \\
		0 \\
		0
	\end{bmatrix}
	\,.
\end{equation}
Notice that $\mathbf{n}=\{1,0,0\}$, and hence, $\left\llbracket \mathbf{P} \mathbf{N} \right\rrbracket_\parallel=\mathbf{0}$. Also note that $ \left \llangle \left\llbracket \mathbf{P} \mathbf{N} \right\rrbracket, \mathbf{n} \right \rrangle_{\mathbf{g}}=\dfrac{r_i^2}{R_i^2}(\sigma_i - e^{-4\Omega_l} p_f)$.

\begin{remark}\label{Rem:Universal-Fluid}
From \eqref{Fluid-Stress}, for a homogeneous compressible isotropic hyperelastic fluid we have $\boldsymbol{\sigma}=\Wf'(J)\,\mathbf{g}^\sharp$ and hence $\operatorname{div}\boldsymbol{\sigma}=\Wf''(J)\,\nabla J$. In the absence of body forces, if the equilibrium equations $\operatorname{div}\boldsymbol{\sigma}=\mathbf{0}$ are to hold for any choice of $\Wf$, then necessarily $\nabla J=\mathbf{0}$, i.e., $J$ must be constant for any universal deformation. Conversely, any deformation with constant $J$ is universal for the class of homogeneous compressible isotropic hyperelastic fluids, since the associated Cauchy stress is a constant hydrostatic pressure field that automatically satisfies the equilibrium equations in the absence of body forces.
\end{remark}

\subsection{The spherical anelastic material surface}

For $\Sur$ and $\sur$ we use the spherical coordinates $(\Theta,\Phi)$ and $(\theta,\phi)$, respectively. The inclusion maps $\iota_{\Sur}:\Sur\hookrightarrow\mathcal{B}$ and $\iota_{\sur}:\sur \hookrightarrow\mathcal{S}$ have the following representations
\begin{equation}
	\iota_{\Sur}(\Theta,\Phi)=(R_i,\Theta,\Phi)\,,\qquad
	\iota_{\sur}(\theta,\phi)=(r_i,\theta,\phi)
	\,.
\end{equation}
Their tangent maps have the representations
\begin{equation}
	\FS=
	\begin{bmatrix}
	0 & 0 \\
	1 & 0 \\
	0 & 1
	\end{bmatrix}
	\,,\qquad
	\Fs=
	\begin{bmatrix}
	0 & 0 \\
	1 & 0 \\
	0 & 1
	\end{bmatrix}
	\,.
\end{equation}
It should be noted that these spherical coordinates are foliation coordinates in both the reference and current configurations, with the first coordinate being along the corresponding normal in each configuration.
The surface spatial and material metrics have the following representations\footnote{A dilatational surface eigenstrain corresponds to a natural (stress-free) surface that is smaller than the actual surface, so that the material surface is in tension in the initial configuration. Thus the material surface metric is written as $\bar{\mathbf{G}}=e^{-2\Omega_s}\mathbf{G}_0$ with $\Omega_s>0$, implying a natural radius $e^{-\Omega_s}R_i<R_i$. Using $e^{2\Omega_s}$ would instead produce a natural radius larger than $R_i$, and therefore an unphysical compressive residual stress.}
\begin{equation} \label{Surface-Material-Metric-Family4}
   \bar{\mathbf{G}}= e^{-2\Omega_s}
   \begin{bmatrix}
    R_i^2 & 0  \\
    0 & R_i^2 \sin^2 \Theta
    \end{bmatrix}\,,\qquad
   \bar{\mathbf{g}}=
   \begin{bmatrix}
    r_i^2 & 0  \\
    0 & r_i^2 \sin^2 \theta
    \end{bmatrix}=\begin{bmatrix}
    r_i^2 & 0  \\
    0 & r_i^2 \sin^2 \Theta
    \end{bmatrix}
	\,,
\end{equation}
where $e^{-\Omega_s}$ ($\Omega_s>0$) is the $2$D dilatational surface eigenstrain.\footnote{The material metric \eqref{Surface-Material-Metric-Family4}$_1$ represents the most general form of surface eigenstrain distribution compatible with the symmetry of Family $4$ universal deformations \citep{Goodbrake2020}.}

Note that 
\begin{equation} 
	\pi_{\SurE}=
	\begin{bmatrix}
	0 & 0 & 0\\
	0 & 1 & 0 \\
	0 & 0& 1
	\end{bmatrix}
	\,,\qquad
	\bar{\pi}_{\Sur}=
	\begin{bmatrix}
	0 & 1 & 0 \\
	0 & 0 & 1
	\end{bmatrix}
	\,.
\end{equation}
The surface restricted, the parallel, and the surface deformation gradients have the following representations
\begin{equation}
	\mathbf{F}_\SurE =
	\begin{bmatrix}
		0 & 0 & 0  \\
		0 & 1 &0 \\
		0 & 0 & 1
	\end{bmatrix}\,,\qquad
	\mathbf{F}_\parallel =
	\begin{bmatrix}
		0 & 0 & 0  \\
		0 & 1 &0 \\
		0 & 0 & 1
	\end{bmatrix}\,,\qquad
	\bar{\mathbf{F}} =
	\begin{bmatrix}
		1 & 0  \\
		0 & 1 
	\end{bmatrix}	
	 \,.
\end{equation}
Thus
\begin{equation}
	\mathbf{C}^\flat_\parallel=
	\begin{bmatrix}
	0 & 0 & 0 \\
	0 & r^{2}(R) & 0 \\
	0 & 0 & r^{2}(R)\sin^{2}\Theta
	\end{bmatrix}\,,\qquad
	\bar{\mathbf{C}}^\flat=
	\begin{bmatrix}
	r^{2}(R) & 0 \\
	0 & r^{2}(R)\sin^{2}\Theta
	\end{bmatrix}
	\,.
\end{equation}
Note that the projected material metric is written as
\begin{equation} \label{Metrics-Sphere}
	\mathring{\mathbf{G}}_\parallel =
	\begin{bmatrix}
	0 & 0 & 0 \\
	0 & R^2 & 0 \\
	0 & 0 & R^{2}\sin^{2}\Theta
	\end{bmatrix}
	\,.
\end{equation}
The surface principal invariants read
\begin{equation} \label{Surface-Principal-Invariants-Sphere}
	\bar{I}_1 =  \frac{2\,e^{2\Omega_s}\,r_i^2}{R_i^{2}}\,, \qquad
	\bar{I}_2 = \bar{J}^{\,2}= \frac{e^{4\Omega_s}\,r_i^4}{R_i^{4}}
	\,.
\end{equation}

Note that
\begin{equation} 
	\frac{\partial \mathbf{g}}{\partial r}\Bigg|_{\surE} = 
	\begin{bmatrix}
	1 & 0 & 0 \\
	0 & 2 r_i & 0 \\
	0 & 0 & 2 r_i \sin^2\theta
	\end{bmatrix}
	\,.
\end{equation}
The second fundamental form $\mathbf{k}$ is calculated by substituting the above expression into \eqref{Second-Fundamental-Form-Coordinates}, which gives us
\begin{equation}
	\mathbf{k} =
	\begin{bmatrix}
	0 & 0 & 0 \\
	0 & r_i & 0 \\
	0 & 0 & r_i \sin^{2}\theta
	\end{bmatrix}=
	\begin{bmatrix}
	0 & 0 & 0 \\
	0 & r_i & 0 \\
	0 & 0 & r_i \sin^{2}\Theta
	\end{bmatrix}=
	\frac{1}{r_i}\begin{bmatrix}
	0 & 0 & 0 \\
	0 & r_i^2 & 0 \\
	0 & 0 & r_i^2 \sin^{2}\Theta
	\end{bmatrix}
	\,.
\end{equation}
This, in particular, implies that 
\begin{equation} \label{2nd-Fundamental-Form-Surface}
   \bar{\mathbf{k}} = \frac{1}{r_i}\bar{\mathbf{g}}=
	\begin{bmatrix}
    r_i & 0  \\
    0 & r_i \sin^2 \Theta
    \end{bmatrix}
	\,.
\end{equation}

From \eqref{Surface-Stress-Isotropic}, the surface Cauchy stress is written as
\begin{equation}
	\bar{\boldsymbol{\sigma}} =
	\begin{bmatrix}
		\dfrac{2\,\Ws_1}{r_i^{2}}
		+ \dfrac{2\,e^{2\Omega_s}\,\Ws_2}{R_i^{2}}
		& 0 \\[8pt]
		0 &
		2\left(
		\dfrac{\Ws_1}{r_i^{2}}
		+ \dfrac{e^{2\Omega_s}\,\Ws_2}{R_i^{2}}
		\right)\csc^{2}\Theta
	\end{bmatrix}
	\,.
\end{equation}
The physical components are
\begin{equation}
	\hat{\bar{\boldsymbol{\sigma}}} =
	\begin{bmatrix}
		2\left(\Ws_1 + \dfrac{r_i^2 }{R_i^2} \,e^{2\Omega_s}\,\Ws_2 \right) & 0 \\[6pt]
		0 & 2\left(\Ws_1 + \dfrac{r_i^2 }{R_i^2} \,e^{2\Omega_s}\,\Ws_2 \right)
	\end{bmatrix}
	=\gamma_0\,\mathbf{I}
	\,,
\end{equation}
where
\begin{equation}
	\gamma_0=2 \left(\Ws_1 + \frac{r_i^2 }{R_i^2} \,e^{2\Omega_s}\,\Ws_2 \right) 	\,,
\end{equation}
is the surface tension.\footnote{Notice that $\bar{\boldsymbol{\sigma}}=\dfrac{\gamma_0}{r_i^2}\mathbf{g}^\sharp$.} As expected, surface tension is deformation dependent.
It should be emphasized that the principal invariants \eqref{Surface-Principal-Invariants-Sphere} are calculated using the non-flat material metric \eqref{Surface-Material-Metric-Family4}$_1$. This implies that the material surface in the initial configuration is residually-stressed.

Note that 
\begin{equation}
	\bar{J} \bar{\boldsymbol{\sigma}} \!:\! \bar{\mathbf{k}}
	=  \frac{2  \gamma_0}{r_i} \bar{J}
	= 2 e^{2\Omega_s} \frac{r_i}{R_i^2}\, \gamma_0
	= 4 e^{2\Omega_s} \frac{r_i}{R_i^2} \left(\Ws_1 + \frac{r_i^2 }{R_i^2} \,e^{2\Omega_s}\,\Ws_2 \right)
	\,.
\end{equation}
The normal equilibrium equation \eqref{Normal-Equilibrium-Equation} is simplified to read
\begin{equation} 
	\frac{r_i^2}{R_i^2} \left(\sigma_i - e^{-4\Omega_l} p_f \right) 	
	- 2 e^{2\Omega_s} \frac{r_i}{R_i^2}\, \gamma_0 = 0
	\,.
\end{equation}
Or
\begin{equation} 
	\sigma_i - e^{-4\Omega_l} p_f 
	- e^{2\Omega_s} \frac{2 \gamma_0}{r_i} = 0
	\,,
\end{equation}
which is a \textit{generalized Laplace’s law}.
In terms of the surface energy function the normal equilibrium equation is written as
\begin{equation} \label{Surface-Equation-General}
	\sigma_i - e^{-4\Omega_l} p_f - \frac{4 e^{2\Omega_s}}{r_i} 
	\left[ \Ws_1 + \frac{r_i^2 }{R_i^2}e^{2\Omega_s}  \,\Ws_2 \right] =0 \,.
\end{equation}

The nonzero connection coefficients of the Levi-Civita connection are $\bar{\gamma}^{\theta}{}_{\phi\phi} =-\sin\theta\cos\theta$ and $\bar{\gamma}^{\phi}{}_{\theta\phi}=\bar{\gamma}^{\phi}{}_{\phi\theta}	=\cot\theta$.
The surface divergence in components is written as
\begin{equation}
	\left(\overline{\operatorname{div}}\bar{\boldsymbol{\sigma}}\right)^{\bar{a}}
	=\bar{\sigma}^{\bar{a}\bar{b}}{}_{|\bar{b}}
	=\bar{\sigma}^{\bar{a}\bar{b}}{}_{,\bar{b}}
	+\bar{\gamma}^{\bar{a}}{}_{\bar{b}\bar{c}}\,\bar{\sigma}^{\bar{c}\bar{b}}
	+\bar{\gamma}^{\bar{b}}{}_{\bar{b}\bar{c}}\,\bar{\sigma}^{\bar{a}\bar{c}}\,.
\end{equation}
For $\bar{a}=\theta$ we have, $\left(\overline{\operatorname{div}}\bar{\boldsymbol{\sigma}}\right)^{\theta}
=\bar{\sigma}^{\theta\theta}{}_{|\theta} +\bar{\sigma}^{\theta\phi}{}_{|\phi}$.
The first term is simplified as 
\begin{equation}
\bar{\sigma}^{\theta\theta}{}_{|\theta}
	=\bar{\sigma}^{\theta\theta}{}_{,\theta}
	+\bar{\gamma}^{\theta}{}_{\theta\bar{c}}\,\bar{\sigma}^{\bar{c}\theta}
	+\bar{\gamma}^{\bar{b}}{}_{\bar{b}\theta}\,\bar{\sigma}^{\theta\bar{b}} 
	=\partial_\theta\bar{\sigma}^{\theta\theta}
	+\bar{\gamma}^{\phi}{}_{\phi\theta}\,\bar{\sigma}^{\theta\theta}
	=\dfrac{\gamma_0}{r_i^{2}}\,\cot\theta\,.
\end{equation}
Similarly, the second term is simplified as 
\begin{equation}
\bar{\sigma}^{\theta\phi}{}_{|\phi}
	=\partial_\phi\bar{\sigma}^{\theta\phi}
	+\bar{\gamma}^{\theta}{}_{\phi\bar{c}}\,\bar{\sigma}^{\bar{c}\phi}
	+\bar{\gamma}^{\phi}{}_{\phi\bar{c}}\,\bar{\sigma}^{\theta\bar{c}} 
	=\bar{\gamma}^{\theta}{}_{\phi\phi}\,\bar{\sigma}^{\phi\phi}
	=-\dfrac{\gamma_0}{r_i^{2}}\,\cot\theta\,.
\end{equation}
Therefore, $\left(\overline{\operatorname{div}}\bar{\boldsymbol{\sigma}}\right)^{\theta}=0$.
For $\bar{a}=\phi$, $\left(\overline{\operatorname{div}}\bar{\boldsymbol{\sigma}}\right)^{\phi}=\bar{\sigma}^{\phi\theta}{}_{|\theta} +\bar{\sigma}^{\phi\phi}{}_{|\phi}$.
Note that $\bar{\sigma}^{\phi\theta}{}_{|\theta} =\bar{\sigma}^{\phi\theta}{}_{,\theta} +\bar{\gamma}^{\phi}{}_{\theta\bar{c}}\,\bar{\sigma}^{\bar{c}\theta}  =0$, and $\bar{\sigma}^{\phi\phi}{}_{|\phi}=\bar{\sigma}^{\phi\phi}{}_{,\phi}	=0$. Thus, $\left(\overline{\operatorname{div}}\bar{\boldsymbol{\sigma}}\right)^{\phi}	=0$.
Therefore, $\overline{\operatorname{div}} \bar{\boldsymbol{\sigma}}=\mathbf{0}$, which is equivalent to $\overline{\operatorname{Div}} \bar{\mathbf{P}}=\mathbf{0}$, i.e., the tangential equilibrium equation \eqref{EL-Compressible-3} is trivially satisfied.

If $\sigma^{rr}(R_o)=-p_o$, the interface equilibrium condition is written as
\begin{equation} \label{Surface-Equation}
	-p_o - \int_{R_i}^{R_o} f(\xi)\,d\xi  - e^{-4\Omega_l} p_f 
	- \frac{4 e^{2\Omega_s}}{r_i} 
	\left[ \Ws_1 + \frac{r_i^2 }{R_i^2}e^{2\Omega_s}  \,\Ws_2 \right] = 0
	\,.
\end{equation}
This nonlinear algebraic equation determines the unknown interfacial radius $r_i$, coupling the internal pressure, surface eigenstrain, and elastic response of the surrounding solid.

In our examples we assume that the bulk body is made of an incompressible neo-Hookean solid, i.e.,
\begin{equation}
	W(I_1,I_2)=\frac{\mu}{2}\,(I_1 - 1)\,,
\end{equation}
where $\mu$ is the shear modulus of the solid. 
Let us also assume that the elastic surface is made of a two-dimensional compressible neo-Hookean material. Its strain energy density is written as
\begin{equation}
	\Wbars(\bar{I}_1,\bar{I}_2)
	= \frac{\mu_s}{2}\,\big(\bar{I}_1 - 2-\ln \bar{I}_2 \big)
	+ \frac{\kappa_s}{2}\left( \bar{I}_2^{\,\frac{1}{2}} - 1\right)^2\,,
\end{equation}
where $\mu_s$ and $\kappa_s$ are the surface shear and bulk moduli, respectively. 
The surface Cauchy stress is written as\footnote{In the initial configuration, $r_i=R_i$ so that $\bar{J}=e^{2\Omega_s}$ and $\gamma_0=(-1 + e^{2\Omega_s})\,\kappa_s + (1 - e^{-2\Omega_s})\,\mu_s$. Clearly, in the absence of surface eigenstrain, $\bar{\boldsymbol{\sigma}}=\mathbf{0}$, as required.}
\begin{equation}
	\bar{\boldsymbol{\sigma}}
	= \frac{\mu_s}{\bar{J}}\,\bar{\mathbf{b}}^\sharp
	+\Big[-\frac{\mu_s}{\bar{J}}+\kappa_s(\bar{J}-1)\Big]\bar{\mathbf{g}}^\sharp
	= \frac{\mu_s}{\bar{J}}\big(\bar{\mathbf{b}}^\sharp -\bar{\mathbf{g}}^\sharp\big)
	+ \kappa_s(\bar{J}-1)\,\bar{\mathbf{g}}^\sharp\,.
\end{equation}
We use the following energy function for the hyperelastic fluid
\begin{equation}
	\Wf(J) = \frac{\kappa_f}{2}\,(J-1)^2\,,
\end{equation}
where $\kappa_f$ is the bulk modulus characterizing compressibility, and
\begin{equation}
	J = J_0 =\frac{r_i^3}{R_i^3}\,e^{-3\Omega_l} \,.
\end{equation}
From \eqref{Fluid-Stress}, we have
\begin{equation} 
	\boldsymbol{\sigma} = \kappa_f (J - 1)\,\mathbf{g}^\sharp
	= \kappa_f\left(\frac{r_i^3}{R_i^3}\,e^{-3\Omega_l} -1 \right) \mathbf{g}^\sharp=p_f \,\mathbf{g}^\sharp
	\,,\qquad 0\leq R <R_i
	\,.
\end{equation}
In the initial configuration, $r_i=R_i$, and hence $\boldsymbol{\sigma} =  \mathring{\boldsymbol{\sigma}}=\kappa_f\left(e^{-3\Omega_l} -1 \right) \mathbf{g}^\sharp$.\footnote{In the literature, the following energy function has been used for a hyperelastic fluid \citep{Huang2006,Javili2013,Ghosh2022}:
\begin{equation}
	\Wf(J) = \mathring{p}_f\,J + \frac{\kappa_f}{2}\,(J-1)^2\,,
\end{equation}
where $\mathring{p}_f<0$ denotes the reference (undeformed) pressure of the liquid. Note that $\mathring{p}_f=\kappa_f\left(e^{-3\Omega_l} -1 \right)$ in our formulation.}
Note that the pressure inside the liquid inclusion is deformation dependent.

For this model we have
\begin{equation}
\begin{aligned}
	& W_1=\dfrac{\mu}{2}\,,\quad	W_2=0\,,\\
	& \Ws_1=\dfrac{\mu_s}{2}\,,\quad 	
	\Ws_2=\dfrac{\kappa_s}{2}\left(1-\bar{J}^{\,-1}\right) -\dfrac{\mu_s}{2} \bar{J}^{\,-2}
	= \dfrac{\kappa_s}{2}\left(1-\dfrac{R_i^{2}}{r_i^{2}}\,e^{-2\Omega_s} \right) 
	-\dfrac{\mu_s}{2} \dfrac{R_i^{4}}{r_i^{4}}\,e^{-4\Omega_s} \,,\\
	& p_f=  \kappa_f \left(\frac{r_i^3}{R_i^3} \,e^{-3\Omega_l} - 1\right)\,.
\end{aligned}
\end{equation}
Note that
\begin{equation}
	\int_{R_i}^{R_o} f(\xi)\,d\xi = 
	\frac{\mu}{2} \left[
	-\frac{4 R_i}{r_i}
	-\frac{R_i^{4}}{r_i^{4}}
	+\frac{R_o\left(4 r_i^{3} - 4 R_i^{3} + 5 R_o^{3}\right)}
	{\left(r_i^{3} - R_i^{3} + R_o^{3}\right)^{\tfrac{4}{3}}}\right]
	\,.
\end{equation}
Eq.~\eqref{Surface-Equation} is now simplified to read
\begin{equation} \label{Surface-Equation-Material}
\begin{aligned}
	& 2 e^{-7\Omega_l}\left(e^{3\Omega_l} - \frac{r_i^{3}}{R_i^{3}}\right)\kappa_f 
	+ \mu \left[
	    \frac{4 R_i}{r_i}
	    + \frac{R_i^{4}}{r_i^{4}}
	    + \frac{R_o\left(4 R_i^{3} - 4 r_i^{3} - 5 R_o^{3}\right)}
	           {\left(r_i^{3} - R_i^{3} + R_o^{3}\right)^{\tfrac{4}{3}}}
	\right] \\[4pt]
	&\quad
	- \frac{4 \,e^{2\Omega_s}}{r_i}\left[
	\kappa_s \left(-1 + e^{2\Omega_s}\,\frac{r_i^{2}}{R_i^{2}}\right)
	+ \mu_s\left(1 - e^{-2\Omega_s} \frac{R_i^{2}}{r_i^{2}}\right)
	\right] = 2\,p_o\,.
\end{aligned}
\end{equation}
The surface stress reads
\begin{equation}
	\gamma_0 = \gamma_0\!\left(\frac{r_i}{R_i},\Omega_s\right)
	=2\left(\Ws_1 + \dfrac{r_i^2 }{R_i^2} \,e^{2\Omega_s}\,\Ws_2 \right) 
	= \left(-1 + \frac{ r_i^2}{R_i^2}\,e^{2\Omega_s} \right) \kappa_s
	+ \left(1 - \frac{ R_i^2}{r_i^2}\, e^{-2\Omega_s} \right) \mu_s
	\,,
\end{equation}
which is deformation dependent. 
In the initial configuration $r_i=R_i$, and hence the initial surface stress is written as
\begin{equation}
	\hat{\gamma}_0
	= \left(-1 + e^{2\Omega_s} \right) \kappa_s	
	+ \left(1 - e^{-2\Omega_s} \right) \mu_s
	\,.
\end{equation}

The (deformation-dependent) \textit{elasto-capillary function} is defined as \citep{Liu2012}
\begin{equation} 
	\mathsf{e}_c \!\left(\frac{r_i}{R_i},\Omega_s\right) = \frac{\gamma_0}{2 R_i \mu}
	= \frac{\kappa_s}{2 R_i \mu} \left(-1 + \frac{e^{2\Omega_s} r_i^2}{R_i^2}\right)
	+\frac{\mu_s}{2 R_i \mu} \left(1 - \frac{e^{-2\Omega_s} R_i^2}{r_i^2}\right)
	\,.
\end{equation}
Surface effects are expected to be significant when $\mathsf{e}_c>1$. 
The \textit{initial elasto-capillary number} is defined as \citep{Ghosh2022}
\begin{equation} \label{Initial-elasto-capillary-number}
	\hat{\mathsf{e}}_c = \frac{\hat{\gamma}_0}{2 R_i \mu}
	= \frac{\kappa_s}{2 R_i \mu} \left(-1 + e^{2\Omega_s} \right) 
	+\frac{\mu_s}{2 R_i \mu} \left(1 - e^{-2\Omega_s} \right)
	\,.
\end{equation}
Let us define the following nondimensional parameters:
\begin{equation} \label{non-dimensionalized-parameters}
	x = \frac{r_i}{R_i}\,, \qquad  
	\alpha = \frac{R_o}{R_i}>1\,,\qquad 
	\hat{p}_o = \frac{p_o}{\mu}\,, \qquad
	\eta_f = \frac{\kappa_f}{\mu}\,, \qquad  
	\xi = \frac{\mu_s}{R_i\,\mu}\,, \qquad  
	\eta = \frac{\kappa_s}{R_i\,\mu}\,.
\end{equation}
Note that
\begin{equation} 
	\mathsf{e}_c(x)= \frac{\eta}{2}\left(e^{2\Omega_s} x^2-1\right)+\frac{\xi}{2}\left(1-e^{-2\Omega_s} x^{-2}\right)
	\,,\qquad
	\hat{\mathsf{e}}_c= \frac{\eta}{2}\left(e^{2\Omega_s}-1\right)+\frac{\xi}{2}\left(1-e^{-2\Omega_s}\right)
	\,.
\end{equation}
In terms of these nondimentional parameters \eqref{non-dimensionalized-parameters}, \eqref{Surface-Equation-Material} is written as
\begin{equation}
\begin{aligned}
	&\frac{1}{x^{4}}
	+ \frac{4}{x}
	+ \frac{\alpha\left(4 - 4 x^{3} - 5 \alpha^{3}\right)}{\left(-1 + x^{3} + \alpha^{3}\right)^{\frac{4}{3}}}
	-\frac{4}{x}\left( e^{2\Omega_s} x^2 -1 \right)\left( \frac{\xi}{x^2}+ \eta \,e^{2\Omega_s} \right)
	+ 2 e^{-7\Omega_l}\left(e^{3\Omega_l} - x^{3}\right)\eta_f = 2 \hat{p}_o \,.
\end{aligned}
\end{equation}

\subsection{Example 1: Spherical cavity without fluid (dry cavity)}

\begin{figure}[hbt!]
	\begin{center}
	\vskip 0.20 in
	\includegraphics[scale=0.65,angle=0]{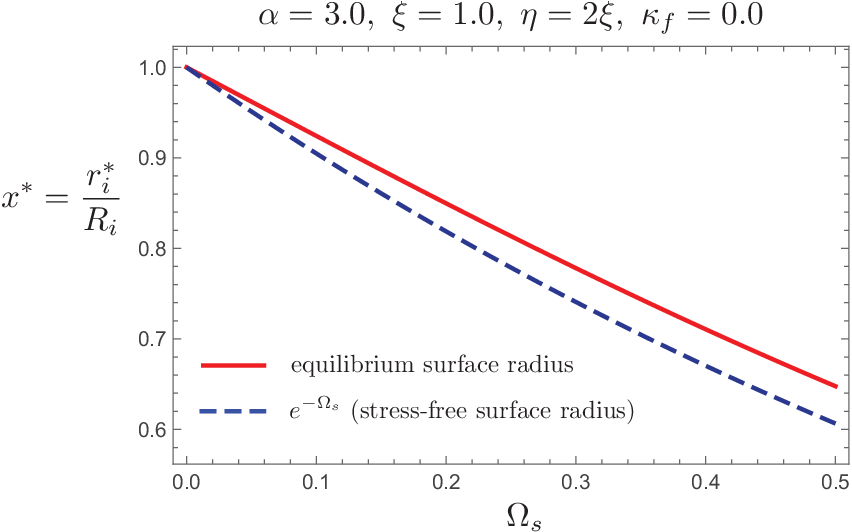}
	\end{center}
	\vskip -0.1 in
	\caption{\footnotesize The solid red curve shows the normalized equilibrium cavity radius $x^*=r_i^*/R_i$ as a function of $\Omega_s$ for $\alpha=3.0$, $\xi=1.0$, and $\eta=2\xi$. The dotted curve is the zero-stress radius $e^{-\Omega_s}R_i/R_i=e^{-\Omega_s}$. When $\Omega_s=0$ and $\hat{p}_o=0$, one has $x=1$ as an equilibrium solution. For $\Omega_s>0$ the material surface prefers the smaller natural radius $e^{-\Omega_s}R_i$ while the surrounding bulk is stress-free at $r(R)=R$. The relaxed radius results from balancing these effects and is always larger than the zero-stress radius.}
	\label{Relaxed-Ball}
\end{figure}

In this example we consider a cavity, i.e., no liquid. In this case $\eta_f=0$ and hence
\begin{equation}
\begin{aligned}
	&\frac{1}{x^{4}}
	+ \frac{4}{x}
	+ \frac{\alpha\left(4 - 4 x^{3} - 5 \alpha^{3}\right)}
	{\left(-1 + x^{3} + \alpha^{3}\right)^{\frac{4}{3}}}
	-\frac{4}{x}\left(e^{2\Omega_s} x^2 - 1\right)
	\left(\frac{\xi}{x^2} + \eta\,e^{2\Omega_s}\right)
	= 2\hat{p}_o\,.
\end{aligned}
\end{equation}
When $\Omega_s=0$ and there is no applied pressure, $x=1$ is an equilibrium solution. For $\hat{p}_o=0$ but $\Omega_s>0$, the initial configuration is not in equilibrium: the material surface has a smaller natural radius $e^{-\Omega_s}R_i<R_i$ and therefore tends to shrink the cavity, whereas the surrounding solid is stress-free when $r(R)=R$. The relaxed radius results from balancing these competing effects. The normalized equilibrium radius $x^*=r^*_i/R_i$ as a function of $\Omega_s$ is shown in Fig.~\ref{Relaxed-Ball}. As expected, the equilibrium radius is larger than the stress-free radius for any value of surface eigenstrain, i.e., $r_i^*> e^{-\Omega_s}\,R_i$.

\begin{figure}[hbt!]
	\begin{center}
	\vskip 0.20 in
	\includegraphics[scale=0.65,angle=0]{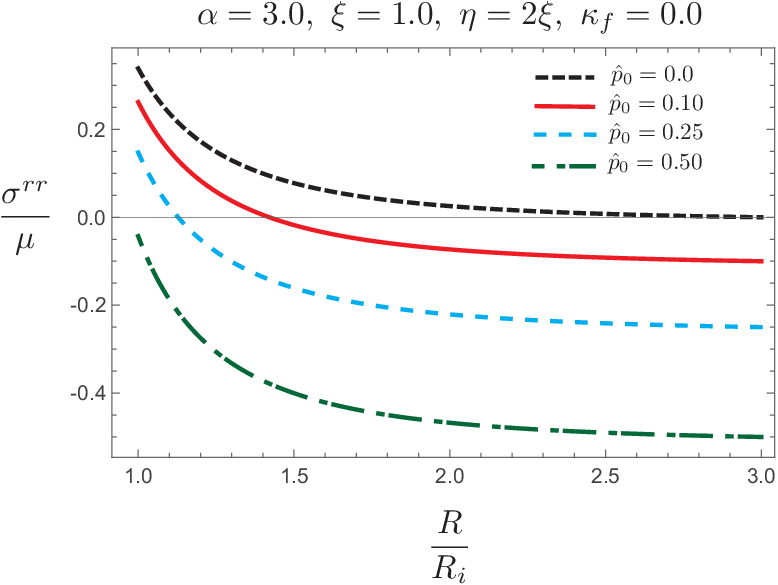}
	\end{center}
	\vskip -0.1 in
\caption{\footnotesize
Normalized radial stress $\sigma^{rr}(R)/\mu$ in the solid for $\alpha=3.0$, $\xi=1.0$, and $\eta=2\xi$, shown for several applied pressures $\hat{p}_o$. Surface eigenstrain induces tensile radial stress at $\hat{p}_o=0$, and increasing $\hat{p}_o$ drives the inner region and eventually the entire shell into compression.}
	\label{Dry-Cavity-Stress}
\end{figure}

From \eqref{Radial-Stress-1} and \eqref{Radial-Stress-2}, the radial stress in the solid has the following distribution
\begin{equation} \label{Radial-Stress}
	\sigma^{rr}(R) =	- p_o-\int_{R_i}^{R_o} f(\xi)\,d\xi
	= - p_o 
	+\frac{R\left(5 R^{3}+4 r_i^{3}-4 R_i^{3}\right)}{2\left(R^{3}+r_i^{3}-R_i^{3}\right)^{\frac{4}{3}}}\,\mu
	+\frac{R_o\left(-4 r_i^{3}+4 R_i^{3}-5 R_o^{3}\right)}{2\left(r_i^{3}-R_i^{3}+R_o^{3}\right)^{\frac{4}{3}}}\,\mu\,.
\end{equation}
Fig.~\ref{Dry-Cavity-Stress} shows the normalized radial stress distribution $\sigma^{rr}(R)/\mu$ in the solid for $\alpha=3.0$, $\xi=1.0$, and $\eta=2\xi$, computed for several values of the applied pressure $\hat{p}_o$. 
For $\hat{p}_o=0$ the surface eigenstrain induces a tensile radial stress in the solid near the cavity, and the entire solid shell carries a residual tensile radial stress that decays toward the outer boundary.
As $\hat{p}_o$ increases, the inner region progressively moves into compression while the outer region may remain in tension. For sufficiently large $\hat{p}_o$ (e.g., $\hat{p}_o=0.50$), the entire solid is under compressive radial stress.

\begin{figure}[hbt!]
\begin{center}
\vskip 0.20 in
\begin{minipage}[c]{0.48\textwidth}
\centering
\includegraphics[width=\textwidth]{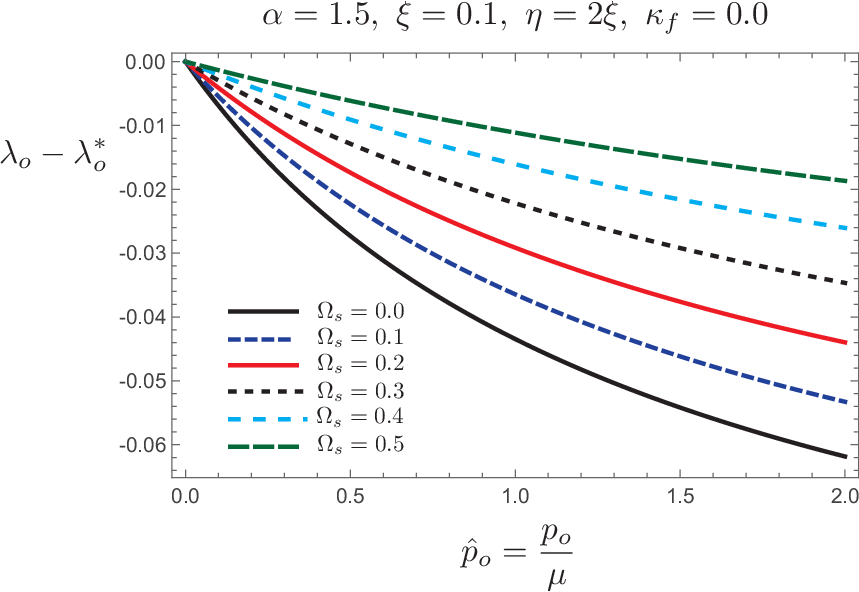}
\end{minipage}
\hskip 0.20 in
\begin{minipage}[c]{0.48\textwidth}
\centering
\includegraphics[width=\textwidth]{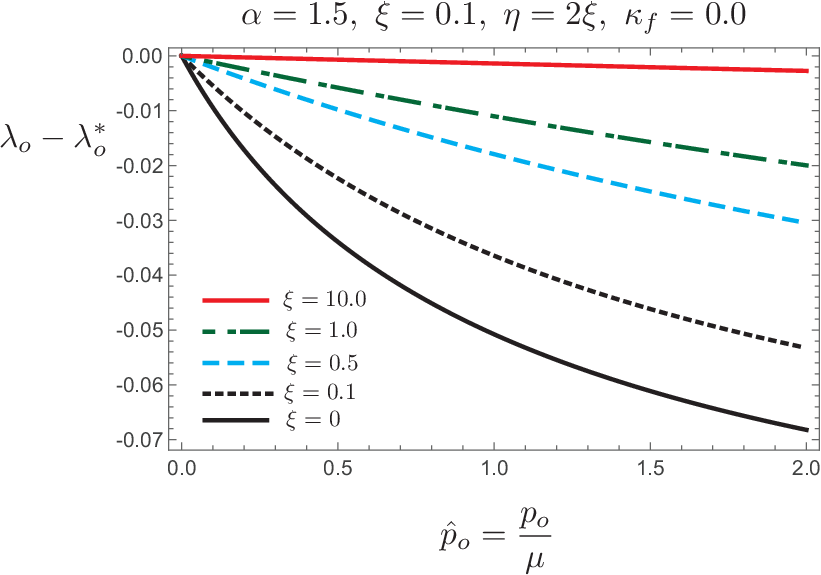}
\end{minipage}

\end{center}
\vskip -0.10 in
\caption{\footnotesize
Left panel: Strain $\lambda_o-\lambda_o^*$ as a function of the normalized applied pressure $\hat{p}_0$ for
$\alpha=1.5$, $\xi=0.1$, and $\eta=2\xi$, shown for several values of the surface eigenstrain $\Omega_s$.
Here $\lambda_o$ denotes the outer radial stretch of the ball and $\lambda_o^*$ is its relaxed value at zero applied pressure. Increasing $\Omega_s$ shifts the relaxed configuration and leads to a stiffer pressure-stretch response. Note that from \eqref{Initial-elasto-capillary-number}, these choices of parameters and surface eigenstrains correspond to the following values of the initial elastic-capillarity number: $\hat{\mathsf{e}}_c=0, 0.13,0.20,0.40,0.60$.
Right panel: Strain $\lambda_o-\lambda_o^*$ as a function of $\hat{p}_0$ for
$\alpha=1.5$ and $\Omega_s=0.1$, for several values of $\xi$ (with $\eta=2\xi$).
For $\xi=0$ the response is governed only by the bulk elasticity and is the most compliant. Increasing $\xi$ strengthens surface effects and results in a progressively stiffer response. For $\xi=10$ the curve is nearly flat at $\lambda_o-\lambda_o^*=0$, corresponding to an effectively rigid response.
Note that these choices of parameters and surface eigenstrains correspond to the following values of the initial elastic-capillarity number: $\hat{\mathsf{e}}_c=0, 0.06,0.31,0.62,6.24$.
}
\label{Surface-Combined}
\end{figure}

Note that
\begin{equation}
	\lambda_o=\frac{1}{\alpha}\left(\alpha^3+x^3-1\right)^{\frac{1}{3}}\,.
\end{equation}
Under zero applied pressure the relaxed stretch is $\lambda_o^*=\frac{1}{\alpha}\left[\alpha^3+(x^*)^3-1\right]^{\frac{1}{3}}$\,.
As the initial configuration is residually stressed, it is natural to regard $\lambda_o-\lambda_o^*$ as the strain, i.e., the stretch measured relative to the relaxed state. For $\alpha=1.5$, $\xi=0.1$, and $\eta=2\xi$, Fig.~\ref{Surface-Combined}\,(left panel) shows the strain as a function of the applied pressure for several values of the surface eigenstrain. As expected, increasing $\Omega_s$ leads to a stiffer response. Next, for fixed $\alpha$ and surface eigenstrain ($\alpha=1.5$, $\Omega_s=0.1$) the strain--pressure plots are shown for different values of $\xi$ (with $\eta=2\xi$) in Fig.~\ref{Surface-Combined}\,(right panel). When $\xi=0$ only the bulk elasticity contributes and the response is the most compliant. As $\xi$ increases surface effects become significant and the ball stiffens. For $\xi=10$ the strain remains essentially zero for all pressures, i.e., the ball behaves effectively as a rigid solid.

\subsection{Example 2: Spherical cavity filled with compressible fluid (wet cavity)}

\begin{figure}[hbt!]
	\begin{center}
	\vskip 0.20 in
	\includegraphics[scale=0.65,angle=0]{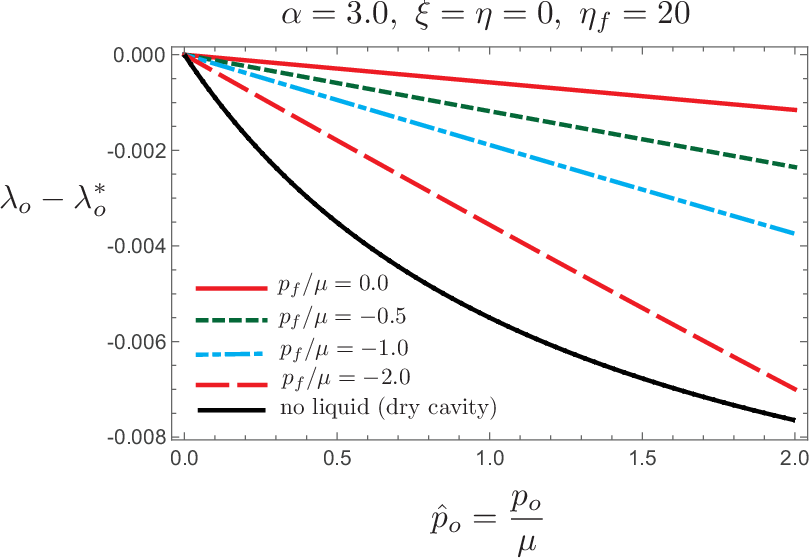}
	\end{center}
	\vskip -0.1 in
\caption{\footnotesize
Normalized pressure-stretch response $\lambda_0(\hat{p}_o)-\lambda_0^{*}$ for a spherical solid ball with a liquid inclusion compared with the corresponding dry cavity, for $\alpha=3.0$, $\xi=\eta=0$ (no surface stress effect), and $\eta_f=20$. $\lambda_0^{*}$ denotes the relaxed stretch at $\hat{p}_o=0$.}
	\label{Pressure-Stretch-Combined-Dry-Cavity}
\end{figure}

In this example we assume that the spherical cavity is filled with a homogeneous isotropic hyperelastic fluid.
We first ignore surface stress effects.
For a liquid inclusion without surface stress the numerical results in Fig.~\ref{Pressure-Stretch-Combined-Dry-Cavity} show a nontrivial dependence of the effective stiffness on the natural fluid pressure.
When the initial internal pressure is small (including $p_f(0)=0$), the inclusion is close to its natural volumetric state and simply contributes an additional compressible phase that resists volumetric changes.
In this regime the ball with a liquid inclusion is stiffer in the pressure-stretch response than the corresponding solid with a dry cavity.
For nonzero values of $p_f$ the fluid prefers a volume larger than that of the actual cavity, and the surrounding solid is prestretched in the relaxed configuration at $\hat{p}_o=0$.
Subsequent loading is measured relative to this residually stressed configuration, and the pressure-stretch response corresponds to the incremental elastic behavior about a prestressed state. For a nonlinear elastic solid, this prestress modifies the tangent stiffness, and in the present case increasing the initial fluid pressure places the material on a softer part of its constitutive response. Consequently, increasing the initial fluid pressure softens the response of the ball.

\begin{figure}[hbt!]
\begin{center}
\vskip 0.20 in
\begin{minipage}[c]{0.47\textwidth}
\centering
\includegraphics[width=\textwidth]{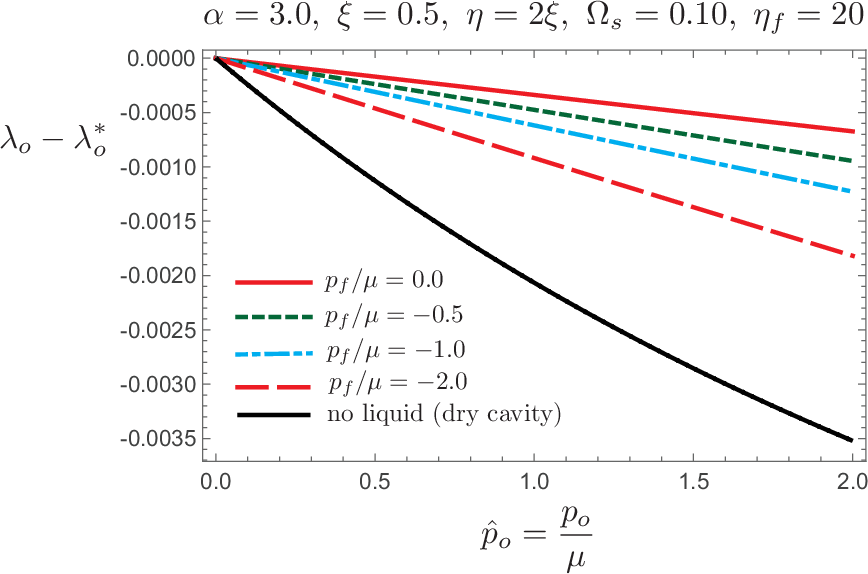}
\end{minipage}
\hskip 0.20 in
\begin{minipage}[c]{0.49\textwidth}
\centering
\includegraphics[width=\textwidth]{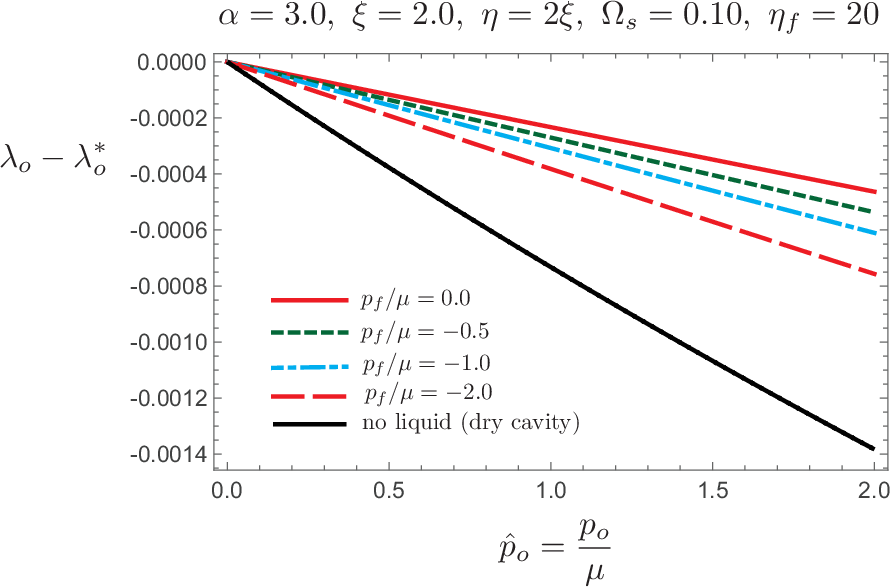}
\end{minipage}

\end{center}
\vskip -0.10 in
\caption{\footnotesize
Left panel: Normalized pressure-stretch response $\lambda_0(\hat{p}_o)-\lambda_0^{*}$ for a spherical solid ball containing a liquid inclusion, compared with the corresponding dry cavity. Results shown for $\alpha=3.0$, $\xi=0.50$, $\eta=2\xi$, $\Omega_s=0.10$, and $\eta_f=20$. Here $\lambda_0^{*}$ is the relaxed stretch at $\hat{p}_o=0$. The liquid-filled configuration is stiffer than the dry cavity for all fluid pressures. Increasing fluid pressures decreases the stiffness of the ball. These parameters correspond to the initial elasto-capillarity number $\hat{\mathsf{e}}_c=0.31$.
Right panel: Normalized pressure-stretch response $\lambda_0(\hat{p}_o)-\lambda_0^{*}$ for the same spherical solid ball containing a liquid inclusion, compared with the corresponding dry cavity. Results shown for $\alpha=3.0$, $\xi=2.0$, $\eta=2\xi$, $\Omega_s=0.10$, and $\eta_f=20$. The pressure-stretch response is similar to the previous case; the liquid-filled configuration is still stiffer than the dry cavity for all fluid pressures. Also. increasing fluid pressures still decreases the stiffness of the ball. The parameters of this case correspond to the initial elasto-capillarity number $\hat{\mathsf{e}}_c=1.25$. We observe that the effect of surface stress in both cases is to stiffen the overall response of the ball.
}
\label{Pressure-Stretch-Combined}
\end{figure}

When surface stress effects are included, the numerical results in Fig.~\ref{Pressure-Stretch-Combined}\,(left panel) show that the relative roles of the fluid and the surface change the pressure-stretch behavior in a systematic manner. 
The following parameters are used in this example: $\alpha=3.0$, $\xi=0.50$, $\eta=2\xi$, $\Omega_s=0.10$, and $\eta_f=20$, which correspond to the initial elasto-capillarity number $\hat{\mathsf{e}}_c=0.31$. 
Here $\lambda_0^{*}$ is the relaxed stretch at $\hat{p}_o=0$. The liquid-filled configuration is stiffer than the dry cavity for all fluid pressures. Increasing fluid pressures decreases the stiffness of the ball. These parameters correspond to the initial elasto-capillarity number $\hat{\mathsf{e}}_c=0.31$.
We next increase the effect of surface stress and use the following parameters: $\alpha=3.0$, $\xi=0.50$, $\eta=2\xi$, $\Omega_s=0.10$, and $\eta_f=20$, which correspond to the initial elasto-capillarity number $\hat{\mathsf{e}}_c=1.25$. In Fig.~\ref{Pressure-Stretch-Combined}\,(right panel) we see a similar trend. Overall, the ball is stiffer than the previous case. However, again including liquid inside the cavity has a stiffening effect and increasing the initial fluid pressure softens the elastic response.

Surface eigenstrain induces an additional residual deformation of the solid even when $p_f(0)=0$, shifting the relaxed configuration and effectively stiffening the response at small applied pressures.
For the liquid-filled configuration the combined effect of surface tension and fluid compressibility produces a response that is stiffer than that of the dry cavity, but as the applied pressure increases the fluid contribution dominates and the configuration eventually becomes more compliant compared to the zero fluid pressure case.
These parameters correspond to the initial elasto-capillarity number $\hat{\mathsf{e}}_c=0.31$.

\begin{remark}[Universality of Radial Deformations of an Isotropic Spherical Ball with a Fluid Inclusion]
For a given class of materials, a deformation is universal if it can be maintained by applying only boundary tractions and in the absence of body forces \citep{Ericksen1954,Ericksen1955}. In our example, we have a spherical ball with a cavity. We know that radial deformations are universal for incompressible isotropic solids---Family $4$ universal deformations \citep{Ericksen1954}, see Footnote~\ref{Footote:Universal-Solid}. We also know that, for compressible isotropic solids---both homogeneous \citep{Ericksen1955} and inhomogeneous \citep{Yavari2021}---the only universal deformations are homogeneous deformations. However, a compressible hyperelastic fluid is special in the sense that its energy function depends only on $J$. We showed that independently of the form of the fluid energy function $r(R)=r_i R/R_i$, i.e., such deformations are universal for the elastic fluid, see Remark~\ref{Rem:Universal-Fluid}. For the spherical material surface, the tangential surface balance is trivially satisfied and only the normal
surface balance \eqref{Surface-Equation} remains.
Thus, for the coupled bulk-surface-fluid system, the only nontrivial equilibrium condition is the generalized Laplace equation \eqref{Surface-Equation} that determines the cavity stretch $x=r_i/R_i$. 
If this scalar equation admits a solution for $x$ for arbitrary incompressible isotropic solid shell and arbitrary compressible isotropic material surface, then the spherically-symmetric deformation would be universal. Whether such a solution exists in general is unknown; the solutions obtained here for the special case of an incompressible neo-Hookean solid and a compressible neo-Hookean material surface do not guarantee universality for broader constitutive classes.
\end{remark}

\begin{remark}
Sphere assemblages were introduced by Hashin \citep{Hashin1962a,Hashin1962b,Hashin1963}.
For finite deformations of incompressible isotropic thick spherical annuli (Family $4$ universal deformations) \citet{Hashin1985} observed that the stress and deformation fields depend only on the ratio $R_i/R_o$ and not on the individual radii. 
This scale invariance is what allows the construction of Hashin's spherical assemblage: starting from a compressible solid subjected to a uniform dilatational strain, one can insert hollow spheres of different sizes but with the same ratio $R_i/R_o$, made of an appropriate second phase, in such a way that the prescribed macroscopic dilatational field in the surrounding material is not perturbed. This observation also plays a central role in the anisotropic extension developed in \citep{GolgoonYavari2021Hashin}. With surface elasticity this invariance is lost. Surface stresses introduce a length scale, and the solution of the spherical cavity problem depends explicitly on $R_i$. Two hollow spheres having the same ratio $R_i/R_o$ but different sizes no longer induce the same stress in the surrounding solid under the same dilatational strain. Consequently, the mechanism that enables Hashin's assemblage is no longer available, and a Hashin-type construction cannot be carried out for materials with elastic or anelastic surfaces. 
\end{remark}

\section{Conclusions} \label{Sec:Conclusions}

In this paper we formulated the mechanics of elastic bodies with material surfaces in the setting of Riemannian geometry of $3$-manifolds and their hypersurfaces. We began with the geometric formulation of bulk elasticity, in which the body is a Riemannian material manifold and all kinematic and kinetic quantities are written in terms of its material metric. This framework, rooted in the earlier work of \citet{MaHu1983}, was then augmented by the differential geometry of embedded material surfaces. Within this setting, kinematics was extended to bodies with elastic material surfaces, and the surface analogs of the classical strain measures of $3$D nonlinear elasticity were constructed using the induced geometry of the material hypersurfaces. Both the bulk and the material surfaces were assumed to be hyperelastic and endowed with their own elastic energies written as functions of the corresponding material metrics.

A key observation is that once bulk elasticity is formulated in the language of Riemannian geometry, the extension to surface elasticity can be carried out in a systematic and transparent manner. Working with hypersurfaces of Riemannian manifolds allows all surface kinematic quantities to be defined unambiguously, including the various surface deformation gradients obtained by restriction and projection. Within this geometric setting the surface strain measures, the surface material metric, and the induced second fundamental form follow naturally from the embedding of the material surface in the material manifold. Surface stresses are then obtained by differentiating the surface energy with respect to the surface material metric, and anisotropy is encoded through surface structural tensors exactly as in the bulk. 

In the geometric formulation, extending bulk elasticity to bulk anelasticity is straightforward: anelastic distortions are encoded by a bulk material metric, following the original ideas introduced by \cite{Eckart1948}. Once surface elasticity is formulated in the setting of hypersurfaces of Riemannian manifolds, the extension to surface anelasticity is equally systematic. In the presence of material surfaces the body carries both bulk and surface anelastic distortions (eigenstrains), each inducing its own material metric. The surface material metric is determined by the surface anelastic distortion and, in general, does not coincide with the first fundamental form of the reference material surface. Thus the geometry of both the bulk and the embedded surface is determined by their respective material metrics, and the balance and constitutive laws retain the same geometric structure after replacing the elastic metrics by the corresponding anelastic ones.

We showed that, for both isotropic and anisotropic elastic surfaces, the constitutive equations retain exactly the same functional form as their bulk counterparts when the first fundamental form and the classical structural tensors are replaced by the surface material metric and the anelastic surface structural tensors. The balance laws were derived variationally using the Lagrange-d'Alembert principle. These include the standard bulk balance of linear momentum together with the surface balance of linear momentum, which has both tangential and normal components---a generalized Laplace's law. The resulting bulk and surface balance equations preserve the familiar structure of classical elasticity with material surfaces, but are expressed entirely in terms of the geometry of the material manifold.

As an example, we analyzed the deformation of a spherical ball containing a concentric inclusion bounded by a material surface. Owing to spherical symmetry, the bulk equilibrium reduces to a single nonlinear equation for the cavity stretch, from which a complete analytical solution was obtained. For a dry cavity with surface eigenstrain we computed the relaxed cavity radius, the residual Cauchy stress field in the solid, and the associated surface stress. The results show that a tensile surface eigenstrain induces a residual tensile Cauchy stress component near the cavity that decays monotonically toward the outer boundary. The pressure-stretch curves further demonstrate that surface stress makes the ball effectively stiffer, shifting the relaxed configuration and increasing the incremental resistance to applied pressure relative to the case with no surface effects.

We then studied a liquid-filled cavity in which the inclusion is a compressible hyperelastic fluid with initial pressure. The fluid contribution enters through an additional eigenstrain parameter and modifies the generalized Laplace's law in a nonlinear manner. Numerical solutions of the exact equilibrium equation showed that an inclusion with a larger initial elasto-capillarity number is stiffer. However, for a fixed initial elasto-capillarity number, a liquid inclusion stiffens the ball. Increasing the initial pressure of the inclusion soften the response. 
The resulting radial stress distributions show that surface or fluid eigenstrains generate tensile residual stress near the cavity and that increasing applied pressure progressively drives the inner region, and eventually the entire shell, into compression.

These examples demonstrate the versatility of the geometric formulation developed in this paper. Bulk and surface elasticity and anelasticity are treated within a single differential-geometric framework in which all kinematic and kinetic quantities are expressed in terms of the material metric and the induced surface geometry. The approach eliminates ad hoc assumptions, clarifies the role of incompatible eigenstrains, and provides a systematic procedure for formulating constitutive equations for both the bulk and the material surfaces. This geometric framework is fully general and extends well beyond spherically symmetric problems, allowing the analysis of inclusions of arbitrary shape, more complex distributions of eigenstrain, and time-dependent processes such as growth, remodeling, or viscoelastic relaxation.

The geometric framework developed here can be extended in several natural directions. Incorporating curvature-dependent surface energies would allow the treatment of bending deformations of material surfaces, in the spirit of Steigmann-Ogden theory but expressed in the present Riemannian setting. Another direction is to include curvature-driven surface remodeling or surface phase transitions within the same geometric structure. Finally, allowing the bulk and surface material metrics to evolve in time would enable a fully geometric treatment of growth, remodeling, and surface viscoelastic relaxation.

\section*{Acknowledgments}

We benefited from discussions with Oscar Lopez-Pamies. This work was partially supported by NSF, Grant No.\ CMMI\,1939901.

\bibliographystyle{plainnat}
\bibliography{ref}

@article{Pradhan2024,
  title={Nonlinear mechanics of phase-change-induced accretion},
  author={Pradhan, Satya Prakash and Yavari, Arash},
  journal={Journal of the Mechanics and Physics of Solids},
  volume={193},
  pages={105888},
  year={2024},
  publisher={Elsevier}
}

@article{Arroyo2009,
  title={Relaxation dynamics of fluid membranes},
  author={Arroyo, Marino and DeSimone, Antonio},
  journal={Physical Review E},
  volume={79},
  number={3},
  pages={031915},
  year={2009},
  publisher={American Physical Society}
}

@article{SozioYavari2020,
  title   = {Riemannian and {E}uclidean material structures in anelasticity},
  author  = {Sozio, Fabio and Yavari, Arash},
  journal = {Mathematics and Mechanics of Solids},
  volume  = {25},
  number  = {6},
  pages   = {1267--1293},
  year    = {2020}
}

@article{SadikYavari2024,
	author = {Sadik, Souhayl and Yavari, Arash},
	journal = {Journal of the Mechanics and Physics of Solids},
	pages = {105461},
	publisher = {Elsevier},
	title = {Nonlinear anisotropic viscoelasticity},
	volume = {182},
	year = {2024}}

@article{Noll1967,
  title={Materially uniform simple bodies with inhomogeneities},
  author={Noll, Walter},
  journal={Archive for Rational Mechanics and Analysis},
  volume={27},
  number={1},
  pages={1--32},
  year={1967},
  publisher={Springer}
}

@incollection{Wang1968,
  title={On the geometric structure of simple bodies, a mathematical foundation for the theory of continuous distributions of dislocations},
  author={Wang, C-C},
  booktitle={Mechanics of Generalized Continua},
  pages={247--250},
  year={1968},
  publisher={Springer}
}

@article{Wang1974,
  title={Material uniformity and inhomogeneity in anelastic bodies},
  author={Wang, C-C and Bloom, F},
  journal={Archive for Rational Mechanics and Analysis},
  volume={53},
  number={3},
  pages={246--276},
  year={1974},
  publisher={Springer}
}

@article{Romano2014,
  title={Geometric continuum mechanics},
  author={Romano, Giovanni and Barretta, Raffaele and Diaco, Marina},
  journal={Meccanica},
  volume={49},
  number={1},
  pages={111--133},
  year={2014}
}

@book{Epstein2011GeometricalLanguage,
  author    = {Morton Epstein},
  title     = {The Geometrical Language of Continuum Mechanics},
  publisher = {Cambridge University Press},
  address   = {Cambridge},
  year      = {2011}
}

@book{Clayton2014,
  title     = {Differential Geometry and Kinematics of Continua},
  author    = {Clayton, John D},
  year      = {2014},
  publisher = {World Scientific}
}

@book{Segev2023,
  author    = {Reuven Segev},
  title     = {Foundations of Geometric Continuum Mechanics},
  series    = {Advances in Mechanics and Mathematics},
  publisher = {Springer},
  year      = {2023}
}

@article{Eremeyev2024,
  author  = {Eremeyev, Victor A.},
  title   = {Surface finite viscoelasticity and surface anti-plane waves},
  journal = {International Journal of Engineering Science},
  volume  = {196},
  pages   = {104029},
  year    = {2024}
}

@article{DianiParks2000,
  title={Problem of an inclusion in an infinite body, approach in large deformation},
  author={Diani, J Lambert and Parks, DM},
  journal={Mechanics of Materials},
  volume={32},
  number={1},
  pages={43--55},
  year={2000},
  publisher={Elsevier}
}

@book{Goldstein2002,
  title={Classical Mechanics},
  author={Goldstein, Herbert and Poole, Charles P. and Safko, John L.},
  edition={3},
  year={2002},
  publisher={Addison-Wesley}
}

@article{Cauchy1828,
	author = {Cauchy, Augustin-Louis},
	journal = {Exercises de Math{\'e}matiques},
	pages = {160--187},
	title = {Sur les {\'e}quations qui expriment les conditions d'{\'e}quilibre ou les lois du mouvement int{\'e}rieur d'un corps solide, {\'e}lastique ou non {\'e}lastique},
	volume = {3},
	year = {1828}}

@article{Truesdell1952,
	author = {Truesdell, C.},
	journal = {Journal of Rational Mechanics and Analysis},
	number = {1},
	pages = {125--300},
	publisher = {Springer},
	title = {The Mechanical foundations of elasticity and fluid dynamics},
	volume = {1},
	year = {1952}}

@article{YavariGoriely2025Cauchy,
  title={Nonlinear {C}auchy Elasticity},
  author={Yavari, Arash and Goriely, Alain},
  journal={Archive for Rational Mechanics and Analysis},
  volume={249},
  number={5},
  pages={57},
  year={2025},
  publisher={Springer}
}

@article{AngoshtariYavari2015,
  title={Differential complexes in continuum mechanics},
  author={Angoshtari, Arzhang and Yavari, Arash},
  journal={Archive for Rational Mechanics and Analysis},
  volume={216},
  number={1},
  pages={193--220},
  year={2015},
  publisher={Springer}
}

@article{Tamim2025,
  title        = {Shaping Capillary Solids from Statics to Dynamics},
  author       = {Tamim, S. I. and Bostwick, J. B.},
  journal      = {Annual Review of Condensed Matter Physics},
  volume       = {16},
  year         = {2025},
  publisher    = {Annual Reviews}
}

@article{Heyden2022,
  title        = {A Robust Method for Quantification of Surface Elasticity in Soft Solids},
  author       = {Heyden, Stefanie and Vlahovska, Petia M. and Dufresne, Eric R.},
  journal      = {Journal of the Mechanics and Physics of Solids},
  volume       = {161},
  pages        = {104786},
  year         = {2022},
  publisher    = {Elsevier}
}

@article{Duan2005,
  title        = {Size-Dependent Effective Elastic Constants of Solids Containing Nano-Inhomogeneities with Interface Stress},
  author       = {Duan, H. L. and Wang, Jian-Xiang and Huang, Z. P. and Karihaloo, Bhushan Lal},
  journal      = {Journal of the Mechanics and Physics of Solids},
  volume       = {53},
  number       = {7},
  pages        = {1574--1596},
  year         = {2005},
  publisher    = {Elsevier}
}

@article{Steigmann1999FluidFilms,
  title   = {Fluid Films with Curvature Elasticity},
  author  = {Steigmann, D. J.},
  journal = {Archive for Rational Mechanics and Analysis},
  volume  = {150},
  pages   = {127--152},
  year    = {1999},
  publisher = {Springer-Verlag}
}

@article{Yavari2021Universal,
  title        = {Universal Deformations in Inhomogeneous Isotropic Nonlinear Elastic Solids},
  author       = {Yavari, Arash},
  journal      = {Proceedings of the Royal Society A},
  volume       = {477},
  number       = {2253},
  pages        = {20210547},
  year         = {2021},
  publisher    = {The Royal Society}
}

@article{YavariGoriely2021Universal,
  title={Universal deformations in anisotropic nonlinear elastic solids},
  author={Yavari, Arash and Goriely, Alain},
  journal={Journal of the Mechanics and Physics of Solids},
  volume={156},
  pages={104598},
  year={2021},
  publisher={Elsevier}
}

@article{Yavari2023universal,
  title={The universal program of nonlinear hyperelasticity},
  author={Yavari, Arash and Goriely, Alain},
  journal={Journal of Elasticity},
  volume={154},
  number={1},
  pages={91--146},
  year={2023},
  publisher={Springer}
}

@article{Ericksen1955,
  title={Deformations possible in every compressible, isotropic, perfectly elastic material},
  author={Ericksen, J. L.},
  journal={Journal of Mathematics and Physics},
  volume={34},
  number={1-4},
  pages={126--128},
  year={1955},
  publisher={Wiley Online Library}
}

@article{Ball1982,
  author       = {Ball, John M.},
  title        = {Discontinuous equilibrium solutions and cavitation in nonlinear elasticity},
  journal      = {Philosophical Transactions of the Royal Society A},
  volume       = {306},
  number       = {1496},
  pages        = {557--611},
  year         = {1982}
}

@article{HorganAbeyaratne1986,
  author       = {Horgan, Cornelius O. and Abeyaratne, Rohan},
  title        = {A bifurcation problem for a compressible hyperelastic sphere},
  journal      = {Journal of Elasticity},
  volume       = {16},
  number       = {2},
  pages        = {189--200},
  year         = {1986}
}

@Article{Hashin1985,
  author    = {Hashin, Zvi},
  title     = {Large isotropic elastic deformation of composites and porous media},
  journal   = {International Journal of Solids and Structures},
  year      = {1985},
  volume    = {21},
  number    = {7},
  pages     = {711--720},
  publisher = {Elsevier},
}

@article{GolgoonYavari2021Hashin,
  title={On {H}ashin’s Hollow Cylinder and Sphere Assemblages in Anisotropic Nonlinear Elasticity},
  author={Golgoon, Ashkan and Yavari, Arash},
  journal={Journal of Elasticity},
  volume={146},
  number={1},
  pages={65--82},
  year={2021},
  publisher={Springer}
}

@article{Hashin1963,
  title={A variational approach to the theory of the elastic behaviour of multiphase materials},
  author={Hashin, Zvi and Shtrikman, Shmuel},
  journal={Journal of the Mechanics and Physics of Solids},
  volume={11},
  number={2},
  pages={127--140},
  year={1963},
  publisher={Elsevier}
}

@article{Hashin1962a,
  title={The elastic moduli of heterogeneous materials},
  author={Hashin, Zvi},
  journal={Journal of Applied Mechanics},
  volume={29},
  pages={143--150},
  year={1962}
}

@article{Hashin1962b,
  title={A variational approach to the theory of the effective magnetic permeability of multiphase materials},
  author={Hashin, Zvi and Shtrikman, Shmuel},
  journal={Journal of Applied Physics},
  volume={33},
  number={10},
  pages={3125--3131},
  year={1962},
  publisher={American Institute of Physics}
}

@book{Laplace1805,
  author       = {Laplace, Pierre-Simon},
  title        = {Trait{\'e} de m{\'e}canique c{\'e}leste, Suppl{\'e}ment au Livre X},
  year         = {1805},
  publisher    = {Courcier},
  address      = {Paris}
}

@book{DeGennes2003,
  author       = {de Gennes, Pierre-Gilles and Brochard-Wyart, Fran{\c c}oise and Qu{\'e}r{\'e}, David},
  title        = {Capillarity and Wetting Phenomena: Drops, Bubbles, Pearls, Waves},
  year         = {2003},
  publisher    = {Springer},
  address      = {New York}
}

@book{Lanczos1962,
  title={The Variational Principles of Mechanics},
  author={Lanczos, Cornelius},
  year={1962},
  publisher={University of Toronto Press}
}

@article{Boehler1979,
  title={A simple derivation of representations for non-polynomial constitutive equations in some cases of anisotropy},
  author={Boehler, Jean-Paul},
  journal={Zeitschrift f{\"u}r Angewandte Mathematik und Mechanik},
  volume={59},
  number={4},
  pages={157--167},
  year={1979},
  publisher={Wiley Online Library}
}

@article{Huang2006,
  title   = {A Theory of Hyperelasticity of Multi-Phase Media with Surface/Interface Energy Effect},
  author  = {Huang, Z. P. and Wang, Jian-Xiang},
  journal = {Acta Mechanica},
  volume  = {182},
  number  = {3},
  pages   = {195--210},
  year    = {2006},
  publisher = {Springer}
}

@article{Shuttleworth1950,
  title   = {The surface tension of solids},
  author  = {Shuttleworth, Richard},
  journal = {Proceedings of the Physical Society. Section A},
  volume  = {63},
  pages   = {444--457},
  year    = {1950},
  publisher = {IOP Publishing}
}

@article{Cammarata1994,
  title   = {Surface and interface stress effects in thin films},
  author  = {Cammarata, Robert C.},
  journal = {Progress in Surface Science},
  volume  = {46},
  pages   = {1--38},
  year    = {1994},
  publisher = {Elsevier}
}

@article{Spaepen2000,
  title   = {Interfaces and stresses in thin films},
  author  = {Spaepen, Frans},
  journal = {Acta Materialia},
  volume  = {48},
  pages   = {31--42},
  year    = {2000},
  publisher = {Elsevier}
}

@article{Krichen2019,
  title   = {Liquid inclusions in soft materials: {C}apillary effect, mechanical stiffening and enhanced electromechanical response},
  author  = {Krichen, Sana and Liu, Liping and Sharma, Pradeep},
  journal = {Journal of the Mechanics and Physics of Solids},
  volume  = {127},
  pages   = {332--357},
  year    = {2019},
  publisher = {Elsevier}
}

@article{Xu2018,
  title   = {Surface elastic constants of a soft solid},
  author  = {Xu, Qin and Style, Robert W. and Dufresne, Eric R.},
  journal = {Soft Matter},
  volume  = {14},
  number  = {6},
  pages   = {916--920},
  year    = {2018},
  publisher = {Royal Society of Chemistry}
}

@article{JaviliSteinmann2010,
  author  = {Javili, A. and Steinmann, P.},
  title   = {A finite element framework for continua and surfaces at finite deformation: {Part I. Continuum and surface kinematics}},
  journal = {Computer Methods in Applied Mechanics and Engineering},
  year    = {2010},
  volume  = {199},
  number  = {9-12},
  pages   = {849--860}
}

@article{Javili2013,
  title   = {Thermomechanics of solids with lower-dimensional energetics: {O}n the importance of surface, interface, and curve structures at the nanoscale. {A} unifying review},
  author  = {Javili, Ali and McBride, Andrew and Steinmann, Paul},
  journal = {Applied Mechanics Reviews},
  volume  = {65},
  number  = {1},
  pages   = {010802},
  year    = {2013},
  publisher = {American Society of Mechanical Engineers}
}

@article{SteigmannOgden1999,
  author  = {Steigmann, D. J. and Ogden, R. W.},
  title   = {Elastic surface--substrate interactions},
  journal = {Proceedings of the Royal Society of London A},
  year    = {1999},
  volume  = {455},
  number  = {1982},
  pages   = {437--474}
}

@book{Lee1997,
	Address = {New York},
	Author = {Lee, J. M.},
	Publisher = {Springer-Verlag},
	Title = {Riemannian Manifold: An Introduction to Curvature},
	Year = {1997}}

@article{SimoMarsden1984,
  title={On the rotated stress tensor and the material version of the {D}oyle-{E}ricksen formula},
  author={Simo, Juan C and Marsden, Jerrold E},
  journal={Archive for Rational Mechanics and Analysis},
  volume={86},
  number={3},
  pages={213--231},
  year={1984}
}

@article{ManCohen1986,
  title={A coordinate-free approach to the kinematics of membranes},
  author={Man, Chi-Sing and Cohen, H},
  journal={Journal of elasticity},
  volume={16},
  pages={97--104},
  year={1986},
  publisher={Springer}
}

@article{Betounes1986,
  title={Kinematics of submanifolds and the mean curvature normal},
  author={Betounes, David E},
  journal={Archive for Rational Mechanics and Analysis},
  volume={96},
  pages={1--27},
  year={1986},
  publisher={Springer}
}

@book{Yano1970,
  title={Integral Formulas in Riemannian Geometry},
  author={Yano, Kentar{\=o}},
  year={1970},
  publisher={Marcel Dekker}
}

@article{Kadianakis2010,
  title={Evolution of Surfaces and the Kinematics of Membranes},
  author={Kadianakis, N},
  journal={Journal of Elasticity},
  volume={99},
  pages={1--17},
  year={2010},
  publisher={Springer}
}

@article{Kadianakis2018,
  title={Infinitesimally affine deformations of a hypersurface},
  author={Kadianakis, Nikos and Travlopanos, Fotios I},
  journal={Mathematics and Mechanics of Solids},
  volume={23},
  number={2},
  pages={209--220},
  year={2018},
  publisher={SAGE Publications Sage UK: London, England}
}

@book{Chen2019,
  title={Geometry of Submanifolds},
  author={Chen, Bang-yen},
  year={2019},
  publisher={Courier Dover Publications}
}

@book{Dajczer2019,
  title={Submanifold Theory},
  author={Dajczer, Marcos and Tojeiro, Ruy},
  year={2019},
  publisher={Springer}
}

@book{Nishikawa2002,
  author    = {Seiki Nishikawa},
  title     = {Variational Problems in Geometry},
  year      = {2002},
  publisher = {American Mathematical Society},
  address   = {Providence, Rhode Island}
}

@article{lu2000covariant,
	author = {Lu, J. and Papadopoulos, P.},
	journal = {Zeitschrift f{\"u}r Angewandte Mathematik und Physik (ZAMP)},
	number = {2},
	pages = {204--217},
	publisher = {Springer},
	title = {A covariant constitutive description of anisotropic non-linear elasticity},
	volume = {51},
	year = {2000}}

@incollection{spencer1982formulation,
	author = {Spencer, A. J. M.},
	booktitle = {Mechanical Behavior of Anisotropic Solids/Comportment M{\'e}chanique des Solides Anisotropes},
	pages = {3--26},
	publisher = {Springer},
	title = {The formulation of constitutive equation for anisotropic solids},
	year = {1982}}

@article{Ericksen1954Anisotropic,
	author = {Ericksen, J. L. and Rivlin, R. S.},
	journal = {Journal of Rational Mechanics and Analysis},
	pages = {281--301},
	title = {Large elastic deformations of homogeneous anisotropic materials},
	volume = {3},
	year = {1954}}

@article{Golgoon2018a,
	author = {Golgoon, A. and Yavari, A.},
	journal = {Journal of Elasticity},
	number = {2},
	pages = {239--269},
	publisher = {Springer},
	title = {Nonlinear elastic inclusions in anisotropic solids},
	volume = {130},
	year = {2018}}

@article{Golgoon2018b,
	author = {Golgoon, A. and Yavari, A.},
	journal = {Zeitschrift f{\"u}r angewandte Mathematik und Physik},
	number = {3},
	pages = {1--28},
	publisher = {Springer},
	title = {Line and point defects in nonlinear anisotropic solids},
	volume = {69},
	year = {2018}}

@incollection{Spencer1986,
	author = {Spencer, A. J. M.},
	booktitle = {Large Deformations of Solids: Physical Basis and Mathematical Modelling},
	pages = {41--52},
	publisher = {Springer},
	title = {Modelling of finite deformations of anisotropic materials},
	year = {1986}}

@incollection{Podio1985,
  title={Cavitation and phase transition of hyperelastic fluids},
  author={Podio-Guidugli, Paolo and Vergara Caffarelli, G and Virga, EG},
  booktitle={Analysis and Thermomechanics: A Collection of Papers Dedicated to W. Noll on His Sixtieth Birthday},
  pages={401--416},
  year={1985},
  publisher={Springer}
}

@book{Tu2017,
  title={Differential Geometry: Connections, Curvature, and Characteristic Classes},
  author={Tu, Loring W},
  volume={275},
  year={2017},
  publisher={Springer}
}

@article{DoyleEricksen1956,
	author = {Doyle, T. C. and Ericksen, J. L.},
	journal = {Advances in Applied Mechanics},
	pages = {53--115},
	publisher = {Academic Press New York},
	title = {Nonlinear elasticity},
	volume = {4},
	year = {1956}}

@article{Kondo1955,
  title={Non-{R}iemannian geometry of imperfect crystals from a macroscopic viewpoint},
  author={Kondo, K},
  journal={Memoirs of the Unifying Study of the Basic Problems in Engineering Science by Means of Geometry},
  volume={1},
  pages={6--17},
  year={1955},
  publisher={Division DI, Gakujutsu Bunken Fukyo-Kai}
}

@article{Bilby1955,
  title={Continuous distributions of dislocations: {A} new application of the methods of non-{R}iemannian geometry},
  author={Bilby, Bruce Alexander and Bullough, R and Smith, Edwin},
  journal={Proceedings of the Royal Society of London. Series A},
  volume={231},
  number={1185},
  pages={263--273},
  year={1955},
  publisher={The Royal Society London}
}

@article{Bilby1956,
  title={Continuous distributions of dislocations. {Ill}},
  author={Bilby, Bruce Alexander and Smith, E},
  journal={Proceedings of the Royal Society of London. Series A.},
  volume={236},
  number={1207},
  pages={481--505},
  year={1956},
  publisher={The Royal Society London}
}

@inproceedings{Bilby1968,
  title={Geometry and continuum mechanics},
  author={Bilby, BA},
  booktitle={Mechanics of Generalized Continua: Proceedings of the IUTAM-Symposium on The Generalized Cosserat Continuum and the Continuum Theory of Dislocations with Applications, Freudenstadt and Stuttgart (Germany) 1967},
  pages={180--199},
  year={1968},
  organization={Springer}
}

@article{Kroner1959,
	Author = {Kr{\"o}ner, E. and Seeger, A.},
	Journal = {Archive for Rational Mechanics and Analysis},
	Number = {1},
	Pages = {97--119},
	Publisher = {Springer},
	Title = {Nicht-lineare elastizit{\"a}tstheorie der versetzungen und eigenspannungen},
	Volume = {3},
	Year = {1959}}

@article{KronerSeeger1959,
  title={Nicht-lineare elastizit{\"a}tstheorie der versetzungen und eigenspannungen},
  author={Kr{\"o}ner, Ekkehart and Seeger, Alfred},
  journal={Archive for Rational Mechanics and Analysis},
  volume={3},
  pages={97--119},
  year={1959},
  publisher={Springer}
}

@book{Schwarz2006,
  title={Hodge Decomposition-A Method for Solving Boundary Value Problems},
  author={Schwarz, G{\"u}nter},
  year={2006},
  publisher={Springer}
}

@book{Nakahara2003,
  title={Geometry, Topology and Physics},
  author={Nakahara, Mikio},
  year={2003},
  publisher={CRC press}
}

@article{MiriRivier2002,
  title={Continuum elasticity with topological defects, including dislocations and extra-matter},
  author={Miri, MirFaez and Rivier, Nicolas},
  journal={Journal of Physics A: Mathematical and General},
  volume={35},
  number={7},
  pages={1727},
  year={2002},
  publisher={IOP Publishing}
}

@article{Kroner1990,
  title={The differential geometry of elementary point and line defects in {B}ravais crystals},
  author={Kr{\"o}ner, E},
  journal={International Journal of Theoretical Physics},
  volume={29},
  pages={1219--1237},
  year={1990},
  publisher={Springer}
}

@article{Grachev1989,
  title={The gauge theory of point defects},
  author={Grachev, AV and Nesterov, AI and Ovchinnikov, SG},
  journal={physica Status Solidi (b)},
  volume={156},
  number={2},
  pages={403--410},
  year={1989},
  publisher={Wiley Online Library}
}

@article{DeWit1981,
  title={A view of the relation between the continuum theory of lattice defects and non-{E}uclidean geometry in the linear approximation},
  author={De Wit, Roland},
  journal={International Journal of Engineering Science},
  volume={19},
  number={12},
  pages={1475--1506},
  year={1981},
  publisher={Elsevier}
}

@article{Falk1981,
  title={Theory of elasticity of coherent inclusions by means of non-metric geometry},
  author={Falk, F},
  journal={Journal of Elasticity},
  volume={11},
  number={4},
  pages={359--372},
  year={1981},
  publisher={Springer}
}

@article{Sadik2016,
  title={A geometric theory of nonlinear morphoelastic shells},
  author={Sadik, Souhayl and Angoshtari, Arzhang and Goriely, Alain and Yavari, Arash},
  journal={Journal of Nonlinear Science},
  volume={26},
  pages={929--978},
  year={2016},
  publisher={Springer}
}

@book{MarsRat2013,
	author = {Marsden, J.E. and Ratiu, T.S.},
	publisher = {Springer New York},
	series = {Texts in Applied Mathematics},
	title = {Introduction to Mechanics and Symmetry: A Basic Exposition of Classical Mechanical Systems},
	year = {2013}	
	}

@book{Goriely2017,
  title={The Mathematics and Mechanics of Biological Growth},
  author={Goriely, Alain},
  volume={45},
  year={2017},
  publisher={Springer}
}

@book{Ogden1984,
  title={Non-Linear Elastic Deformations},
  author={Ogden, Raymond W},
  year={1997},
  publisher={Dover}
}

@article{YavariSozio2023,
	author = {Yavari, Arash and Sozio, Fabio},
	journal = {Journal of the Mechanics and Physics of Solids},
	pages = {105101},
	publisher = {Elsevier},
	title = {On the direct and reverse multiplicative decompositions of deformation gradient in nonlinear anisotropic anelasticity},
	volume = {170},
	year = {2023}}

@article{OzakinYavari2014,
  title={Affine development of closed curves in {W}eitzenb{\"o}ck manifolds and the {B}urgers vector of dislocation mechanics},
  author={Ozakin, Arkadas and Yavari, Arash},
  journal={Mathematics and Mechanics of Solids},
  volume={19},
  number={3},
  pages={299--307},
  year={2014},
  publisher={SAGE Publications Sage UK: London, England}
}

@article{YavariGoriely2014,
  title={The geometry of discombinations and its applications to semi-inverse problems in anelasticity},
  author={Yavari, Arash and Goriely, Alain},
  journal={Proceedings of the Royal Society A},
  volume={470},
  number={2169},
  pages={20140403},
  year={2014},
  publisher={The Royal Society Publishing}
}

@book{Nicolaescu2020,
  title={Lectures on the Geometry of Manifolds},
  author={Nicolaescu, Liviu I},
  year={2020},
  publisher={World Scientific}
}

@book{Spivak1970II,
  title={A Comprehensive Introduction to Differential Geometry},
  author={Spivak, Michael},
  volume={2},
  year={1970},
  publisher={Publish or Perish, Incorporated}
}

@book{Spivak1970III,
  title={A Comprehensive Introduction to Differential Geometry},
  author={Spivak, Michael},
  volume={3},
  year={1970},
  publisher={Publish or Perish, Incorporated}
}

@article{Gordeeva2010,
  title={Riemann--{C}artan manifolds},
  author={Gordeeva, IA and Pan’zhenskii, VI and Stepanov, SE},
  journal={Journal of Mathematical Sciences},
  volume={169},
  number={3},
  pages={342--361},
  year={2010},
  publisher={Springer}
}

@book{Sternberg1999,
  title={Lectures on Differential Geometry},
  author={Sternberg, Shlomo},
  volume={316},
  year={1999},
  publisher={American Mathematical Society}
}

@book{Sternberg2013,
  title={Curvature in Mathematics and Physics},
  author={Sternberg, Shlomo},
  year={2013},
  publisher={Courier Corporation}
}

@book{Hehl2003,
  title={Foundations of classical electrodynamics: Charge, flux, and metric},
  author={Hehl, Friedrich W and Obukhov, Yuri N},
  volume={33},
  year={2003},
  publisher={Springer Science \& Business Media}
}

@article{CapovillaGuven1995,
	Author = {Capovilla, R. and Guven, J.},
	Journal = {Physical Review D},
	Number = {12},
	Pages = {6736-6743},
	Title = {Geometry of deformations of relativistic membranes},
	Volume = {51},
	Year = {1995}}

@book{do1992riemannian,
	Author = {do Carmo, M.P.},
	Isbn = {1584883553},
	Lccn = {2003058473},
	Publisher = {Birkh{\"a}user Boston},
	Series = {Mathematics: Theory \& Applications},
	Title = {Riemannian Geometry},
	Year = {1992}
}

@article{Kuchar1976,
	Author = {Kucha\v{r}, Karel},
	Journal = {Journal of Mathematical Physics},
	Number = {5},
	Pages = {777-791},
	Title = {Geometry of hyperspace. I},
	Volume = {17},
	Year = {1976}
}

@book{Straumann2012,
  title={General Relativity},
  author={Straumann, Norbert},
  year={2012},
  publisher={Springer Science \& Business Media}
}

@article{YavariOzakinSadik2016,
  title={Nonlinear elasticity in a deforming ambient space},
  author={Yavari, Arash and Ozakin, Arkadas and Sadik, Souhayl},
  journal={Journal of Nonlinear Science},
  volume={26},
  pages={1651--1692},
  year={2016},
  publisher={Springer}
}

@article{YavariGoriely2013Inclusions,
  title={Nonlinear elastic inclusions in isotropic solids},
  author={Yavari, Arash and Goriely, Alain},
  journal={Proceedings of the Royal Society A},
  volume={469},
  number={2160},
  pages={20130415},
  year={2013},
  publisher={The Royal Society Publishing}
}

@article{Ghosh2022,
  title={Elastomers filled with liquid inclusions: {T}heory, numerical implementation, and some basic results},
  author={Ghosh, Kamalendu and Lopez-Pamies, Oscar},
  journal={Journal of the Mechanics and Physics of Solids},
  pages={104930},
  year={2022},
  publisher={Elsevier}
}

@article{Liu2012,
  title={On elastocapillarity: {A} review},
  author={Liu, Jian-Lin and Feng, Xi-Qiao},
  journal={Acta Mechanica Sinica},
  volume={28},
  number={4},
  pages={928--940},
  year={2012},
  publisher={Springer}
}

@article{Bico2018,
  title={Elastocapillarity: {W}hen surface tension deforms elastic solids},
  author={Bico, Jos{\'e} and Reyssat, {\'E}tienne and Roman, Beno{\^\i}t},
  journal={Annual Review of Fluid Mechanics},
  volume={50},
  number={1},
  pages={629--659},
  year={2018}
}

@article{Style2015a,
  title={Stiffening solids with liquid inclusions},
  author={Style, Robert W and Boltyanskiy, Rostislav and Allen, Benjamin and Jensen, Katharine E and Foote, Henry P and Wettlaufer, John S and Dufresne, Eric R},
  journal={Nature Physics},
  volume={11},
  number={1},
  pages={82--87},
  year={2015},
  publisher={Nature Publishing Group}
}

@article{Style2015b,
  title={Surface tension and the mechanics of liquid inclusions in compliant solids},
  author={Style, Robert W and Wettlaufer, John S and Dufresne, Eric R},
  journal={Soft Matter},
  volume={11},
  number={4},
  pages={672--679},
  year={2015},
  publisher={Royal Society of Chemistry}
}

@article{Style2017,
  title={Elastocapillarity: {S}urface tension and the mechanics of soft solids},
  author={Style, Robert W and Jagota, Anand and Hui, Chung-Yuen and Dufresne, Eric R},
  journal={Annual Review of Condensed Matter Physics},
  volume={8},
  pages={99--118},
  year={2017},
  publisher={Annual Reviews}
}

@article{Javili2018,
  title={Aspects of interface elasticity theory},
  author={Javili, Ali and Ottosen, Niels Saabye and Ristinmaa, Matti and Mosler, J{\"o}rn},
  journal={Mathematics and Mechanics of Solids},
  volume={23},
  number={7},
  pages={1004--1024},
  year={2018},
  publisher={SAGE Publications Sage UK: London, England}
}

@article{Gurtin1975,
  title={A continuum theory of elastic material surfaces},
  author={Gurtin, Morton E and Murdoch, A I},
  journal={Archive for Rational Mechanics and Analysis},
  volume={57},
  number={4},
  pages={291--323},
  year={1975},
  publisher={Springer}
}

@article{Gurtin1978,
  title={Surface stress in solids},
  author={Gurtin, Morton E and Murdoch, A Ian},
  journal={International Journal of Solids and Structures},
  volume={14},
  number={6},
  pages={431--440},
  year={1978},
  publisher={Elsevier}
}

@article{Gurtin1998,
  title={A general theory of curved deformable interfaces in solids at equilibrium},
  author={Gurtin, Morton E and Weissm{\"u}ller, Jorg and Larche, Francis},
  journal={Philosophical Magazine A},
  volume={78},
  number={5},
  pages={1093--1109},
  year={1998},
  publisher={Taylor \& Francis}
}

@book{MaHu1983,
	Address = {New York},
	Author = {Marsden, J. E. and T. J. R. Hughes},
	Isbn = {9780486678658},
	Lccn = {93042631},
	Publisher = {Dover},
	Series = {Dover Civil and Mechanical Engineering Series},
	Title = {Mathematical Foundations of Elasticity},
	Year = {1983}}

@article{zheng1993,
	Author = {Zheng, Q. S. and Spencer, A J M},
	Journal = {International Journal of Engineering Science},
	Number = {5},
	Pages = {679--693},
	Publisher = {Elsevier},
	Title = {Tensors which characterize anisotropies},
	Volume = {31},
	Year = {1993}}

@article{lu2000,
	Author = {Lu, Jia and Papadopoulos, Panayiotis},
	Journal = {Zeitschrift f{\"u}r Angewandte Mathematik und Physik (ZAMP)},
	Number = {2},
	Pages = {204--217},
	Publisher = {Springer},
	Title = {A covariant constitutive description of anisotropic non-linear elasticity},
	Volume = {51},
	Year = {2000}}

@article{zheng1994theory,
	Author = {Zheng, Q S},
	Journal = {Applied Mechanics Reviews},
	Number = {11},
	Pages = {545--587},
	Title = {Theory of representations for tensor functions},
	Volume = {47},
	Year = {1994}}

@article{MazzucatoRachele2006,
	Author = {Mazzucato, Anna L and Rachele, Lizabeth V},
	Journal = {Journal of Elasticity},
	Number = {3},
	Pages = {205--245},
	Publisher = {Springer},
	Title = {Partial uniqueness and obstruction to uniqueness in inverse problems for anisotropic elastic media},
	Volume = {83},
	Year = {2006}}

@book{boehler1987,
	Author = {Boehler, Jean-Paul},
	Publisher = {Springer},
	Title = {Applications of Tensor Functions in Solid Mechanics},
	Volume = {292},
	Year = {1987}}

@article{liu1982,
	Author = {Liu, I},
	Journal = {International Journal of Engineering Science},
	Number = {10},
	Pages = {1099--1109},
	Publisher = {Elsevier},
	Title = {On representations of anisotropic invariants},
	Volume = {20},
	Year = {1982}}

@article{Spencer1971,
	Author = {Spencer, A J M},
	Journal = {Continuum Physics},
	Pages = {239--353},
	Title = {Part {III}. {T}heory of {I}nvariants},
	Volume = {1},
	Year = {1971}}

@book{Eisenhart1926,
  title={Riemannian Geometry},
  author={Eisenhart, Luther Pfahler},
  year={1926},
  publisher={Princeton university press}
}

@Article{kinoshita1971,
  author    = {Kinoshita, N and Mura, T},
  title     = {Elastic fields of inclusions in anisotropic media},
  journal   = {Physica Status Solidi (a)},
  year      = {1971},
  volume    = {5},
  number    = {3},
  pages     = {759--768},
  publisher = {Wiley Online Library},
}

@article{Mindlin1950,
  title={Nuclei of strain in the semi-infinite solid},
  author={Mindlin, Raymond D and Cheng, David H},
  journal={Journal of Applied Physics},
  volume={21},
  number={9},
  pages={926--930},
  year={1950},
  publisher={American Institute of Physics}
}

@Article{Ueda1975,
  Title                    = {A New Measuring Method of Residual Stresses with the Aid of Finite Element Method and Reliability of Estimated Values},
  Author                   = {Ueda, Yukio and Fukuda, Keiji and Nakacho, Keiji and Endo, Setsuo},
  Journal                  = {Transactions of JWRI},
  Year                     = {1975},
  Number                   = {2},
  Pages                    = {123-131},
  Volume                   = {4},
  Publisher                = {Osaka University},
}

@article{Reissner1931,
  title={Eigenspannungen und eigenspannungsquellen},
  author={Reissner, H},
  journal={Zeitschrift f{\"u}r Angewandte Mathematik und Mechanik},
  volume={11},
  number={1},
  pages={1--8},
  year={1931},
  publisher={Wiley Online Library}
}

@Book{Mura1982,
  Title                    = {Micromechanics of Defects in Solids},
  Author                   = {Mura, Toshio},
  Publisher                = {Martinus Nijhoff},
  Year                     = {1982},
}

@article{Yavari2006,
	author = {A. Yavari and J. E. Marsden and M. Ortiz},
	journal = {Journal of Mathematical Physics},
	pages = {85--112},
	title = {On the spatial and material covariant balance laws in elasticity},
	volume = {47},
	year = {2006}}

@article{Ericksen1954,
	Author = {Ericksen, J L},
	Journal = {Zeitschrift f{\"u}r Angewandte Mathematik und Physik (ZAMP)},
	Number = {6},
	Pages = {466--489},
	Publisher = {Springer},
	Title = {Deformations possible in every isotropic, incompressible, perfectly elastic body},
	Volume = {5},
	Year = {1954}}

@article{Goodbrake2020,
  title={The Anelastic {E}ricksen Problem: {U}niversal Deformations and Universal Eigenstrains in Incompressible Nonlinear Anelasticity},
  author={Goodbrake, Christian and Yavari, Arash and Goriely, Alain},
  journal={Journal of Elasticity},
  volume={142},
  number={2},
  pages={291--381},
  year={2020},
  publisher={Springer}
}

@article{Yavari2013,
	Author = {Yavari, A. and Goriely, A.},
	Journal = {Proceedings of the Royal Society A},
	Number = {2160},
	Pages = {20130415},
	Title = {Nonlinear elastic inclusions in isotropic solids},
	Volume = {469},
	Year = {2013}}

@article{Yavari2021,
  title={On {E}shelby’s inclusion problem in nonlinear anisotropic elasticity},
  author={Yavari, Arash},
  journal={Journal of Micromechanics and Molecular Physics},
  volume={6},
  number={01},
  pages={2150002},
  year={2021},
  publisher={World Scientific}
}

@article{Eckart1948,
  title={The thermodynamics of irreversible processes. {IV}. {T}he theory of elasticity and anelasticity},
  author={Eckart, Carl},
  journal={Physical Review},
  volume={73},
  number={4},
  pages={373},
  year={1948},
  publisher={APS}
}

@article{ambrosi2019growth,
	Author = {Ambrosi, Davide and Ben Amar, Martine and Cyron, Christian J and DeSimone, Antonio and Goriely, Alain and Humphrey, Jay D and Kuhl, Ellen},
	Journal = {Journal of the Royal Society Interface},
	Number = {157},
	Pages = {20190233},
	Publisher = {The Royal Society},
	Title = {Growth and remodelling of living tissues: perspectives, challenges and opportunities},
	Volume = {16},
	Year = {2019}}

@article{Eshelby1957,
  Title                    = {The determination of the elastic field of an ellipsoidal inclusion, and related problems},
  Author                   = {Eshelby, John D},
  Journal                = {Proceedings of the Royal Society of London A},
  Year                     = {1957},
  Number                   = {1226},
  Organization             = {The Royal Society},
  Pages                    = {376--396},
  Volume                   = {241},
}

@article{YavariGoriely2012b,
	Author = {Yavari, A. and Goriely, A.},
	Journal = {Proceedings of the Royal Society A},
	Number = {2148},
	Pages = {3902--3922},
	Title = {Weyl geometry and the nonlinear mechanics of distributed point defects},
	Volume = {468},
	Year = {2012}}

@article{YavariGoriely2012a,
	Author = {Yavari, A. and Goriely, A.},
	Journal = {Archive for Rational Mechanics and Analysis},
	Number = {1},
	Pages = {59--118},
	Title = {Riemann--{C}artan geometry of nonlinear dislocation mechanics},
	Volume = {205},
	Year = {2012}}

@article{YavariGoriely2013a,
	Author = {Yavari, A and Goriely, A},
	Journal = {Mathematics and Mechanics of Solids},
	Number = {1},
	Pages = {91--102},
	Publisher = {Sage Publications Sage UK: London, England},
	Title = {Riemann--{C}artan geometry of nonlinear disclination mechanics},
	Volume = {18},
	Year = {2013}}

@article{Sozio2019,
  title = {Nonlinear mechanics of accretion},
  author = {Sozio, Fabio and Yavari, Arash},
  journal = {Journal of Nonlinear Science},
  volume = {29},
  number = {4},
  pages = {1813--1863},
  year = {2019},
  publisher = {Springer}
}

@article{Sadik2017,
  title={On the origins of the idea of the multiplicative decomposition of the deformation gradient},
  author={Sadik, Souhayl and Yavari, Arash},
  journal={Mathematics and Mechanics of Solids},
  volume={22},
  number={4},
  pages={771--772},
  year={2017}
}

@Article{Kondo1949,
  author    = {Kondo, Kazuo},
  title     = {A Proposal of a New Theory concerning the Yielding of Materials based on {R}iemannian Geometry},
  journal   = {The Journal of the Japan Society of Aeronautical Engineering},
  year      = {1949},
  volume    = {2},
  number    = {8},
  pages     = {29-31},
  publisher = {THE JAPAN SOCIETY FOR AERONAUTICAL AND SPACE SCIENCES},
}

\end{document}